\documentclass[pdftex,twocolumn,epjc3_preprint,runningheads]{svjour3}

\usepackage[T1]{fontenc}
\usepackage{lmodern}
\usepackage{calc}
\usepackage{graphicx}
\usepackage{booktabs}
\usepackage{textcomp}
\usepackage{xspace}
\usepackage{relsize}
\usepackage{amssymb}
\usepackage{amsmath}
\usepackage{listings}
\usepackage{microtype}
\usepackage{multirow}
\usepackage{tabularx}
\usepackage{array}
\usepackage{placeins}
\usepackage{cuted}
\usepackage{soul}
\usepackage{fixltx2e}
\usepackage[numbers,sort&compress]{natbib}
\usepackage[labelfont=bf,font=small]{caption}
\usepackage[skip=-2pt]{subcaption}
\usepackage[colorlinks,citecolor=blue,urlcolor=blue,linkcolor=blue,breaklinks=breakall]{hyperref}
\usepackage{breakurl}
\usepackage[dvipsnames]{xcolor}
\usepackage[clockwise,figuresright]{rotating}
\usepackage{siunitx}
\usepackage{tikz}

\usepackage{etoolbox}
\AfterEndEnvironment{strip}{\leavevmode}

\allowdisplaybreaks

\newcolumntype{L}{>{\raggedright\let\newline\\\arraybackslash\hspace{0pt}}X}
\newcolumntype{R}{>{\raggedleft\let\newline\\\arraybackslash\hspace{0pt}}X}
\newcolumntype{C}{>{\centering\let\newline\\\arraybackslash\hspace{0pt}}X}

\newcommand{\imperial}{Department of Physics, Imperial College London, Blackett Laboratory, Prince Consort Road, London SW7 2AZ, UK}

\newcommand{\okc}{Oskar Klein Centre for Cosmoparticle Physics, AlbaNova University Centre, SE-10691 Stockholm, Sweden}
\newcommand{\su}{Department of Physics, Stockholm University, SE-10691 Stockholm, Sweden}
\newcommand{\mcgill}{Department of Physics, McGill University, 3600 rue University, Montr\'eal, Qu\'ebec H3A 2T8, Canada}
\newcommand{\ucla}{Physics and Astronomy Department, University of California, Los Angeles, CA 90095, USA}
\newcommand{\annecy}{LAPTh, Universit\'e de Savoie, CNRS, 9 chemin de Bellevue B.P.110, F-74941 Annecy-le-Vieux, France}

\newcommand{\gambitacknos    }{We warmly thank the Casa Matem\'aticas Oaxaca, affiliated with the Banff International Research Station, for hospitality whilst part of this work was completed, and the staff at Cyfronet, for their always helpful supercomputing support.  \GB has been supported by STFC (UK; ST/K00414X/1, ST/P000762/1), the Royal Society (UK; UF110191), Glasgow University (UK; Leadership Fellowship), the Research Council of Norway (FRIPRO 230546/F20), NOTUR (Norway; NN9284K), the Knut and Alice Wallenberg Foundation (Sweden; Wallenberg Academy Fellowship), the Swedish Research Council (621-2014-5772), the Australian Research Council (CE110001004, FT130100018, FT140100244, FT160100274), The University of Sydney (Australia; IRCA-G162448), PLGrid Infrastructure (Poland), Polish National Science Center (Sonata UMO-2015/17/D/ST2/03532), the Swiss National Science Foundation (PP00P2-144674), the European Commission Horizon 2020 Marie Sk\l{}odowska-Curie actions (H2020-MSCA-RISE-2015-691164), the ERA-CAN+ Twinning Program (EU \& Canada), the Netherlands Organisation for Scientific Research (NWO-Vidi 680-47-532), the National Science Foundation (USA; DGE-1339067), the FRQNT (Qu\'ebec) and NSERC/The Canadian Tri-Agencies Research Councils (BPDF-424460-2012).}

\makeatletter

\newcommand{\preprintnumber}[1]{\gdef\@preprintnumber{\begin{flushright}{#1}\end{flushright}}}

\g@addto@macro\bfseries{\boldmath}
\makeatother

\newcommand{\subparagraph}{} 
\usepackage{titlesec}
\titleformat*{\paragraph}{\bfseries}

\journalname{Eur. Phys. J. C}
\smartqed
\let\underscore\_
\renewcommand{\_}{\discretionary{\underscore}{}{\underscore}}

\makeatletter
\let\orgdescriptionlabel\descriptionlabel
\renewcommand*{\descriptionlabel}[1]{%
  \let\orglabel\label
  \let\label\@gobble
  \phantomsection
  \protected@edef\@currentlabel{#1}%
  \let\label\orglabel
  \orgdescriptionlabel{#1}%
}
\makeatother

\newcommand\postnewlinemarker{\hbox{\ensuremath{\hookrightarrow}}}
\newcommand\cpp[1]{{\lstinline!#1!}}  
\newcommand\cpppragma[1]{{\CPPcommentstyle#1}}
\newcommand\yaml[1]{{\lstset{style=yaml}\lstinline!#1!\lstset{style=cpp}}}
\newcommand\yamlvalue[1]{{\YAMLvaluestyle\ttfamily#1}}
\newcommand\term[1]{{\lstset{style=terminal}\lstinline!#1!\lstset{style=cpp}}}
\newcommand\fortran[1]{{\lstset{style=fortran}\lstinline!#1!\lstset{style=cpp}}}
\newcommand\py[1]{{\lstset{style=python}\lstinline!#1!\lstset{style=cpp}}}
\newcommand\customtilde{{\raisebox{0.2ex}{\scalebox{0.6}{\boldmath$\sim$}}}}
\newcommand\mathematica[1]{{\lstset{style=Mathematica}\lstinline!#1!\lstset{style=cpp}}}

\lstnewenvironment{lstlistingyaml}{\lstset{style=yaml}}{\lstset{style=cpp}}
\lstnewenvironment{lstlistingterm}{\lstset{style=terminal}}{\lstset{style=cpp}}
\lstnewenvironment{lstlistingfortran}{\lstset{style=fortran}}{\lstset{style=cpp}}
\lstnewenvironment{lstcpp}{\lstset{style=cpp}}{\lstset{style=cpp}}
\lstnewenvironment{lstcppalt}{\lstset{style=cppalt}}{\lstset{style=cpp}}
\lstnewenvironment{lstcppnum}{\lstset{style=cppnum}}{\lstset{style=cpp}}
\lstnewenvironment{lstyaml}{\lstset{style=yaml}}{\lstset{style=cpp}}
\lstnewenvironment{lstterm}{\lstset{style=terminal}}{\lstset{style=cpp}}
\lstnewenvironment{lsttermalt}{\lstset{style=terminalalt}}{\lstset{style=cpp}}
\lstnewenvironment{lsttext}{\lstset{style=text}}{\lstset{style=cpp}}
\lstnewenvironment{lstfortran}{\lstset{style=fortran}}{\lstset{style=cpp}}
\lstnewenvironment{lstpy}{\lstset{style=python}}{\lstset{style=cpp}}
\lstnewenvironment{lstmathematica}{\lstset{style=mathematica}}{\lstset{style=cpp}}

\newcommand{\tmpname}{}
\newcommand{\tmplistingname}{}
\makeatletter
\newif\ifATOlabelname
\lst@Key{labelname}{Listing}{\def\ATOlabelname{#1}\global\ATOlabelnametrue}
\makeatother
\lstnewenvironment{lstcpplabel}[1][]{
  \lstset{style=cpp,#1} 
  \ifATOlabelname
    \renewcommand{\tmpname}{\lstlistingname}
    \renewcommand{\tmplistingname}{\lstlistlistingname}
    \renewcommand{\lstlistingname}{\ATOlabelname}
    \renewcommand{\lstlistlistingname}{List of \lstlistingname s}
  \fi
}{
  \renewcommand{\lstlistingname}{\tmpname}
  \renewcommand{\lstlistlistingname}{\tmplistingname}
  \lstset{style=cpp}
}
\definecolor{solarized@base03}{HTML}{002B36}
\definecolor{solarized@base02}{HTML}{073642}
\definecolor{solarized@base01}{HTML}{586e75}
\definecolor{solarized@base00}{HTML}{657b83}
\definecolor{solarized@base0}{HTML}{839496}
\definecolor{solarized@base1}{HTML}{93a1a1}
\definecolor{solarized@base2}{HTML}{EEE8D5}
\definecolor{solarized@base3}{HTML}{FDF6E3}
\definecolor{solarized@yellow}{HTML}{B58900}
\definecolor{solarized@orange}{HTML}{CB4B16}
\definecolor{solarized@red}{HTML}{DC322F}
\definecolor{solarized@magenta}{HTML}{D33682}
\definecolor{solarized@violet}{HTML}{6C71C4}
\definecolor{solarized@blue}{HTML}{268BD2}
\definecolor{solarized@cyan}{HTML}{2AA198}
\definecolor{solarized@green}{HTML}{859900}
\definecolor{darkred}{HTML}{550003}
\definecolor{darkgreen}{HTML}{00AA00}

\newcommand\YAMLstringstyle{\footnotesize\color{solarized@green}\mdseries}
\newcommand\YAMLkeystyle{\footnotesize\color{solarized@blue}\ttfamily}
\newcommand\YAMLvaluestyle{\footnotesize\color{blue}\mdseries}
\newcommand\ProcessThreeDashes{\llap{\color{cyan}\mdseries-{-}-}}

\newcommand\CPPidentifierstyle{\color{solarized@blue}\footnotesize\ttfamily}
\newcommand\CPPcommentstyle{\color{solarized@violet}\footnotesize\ttfamily}
\newcommand\CPPdirectivestyle{\color{solarized@magenta}\footnotesize\ttfamily}
\newcommand\termplainstyle{\footnotesize\ttfamily}

\newcommand\processLongMacroDelimiter
{%
\CPPdirectivestyle%
\#define%
}

\lstdefinestyle{cpp}
{
  language=C++,
  basicstyle=\footnotesize\ttfamily,
  basewidth={0.53em,0.44em},
  numbers=none,
  tabsize=2,
  breaklines=true,
  escapeinside={@}{@},
  showstringspaces=false,
  numberstyle=\tiny\color{solarized@base01},
  keywordstyle=\color{solarized@orange},
  stringstyle=\color{solarized@red}\ttfamily,
  identifierstyle=\color{solarized@blue},
  commentstyle=\CPPcommentstyle,
  directivestyle=\CPPdirectivestyle,
  emphstyle=\color{solarized@green},
  frame=single,
  rulecolor=\color{solarized@base2},
  rulesepcolor=\color{solarized@base2},
  literate={~} {\customtilde}1,
  moredelim=*[directive]\ \ \#,
  moredelim=*[directive]\ \ \ \ \#
}

\lstdefinestyle{cppalt}
{
  language=C++,
  basicstyle=\footnotesize\ttfamily,
  basewidth={0.53em,0.44em}, 
  numbers=none,
  tabsize=2,
  breaklines=true,
  escapeinside={*@}{@*},
  showstringspaces=false,
  numberstyle=\tiny\color{solarized@base01},
  keywordstyle=\color{solarized@orange},
  stringstyle=\color{solarized@red}\ttfamily,
  identifierstyle=\color{solarized@blue},
  commentstyle=\CPPcommentstyle,
  directivestyle=\CPPdirectivestyle,
  emphstyle=\color{solarized@green},
  frame=single,
  rulecolor=\color{solarized@base2},
  rulesepcolor=\color{solarized@base2},
  literate={~}{\customtilde}1,
  moredelim=**[is][\processLongMacroDelimiter]{BeginLongMacro}{EndLongMacro} 
}

\lstdefinestyle{cppnum}
{
  language=C++,
  basicstyle=\footnotesize\ttfamily,
  basewidth={0.53em,0.44em},
  numbers=none,
  tabsize=2,
  breaklines=true,
  escapeinside={@}{@},
  numberstyle=\tiny\color{solarized@base01},
  showstringspaces=false,
  numberstyle=\tiny\color{solarized@base01},
  keywordstyle=\color{solarized@orange},
  stringstyle=\color{solarized@red}\ttfamily,
  identifierstyle=\color{solarized@blue},
  commentstyle=\CPPcommentstyle,
  directivestyle=\CPPdirectivestyle,
  emphstyle=\color{solarized@green},
  frame=single,
  rulecolor=\color{solarized@base2},
  rulesepcolor=\color{solarized@base2},
  literate={~} {\customtilde}1,
  moredelim=*[directive]\ \ \#,
  moredelim=*[directive]\ \ \ \ \#
}

\lstdefinestyle{python}
{
  language=Python,
  basicstyle=\footnotesize\ttfamily,
  basewidth={0.53em,0.44em},
  numbers=none,
  tabsize=2,
  breaklines=true,
  escapeinside={@}{@},
  showstringspaces=false,
  numberstyle=\tiny\color{solarized@base01},
  keywordstyle=\color{blue},
  stringstyle=\color{orange}\ttfamily,
  identifierstyle=\color{darkred},
  commentstyle=\color{purple},
  emphstyle=\color{green},
  frame=single,
  rulecolor=\color{solarized@base2},
  rulesepcolor=\color{solarized@base2},
  literate = {~}{\customtilde}1
             {\ as\ }{{\color{blue}\ as\ \color{black}}}3
}

\lstdefinestyle{fortran}
{
  language=Fortran,
  basicstyle=\footnotesize\ttfamily,
  basewidth={0.53em,0.44em},
  numbers=none,
  tabsize=2,
  breaklines=true,
  escapeinside={@}{@},
  showstringspaces=false,
  numberstyle=\tiny\color{solarized@base01},
  keywordstyle=\color{blue},
  stringstyle=\color{orange}\ttfamily,
  identifierstyle=\color{Periwinkle},
  commentstyle=\color{purple},
  emphstyle=\color{green},
  morekeywords={and, or, true, false},
  frame=single,
  rulecolor=\color{solarized@base2},
  rulesepcolor=\color{solarized@base2},
  literate={~}{\customtilde}1
}

\lstdefinestyle{terminal}
{
  language=bash,
  basicstyle=\termplainstyle,
  numbers=none,
  tabsize=2,
  breaklines=true,
  escapeinside={@}{@},
  frame=single,
  showstringspaces=false,
  numberstyle=\tiny\color{solarized@base01},
  keywordstyle=\color{solarized@orange},
  stringstyle=\color{solarized@red}\ttfamily,
  identifierstyle=\color{black},
  commentstyle=\color{solarized@violet},
  emphstyle=\color{solarized@green},
  frame=single,
  rulecolor=\color{solarized@base2},
  rulesepcolor=\color{solarized@base2},
  morekeywords={gambit, cmake, make, mkdir},
  deletekeywords={test},
  literate = {\ gambit}{{\ }{\color{black}}gambit}7
             {/gambit}{{/}{\color{black}}gambit}6
             {gambit/}{{\color{black}}gambit{/}}6
             {/include}{{/}{\color{black}}include}8
             {cmake/}{{\color{black}}cmake/}6
             {.cmake}{{.}{\color{black}}cmake}6
             {~}{\customtilde}1
}

\lstdefinestyle{terminalalt}
{
  language=bash,
  basicstyle=\footnotesize\ttfamily,
  numbers=none,
  tabsize=2,
  breaklines=true,
  escapeinside={*@}{@*},
  frame=single,
  showstringspaces=false,
  numberstyle=\tiny\color{solarized@base01},
  keywordstyle=\color{solarized@orange},
  stringstyle=\color{solarized@red}\ttfamily,
  identifierstyle=\color{black},
  commentstyle=\color{solarized@violet},
  emphstyle=\color{solarized@green},
  frame=single,
  rulecolor=\color{solarized@base2},
  rulesepcolor=\color{solarized@base2},
  morekeywords={gambit, cmake, make, mkdir},
  deletekeywords={test},
  literate = {\ gambit}{{\ }{\color{black}}gambit}7
             {/gambit}{{/}{\color{black}}gambit}6
             {gambit/}{{\color{black}}gambit{/}}6
             {/include}{{/}{\color{black}}include}8
             {cmake/}{{\color{black}}cmake/}6
             {.cmake}{{.}{\color{black}}cmake}6
             {~}{\customtilde}1
}

\lstdefinestyle{text}
{
  language={},
  basicstyle=\footnotesize\ttfamily,
  identifierstyle=\color{black},
  numbers=none,
  tabsize=2,
  breaklines=true,
  escapeinside={*@}{@*},
  showstringspaces=false,
  frame=single,
  rulecolor=\color{solarized@base2},
  rulesepcolor=\color{solarized@base2},
  literate={~}{\customtilde}1
}

\lstdefinestyle{yaml}
{
  language=bash,
  escapeinside={@}{@},
  keywords={true,false,null},
  otherkeywords={},
  keywordstyle=\color{solarized@base0}\bfseries,
  basicstyle=\footnotesize\color{black}\ttfamily,
  identifierstyle=\YAMLkeystyle,
  sensitive=false,
  commentstyle=\color{solarized@orange}\ttfamily,
  morecomment=[l]{\#},
  morecomment=[s]{/*}{*/},
  stringstyle=\YAMLstringstyle\ttfamily,
  moredelim=**[s][\YAMLkeystyle]{,}{:},   
  moredelim=**[l][\YAMLvaluestyle]{:},    
  morestring=[b]',
  morestring=[b]",
  literate =    {---}{{\ProcessThreeDashes}}3
                {>}{{\textcolor{solarized@red}\textgreater}}1
                {|}{{\textcolor{solarized@red}\textbar}}1
                {\ -\ }{{\mdseries\color{black}\ -\ \negmedspace}}3
                {\}}{{{\color{black} \}}}}1
                {\{}{{{\color{black} \{}}}1
                {[}{{{\color{black} [}}}1
                {]}{{{\color{black} ]}}}1
                {~}{\customtilde}1,
  breakindent=0pt,
  breakatwhitespace,
  columns=fullflexible
}

\lstdefinestyle{mathematica}
{
  language={Mathematica},
  basicstyle=\footnotesize\ttfamily,
  basewidth={0.53em,0.44em},
  numbers=none,
  tabsize=2,
  breaklines=true,
  escapeinside={@}{@},
  numberstyle=\tiny\color{black},
  showstringspaces=false,
  numberstyle=\tiny\color{solarized@base01},
  keywordstyle=\color{solarized@orange},
  stringstyle=\color{solarized@red}\ttfamily,
  identifierstyle=\color{solarized@orange}\ttfamily,
  commentstyle=\color{solarized@gray}\ttfamily,
  directivestyle=\color{solarized@orange}\ttfamily,
  emphstyle=\color{solarized@green},
  frame=single,
  rulecolor=\color{solarized@base2},
  rulesepcolor=\color{solarized@base2},
  literate={~} {\customtilde}1,
  moredelim=*[directive]\ \ \#,
  moredelim=*[directive]\ \ \ \ \#,
  mathescape=true
}

\newcommand{\cross}[1]{\ref{#1}}
\newcommand{\doublecross}[2]{\hyperref[#2]{\textbf{#1}}}
\newcommand{\doublecrosssf}[2]{\hyperref[#2]{\textbf{\textsf{#1}}}}
\newcommand{\gitem}[1]{\item[\textbf{#1}\label{#1}]}

\newcommand{\startglossary}{\section{Glossary}\label{glossary}Here we explain some terms that have specific technical definitions in \GB.\begin{description}}
\newcommand{\finishglossary}{\end{description}}

\newcommand{\bcode}{\begin{lstlisting}}
\newcommand{\ecode}{\end{lstlisting}}
\newcommand{\metavarf}[1]{\textit{\color{darkgreen}\footnotesize\textrm{#1}}}

\newcommand{\metavar}{\metavarf}

\newcommand{\sss}{\scriptscriptstyle}
\newcommand{\ms}{m_{\sss S}}
\newcommand{\lhs}{\lambda_{h\sss S}}

\newcommand{\gambit}{\textsf{GAMBIT}\xspace}

\newcommand{\precisionbit}{\textsf{PrecisionBit}\xspace}
\newcommand{\scannerbit}{\textsf{ScannerBit}\xspace}

\newcommand{\GB}{\gambit}

\newcommand{\omp}{\textsf{OpenMP}\xspace}
\newcommand{\mpi}{\textsf{MPI}\xspace}

\newcommand\flexiblesusy{\FlexibleSUSY}
\newcommand\FlexibleSUSY{\textsf{FlexibleSUSY}\xspace}

\newcommand\SOFTSUSY{\textsf{SOFTSUSY}\xspace}

\newcommand\pippi{\textsf{pippi}\xspace}
\newcommand\MultiNest{\textsf{MultiNest}\xspace}
\newcommand\multinest{\MultiNest}
\newcommand\great{\textsf{GreAT}\xspace}
\newcommand\twalk{\textsf{T-Walk}\xspace}
\newcommand\diver{\textsf{Diver}\xspace}

\newcommand\xx{\raisebox{0.2ex}{\smaller ++}\xspace}
\newcommand\Cpp{\textsf{C\xx}\xspace}

\newcommand\plainC{\textsf{C}\xspace}

\newcommand\Fortran{\textsf{Fortran}\xspace}
\newcommand\YAML{\textsf{YAML}\xspace}

\newcommand\beq{\begin{equation}}
\newcommand\eeq{\end{equation}}

\renewcommand{\url}[1]{\href{#1}{#1}}

\begin{document}

\preprintnumber{}

\title{Comparison of statistical sampling methods with ScannerBit, the GAMBIT scanning module}

\author
{
  The GAMBIT Scanner Workgroup:
  Gregory D.\ Martinez\thanksref{inst:a,e1} \and
  James McKay\thanksref{inst:b,e2} \and
  Ben Farmer\thanksref{inst:c,inst:d,e3} \and
  Pat Scott\thanksref{inst:b,e4} \and
  Elinore Roebber\thanksref{inst:e} \and
  Antje Putze\thanksref{inst:f} \and
  Jan Conrad\thanksref{inst:c,inst:d}
}

\institute{%
  \ucla\label{inst:a} \and
  \imperial\label{inst:b} \and
  \okc\label{inst:c} \and
  \su\label{inst:d} \and
  \mcgill\label{inst:e} \and
  \annecy\label{inst:f}
}

\thankstext{e1}{gmartine@astro.ucla.edu}
\thankstext{e2}{j.mckay14@imperial.ac.uk}
\thankstext{e3}{benjamin.farmer@fysik.su.se}
\thankstext{e4}{p.scott@imperial.ac.uk}

\titlerunning{ScannerBit}
\authorrunning{GAMBIT Scanner Workgroup}

\date{Received: date / Accepted: date}

\maketitle

\begin{abstract}
We introduce \scannerbit, the statistics and sampling module of the public, open-source global fitting framework \gambit. \scannerbit provides a standardised interface to different sampling algorithms, enabling the use and comparison of multiple computational methods for inferring profile likelihoods, Bayesian posteriors, and other statistical quantities.  The current version offers random, grid, raster, nested sampling, differential evolution, Markov Chain Monte Carlo (MCMC) and ensemble Monte Carlo samplers.  We also announce the release of a new standalone differential evolution sampler, \diver, and describe its design, usage and interface to \scannerbit.  We subject \diver and three other samplers (the nested sampler \multinest, the MCMC \great, and the native \scannerbit implementation of the ensemble Monte Carlo algorithm \twalk) to a battery of statistical tests.  For this we use a realistic physical likelihood function, based on the scalar singlet model of dark matter.  We examine the performance of each sampler as a function of its adjustable settings, and the dimensionality of the sampling problem.  We evaluate performance on four metrics: optimality of the best fit found, completeness in exploring the best-fit region, number of likelihood evaluations, and total runtime.  For Bayesian posterior estimation at high resolution, \twalk provides the most accurate and timely mapping of the full parameter space.  For profile likelihood analysis in less than about ten dimensions, we find that \diver and \multinest score similarly in terms of best fit and speed, outperforming \great and \twalk; in ten or more dimensions, \diver substantially outperforms the other three samplers on all metrics.
\end{abstract}

\tableofcontents

\section{Introduction}

Science has entered an era of increasing computational complexity.  Large data sets and burgeoning model complexity have necessitated the development of increasingly sophisticated and efficient analysis techniques.  As datasets and theories in particle physics and cosmology have become more computationally expensive to work with, the problem of efficiently and comprehensively sampling model parameter spaces has become steadily more challenging.  Simple random parameter sampling (e.g.\ \cite{Berger09,ATLAS15}) has gradually proven more inadequate as time goes on, as it typically leads to incomplete and biased inferences when applied to all but the simplest problems.

Workers in various fields have employed increasingly advanced numerical and statistical methods to deal with this challenge.  Bayesian numerical techniques such as Markov Chain Monte Carlos (MCMCs) became particularly popular in cosmology, because of their theoretical near-linear scalability with parameter dimensionality.  Cosmic Microwave Background (CMB) analyses were amongst the first such applications of MCMCs \citep{Christensen01}, with later improvements and optimisations brought about through the use of adaptive techniques and robust convergence criteria \citep{Dunkley05, CosmoMC, CosmoMC++}.  MCMCs also proved popular in particle physics, for the exploration of moderately complex supersymmetric model parameter spaces \citep{Baltz04, Allanach06, Fittino06, Ruiz06, Buchmueller08}.  Nested sampling \cite{Skilling04} gradually displaced MCMCs in many such applications \cite{Trotta08,Scott09c,Planck15cosmo,MasterCodeMSSM10}, owing to its efficiency for mapping posterior distributions and calculating the Bayesian evidence, especially when dealing with multi-modal likelihoods \cite{MultiNest}.

Because the likelihood functions involved are computationally expensive (see e.g.\ \cite{IC79_SUSY,ColliderBit}), fully frequentist Neyman constructions are typically not possible.  A popular alternative to Bayesian inference is to examine the prior-independent profile likelihood.  However, Bayesian methods such as MCMCs and nested sampling are not necessarily optimal sampling strategies in this case \cite{Akrami09}.  Estimating the Bayesian posterior requires integrating the likelihood in various directions of the parameter space, whereas the profile likelihood relies instead on maximising it in those directions. From the perspective of numerical analysis, to a first approximation Bayesian sampling is an integration problem, whereas profile likelihood estimation is an optimisation problem.  It is therefore unsurprising that modern multi-modal optimisation strategies such as genetic algorithms and differential evolution have proven more efficient than Bayesian methods in some applications of the profile likelihood \cite{Akrami09,DEnu}.

This picture is further complicated by additional requirements not present in traditional optimisation problems.  To be able to infer reliable confidence intervals on parameters, the likelihood function must be sampled sufficiently well around the maximum to allow isolikelihood contours to be inferred.  Unfortunately, determination of the global best-fit point does not necessarily guarantee that this will be the case. In this respect, some Bayesian methods can in fact be more efficient than optimisers, even if they are less efficient at finding the global maximum \cite{SBSpike}.   Another issue is the degree to which the resulting confidence intervals achieved in the profile likelihood analysis have the expected statistical coverage properties \cite{Akrami11coverage, SBcoverage, Strege12, Fittinocoverage}; this can be strongly influenced by the choice of scanning algorithm.

In this paper we provide a detailed manual for \scannerbit, a package designed to provide a common interface to a range of different sampling algorithms, so that the performance of the different algorithms can be easily compared, and the most appropriate algorithm (or combination thereof) chosen for the problem at hand.  We also carry out some such comparisons of sampling algorithms, and provide recommended settings for different samplers.

\scannerbit is designed to be modular and expandable, allowing it to access a multitude of different samplers in a plug and play fashion.  As the \gambit project grows, we will continually add scanners to the \scannerbit suite.  Users can also easily implement various scanners to meet their personal needs.  \scannerbit initially ships with four production-quality scanners: an adaptive MCMC (\great), an ensemble MCMC (\twalk), a nested sampler (\multinest) and a differential evolution sampler (\diver).  \great \cite{great} and \multinest \cite{MultiNest} are existing external packages. \diver is a new external package that we describe for the first time here. \twalk is implemented natively in \scannerbit.  The \scannerbit package also contains a postprocessor and a series of simple scanners, including random, grid and list-based samplers and a more basic toy MCMC (for tutorial purposes).

All the scanners initially accessible from \scannerbit are designed for the calculation of profile likelihoods or Bayesian posteriors, such that they select optimal parameter combinations for which to perform likelihood calculations. These samplers therefore require the likelihood to be explicitly calculable for any parameter combination, either by parametrisation or numerical approximation.  The design of \scannerbit is not limited to this operation mode, however, and can easily support methods that do not require explicit calculation of a likelihood, such as Approximate Bayesian Computation \cite{cosmoabc}.

\scannerbit can either be used within its parent code \gambit \cite{gambit}, or as a standalone package, or simply interfaced directly to an external likelihood function.

We begin by describing the \scannerbit package in Sec.\ \ref{sec:description}, before giving the implementation details and the underlying statistical methods that we employ in Sec.\ \ref{sec:stats}.  The user interface is covered in Sec.\ \ref{sec:interface}, and the simple scanners in Sec.\ \ref{sec:simple}.  Secs.\ \ref{sec:postprocessor}--\ref{sec:de} respectively describe the postprocessor, MCMC, ensemble MCMC, nested sampler and differential evolution samplers.  In Sec.\ \ref{sec:comparison} we perform a detailed comparison of the different algorithms implemented in \scannerbit, and their available parameters and options.  We summarise in Sec. \ref{sec:conclusions}, then provide an extensive set of appendices.  These cover the sources, options and outputs of our differential evolution sampler \diver (Appendix \ref{app:diver}), \scannerbit options and outputs for all five major scanners (Appendix \ref{app:options}), examples of how to implement new priors (Appendix \ref{app:priors}), examples of adding new scanners and objective functions (Appendix \ref{app:plugin_decl}), some supplementary comparisons of scanner performance (Appendix \ref{app:scanner_comparisons}),
a minimal example input file (Appendix \ref{app:inifile}), and a glossary of the most commonly-used \GB terms (Appendix \ref{glossary}).

More details on \gambit itself can be found in Ref.\ \cite{gambit}, on its various physics modules in Refs.\ \cite{DarkBit,ColliderBit,SDPBit,FlavBit}, and on first physics results in Refs.\ \cite{CMSSM,MSSM,SSDM}.

\section{Package description}
\label{sec:description}

To efficiently sample an $n$-dimensional parameter space, \scannerbit works by separating the sampling problem into three distinct steps:\begin{enumerate}
\item Choosing $n$ values in the interval between 0 and 1.  Taken together, these values constitute a `point' in the $n$-dimensional `unit hypercube'.
\item Transforming the point in the unit hypercube into a point in the physical $n$-dimensional parameter space.
\item Passing the values of the physical parameters to a user-specified function, which may compute a number of things from the parameter values, including theoretical predictions of different observable quantities and corresponding likelihoods, based on e.g.\ comparison with experimental data.
\end{enumerate}
The steps then repeat until some convergence criterion is satisfied, with the results of Step 3 used to help choose the next point in the unit hypercube in the subsequent iteration of Step 1.

The results of Step 3 are output in each iteration, in a format of the user's choice.  Any output format supported by the \GB printer system can be chosen, including \textsf{ASCII} and \textsf{HDF5} database files. The \GB printers are described in detail in Sec.\ 9 of Ref.\ \cite{gambit}, and \scannerbit's interface to them is described in Sec.\ \ref{sec:printers}.  Both \textsf{ASCII} and \textsf{HDF5} outputs can be parsed and plotted as profile likelihoods with \pippi \cite{pippi}, which can be installed automatically from within \GB (or \scannerbit) by typing \term{make get-pippi} from within the build directory.  As the printer output is handled in Step 3, independent of the sampling algorithms responsible for Step 1, every parameter set tested is printed, along with every quantity derived from this set, regardless of whether the scanner accepts the tested point or not.  This provides the maximum information possible for profile likelihood and post-processing analyses.  The \GB printers can also be sent additional supplementary data computed by the samplers themselves, such as the posterior weights needed for plotting posterior probability densities with \pippi.

\subsection{\scannerbit plugins}
\label{sec:plugins}

\scannerbit is designed to be completely modular and expandable.  It achieves this via a plugin interface, which allows various scanners and likelihood functions to be connected at will.  Plugins are either \doublecross{scanner plugins}{scanner plugin}, which each contain code implementing a single sampling algorithm, or objective plugins, also known as \doublecross{test function plugins}{test function plugin}, which contain specific objective functions to be scanned (such as simple test functions and likelihoods).  Scanner plugins are responsible for efficiently navigating the unit cube in Step 1, whereas objective plugins provide the user-specified function in Step 3.  Each plugin is compiled into an independent library with a common interface to \scannerbit, so that at runtime it can be passed necessary information like the dimensionality of the space being scanned and the user's preferred method of outputting the results.

The transformation that must be applied in Step 2 constitutes a sampling prior.  This is relevant both for Bayesian analyses, where the final posterior probability of the model parameters is directly proportional to the prior, and for profile likelihood analyses, where the sampling prior can have an impact on how efficiently the likelihood function can be sampled.  \scannerbit implements priors as transformations of the uniform probability distribution, as it instructs all scanner plugins to carry out Step 1 by sampling the unit hypercube using a uniform sampling prior. \scannerbit transforms the samples generated from the unit hypercube into actual model-space parameter values by requiring the user to select a prior transformation to apply to each parameter.  This allows scanner plugins to operate completely independently of priors.  Sampler implementations are kept entirely independent of prior implementations, allowing any scanner to be used with any prior.\footnote{Although scanning the unit hypercube is the default, \scannerbit does also permit special scanners developed for specific models to choose to bypass the prior transformation entirely, in order to work directly with model parameter values.  Users are advised to avoid this unless strictly necessary though, as the resulting scanner is neither usable with other models nor other priors.}  Priors can be added to \scannerbit in a similarly modular way to scanner and test function plugins (see Sec.\ \ref{sampling_distributions}).

\scannerbit grants scanner plugins access to specific functions necessary for them to perform their sampling task.  At the simplest level, the only such functions are the prior transformation of Step 2, and a log-likelihood function for Step 3, allowing the likelihood to be evaluated for any given point in the hypercube.  The function(s) provided to a scanner plugin at runtime are selected by assigning \doublecross{purposes}{purpose} (such as ``\yaml{LogLike}'') to different objective plugins or results provided by \gambit, and then telling each scanner which purpose(s) corresponding to the inputs it should collect.  The purposes are specified in the input file for a \scannerbit run, which should be written in \YAML format.\footnote{\href{http://www.yaml.org/}{www.yaml.org}}  All \scannerbit objective functions tagged for a common purpose are combined into a single function, and provided to the scanner as a function pointer.  In a regular \gambit scan, this is the total log-likelihood function provided by the \cross{likelihood container}, which combines \GB functions tagged with a common purpose, according to the specific function \doublecross{capabilities}{capability} requested by the user in their input \YAML file.

Generically, objective plugins take model parameter values as inputs, and return some quantity useful to \scannerbit for performing a scan.  Each objective can be individually assigned a purpose to enable its output to be assigned appropriately in a scanner plugin.  The canonical example of an objective plugin is the merit function to be used in a given scan, allowing \scannerbit to determine which parameter combinations are better than others, and to make informed choices about which combinations to sample next.  This function might be a complicated likelihood (as in the case of the \gambit \cross{likelihood container}), or just a simple test function for evaluating the performance of a new scanner.  A more advanced example of an objective plugin would be one that provides the derivative of a merit function, for use with e.g.\ optimisers that use derivatives to accelerate their searches.  Whilst each objective plugin is automatically given access to the prior chosen for a given scan, objective plugins can in fact \textit{also} be employed to provide the underlying transformation function used in a prior (although this method is not mandatory for defining a new prior -- see Sec.\ \ref{sampling_distributions}).

Each plugin's source code is placed in its own subdirectory within \term{ScannerBit/src/}\metavar{plugin\_kind}, where \metavar{plugin\_kind} is either \term{scanners} or \term{objectives}.  The plugin headers reside in their own subdirectory within \term{ScannerBit/headers/gambit/ScannerBit/}\metavar{plugin\_kind}.  Each plugin's compilation and linkage is handled by the \scannerbit \textsf{CMake} script.

\section{Statistics and scanning}
\label{sec:stats}

To run a parameter scan in \GB, the user writes an input \YAML file specifying that they want to analyse a particular model.  They indicate the parameter ranges and priors over which \GB should sample that model, how that sampling should be done, and what quantities should be computed for each parameter combination.  \GB activates the model in its model database, along with all other models that the model in question is a subspace of.  The dependency resolver uses the activated model hierarchy and the list of the user's requested quantities to activate and connect various module functions into a dependency graph   (see Ref.\ \cite{gambit}).  \scannerbit is then responsible for determining which parameter combinations to run through the dependency graph.

Every quantity requested for calculation in a scan must be assigned a \cross{purpose} in the input \YAML file, using the eponymous option \yaml{purpose}.  This can be set to \yamlvalue{Test} or \yamlvalue{Observable}, to flag that the quantity must be computed and output for each sample.  To include the quantity in the function that actually drives a sampler, the user must match the \yaml{purpose} of the quantity to whatever \yaml{purpose} he or she instructs the sampler to seek out in order to define its objective function. Once dependency resolution has been completed, \GB constructs a \cross{likelihood container}, which consists of the dependency tree of all module functions assigned the purpose sought by the sampler.  This container essentially packages the results of the different module functions into a single function that can be called by any sampling algorithm.

Conventionally, \GB example \YAML files assign \yaml{purpose: LogLike} to any quantity that should enter the fit as a likelihood component, and expects such functions to return the natural logarithm of the likelihood $\log\mathcal{L}$.  By simply summing their return values, the likelihood container combines the results of all log-likelihood functions and returns the result to \scannerbit as the total log-likelihood.  At present, the sampling algorithms callable by \scannerbit allow only a single \yaml{purpose} to dictate the behaviour of a scan, although future scanners are anticipated to make use of two or more distinct purposes in a single scan (as in e.g. in multi-objective optimisation).

\subsection{Priors and sampling distributions}
\label{sampling_distributions}

Most samplers are driven by \scannerbit to draw from the unit interval $[0,1]$.  The sampled values are then converted to real physical parameters internally, using whatever prior the user has chosen when launching the scan. In the simplest cases, this occurs by applying the transformation method, where samples from the unit interval are converted to samples from the desired sampling distribution (i.e.~prior), by applying the inverse of the cumulative distribution function (CDF) of the desired distribution. Here, a uniform random deviate $x$ is transformed into a random deviate $y$ sampled from a target distribution $D$ with cumulative distribution function $F(y)$, by computing
\begin{equation}
  y = F^{-1}(x).
\end{equation}
Take as an example the case where a user requests a flat `prior' over the range $[a,b]$ for some parameter. \scannerbit expects the underlying sampler to provide a number $x$ in the interval $[0,1]$, and then applies the transformation
\begin{equation}
  y = F^{-1}(x) = (b-a) x + a ,
\end{equation}
in order to obtain a sample in the range $[a,b]$.  Here $F^{-1}(x)$ is the inverse of
\begin{align}
F(y) &\equiv \int_a^y P(x)\,dx\nonumber\\
     &= \int_a^y \frac{dx}{b-a}\nonumber\\
     & = \frac{y-a}{b-a},
\end{align}
which is the CDF of $P(x)\equiv1/(b-a)$, the uniform distribution over $[a,b]$. Thus, although the underlying sampler chooses uniform random numbers for $x$ from the interval $[0,1]$, the final `physical' parameter $y$ will be sampled uniformly from the interval $[a,b]$. Similarly, if the user requests a `Gaussian' prior (with mean $\mu$ and standard deviation $\sigma$) for parameter $y$, then \scannerbit will apply the transformation
\begin{equation}
  y = \mu + \sigma \sqrt{2} \,\, \text{erf}^{-1}\left( 2 x - 1 \right),
\end{equation}
so that uniform samples from the unit interval are transformed into samples from the normal distribution $\mathcal{N}(\mu,\sigma)$.

It is important to note that the actual sampling distribution of a scan only follows these transformed distributions in the special case where the underlying unit-interval sampling is actually uniform. This corresponds to the case of a purely random sampling algorithm (implemented as the \textsf{random} sampler in \scannerbit; see Sec.\ \ref{sec:random_sampler}).

If the underlying sampling is driven, for example, by a Metropolis-Hastings algorithm, or an evolutionary sampler, then the final samples will of course not be drawn directly from the user-requested distribution. In this case the user-requested sampling distribution still has statistical implications, particularly for the Bayesian interpretation of results, where it plays the role of the prior probability distribution. For example, if the user requests that a parameter have a Gaussian prior $\pi(y)$, and chooses to draw samples with a Metropolis-Hastings algorithm, then the final density of points will be proportional to the posterior probability density $p(y)$
\begin{equation}
  p(y)\propto\mathcal{L}(y)\pi(y).
\end{equation}
This is because it is a property of the Metropolis-Hastings algorithm that the density of sample points is proportional to $\mathcal{L}$ in the unit-interval parameter space -- which is then distorted to the physical parameter space density $d(y)$ under the mapping $y = F^{-1}(x)$
\begin{align}
  d(y) &= \mathcal{L}(y) \left| \frac{\mathrm{d} F(y)}{\mathrm{d}y} \right| \\
       &= \mathcal{L}(y) f(y).
\end{align}
Here $f(y)$ is the probability distribution function (PDF) corresponding to the CDF $F(y)$, and is therefore the user-requested `prior', and $d(y)$ is proportional to the posterior probability density $p(y)$.

\scannerbit makes a wide range of possible prior transformations available.  These priors are separated into three groups: one-dimensional (\textsf{flat}, \textsf{log}, \textsf{double\_log\_flat\_join}, \textsf{sin}, \textsf{cos}, \textsf{tan}, \textsf{cot}),  multi-dimensional (\textsf{gaussian}, \textsf{cauchy}), and others (\textsf{same\_as}, \textsf{fixed\_value}, \textsf{none}, \textsf{plugin}).  These priors, and their corresponding options, can be specified in the \yaml{Priors} section of the \YAML input file that defines a scan, or, in the case of one-dimensional priors, also in the \yaml{Parameters} section (see Section \ref{sec:interface}). Users can also define custom priors, which can be added to the set of priors available to \scannerbit (see Appendix\ \ref{app:priors}).

\subsubsection{Built-in one-dimensional priors}
\label{sec:1Dpriors}

\scannerbit currently includes six one-dimensional priors:
\begin{description}
 \item \textsf{sin}: $\mathcal{P}(x) \varpropto \sin(x)$
 \item \textsf{cos}: $\mathcal{P}(x) \varpropto \cos(x)$
 \item \textsf{tan}: $\mathcal{P}(x) \varpropto \tan(x)$
 \item \textsf{cot}: $\mathcal{P}(x) \varpropto \cot(x)$
 \item \textsf{flat}: Uniform in $x$, i.e.~$\mathcal{P}(x) \varpropto \mathrm{const}$.
 \item \textsf{log}: Uniform in $\log x$, i.e.~$\mathcal{P}(x) \varpropto 1/x$.
 \item \textsf{double\_log\_flat\_join}: A piecewise prior that patches together sections uniform in $\log (-x)$, uniform in $x$, and uniform in $\log x$.  Useful when the desired prior density is positive at zero, but logarithmic at large absolute values of the parameter. i.e.
 \begin{equation}
  \mathcal{P}(x) \varpropto \left\{
  \begin{array}{lcrcl}
  \nonumber
   1/|x| & : & \text{\cpp{lower}} &< x <& \text{\cpp{flat\_start}} \\
   \mathrm{const} & : & \text{\cpp{flat\_start}} &\leq x \leq& \text{\cpp{flat\_end}} \\
   1/x & : & \text{\cpp{flat\_end}} &< x <& \text{\cpp{upper}}
  \end{array}
  \right.
 \end{equation}
\end{description}
Each prior has a number of configurable options.  These may be entered as key-value entries for the parameter in question, in the input \YAML file.  For one-dimensional priors, the options can be entered in either the \yaml{Priors} or the \yaml{Parameters} section of the \YAML file (further details on the input file format can be found in Sec.\ \ref{sec:interface}).  The following options are available for all 1D priors except \textsf{double\_log\_flat\_join}:
\begin{description}
 \item\yaml{range:} Specifies the range in the form [low, high].
 \item\yaml{shift:} Shifts all parameter samples by the specified value.  Defaults to \yamlvalue{0} if absent.
 \item\yaml{scale:} Multiplies all parameter samples by the specified value.  If set to \yamlvalue{degrees}, will convert degrees to radians.  Defaults to \yamlvalue{1} if absent.
 \item\yaml{output\_scaled\_values:} If \yamlvalue{true}, any scale and/or shift applied to the parameter during a scan is also applied to the printed value of the parameter.  Defaults to \yamlvalue{true} if absent.
\end{description}
The \textsf{double\_log\_flat\_join} prior also accepts the same \yaml{range} option, as well as
\begin{description}
 \item\yaml{ranges:} An extended version of \yaml{range}, taking the form [lower, flat\_start, flat\_end, upper].  The negative log prior is applied over parameter values ranging from the first to the second entry, the flat prior is applied from the second to the third entry, and the positive log prior is applied between the third and fourth entries.  This option takes precedence over \yaml{range}.
 \item\yaml{flat\_start, flat\_end:} The boundaries of the interior region over which to apply the flat prior; these options are expected whenever the 4-component \yaml{ranges} option is not in use.
 \item\yaml{lower, upper:} The outer boundaries of the logarithmic prior sections. These options are only used if neither \yaml{ranges} nor \yaml{range} is present. They require the presence of \yaml{flat\_start} and \yaml{flat\_end}.
\end{description}

\subsubsection{Built-in multi-dimensional priors}
\label{sec:2Dpriors}

\scannerbit presently ships with two real multi-dimensional priors, and one example function:
\begin{description}
 \item \textsf{gaussian}: Gaussian distribution of the form\\
  $\mathcal{P}(\vec{x}) \varpropto \exp[-(\vec{x}-\bar{\vec{x}}) \cdot C^{-1} \cdot (\vec{x}-\bar{\vec{x}})/2]$,\\with $C$ a covariance matrix.
 \item \textsf{cauchy}: Cauchy distribution of the form\\
  $\mathcal{P}(\vec{x}) \varpropto \left[1 + (\vec{x}-\bar{\vec{x}}) \cdot C^{-1} \cdot (\vec{x}-\bar{\vec{x}})\right]^{-1}$,\\with $C$ a covariance matrix.
 \item\textsf{dummy:} Performs a dummy transformation of the unit hypercube parameters back to themselves; included as a simple example of the code needed to define a new multidimensional prior (see Appendix\ \ref{app:priors}).
\end{description}
The \textsf{gaussian} and \textsf{cauchy} priors have options:
\begin{description}
 \item\yaml{cov:} Full covariance matrix.  Off-diagonal elements default to zero if this option is omitted.
 \item\yaml{sigs:} A vector containing the square root of each of the diagonal components of the covariance matrix. Defaults to \yamlvalue{1} if absent.
 \item\yaml{mean:} A vector containing the mean (for \textsf{gaussian}) or median (for \textsf{cauchy}) of each parameter.  Defaults to \yamlvalue{0} if absent.
\end{description}

\subsubsection{Additional built-in priors}
\label{sec:other_priors}

\scannerbit is also equipped with some useful non-standard priors:
\begin{description}
 \item\textsf{ same\_as}: Specifies that some parameter is the same as another parameter.  The net effect is to make both parameters appear as a single parameter to the scanner, but as two distinct parameters to the objective function.  This prior accepts an eponymous option \yaml{same_as}, which is used to choose which parameter to shadow.  It also optionally accepts the \yaml{scale} and \yaml{shift} keywords described in Sec.\ \ref{sec:1Dpriors}, allowing the parameter to be presented to the objective function as a rescaled, shifted version of the parameter it has been set up to shadow.
 \item\textsf{ fixed\_value}: Fixes this parameter to a specified value, with the actual value set by the option of the same name.  If a sequence of values is given, the values are simply iterated over in each subsequent point, repeating from the beginning once exhausted.  This prior also accepts the \yaml{scale} and \yaml{shift} keywords.
 \item\textsf{ none}: Specifies that this parameter will be directly set by the scanner.  If the scanner does not do so, \scannerbit will throw an error.
 \item\textsf{ plugin}: Uses a plugin function as the prior.  The plugin to be used is set with an option of the same name (i.e., \yaml{plugin}), and must be defined as an objective plugin under the \yaml{objectives} tag in the \yaml{Scanner} section of the input \YAML file.  Note that in the current version of \scannerbit, using the same plugin more than once in a given scan is not supported, e.g. as two separate applications of a one-dimensional prior to two different parameters.
\end{description}

\subsection{Plugins}\label{sec:plugins}

\scannerbit plugins are independent code snippets, separate from the main \scannerbit code.  \doublecross{Scanner plugins}{scanner plugin} provide a standard interface between \scannerbit and sampling algorithms (whether external libraries or native \scannerbit implementations).  Objective plugins (otherwise known as \doublecross{test function plugins}{test function plugin}) provide an interface between \scannerbit and external objective or test functions.

Plugin functionality falls into three main categories: loading, unloading, and the main function provided to \scannerbit by the plugin.

\begin{description}
 \item \textbf{loading}: When a plugin is loaded, it is provided with some generic information needed for running any plugin, as well as specific information relevant to its plugin type. The generic information includes a list of expected input file options, as well as interfaces to the \cross{printer} and prior transform.  Plugin-specific information may include likelihood functor access, hypercube parameter dimension, and parameter key names.  Each plugin has a constructor that runs when the plugin is loaded, allowing it to perform startup operations such as variable initialisation.
 \item \textbf{unloading}: When a plugin is no longer needed, any shared libraries it has loaded are unloaded, and the plugin deconstructor runs.  This typically performs any plugin-specific shutdown operations, such as closing files or releasing memory.
 \item \textbf{main function}: Every plugin has some core functionality, provided by its \cpp{plugin_main} function.  For example, a scanner plugin's \cpp{plugin_main} should contain code that samples an objective function over a specified parameter space --- whereas an objective plugin to be used as a likelihood function would provide functionality necessary for likelihood evaluations.  This functionality may have any interface, but it must be consistent with the goal of the plugin.  For example, a likelihood plugin should accept a map of parameters and return a likelihood value, whereas a scanner plugin would not accept inputs.
\end{description}
Because of this general format, plugins can be used for a wide range of tasks.  Scanner plugins specifically contain code to perform parameter scans of various models, do not require inputs, and simply return an integer indicating the success or failure of the scan.  Objective plugins are for more general use, and may provide functions that can be used as likelihoods, observable functions, prior transforms, or in fact any other quantity that might need to be computed for each point in parameter space (e.g.\ likelihood gradients).  Objective plugins are not required to have any specific interface, but are all granted access to the same information and utility functions by \scannerbit.  Detailed information about definition, design and operation of \scannerbit plugins can be found in Appendix\ \ref{app:plugin_decl}.

\section{Setup and input file options}
\label{sec:interface}

\scannerbit scans are specified and initiated using an input file written in \YAML.  This file must contain at least four sections: \yaml{Parameters}, \yaml{Scanner}, \yaml{Printers} and \yaml{KeyValues}.  It may also optionally contain a \yaml{Priors} section.  We do not deal with the \yaml{Printers} and \yaml{KeyValues} sections in this paper, as they refer to \GB features described in detail in Ref.\ \cite{gambit}; minimal working entries for these sections can be found in the example input \YAML file given in Appendix\ \ref{app:inifile}.  Additionally, \scannerbit includes an example \YAML file, \term{ScannerBit.yaml}, in the \term{yaml_files} folder.  The \yaml{Parameters} section indicates which models and parameters to scan, as well as (optionally) simple prior definitions for individual parameters.  The \yaml{Priors} section contains additional --- potentially more complicated --- prior definitions not included in the \yaml{Parameters} section.  The \yaml{Scanner} section contains all scanner and plugin options and definitions.

\subsection{Input file {\YAMLkeystyle Parameters} section}

The \yaml{Parameters} section contains information about the models and their associated parameters, and follows the basic format:
\begin{lstyaml}
Parameters:
  @\metavar{model}@:
    @\metavar{parameter\_name1}@:
      @\metavar{...options...}@
    @\metavar{parameter\_name2}@:
      @\metavar{...options...}@
    ...
\end{lstyaml}
The \yaml{Parameters} section can contain several models, where each model contains several parameters.  Each declared parameter can have the following options, associated with the prior to be applied to the parameter:
\begin{description}
 \item\yaml{ prior\_type}: Specifies a one-dimensional prior to be applied to the parameter.  If this option is absent but either the \yaml{range}, \yaml{same\_as} or \yaml{fixed\_value} option is given, \scannerbit will deduce the prior type from the presence of the other option.
 \item\yaml{ range}: Specifies the range of parameter values to be sampled.  In the absence of an entry for \yaml{prior\_type}, specifying a \yaml{range} causes a \textsf{flat} prior to be adopted.
 \item\yaml{ shift}: Adds the given value to the parameter.
 \item\yaml{ scale}: Multiplies the parameter by the given amount.
 \item\yaml{ same\_as}: Indicates that this prior is the same as another parameter.  Note that \scannerbit parameters are denoted by a string of the form \metavar{model}::\metavar{parameter\_name}.
 \item\yaml{ fixed\_value}: Fixes the parameter to the given value.  The same effect can be achieved in even more compact form, by giving no options for a parameter except a value or sequence of values, in the form \metavar{parameter\_name}\yaml{:} \metavar{value}.
 \item\yaml{ lower, flat\_start, flat\_end, upper}: for the \textsf{double\_log\_flat\_join} prior (see Sec.\ \ref{sec:1Dpriors}).

\end{description}
Each of these options are optional.  If none of them is set, the prior must be specified in the \yaml{Priors} section.  Like the \textsf{flat} prior, the \textsf{fixed\_value} and \textsf{same\_as} priors do not need to be specifically indicated with \yaml{prior\_type}, as they are implicitly defined by the declaration of their options.  More details can be found in the subsection dealing specifically with one-dimensional priors (Sec.\ \ref{sec:1Dpriors}).

\subsection{Input file {\YAMLkeystyle Priors} section}

Any parameter lacking a specified one-dimensional prior in the \yaml{Parameters} section must be associated with a sampling range and prior in the \yaml{Priors} section.  A prior definition in this section takes the form:
\begin{lstyaml}
Priors:
  @\metavar{prior\_name}@:
    parameters: [@\metavar{parameter\_list}@]
    prior_type: @\metavar{type}@
    @\metavar{options}@
\end{lstyaml}
Here, \metavar{prior\_name} can be any unique identifier, and need not map to any particular name within \scannerbit.  The \metavar{parameter\_list} is a sequence of parameters to apply the prior to.  The \metavar{type} of the prior must match one of the known \scannerbit priors listed in Sec.~\ref{sampling_distributions}.  This should be followed by any additional key-value pairs needed to set the desired \metavar{options} of the chosen prior.

\subsection{Input file {\YAMLkeystyle Scanner} section}

The \yaml{Scanner} section defines the scanners, objectives and their options.  It has the general form:
\begin{lstyaml}
Scanner:
  use_objectives: [@\metavar{objective1}@, @\metavar{objective2}@, ...]
  use_scanner: @\metavar{chosen\_scanner}@

  scanners:
    @\metavar{scanner1}@:
      plugin: @\metavar{plugin1}@
      @\metavar{options}@

    @\metavar{scanner2}@:
      plugin: @\metavar{plugin2}@
      @\metavar{options}@

    ...

  objectives:
    @\metavar{objective1}@:
      purpose: @\metavar{purpose1}@
      plugin: @\metavar{plugin3}@
      @\metavar{options}@

    @\metavar{objective2}@:
      purpose: @\metavar{purpose2}@
      plugin: @\metavar{plugin4}@
      @\metavar{options}@

    ...

\end{lstyaml}
All scanners that a user wishes to make available for a given scan must be listed in the \yaml{scanners} node, and all objectives in the \yaml{objectives} node.  Each scanner or objective must be given a local name (\metavar{scanner1}, \metavar{scanner2}, \metavar{objective1}, etc), and a plugin and any relevant options must be associated with that name.  Objectives also need to be assigned a \yaml{purpose}, which tells \scannerbit and its scanner plugins how the objective plugin should be used.   Exactly one of the scanners under the \yaml{scanner} node can be chosen as the sampling algorithm for the scan, by setting \yaml{use_scanner} to the name of the block that defines the preferred scanner.  Arbitrarily many objectives can be activated with the \yaml{use_objectives} directive.

\subsection{\scannerbit standalone executable}

Like other \GB modules, \scannerbit can be compiled into a standalone executable, and used independently of \GB.  This can be useful for sampling external objective functions that do not come from \GB.  The build command is simply
\begin{lstterm}
make ScannerBit_standalone
\end{lstterm}
which creates the executable \textsf{ScannerBit\_standalone} and places it in the main \GB directory.

The interface of the \scannerbit\textsf{\_standalone} is similar to that of \GB itself.  Launching \term{ScannerBit_standalone -f} \metavar{yaml\_file} runs a scan defined in the file \metavar{yaml\_file}.  To replace rather than resume from any existing files when beginning a scan, use the \term{-r} option.

\textsf{ScannerBit\_standalone} also provides a diagnostic list of recognised scanners and objective plugins \`a la \GB, using the commands \term{ScannerBit_standalone} \term{scanners} and \term{ScannerBit_standalone objectives} (or simply \term{ScannerBit_standalone plugins} to see both together).  These commands list the name, version, and status of all the plugins that \scannerbit is aware of.

The standalone can also provide diagnostic information on a specific plugin, using the command \term{ScannerBit_standalone} \metavar{plugin\_name}.  Individual plugin diagnostics contain three sections.  The \textit{General Plugin Information} section displays the name, type, version, and status of the plugin.  The status \texttt{\color{darkgreen} ok} indicates that a plugin is properly linked.  The status \texttt{\color{red} reqd lib(s) not found} indicates that a library requested by the \cpp{reqd\_libraries} macro cannot be found.  A status of \texttt{\color{red} invalid lib path(s) in locations file} indicates that a library specified in \term{config/scanner_locations.yaml} or \term{config/objective_locations.yaml} (or their default equivalents; see Sec. \ref{plugin_setup}) cannot be found at the specified location. Similarly, \texttt{\color{red} reqd header file(s) not found} occurs when a header listed under \cpp{reqd\_headers} cannot be located, and \texttt{\color{red} invalid include dir(s) in locations file} indicates that an include folder that was specified in the \term{scanner_locations.yaml} or \term{objective_locations.yaml} files cannot be located. Finally, a status of \texttt{\color{red} excluded} indicates that the plugin was \term{-Ditch}ed from the configuration of the code when \textsf{CMake} was invoked.  The \textit{Header \& Link Info} section contains include and link paths of headers and libraries requested by the plugin, information about which of them were found, and a list of all input file options that the plugin requires to be defined in order to run.  Finally, the \textit{Description} section contains a short description of the plugin.  This typically includes recognised input file options and a description of the algorithm or function that the plugin provides.

\section{Simple scanners}
\label{sec:simple}

\scannerbit includes four simple scanners, all found in \term{ScannerBit/src/scanners/simple/}: a \textsf{random} sampler, a \textsf{grid} sampler, a list-based \textsf{raster} sampler, and a simple toy Metropolis MCMC \textsf{toy\_mcmc}.  These are all parallelised with \mpi, using a simple prescription that simply distributes objective calculations evenly among the available processes.  Below we give the available options for each simple scanner, and default values in square brackets (where defaults exist).

\subsection{The \textsf{random} sampler}
\label{sec:random_sampler}

The \textsf{random} sampler draws a user-defined number of random points from the specified prior.  The only available option is
\begin{description}
 \item \yaml{point\_number[10]}: The number of random samples desired.
\end{description}

\subsection{The \textsf{grid} and \textsf{square{\underscore}grid} scanners}

These scanners calculate likelihoods at points on a uniform, user-defined grid in the unit hypercube.  The \textsf{grid} scanner allows the grid resolution be specified separately for each parameter, whereas \textsf{square\_grid} is simply a shortcut for the special case where the grid has the same number of points in every dimension.  The grid resolution is set with the option
\begin{description}
\item \yaml{grid\_pts[2]}: For the \textsf{grid} scanner, a vector of integers that specifies the number of grid points in each dimension of the parameter space.  For the \textsf{square\_grid} scanner, a single integer.
\end{description}

\subsection{The \textsf{raster} scanner}

This scanner computes an objective over a user-defined list of parameter points.  The available options are:
\begin{description}
 \item\yaml{like}: The purpose to use as the objective.
 \item\yaml{parameters}: The parameters specified by the user.
\end{description}
The \yaml{parameters} option should contain a list of parameters, with a number or sequence that specifies the user-defined values, e.g.
\begin{lstyaml}
  raster_example:
    plugin: raster
    like: LogLike
    parameters:
      "model::param_1": [0@\yamlvalue{, 1}@]
      "model::param_2": 0.5
      "model::param_3": [2@\yamlvalue{, 3, 4}@]
\end{lstyaml}

To obtain sensible results, the \textsf{none} prior should be employed for any parameters where values are given via the \yaml{parameters} option. Any parameters not specified are chosen randomly, and transformed by the chosen priors.  Parameters can be specified with a single number to apply to all points in the list, or as a vector of values.  Different parameters can be assigned lists of different lengths, which simply repeat once they are exhausted.  In the example above, \scannerbit will run the points $(0, 0.5, 2) \rightarrow (1, 0.5, 3) \rightarrow (0, 0.5, 4)$, and then terminate.

\subsection{The \textsf{toy{\underscore}mcmc} scanner}

This the simplest possible implementation of the Metropolis algorithm \cite{Metropolis}, with the proposal distribution set to the prior.  Given a randomly drawn initial point $x_i$, a candidate point $x_i'$ is randomly selected from the unit hypercube.  The candidate is then accepted with probability
\begin{equation}
 \alpha = \min[1, \mathcal{L}(x_i')/\mathcal{L}(x_i)].
\end{equation}
If a point is accepted, it becomes the comparison point in the next iteration.  If it is rejected, the previous point is retained.  The scanner keeps track of the number of times a given point is retained, and the resulting point multiplicities can then be used as weights in subsequent analysis, in particular for computing Bayesian posterior probability densities.  There is no convergence criterion implemented in the \textsf{toy\_mcmc}; the scanner simply runs for a fixed number of points given by the user:
\begin{description}
 \item\yaml{ point\_number[1000]}: The number of distinct (accepted) points to be computed in the chain.
\end{description}

\section{The \textsf{postprocessor}}
\label{sec:postprocessor}

This plugin reads a series of samples computed in some previous scan, and computes additional likelihoods or observables for them. Log-likelihoods for the original samples may be added to or subtracted from a newly-computed contribution, allowing existing likelihood constraints to be replaced or new ones added to previously-completed scans.  Like the simple scanners, the \textsf{postprocessor} uses \mpi to divide its objective calculations evenly between available processes.

The \textsf{postprocessor} operates as a scanner plugin.  From the perspective of \scannerbit and \GB, it {\em is} a scanning algorithm.  However, it does not generate sample points for itself, but instead obtains them directly from previous scan output. When running from \GB, this means that the \cross{likelihood container} then operates using the parameter values from the previous scan as input, and the output likelihood and observables are added to the existing data from the previous scan. A new set of output files is created, just as they are when running a `true' scan. All data from the original output that does not conflict with new output is copied to the new output files, leaving the original files unchanged.

In most respects, the \textsf{postprocessor} operates as a standard \GB scanner: it can be run via the standard \GB interface, it can be run in parallel via \mpi, it can be stopped and resumed, and all \cross{printer} output from the likelihood container is treated the same as it would be during a `normal' scan.  The options and particulars of the \textsf{postprocessor} are given in Appendix\ \ref{app:options:postprocessor}.

\section{Markov Chain Monte Carlo}
\label{sec:mcmc}

In Bayesian parameter estimation and model comparison, calculating evidence values or one-dimensional posterior PDFs for individual parameters or observables requires the ability to integrate the full multi-dimensional posterior density.  An efficient sampling method for the posterior PDF is therefore mandatory. Of the methods proposed for this task, Markov Chain Monte Carlo (MCMC) algorithms are amongst the most tried and tested \cite{2003it...book....M, 1993Neal}.

In general, MCMC methods allow one to study any target distribution of a vector of parameters $\boldsymbol{\theta}$, by generating a sequence of $n$ parameter combinations (a `chain') $\{\boldsymbol{\theta}_i\}_{i =1, \ldots, n} = \{\boldsymbol{\theta}_1, \boldsymbol{\theta}_2, \ldots, \boldsymbol{\theta}_n\}$. The chain constitutes a Markov process, because each $\boldsymbol{\theta}_{i+1}$ is drawn from a proposal distribution that is fully determined by the previous point $\boldsymbol{\theta}_i$. MCMC algorithms are designed to ensure that the time spent by the Markov chain in a region of the parameter space is proportional to the target posterior PDF value in this region. Hence, from such a chain, one can obtain a series of independent samples from the posterior PDF.  Up to a common normalisation constant (the evidence), both the target posterior PDF and any marginalised versions of it can be estimated by simply counting the number of samples within the relevant region of parameter space.

\subsection{The \great software}

The Grenoble Analysis Toolkit (\great) \cite{great} is a modular, user-friendly, object-oriented \Cpp MCMC framework for sampling user-defined parameter spaces. It uses the Metropolis-Hastings algorithm \citep{Metropolis,Hastings, 1993Neal, 2003it...book....M} to generate Markov chains. This prescription ensures that the stationary distribution of the chain asymptotically tends to the target distribution (typically the posterior PDF), by generating a candidate state $\boldsymbol{\theta}_\mathrm{trial}$ picked at random from a proposal distribution $q(\boldsymbol{\theta}_\mathrm{trial} | \boldsymbol{\theta}_{i})$ and accepting the candidate with probability $a$,
\begin{equation}
	a (\boldsymbol{\theta}_\mathrm{trial} | \boldsymbol{\theta}_{i}) = \text{min} \left( 1,  \frac{p (\boldsymbol{\theta}_\mathrm{trial}) }{p (\boldsymbol{\theta}_i)} \frac{q(\boldsymbol{\theta}_i | \boldsymbol{\theta}_\mathrm{trial})}{q(\boldsymbol{\theta}_\mathrm{trial} | \boldsymbol{\theta}_{i})} \right).
\end{equation}
Here, the target distribution $p (\boldsymbol{\theta})$ can be reduced to the likelihood function $\mathcal{L}$ assuming a flat prior for $\boldsymbol{\theta}$. If the trial is accepted, it becomes the new state, whereas if it is rejected, the current state is retained. This criterion ensures that once at its equilibrium, the chain samples the target distribution $p (\boldsymbol{\theta})$.

To optimise the efficiency of an MCMC, the proposal distribution should be as close as possible to the true distribution. The MCMC implemented in \great uses a multivariate Gaussian distribution, accounting for possible correlations between the parameters of the model.  \great runs multiple MCMC chains, either sequentially or in parallel depending on the user's \mpi configuration.  At the termination of each chain, based on the samples contained in all chains completed so far, i.e.\ minus a `burn-in' period at the beginning of each chain and after the removal of correlated samples by thinning the chains, \great updates the covariance matrix to be used to define the proposal distribution in subsequent chains.  The updated covariance matrix is saved externally, in order to allow chains running in parallel to always use the latest version.

To obtain a reliable estimate of the target distribution, \great bases its analysis of a chain on a selected subset of its points. Burn-in points are discarded, to avoid the random starting point of the chain biasing the sampling. By construction, each step of the chain is correlated with the previous steps: \great obtains sets of independent samples by thinning the chain over its autocorrelation length $l$.  The single-parameter autocorrelation on length scale $k$, in a chain of total length $N$ and for parameter $\theta$, is
\begin{equation}
\label{autocorr}
r(k) = \frac{\sum_{i=1}^{N-k} (\theta_i - \bar\theta)(\theta_{i+k} - \bar\theta)}{\sum_{i=1}^{N} (\theta_i - \bar\theta)^2}.
\end{equation}
\great defines the correlation length $l_j$ for the $j$th parameter to be the smallest inter-sample interval such that $r(l_j) \le 0.5$; samples separated by scales larger than this are considered independent. The overall correlation length $l$ for the chain is defined as the maximum correlation length across all $m$ parameters, i.e.\ $l \equiv \max_{j=1..m} l_j$.

The fraction of independent samples measuring the efficiency of the MCMC is defined to be the fraction of samples remaining after discarding the burn-in steps and thinning the chain. The final results of the MCMC analysis are the target distribution and all marginalised distributions, obtained by counting the number of samples within the relevant region of parameter space.

\subsection{\great--\scannerbit interface}

As implemented in \great, the Metropolis-Hastings algorithm has no default convergence criterion. The user is required to specify a chain length, i.e.~a number of steps, for each Markov chain. These options are given in Appendix\ \ref{app:options:great}

\great also extracts the relevant trials for further analysis. It first calculates the burn-in length $b$ corresponding to the first sample $\boldsymbol{\theta}_b$ for which $p (\boldsymbol{\theta}_b) > p_{1/2}$, where $p_{1/2}$ is the median of the target distribution obtained from the entire chain (i.e.\ the median posterior density, at least in standard applications). To obtain uncorrelated samples within each chain, it then computes the autocorrelation function for each parameter (Eq.\ \ref{autocorr}).  If the chain does not have any samples for which $p(\boldsymbol{\theta}_b) > p_{1/2}$, then a burn-in length cannot be defined.  This can happen if e.g.\ every sample has the same likelihood, as can occur if every sample in the chain is deemed invalid by \scannerbit, and assigned the default minimum log-likelihood (set by the \YAML entry \yaml{KeyValues::} \yaml{likelihood::}\yaml{model_invalid_for_lnlike_below}).

\great performs these operations after computing each chain, before using the results to update the covariance matrix. If the burn-in length of the last chain is undefined, a warning message is printed, and a new chain is started using the old covariance matrix.  Chains for which the burn-in length is undefined are not retained for any further analysis, and are considered invalid. At the end of the run, the complete statistics for all valid chains (burn-in length, correlation length, number of independent samples) are printed out in \great's native format. The independent samples and their multiplicities are stored in whatever output format the user has instructed \gambit to use for printing results.

\section{Ensemble MCMC}
\label{sec:ensemble}

Standard MCMC algorithms are traditionally somewhat problematic in large or highly multi-modal parameter spaces, as their efficient operation requires a well-tuned proposal density.  Some modern MCMC samplers (such as \great) address this by adaptively varying the proposal distribution based on samples from previous runs.  Other successful strategies use multiple concurrent MCMC chains as the basis of the proposal distribution.  These are commonly referred to as ensemble samplers.

In an ensemble MCMC, each chain is individually advanced by constructing a proposal PDF from the set of all current points across the full set of concurrent chains.  Procedurally, this equates to exploring an augmented parameter space consisting of $n$ copies of the original space, $\{\vec{\theta}_{(0)}, \vec{\theta}_{(1)}, \dots, \vec{\theta}_{(n)}\}$ corresponding to a composite posterior distribution $\mathcal{P}(\vec{\theta}_{(0)}, \vec{\theta}_{(1)}, \dots, \vec{\theta}_{(n)}) = \prod_{i=0}^n\mathcal{P}(\vec{\theta}_{(i)})$ where $\mathcal{P}(\vec{\theta})$ is the actual target distribution of interest.  These algorithms are able to easily adapt their proposal densities to the target distribution, and exhibit performance that is generally invariant under affine transforms (e.g.\ $\vec{\theta} \rightarrow \phi \vec{\theta}$).  Unfortunately, the performance of these algorithms is highly sensitive to the number of concurrent chains, with the number of chains required typically scaling linearly with the parameter dimension; this makes the overall number of likelihood evaluations needed for convergence proportional to the square of the parameter dimension.

\subsection{\twalk}

In the serial version of the \twalk algorithm \cite{ChristenFox10}, chains are advanced one at a time, with the proposal density based on the current parameter points of all chains not chosen for advancement, and the chain to be advanced chosen randomly at each iteration.  In the parallel version, each \mpi process randomly selects a chain for advancement at each iteration, and the proposal distribution used for advancing all chains is based only on the state of the remaining chains not chosen for advancement by any process in that iteration.  In what follows, we refer to chains that are being advanced in a given iteration as the advancing chains, and the others (those contributing to the proposal distribution) as the proposal chains.

\twalk uses one of four movement strategies when advancing a chain, choosing randomly between them at each iteration.  Two of these strategies, the \textsf{walk} and \textsf{traverse} moves, shift the current chain position ($\vec{\theta}_i$) by some multiple of the distance between it and the current point in a randomly-selected proposal chain.  The remaining two moves, \textsf{hop} and \textsf{blow}, cause advancing chains to perform different random Gaussian jumps, with covariance matrices calculated from the full set of current points in the proposal chains.

\begin{description}
 \vspace{2mm}
 \item \textsf{Walk}: advances the current chain $\vec{\theta_i}$ by jumping either towards or away from a randomly selected proposal chain $\vec{\theta}_j, i \neq j$.  This move produces a candidate point $\vec{\theta}_i'$,
 \begin{equation}\label{eq:walk}
 \vec{\theta}_i' =
 \begin{array}{lcr}
   \vec{\theta}_i + (1-\alpha)(\vec{\theta}_j - \vec{\theta}_i),
 \end{array}
 \end{equation}
 where $\alpha$ is a parameter drawn from a distribution $\mathcal{G}(\alpha)$.  For distributions satisfying
 \begin{equation}\label{eq:twalkalphag}
  \mathcal{G}\left(\frac{1}{\alpha}\right) = \alpha \mathcal{G}(\alpha),
 \end{equation}
 detailed balance is satisfied if the candidate point is accepted with probability
 \begin{equation}
  p = \min\left[{1, \alpha^{n-1}\frac{\mathcal{P}(\vec{\theta}_i')}{\mathcal{P}(\vec{\theta}_i)}}\right].
 \end{equation}
 Here $n$ is the dimension in which the \twalk moves are being performed (the so-called `projection dimension', described later in this subsection).  \scannerbit's implementation of \twalk uses the distribution
 \begin{equation}
  \mathcal{G}(\alpha) = \frac{\sqrt{a_w}}{2(a_w - 1)} \times \left\{
  \begin{array}{ccc}
   \frac{1}{\sqrt{\alpha}}, & \textrm{for} & \frac{1}{a_w} \leq \alpha \leq a_w \\
   0 & & \textrm{otherwise},
  \end{array}\right.
 \end{equation}
 where $a_w$ is a user-configurable input parameter of the algorithm.\vspace{2mm}

 \item \textsf{Traverse}: similar to \textsf{walk}, but the chain is advanced by jumping \textit{over} the point in the proposal chain.  The candidate point is
 \begin{equation}\label{eq:traverse}
 \vec{\theta}_i' =
 \begin{array}{lcr}
   \vec{\theta}_i + (1+\beta)(\vec{\theta}_j - \vec{\theta}_i),
 \end{array}
 \end{equation}
 where $\beta$ can take any positive value.  Detailed balance is satisfied if $\beta$ follows a distribution $\mathcal{H}(\beta)$ that satisfies
 \begin{equation}\label{eq:twalkalphah}
  \mathcal{H}\left(\frac{1}{\beta}\right) = \mathcal{H}(\beta),
 \end{equation}
 and the Metropolis-Hastings acceptance probability is modified as
 \begin{equation}
  p = \min\left[{1, \beta^{n-2}\frac{\mathcal{P}(\vec{\theta}_i')}{\mathcal{P}(\vec{\theta}_i)}}\right],
 \end{equation}
 where $n$ is again the projection dimension.  \scannerbit's implementation of \twalk uses
 \begin{equation}
  \mathcal{H}(\beta) = \frac{a^2_t - 1}{2 a_t} \times \left\{
  \begin{array}{lcc}
   \beta^{a_t} & \textrm{for} & 0 < \beta \leq 1\\
   \beta^{-a_t} & \textrm{for} & \beta > 1,
  \end{array}\right.
 \end{equation}
 where $a_t$ is another parameter of the algorithm, configurable by the user.\vspace{2mm}

 \item \textsf{Hop} and \textsf{blow}: In general, the \textsf{walk} and \textsf{traverse} moves available to the advancing chains only form a basis for some smaller-dimensional subspace of the full parameter space. With only these moves available, if the current chain positions are co-planar or are sufficiently clustered, mixing between chains can be low, and infinite loops of identical repeated and reversed jumps can occur.  For this reason, traditional MCMC jumps are mixed into the proposal distribution.  These moves use the total set of current points in the proposal chains to infer a covariance matrix $C$.  The \textsf{hop} and \textsf{blow} moves use $C$ to construct a Gaussian proposal function and perform an MCMC jump based on the resulting conditional PDF; \textsf{hop} centers the proposal on the current point of the chain being advanced, whereas \textsf{blow} centers it on the current point of one of the proposal chains.

 \qquad\scannerbit's implementations of \textsf{hop} and \textsf{blow} advance a chain some distance $r$ in a chosen direction $\hat{\vec{r}}$ from the center of the proposal distribution.  Following \cite{CosmoMC++}, $r$ is drawn from the distribution
 \begin{equation}\label{eq:hopblow}
  \mathcal{P}(r) = \frac23\mathcal{P}_2(r/d) + \frac13 e^{-r},
 \end{equation}
 where $\mathcal{P}_n(x)$ is the distribution of radii arising from an $n$-dimensional normal distribution centred at the origin, and $d$ is the user-configurable Gaussian jump parameter.  The distance $r$ is related to the hypercube parameters $\vec{\theta}$ via a Cholesky decomposition $C = \vec{L}\vec{L}^T$,
 \begin{equation}
  \vec{\theta}'_{i} = \vec{\theta}_k + r \vec{L} \cdot \hat{\vec{r}}.
 \end{equation}
 The starting point $\vec{\theta}_k$ of the jump for the \textsf{hop} move is the current point of the chain to be advanced, whereas the starting point of the \textsf{blow} move is the current point of any other advancing or proposal chains.

 \qquad In order to promote exploration of the parameter space in scenarios where the best-fit regions are highly degenerate in the parameters, \twalk chooses the direction of propagation $\hat{\vec{r}}$ by first choosing a random orthonormal basis for the parameter space.  It then chooses $\hat{\vec{r}}$ in successive \textsf{hop} and \textsf{blow} moves by cycling through the basis vectors in random order.  Once it has used all basis vectors once, it generates a new random orthonormal basis.

 \qquad\twalk calculates $C$ directly from the current points of the proposal chains,
 \begin{equation}
  C = \sum_j \left(\vec{\theta}_{(j)} - \bar{\vec{\theta}}\right)\left(\vec{\theta}_{(j)} - \bar{\vec{\theta}}\right)^T,
 \end{equation}
 where $j$ indexes the proposal chains, and $\bar{\vec{\theta}}$ gives the mean current point across them.  If this matrix is not positive-definite, then \twalk approximates it as
 \begin{equation}
  C_{l, l} = \left(\max_{j,k}\left[\theta_{l(j)} - \theta_{l(k)}\right]\right)^2/12
 \end{equation}
 where $j$ and $k$ run over all proposal chains.

\end{description}

\scannerbit's implementation performs each \textsf{walk} and \textsf{traverse} step within a randomly chosen subspace of lower dimensionality, known as the projection subspace. This encourages chain movement by avoiding a narrow distribution, which is endemic to higher-dimensional proposal distributions.  The relative probabilities of \textsf{walk} and \textsf{traverse} moves are set equal, as are those of \textsf{hop} and \textsf{blow}.  The ratio of \textsf{walk}+\textsf{traverse} to \textsf{hop}+\textsf{blow} moves, and the dimension of the projection subspace, are user-configurable.

The version of the \twalk algorithm described above, and implemented in \scannerbit, differs slightly from the original algorithm \cite{ChristenFox10} in two ways.  The first is the use of the full concurrent covariance matrix for the Gaussian jumps in the \textsf{hop} and \textsf{blow} moves, making them similar to the ``walk'' move of Ref.\ \cite{GoodmanWeare10}. Second, the algorithm is formulated to work with any number of chains greater than one, rather than just a pair (making the \textsf{walk} and \textsf{traverse} moves described here similar to the ``stretch'' move in Ref.\ \cite{GoodmanWeare10}).

The version of \twalk in \scannerbit uses the Gelman-Rubin convergence diagnostic $\sqrt{R}$ \cite{GelmanRubin} to determine convergence. This statistic compares the inter-chain dispersion to the total dispersion of each parameter.

See Appendix\ \ref{app:options:twalk} for the available options and outputs of \twalk.

\section{Nested sampling}
\label{sec:nested sampling}

Nesting sampling is a method designed for efficient calculation of the Bayesian evidence.  As a byproduct, it also produces samples from the posterior.  The algorithm samples the posterior in nested shells of probability, by continually updating a set of ``live'' points, replacing the lowest-likelihood live point in each iteration with a better point.  As the algorithm progresses, the set of live points naturally splits into clusters that shrink around the peaks of the posterior, making the algorithm well-suited to efficiently sampling multimodal distributions. \MultiNest \cite{MultiNest} is a \Fortran library that implements the nested sampling algorithm, with the addition of a clustering algorithm to estimate bounding ellipsoids for the the live points.  These bounding ellipsoids are used to approximate the iso-likelihood contours of the function being explored, allowing the algorithm to efficiently propose new live points when scanning parameter spaces of low to moderate dimension. For large dimensionalities the \MultiNest algorithm is computationally expensive, as the bounding ellipsoids typically encompass large swathes of uninteresting parameter space -- but for small and moderate-size parameter spaces it usually offers quite competitive efficiency.  The \scannerbit plugin runs the \MultiNest sampler developed by Feroz et. al. \cite{MultiNest}.  Its options and outputs are listed in Appendix\ \ref{app:options:multinest}.

\section{Differential evolution}
\label{sec:de}

Differential evolution \cite{StornPrice95, Price05wholebook, DasSuganthan11, Price13} (DE) is an efficient algorithm for global optimisation, with similarities to both genetic algorithms and the Nelder-Mead simplex method \cite{DasSuganthan11}.  It has been found to be quite robust, and is often the algorithm of choice for for multimodal, high-dimensional problems.

DE works by evolving a population of points in parameter space, with successive generations chosen by a form of vector addition between members of the current population.  The vector addition step gives the algorithm the character of a random walk with a step size provided by the population.  This makes it highly adaptive, and helps to limit the number and tuning of control parameters required.  In its simplest form, DE requires only three controlling parameters; this can be reduced even further in variants that allow self-adaptation of parameters.  It is straightforward and efficient to parallelise, as each member of the population can be simultaneously and independently evaluated against a replacement candidate.

DE's population-based mutation also leads to \emph{contour matching} \cite{Price05chp2}, where members of a population will tend to be at similar likelihood values, with the worst individuals improving the fastest, allowing the algorithm to trace out contours of the objective function rather effectively.  This not only allows good mapping of likelihood contours, but further aids with adaptive stepping from one generation to the next, and promotes transfer of population members between local minima, improving the overall convergence towards the global minimum.

\subsection{Algorithmic details}
\label{subsec:de_algorithm}

All variants of DE consist of three main steps: mutation, crossover, and selection.  These are controlled by three parameters: the population size $NP$, the mutation scale factor $F$, and the crossover rate $Cr$.  The simplest form of DE, known as `rand/1/bin', was first described in 1995 \cite{StornPrice95}, and continues to be widely used.  The first two parts of the name refer to the strategy for mutation, and the the third refers to the crossover; these are described in detail below.

\begin{figure}
	\centering
	\includegraphics{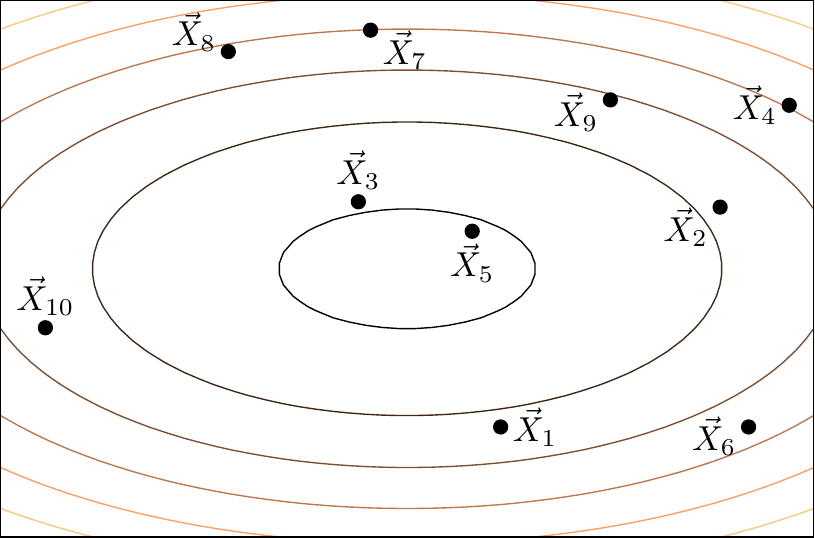}
	\caption{A simple example of differential evolution in two dimensions.  This
				figure shows the likelihood function represented by contours (with more central contours corresponding to higher likelihood values), and an initial random
				population of \protect\fortran{NP=10} vectors $\{\vec{X}_i^0\}$.  Subsequent figures illustrate the remaining
				steps of the algorithm.}
	\label{fig:initial_pop}
\end{figure}

The algorithm begins by initialising the population to a random selection of points within the allowed parameter space (Fig.~\ref{fig:initial_pop}).
We will denote the population of points (also referred to as \emph{target vectors}) as $\{\vec{X}_i^g\},$ with $i$ indexing the members
of the population, and $g$ indexing the generation.  Each subsequent generation of the population is chosen by performing mutation,
crossover and selection on the previous generation.

\begin{figure}
	\centering
	\includegraphics{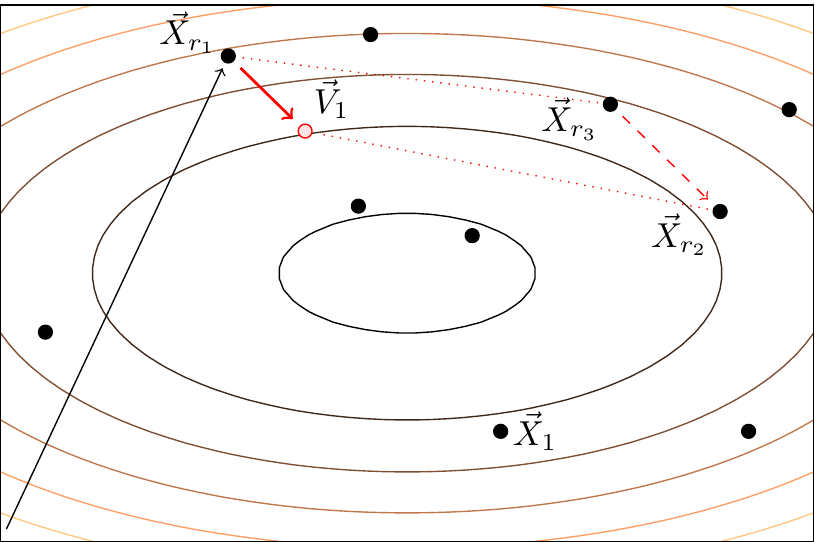}
	\caption{The process of creating the first donor vector during mutation in the simple `rand/1' variant of this step. The difference vector
		between two randomly chosen points is shown as a dashed red line, and the scaled
		difference vector (thick red line) is shown added to another randomly chosen point
		to create the donor vector $\vec{V}_1$.  Note that the current target vector $\vec{X}_1$
		is not used during rand/1 mutation.  The scale factor in this example is \protect\fortran{F=0.7}.  Ellipses are isolikelihood contours, with more central contours corresponding to higher likelihood values.}
	\label{fig:mutation}
\end{figure}

\subsubsection{Mutation}
\label{subsec:mutation}

The first step in DE is mutation, which will produce the \emph{donor vectors} $\{\vec{V}_i\}$ from
the current population of target vectors $\{\vec{X}_i^0\}$.  This step is illustrated in
Fig.~\ref{fig:mutation}.  In the rand/1 mutation scheme, a random vector is combined with a single
difference vector scaled by the mutation scale factor $F$.  To produce each donor vector
$\vec{V}_i$, three random vectors $\vec{X}_{r1}$, $\vec{X}_{r2}$ and $\vec{X}_{r3}$
are chosen from the current population, such that none of the  $\vec{X}_k$ are the same, and none matches the current target vector
$\vec{X}_i$.  The vectors are then combined using vector addition to produce the donor vector:
\begin{equation}
\vec{V}_i = \vec{X}_{r1} + F(\vec{X}_{r2} - \vec{X}_{r3}).
\label{eq:mutation}
\end{equation}
This name rand/1 refers specifically to the fact that the donor is formed by choosing a \textit{random} base vector from the population, and vector-adding it
to \emph{one} scaled difference vector between population members.  The combination of a single target vector (referred to as the \textit{base vector}) with a donor vector constructed from scaled differences between other population members is a general feature of DE.  Further variants are detailed in section~\ref{subsec:mutation_strategies}.

The usage of this vector addition strategy allows DE to explore a function dynamically, based on the size and shape of the evolving population (which reflects the size and shape of the contours of the objective function).  The value of $F$ is the main determinant of how broad this search is.  In general, $F$ is required to be less than 1 for convergence to be achievable -- but too low a value can lead to insufficient exploration, and premature convergence \cite{Price05chp2}.

\begin{figure}
	\centering
	\includegraphics{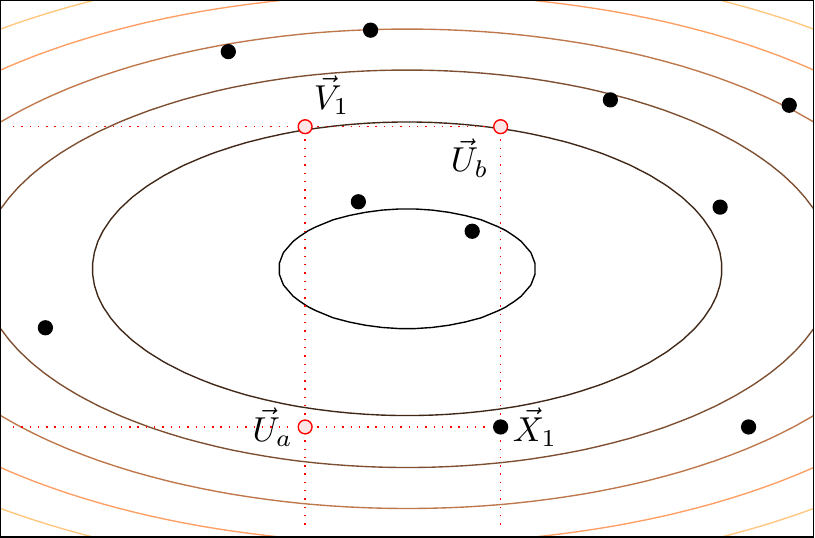}
	\caption{Binomial crossover between the donor vector $\vec{V}_1$ and the target vector
				$\vec{X}_1$ in rand/1/bin differential evolution.  This produces three possible trial vectors, shown in lightly-filled red
				circles.  Because at least one component of the donor vector always goes into the trial vector, but no
				components are guaranteed to come from the target vector, $\vec{V}_1$ is
				a possible trial vector (in the case where both components have been taken from
				the donor vector), as are $\vec{U}_a$ and $\vec{U}_b$ (where only one
				component has been chosen from the donor vector).  The target vector
				$\vec{X}_1$ itself is not a possible trial vector. Ellipses are isolikelihood contours, with more central contours corresponding to higher likelihood values.}
	\label{fig:crossover}
\end{figure}

\subsubsection{Crossover}
\label{subsec:crossover}

The second step in DE is crossover, also called recombination.  This is illustrated in Fig.~\ref{fig:crossover}.  Crossover combines
the donor vectors produced by mutation with the original population of target vectors to produce the \emph{trial vectors} $\vec{U}_i$.  The trial vectors
will potentially form the next generation of vectors.  The degree to which the trial vectors are composed of components of the donor vectors
rather than components of target vectors is influenced by the parameter $Cr$, which takes a value between $0$ and $1$.  In binomial
crossover (the `bin' of rand/1/bin DE), the trial vector is chosen according to the following procedure:
\begin{enumerate}
	\item For the $k$th component of the trial vector $\vec{U}_i$, denoted $\vec{U}_{i,k}$, a random number $r_k$ is chosen such that $0<r<1$.
	\item If $r_k \leq Cr$, the component is taken from the donor vector: $\vec{U}_{i,k} = \vec{V}_{i,k}$.
	\item If $r_k > Cr$, it is taken from the target vector instead: $\vec{U}_{i,k} = \vec{X}_{i,k}$.
	\item After all components of $\vec{U}_i$ have been chosen in this fashion, one component is reassigned in order to ensure that
		trial vectors are always different from their parent target vectors.  A dimension $l$ is chosen randomly for each member of the
		population.  The corresponding component of the donor vector is then assigned to the target vector: $\vec{U}_{i,l} = \vec{V}_{i,l}$,
		irrespective of its previous value.
\end{enumerate}

As $Cr$ increases, the probability that components are chosen from the donor vector increases: for many-dimensional problems, the
percentage of components taken from the donor vector is approximately $Cr$ (see Ref.\ \cite{Zaharie07} for a full analysis).  High values
of $Cr$ therefore lead to increased exploration, as the trial vectors will differ from the target vectors along many dimensions.  Low
values of $Cr$ are primarily effective for the special case where the likelihood function is a separable function of the parameters, because this
allows the algorithm to explore along individual dimensions \cite[e.g.][]{Price05chp2}.  In the more general case, where the objective
function is non-separable, $Cr$ should be kept high to allow better exploration. A small amount of crossover with the target vectors remains useful, however, as
it improves the diversity of the population of trial vectors \cite{Price05chp2}.

\begin{figure}
	\centering
	\includegraphics{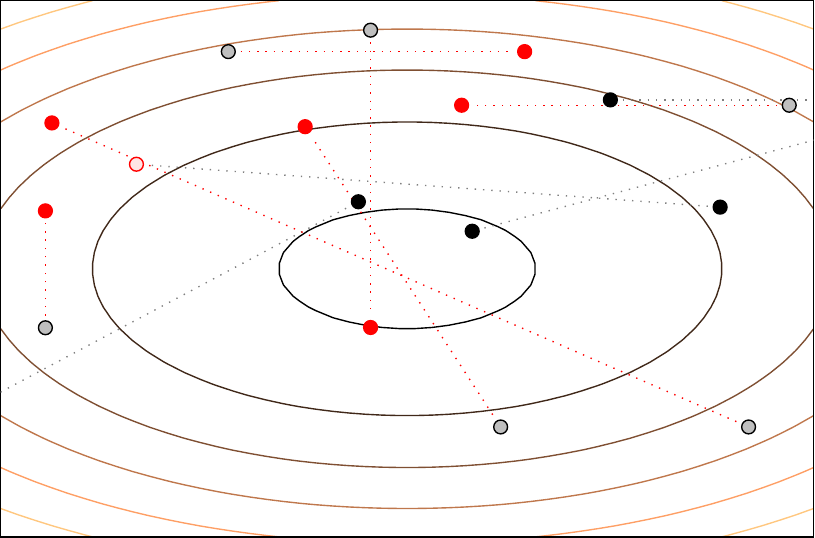}
	\caption{The last step in a generation of differential evolution.  This shows the process of
				selection after trial vectors have been chosen for the entire population.  Each
				target vector is compared with its associated trial vector, and the better one is
				retained for the next generation.  Here red indicates trial vectors and black indicates target vectors. Filled circles have been kept for the next generation, whereas open circles have been rejected.  Note that several points have trial vectors outside the allowed	boundaries; these are rejected automatically. Ellipses are isolikelihood contours, with more central contours corresponding to higher likelihood values.}
	\label{fig:selection}
\end{figure}

\subsubsection{Selection}
\label{subsec:selection}

The final step in DE is selection, which generates the next population of vectors. This step is shown in Fig.~\ref{fig:selection}.  The value of the objective function (typically the likelihood) for
each target vector $\vec{X}^g_i$ (the previous population) is compared with the trial vector  $\vec{U}_i$ constructed from it using
mutation and crossover. The point with the better likelihood is retained as a member of the next generation, and becomes one of the new target vectors $\vec{X}^{g+1}_i$.
If both have the same likelihood, the trial vector $\vec{U}_i$ is preferred, in order to allow the population to move across flat surfaces.

Selection makes DE what is known as a \textit{greedy} algorithm: it takes any improvement offered, and never accepts steps that would lead to a poorer fit.  This allows faster convergence, but unlike non-greedy sampling methods (e.g.\ MCMCs), where poorer fits are sometimes accepted, discovery of the global minimum is not guaranteed even for infinite running time.

It is possible for trial vectors to be located outside of the allowed parameter space boundary.  This is most common during the first few generations of the algorithm, when the population is spread out, allowing very large difference vectors to be produced.  However, if a local or global minimum is located near the edges of parameter space, out of bounds vectors can occur throughout the minimisation process.  The simplest way to enforce parameter boundaries is to reject any trial points that lie outside them; for alternatives see Sec.~\ref{sec:options}.

\subsubsection{Advanced mutation and crossover strategies}
\label{subsec:mutation_strategies}

Although rand/1/bin DE is simple and popular, many other variants have been
proposed.  The simplest variations involve either a different choice of base vector, or a different
method to calculate the difference vector.  The name of the DE strategy is typically written in the
form \emph{base/difference number/crossover}, where
\begin{description}
	\item[\textbf{base}:] how the base vector, $\vec{X}_{r1}$ in equation~\ref{eq:mutation} and
			Fig.~\ref{fig:mutation}, is chosen for mutation.
	\item[\textbf{difference number}:] the number of difference vectors $F\left(\vec{X}_{r2} -\vec{X}_{r3}\right)$ in
			equation~\ref{eq:mutation} and Fig.~\ref{fig:mutation} that are used in mutation.
	\item[\textbf{crossover}:] the form of crossover used.
\end{description}
Some options for the base vector beyond a random choice from the population include the current
target vector (`current'), the best vector in the population (`best'), or a
base vector made up of a combination of these (e.g. `rand-to-best').

A `general' mutation strategy encompassing several possible mutation strategies can be written as follows \cite{Zaharie08}:
\begin{equation}
	\vec{V}_i = \lambda \vec{X}_\text{best} + (1-\lambda) \vec{X}_1 + \sum_{q=0}^Q F_q(\vec{X}_{2_q} - \vec{X}_{3_q}), 
\end{equation}
where $\vec{X}_1$ is the current vector or is chosen randomly as before and $\vec{X}_{2,3}$ are chosen
randomly from the population.  No vectors may be used twice.  This form allows rand base vectors
($\vec{X}_1 = \vec{X}_\text{rand}$ and $\lambda=0$), current base vectors
($\vec{X}_1 = \vec{X}_i$ and $\lambda=0$), best base vectors ($\lambda = 1$),
rand-to-best base vectors ($\vec{X}_1 = \vec{X}_\text{rand}$ and $0<\lambda<1$), and
current-to-best base vectors ($\vec{X}_1 = \vec{X}_i$ and $0<\lambda<1$). It also allows
for the use of $Q$ difference vectors along with a corresponding set $\{F_q\}$ of scale factors.  Note that there are other forms of mutation that are not described by this equation.

Using the best individual in the population as the base vector (e.g.\ best/1/bin) speeds up convergence, as it reduces stagnation in the population -- but it makes DE less likely to find the global minimum compared to simply choosing the base randomly.  This tends to be a good choice for near-unimodal functions, but poor for highly multimodal functions \cite{Price05chp2, Mezura-Montes06}. Using the current vector as the base can slow convergence because it reduces the diversity of the resulting population \cite{Price05chp2}, but can be more efficient than randomly choosing the base because it reduces so-called `selection drift bias' \cite{Price13}.  Combining multiple difference vectors can help combat the loss of diversity induced by using either the best or current vector as the base, but may hamper contour-matching \cite{Price05chp2}.

In contrast to the proliferation of mutation strategies, binomial crossover has only one main competitor,
exponential crossover (`exp').  The lack of additional recipes is mostly a result of the lesser impact of crossover on performance than mutation \cite{Zaharie09}.  Exponential crossover was used in the original DE algorithm \cite{StornPrice95}, but is generally less popular than binomial crossover.

In exponential crossover, a length $L$ to be crossed over is chosen by drawing random numbers between 0 and 1 until one of them exceeds $Cr$.  $L$ is then set to the total number of draws required, with
the provision that it must be less than the dimensionality of the parameter space $D$.  A random dimension $d$ is then chosen from $[1, D]$, and
the next $L$ entries in the donor vector (wrapping around to the first if necessary) are chosen to contribute to the trial vector.  The remaining $D-L$ components are taken from the target vector.

Exponential crossover is generally considered to perform less well than binomial crossover.  This has been suggested \cite{Mezura-Montes06}
to be due to the requirement in exponential crossover that dimensions taken from the target vector must be adjacent, whereas in binomial
crossover all combinations are possible.  Both forms of crossover suffer from the fact that the process is not rotationally invariant, as it preferentially acts along dimensions, and therefore cannot perform identically on separable and inseparable functions, decreasing efficiency when working with parameterisations that induce correlations between parameters \cite{Price05chp2,Zaharie09}.  This is a common feature of evolutionary algorithms, including e.g. genetic algorithms.

\subsubsection{Self-adaptive differential evolution}
\label{subsec:adaptiveDE}

As with all optimisation strategies, the ideal choice of parameters for DE depends on the type of problem to be solved, and is frequently unclear \textit{a priori}.  The ability for the algorithm to adapt its parameters in real time is therefore advantageous.  One example of self-adaptive differential evolution is known as jDE \cite{Brest06}, which compares favourably with classic DE and other modifications of DE across problem types and in high-dimension parameter spaces \cite{NeriTirronen10, DasSuganthan11}.

The jDE algorithm is based on classic rand/1/bin DE but adapts the values of $F$ and $Cr$ as the run progresses.  Each vector in the population is associated with personal values of $F$ and $Cr$, which are then used to generate the next generation of vectors.  Before
mutation occurs for the $i$th member of the population, $F_i$ has a chance to change.  The same is true of $Cr_i$ immediately before crossover.  During selection, the values of $F$ and $Cr$ belonging to successful vectors are retained in the next generation of the population.  Variants on the jDE algorithm can extend the self-adaptive behaviour to other mutation or crossover strategies.  We introduce one such variant, $\lambda$jDE, which dynamically modifies $\lambda$ in a similar way over the course of the run. We describe the jDE and $\lambda$jDE algorithms, as well as our implementations and variations of them, in greater detail in section~\ref{subsec:jDE}.

\subsection{The \diver package}

In this section, we introduce \diver, an open-source differential evolution sampler intended for use in optimisation problems in physics and astronomy.  \diver can be downloaded either as a source tarball or a \textsf{git} repository from \href{http://diver.hepforge.org}{http://diver.hepforge.org}.  It is released under an academic use license.

\subsubsection{Design and invocation}

\diver is a fully-featured, standalone parallel implementation of differential evolution.  Its default mode is to perform self-adaptive $\lambda$jDE optimisation, with jDE, rand/1/bin and all mutation and crossover strategies in between available through an extensive set of runtime options.  It also includes additional options for outputting derived parameters, stopping and restarting scans, computing approximations to various Bayesian quantities, and dealing with discrete parameters.

\diver is written in \Fortran, and includes wrappers for calling it from \plainC/\Cpp.  It is compatible with \textsf{gcc 4.4} and later, and version 11 and later of the intel compiler suite.  Parallelism in \diver makes use of \mpi, and works by simply dividing each generation up evenly across all \mpi processes.   It is invoked by calling the \Fortran function \fortran{diver()} or its \plainC equivalent \cpp{cdiver()} from some user-supplier driver program.  When calling these functions, the driver program must pass the address of another, user-supplied, likelihood/objective function, which \diver then minimises.  The package includes example driver programs and objective functions in \Fortran, \plainC and \Cpp; these can be respectively found in the \term{example\_f}, \term{example\_c}, and \term{example\_cpp} subdirectories of the main \diver installation directory.

Synopses of the different source files in \diver, the various run options it offers, and the format of its outputs can be found in Appendices\ \ref{sec:sources}, \ref{sec:options} and \ref{sec:outputs}, respectively.

\subsubsection{Adaptive differential evolution: jDE and \texorpdfstring{$\lambda$}{lambda}jDE}
\label{subsec:jDE}

We include two options to use self-adaptive evolution, based on the jDE algorithm initially proposed by Brest et al.~\cite{Brest06}.
In regular jDE (accessed by setting \fortran{jDE} \fortran{=} \fortran{true}), rand/1/bin evolution is used, but each vector has unique values for $F$ and $Cr$, which evolve along with the population.

The evolution of $F$ is controlled by a value $\tau_1$, which we take to be 0.1 throughout.  The permissible range for $F$ extends from
$F_l=0.1$ to $F_u=0.9$, as values of $F$ too close to zero imply no evolution, whereas values too close to 1 prevent convergence.
We choose the initial value of $F$ for each vector randomly from a uniform distribution between $F_l$ and $F_u$.  Before mutating the the vectors, we draw a random number and compare it to $\tau_1$.  If it less than $\tau_1$, we update $F$ to a new random value between $F_l$ and $F_u$, and the new value is used for mutation.  Then, during selection, if the trial vector is accepted, the new value for $F$ is kept as well.  If the trial vector is rejected, the previous value for $F$ is kept instead.

Similarly, the evolution of $Cr$ is controlled by a value $\tau_2$, also taken to be 0.1.  Unlike $F$, $Cr$ is allowed to vary between 0 and 1 inclusive, as crossover does not exhibit any pathological behaviour in either limit.  For each member of the population, we initialise $Cr$ to a random value between 0 and 1.  For each generation, before crossover we then choose a trial value for $Cr$.  As for $F$, we draw a uniform random deviate and compare it to $\tau_2$; if it is larger than $\tau_2$, the trial value for $Cr$ remains unchanged; if it is smaller, we choose a random new value for $Cr$ and use it during crossover.  During selection, if the trial vector is kept, the new crossover parameter is kept as well; if not, the value of $Cr$ reverts to the previous value.

The justification for this process is that different values of $F$ and $Cr$ are useful for different classes of problems, but the preferred values are usually not known.  It is presumed that successful choices of  $F$ or $Cr$ are more likely to lead to successful trial vectors, and so by tying the evolution of $F$ and $Cr$ to the evolution of the vectors, desirable values of $F$ and $Cr$ will be preferentially propagated.

In addition to the standard jDE, we offer the possibility to use self-adaptive rand-to-best/1/bin evolution.  This works just as in jDE, but with the addition of an adaptive $\lambda$ mutation parameter, which evolves via a scheme that mirrors the way $Cr$ is evolved.  The addition of this parameter harnesses the benefits of jDE, while allowing for more aggressive optimisation, since information about the position of the best member of the current generation is used.  This option is accessed by setting \fortran{lambdajDE = true}.

\subsubsection{Discrete parameters and parameter-space partitioning}
\label{subsec:discrete}

\diver offers the ability to label one or more parameters as discrete rather than continuous, using the \fortran{discrete} keyword.  This may be desirable because some parameter(s) are indeed discrete at some fundamental level, or simply as a means of labelling a set of individual fits that are interrelated in some way.

The main complication when working with discrete parameters is that mutation must be a floating-point operation in DE, in order to ensure that the donor vectors are valid, to allow for enough variety in potential donor vectors, and to ensure proper convergence.  When treating a parameter as discrete in \diver, we deal with this by storing the values of the discrete parameter internally as floating-point values, so that mutation works as normal, but evaluation of the likelihood is done by rounding the parameter to the closest integer.  The output \term{.raw} file stores the underlying floating-point representation of the parameters (to allow runs to be properly resumed), whereas the desired integer values are output in a \term{.sam} file (we discuss output formats in more detail in Appendix\ \ref{sec:outputs}).

The \fortran{partitionDiscrete} option can also be used to partition the DE population evenly into the allowed values of the discrete parameters.  With this option, no vector is allowed to change its discrete value.  This mode allows simultaneous fitting of multiple objective functions, with the discrete dimension simply treated as a label for assigning subpopulations to the different problems.  One useful application of this option is to perform multi-objective optimisation where the value of each fitness function depends (preferably only weakly) on the best-fit parameters of the other subpopulations.

\subsubsection{Population diversity and duplicate individuals}
\label{subsec:duplicates}
In order for DE to converge appropriately, it is necessary to retain sufficient population diversity.  Duplicate vectors in the population lead to artificial drops in diversity.  Duplicate vectors can arise naturally in rand/ or best/ mutation if two separate vectors in the population are updated using the same combination of random vectors.  Once there are multiple identical vectors in a population, the diversity of the population will decrease, making premature convergence more likely.

Even more problematically, duplicate vectors have a tendency to infect the rest of the population: whenever a pair of duplicates is chosen to create the difference vector during mutation, the resulting donor vector will match the third vector chosen, possibly creating another duplicate.  In best/ mutation, such a process can rapidly lead to an entire population matching the `best' vector.

\diver includes a facility for weeding out duplicate vectors as soon as they arise to prevent these problems.  When \fortran{removeDuplicates = true}, the population is examined after selection.  If a set of duplicates is discovered, one is modified,
according to the following rules:
\begin{enumerate}
	\item If one vector was inherited from the previous generation, and the other is new, the new vector is reverted to its previous value.
	\item If both vectors are new, the one that improved the most is kept and the other is reverted to its previous value.
	\item The appearance of duplicate vectors in the initial population, or inheritance of multiple copies of the same vector from a previous generation,
        are strong indications of coding errors.	In these cases, a warning is printed and one vector is re-initialised to a random point in the parameter space.
\end{enumerate}

Duplicate removal is disabled by default for current/ mutation (\fortran{current = true}), jDE (\fortran{jDE = true}), and $\lambda$jDE (\fortran{lambdajDE = true}), as the presence of duplicates in the results of these algorithms would be surprising.  It is enabled by default for all other settings, i.e.\ rand/, best/, or rand-to-best/ mutation, as these forms of mutation are susceptible to duplicate creation.  If \diver is compiled with \mpi support, duplicate removal is enabled by default regardless of any other settings, and is recommended as a useful diagnostic for insuring against \mpi library issues.

\subsubsection{Approximate posterior and evidence estimates}
\label{subsec:posterior}

\diver can compute the Bayesian posterior and evidence from its samples when using a negative log-likelihood function as the objective, by using the likelihood samples to perform Monte Carlo integration of the (prior-weighted) likelihood.  These calculations can be activated by setting \fortran{doBayesian = true} and specifying a \fortran{prior} function.

Because DE does not share the property of Bayesian algorithms that the sampling distribution is proportional to the posterior, this requires a bootstrap estimate of the actual sampling distribution produced in a DE run.  This invariably leads to fairly rough estimates of Bayesian quantities, especially when the likelihood function is multimodal and/or highly non-Gaussian, but the results can be useful for some quick estimates before deploying more expensive algorithms optimised for Bayesian inference.

\diver obtains a bootstrap estimate of its sampling density by performing a binary space partitioning on the parameter space being scanned, using the actual samples obtained in a scan.  Each sample is sorted into a cell in the partitioned parameter space, with cells partitioned further as soon as their populations exceed \fortran{maxNodePop}.  The partitioning is done alternately in each direction of the parameter space, so that each cell remains rectangular in the parameters.

The resulting posterior weight for a sample $\theta$ can then be estimated as
\begin{equation}
\label{posterior}
P(\theta) \approx \frac{N_\mathrm{c}}{N_\mathrm{s}} V(\theta) \Pi(\theta) \mathcal{L}(\theta),
\end{equation}
where $N_\mathrm{c}$ is the number of cells, $N_\mathrm{s}$ the total number of samples, $V(\theta)$ is the parameter volume occupied by the cell containing the sample $\theta$, $\Pi(\theta)$ is the prior function (provided explicitly by \fortran{prior} -- note that this is \textit{not} the prior transform, but the prior itself), and $\mathcal{L}(\theta)$ is the likelihood, i.e.~$\exp(-x)$, where $x \equiv -\ln\mathcal{L}$ is the objective function being sampled.  The corresponding Monte Carlo estimate of the Bayesian evidence is then
\begin{equation}
\mathcal{Z} \approx \sum_{i=1}^{N_\mathrm{s}} P(\theta_i).
\end{equation}
Taking the estimate to be Gaussianly distributed, the $1\sigma$ uncertainty on the evidence can be approximated from its variance,
\begin{equation}
\Delta \mathcal{Z} \approx \sqrt{(\langle P^2 \rangle - \mathcal{Z}^2)/N_\mathrm{s}},
\end{equation}
where
\begin{equation}
\langle P^2 \rangle \equiv \frac{1}{N_\mathrm{s}} \sum_{i=1}^{N_\mathrm{s}} P^2(\theta_i)
\end{equation}
is the mean square posterior.

If \fortran{doBayesian = true}, \diver will continue to sample until the logarithmic uncertainty on $\mathcal{Z}$ reaches or passes below \fortran{Ztolerance}, i.e.
\begin{equation}
\ln\left(\frac{\mathcal{Z}}{\mathcal{Z} - \Delta\mathcal{Z}}\right) \le \text{\fortran{Ztolerance}}.
\end{equation}
Once this convergence criterion has been satisfied, \diver then further polishes its posterior and evidence estimates by taking the final binary spanning tree so generated during the scan, and re-calculating Eq.\ \ref{posterior} for each individual of every population. This improves the final posterior and evidence estimates because the resulting weights for all individuals get computed on the basis of the complete tree, rather than the tree as it was at the time each individual was initially created.

\subsubsection{\scannerbit interface}

Because \diver is specifically designed to minimise positive-definite fitness functions, the \diver plugin for \scannerbit uses the \textit{negative} of the composite log-likelihood function provided by \GB as its fitness function.  If desired, \scannerbit will also apply an offset to the log-likelihood passed to \diver, and have the printer remove that offset again before printing.  This can be useful in cases where the likelihood normalisation leads to positive total log-likelihoods; taken without an offset, these likelihoods would prevent the fitness passed to \diver from remaining positive definite.  The offset can be specified with the \yaml{lnlike_offset} option in the \yaml{likelihood} node of the \yaml{KeyValues} section of a run's main \YAML file.  If this option is absent, the offset will default to $10^{-4}$ times the value of \yaml{model_invalid_for_lnlike_below} (also in \yaml{KeyValues::}\yaml{likelihood}).  The full range of \diver options available from the \YAML file is given in Appendix\ \ref{app:options:diver}.

The \diver interface in \scannerbit does not yet make use of the ability of \diver to scan discrete parameters, as doing so is not yet supported by \scannerbit itself; this feature is slated for inclusion in a future revision of \GB.

\section{Scanner performance comparisons}\label{sec:comparison}

By offering the capacity to vary the scanning algorithm and its operating parameters --- whilst keeping all other aspects of a scan identical --- \scannerbit provides a unique testbed for comparing sampling algorithms.  In this section we present an exploration of the performance of the four major scanners available in \GB \textsf{1.0.0}, when applied to a physically realistic likelihood function.  The modularity of the scanner interface allows consistent comparison between both the algorithms themselves, and between different choices of algorithm parameters.

This investigation is intended to reveal the strengths and weaknesses of different sampling algorithms with respect to typical user requirements.  These requirements can be quite varied, and may include the choice of statistical approach (frequentist or Bayesian), the time taken for a scan to converge, the reliability of the results, or some combination of the three.  However, for any thorough investigation, the user should typically take advantage of the unique flexibility offered by \scannerbit to employ a range of algorithms, statistical methods, and scanner parameters in order to obtain the most complete and robust sampling possible.

For this demonstration, we work with the scalar singlet dark matter model.  This model has two parameters beyond the Standard Model (SM): the Higgs portal coupling $\lhs$, and the singlet Lagrangian mass parameter $\mu_S$.  We present the results in the effective parameter space of $\lhs$ and $\ms$, where the physical singlet mass $\ms$ is given by
\begin{align}
\ms = \sqrt{\mu_S^2+\frac{1}{2}\lhs v_0^2}
\end{align}
where $v_0 = 246$ GeV is the vacuum expectation value of the Higgs field.  The likelihood and posterior are both multimodal and highly degenerate across several orders of magnitude in the values of these parameters.

To investigate how performance scales with dimensionality, we introduce additional parameters that enter into the combined likelihood function. These parameters are well constrained by unimodal likelihood functions, but still create a significant challenge for any sampling algorithm due to the increase in the dimensionality of the parameter space.  In particular, we carry out detailed tests in two, seven and fifteen dimensions, and one scan with each sampler for dimensionalities between two and fifteen. We list the free parameters for each scan in Table \ref{table:params}.  For all test scans, we apply a logarithmic prior to the singlet parameters $\lhs$ and $\ms$, and flat priors to the additional parameters.

In the following, we only show full results from the fifteen-dimensional scans.  Increasing the dimensionality of the problem across this particular parameter space does not substantially shift the location nor shape of the final likelihood with respect to $\lhs$ and $\ms$.  As a result, the best-fit point and regions of maximum likelihood remain similar. For comparison, in Appendix \ref{app:scanner_comparisons}, we give additional detailed results in two dimensions.  The inclusion of additional parameters does significantly increase the runtime for the scanning algorithms, and degrades their ability to locate the maximum likelihood point.  Note that choosing a more complicated model, with more complicated parameters in the `higher' dimensions, would only increase the required computing time, making such an extensive comparison study infeasible.  We refer the interested reader to the companion papers on supersymmetric models \cite{CMSSM,MSSM} for applications of \diver and \multinest to higher-dimensional multimodal parameter spaces.

The dominant physical constraints on the model that we consider here come from experiments searching for dark matter via direct and indirect detection, the observed limit on the thermal relic abundance of dark matter, and constraints on the rate of invisible Higgs decays at the Large Hadron Collider.  We also apply the constraint $\lhs <10$, as larger values would violate perturbative unitarity and are therefore not physically interesting.  More details on the model can be found in accompanying and earlier papers \cite{gambit,SSDM,Cuoco:2016jqt,Beniwal,Cline13b,Cheung:2012xb,Mambrini11,Burgess01,McDonald94,SilveiraZee}.  Here our test function consists of the same likelihood components as in Ref.\ \cite{SSDM}.  Although this is a simple, well-studied extension of the SM, the parameter space is still sufficiently non-trivial that it constitutes an illustrative test of scanner performance.

\begin{table}[tp]
\centering
\caption{Parameters, ranges and central values of the test scans of this section, for each scan dimensionality.  The ranges for most SM parameters correspond to $\pm3\sigma$ variations around the 2014 PDG central values \cite{PDB}.  For the Higgs, the range is $\pm4\sigma$ about the 2014 central value (which encompasses the 2015 $4\sigma$ range \cite{PDG15}).  For the up and down quark masses, we take the central values from the 2014 review, and scan over a range of $\pm20\%$ around the central values.  This is intended to capture the $\pm3\sigma$ range implied by the likelihoods in \precisionbit \cite{SDPBit}, which deal with correlated mass-ratio measurements. The nuclear couplings also incorporate a range of $\pm3\sigma$ around the best estimates.  The dark matter density has an asymmetric range about the central value, as the likelihood that we apply to this parameter is log-normal rather than Gaussian. We refer the reader to Refs.\ \cite{SSDM,Cline13b} for further details and references on the central values and uncertainties associated with the local density and nuclear parameters.} \label{table:params}
\begin{tabular}{l@{\hspace{-4mm}}c@{}c}
\hline
Parameter & & Values \\
\hline
Scalar pole mass  & $\ms$ & $[45,  10^4]$\,GeV \\
Higgs portal coupling & $\lhs$ &$[10^{-4}, 10]$\vspace{2mm}\\
\multicolumn{2}{c}{Varied in 7 and 15-dimensional scans}\\
\hline
Electromagnetic coupling & $1/\alpha^{\overline{MS}}(m_Z)$        & $127.940(42)$       \\
Strong coupling & $\alpha_s^{\overline{MS}}(m_Z)$      & $0.1185(18)$   \\
Top pole mass  & \phantom{$^{\overline{MS}}$}$m_t$\phantom{$^{\overline{MS}}$}  &  $173.34(2.28)$\,GeV\\
Higgs pole mass & \phantom{$^{\overline{MS}}$}$m_h$\phantom{$^{\overline{MS}}$}  & $125.7(1.6)$\,GeV \\
Local dark matter density & \phantom{$^{\overline{MS}}$}$\rho_0$\phantom{$^{\overline{MS}}$} &  $0.4^{+0.4}_{-0.2}$\,GeV\,cm$^{-3}$\vspace{2mm}\\
\multicolumn{2}{c}{Varied in 15-dimensional scans}\\
\hline
Nuclear matrix el. (strange)  & \phantom{$^{\overline{MS}}$}$\sigma_s$\phantom{$^{\overline{MS}}$} & $43(24)$\,MeV \\
Nuclear matrix el. (up + down) & \phantom{$^{\overline{MS}}$}$\sigma_l$\phantom{$^{\overline{MS}}$} & $58(27)$\,MeV \\
Fermi coupling $\times$ $10^{5}$ & \phantom{$^{\overline{MS}}$}$G_{F,5}$\phantom{$^{\overline{MS}}$} & $1.1663787(18)$ \\
Down quark mass & $m_d^{\overline{MS}}(2\,\text{GeV})$  &   $4.80(96)$\,MeV  \\
Up quark mass & $m_u^{\overline{MS}}(2\,\text{GeV})$     & $2.30(46)$\,MeV \\
Strange quark mass & $m_s^{\overline{MS}}(2\,\text{GeV})$  & $95(15)$\,MeV\\
Charm quark mass & $m_c^{\overline{MS}}(m_c)$ & $1.275(75)$\,GeV \\
Bottom quark mass & $m_b^{\overline{MS}}(m_b)$    & $4.18(9)$\,GeV \\
\hline\end{tabular}
\end{table}

\begin{figure*}[tp]
  \centering
  \includegraphics[width=1\linewidth]{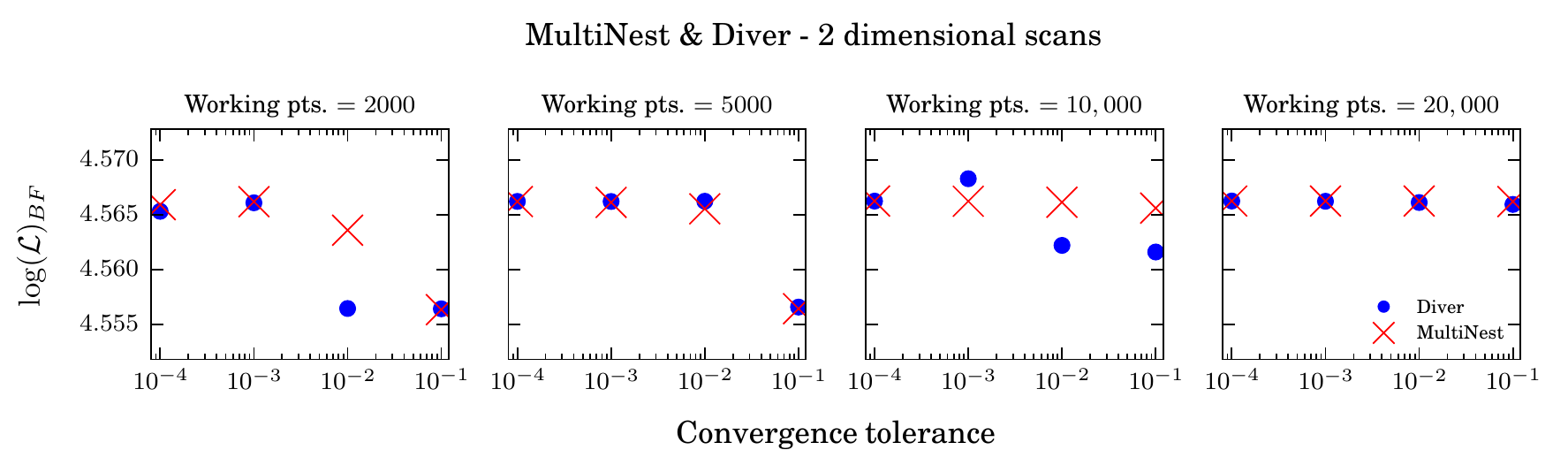}
  \includegraphics[width=1\linewidth]{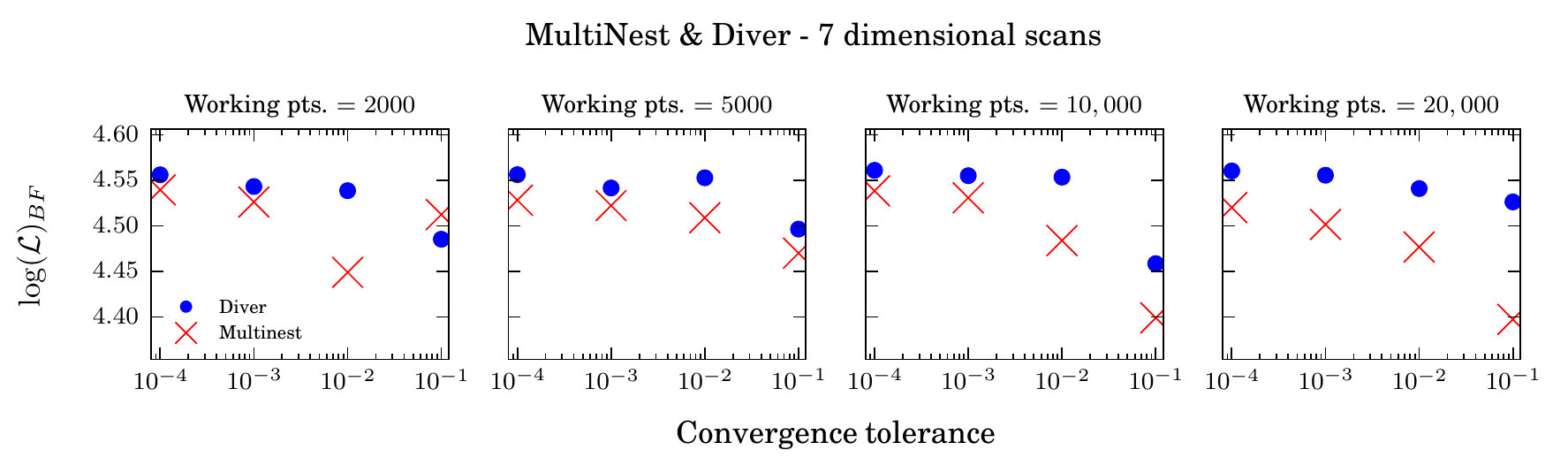}
  \includegraphics[width=1\linewidth]{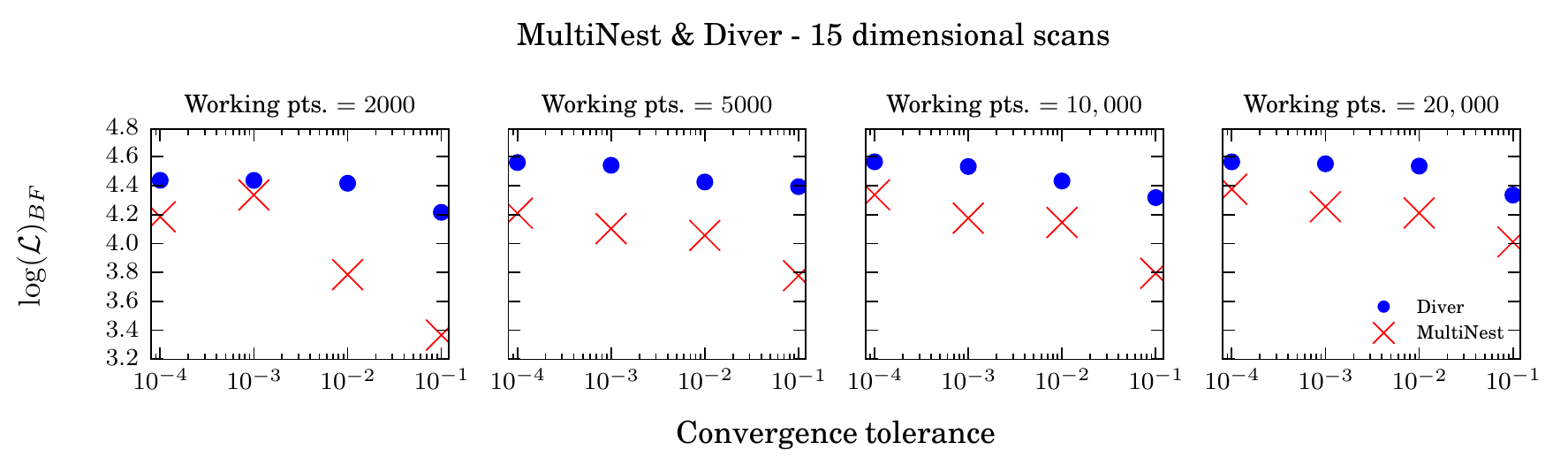}
  \caption{Best-fit log-likelihoods in scans of the scalar singlet space using the \diver and \multinest scanners, for a range of convergence tolerances and a fixed number of working points.  Tolerances correspond to the parameter \lstinline{tol} for \multinest and the parameter \protect\fortran{convthresh} for \diver.  Working points correspond to the parameter $N_\mathrm{live}$ for \multinest and the parameter \protect\fortran{NP} for \diver.  Note that the likelihood is dimensionful, leading to $\mathcal{L_{\rm BF}}>1$ \cite{gambit}.}
  \label{fig:Diver_MultiNest_1}
\end{figure*}

\begin{figure*}[tp]
  \centering
  \includegraphics[width=1\linewidth]{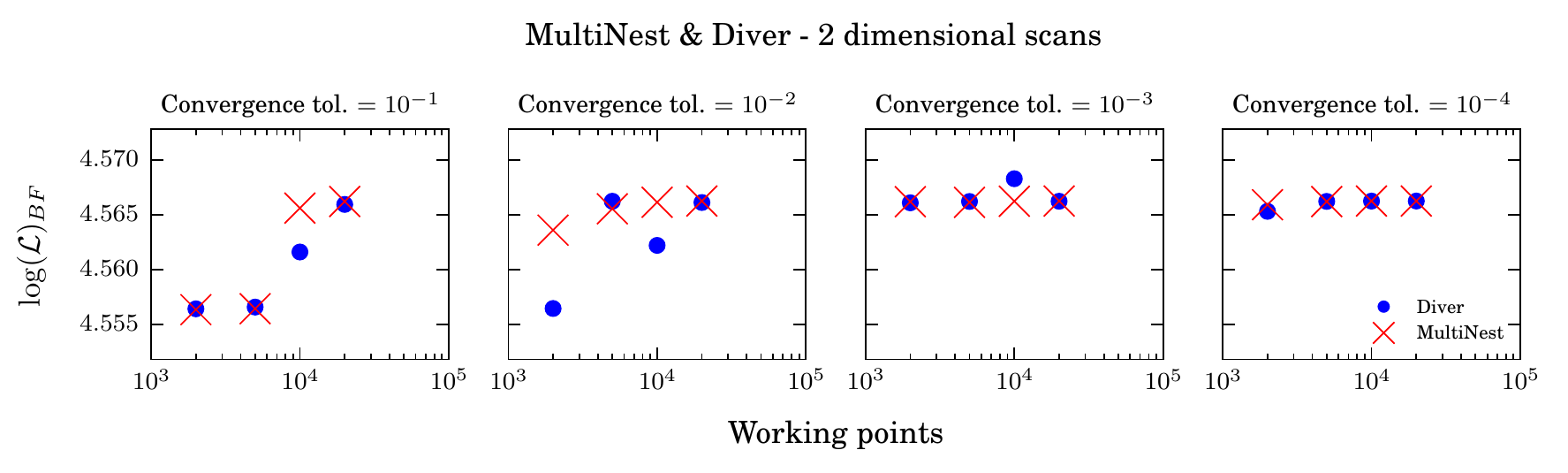}
  \includegraphics[width=1\linewidth]{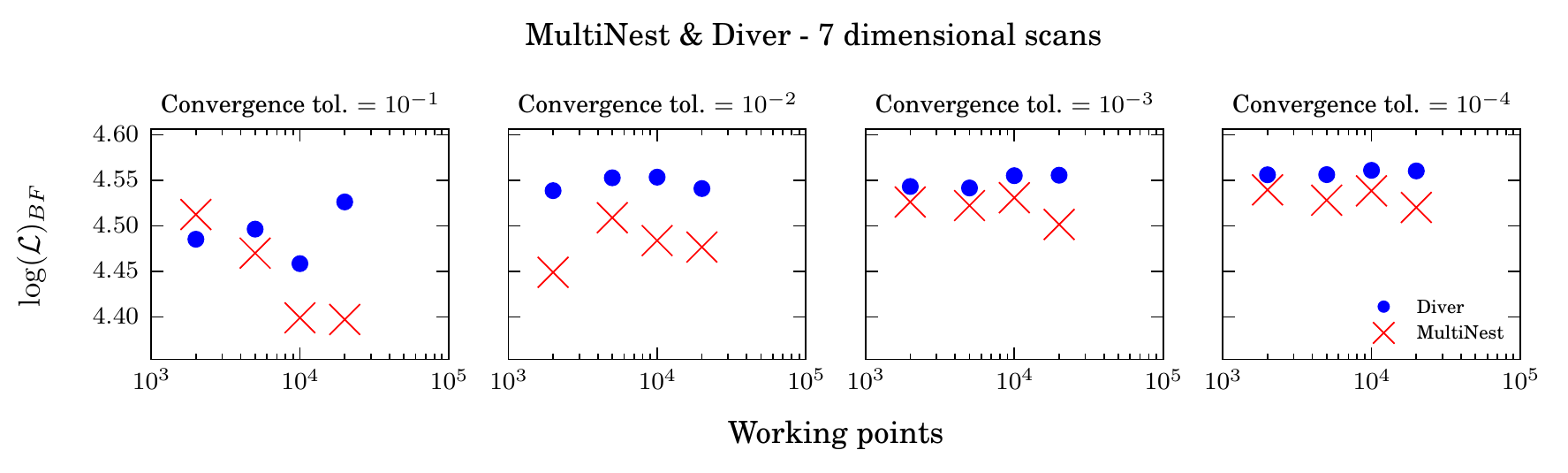}
  \includegraphics[width=1\linewidth]{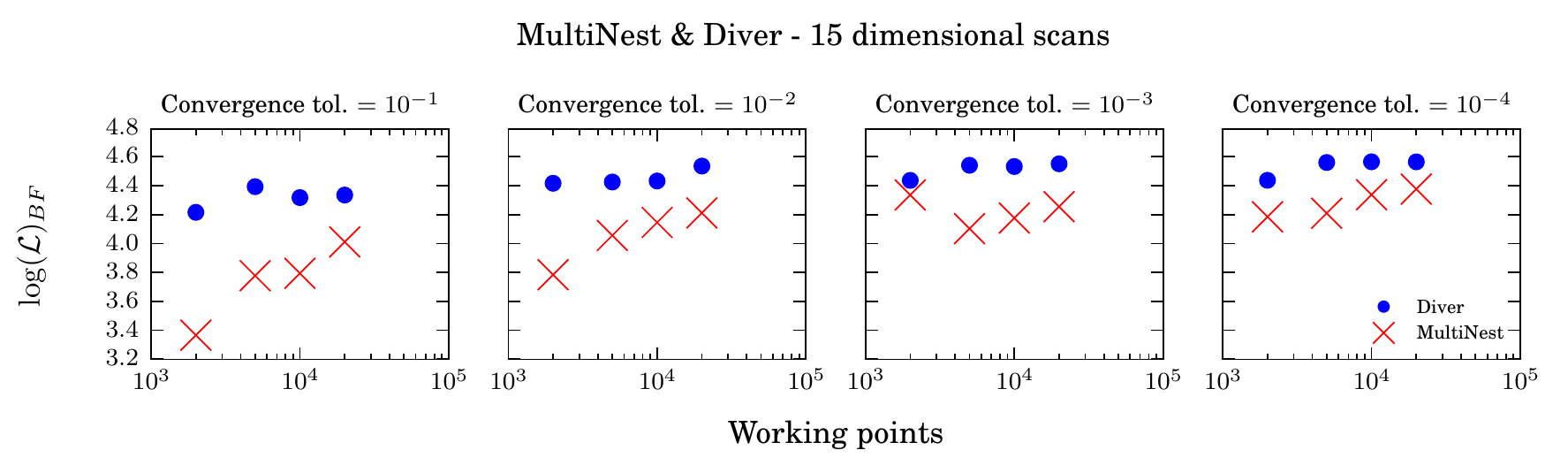}
  \caption{Best-fit log-likelihoods in scans of the scalar singlet space using the \diver and \multinest scanners, for different numbers of working points and fixed convergence tolerances.  Working points correspond to the parameter $N_\mathrm{live}$ for \multinest and the parameter \protect\fortran{NP} for \diver.  Tolerances correspond to the parameter \lstinline{tol} for \multinest and the parameter \protect\fortran{convthresh} for \diver.  Note that the likelihood is dimensionful, leading to $\mathcal{L_{\rm BF}}>1$ \cite{gambit}.}
  \label{fig:Diver_MultiNest_2}
\end{figure*}

\begin{figure*}[tp]
  \centering
  \includegraphics[height=0.234\linewidth]{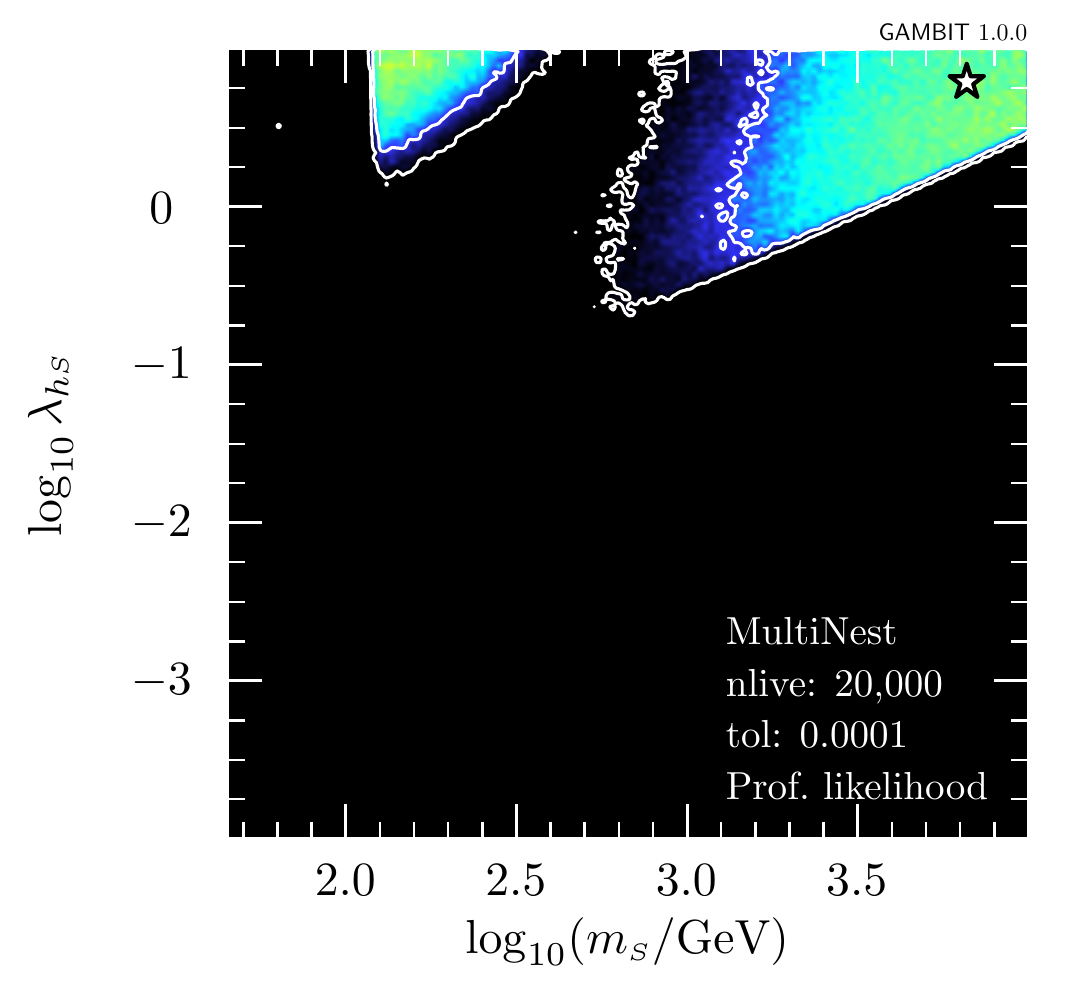}\hspace{-3mm}%
  \includegraphics[height=0.234\linewidth]{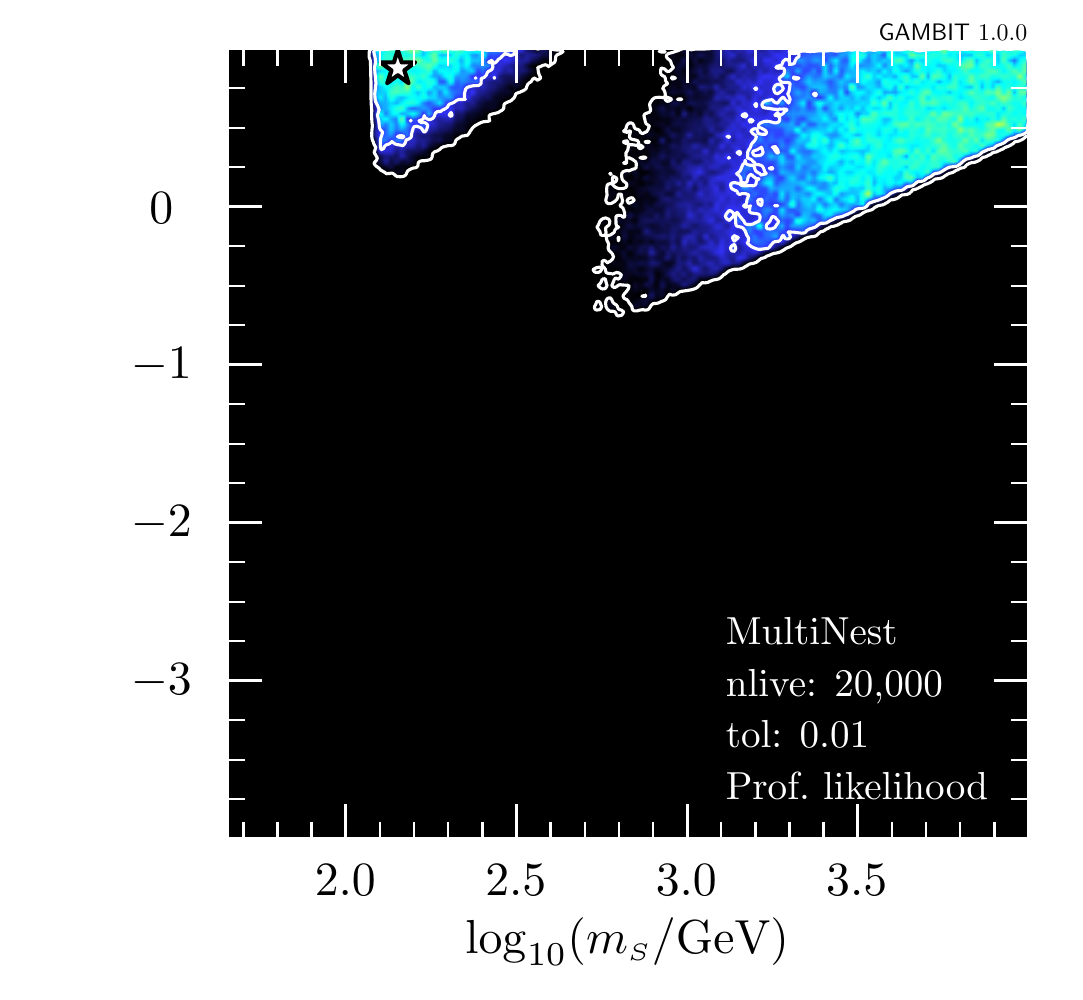}\hspace{-3mm}%
  \includegraphics[height=0.234\linewidth]{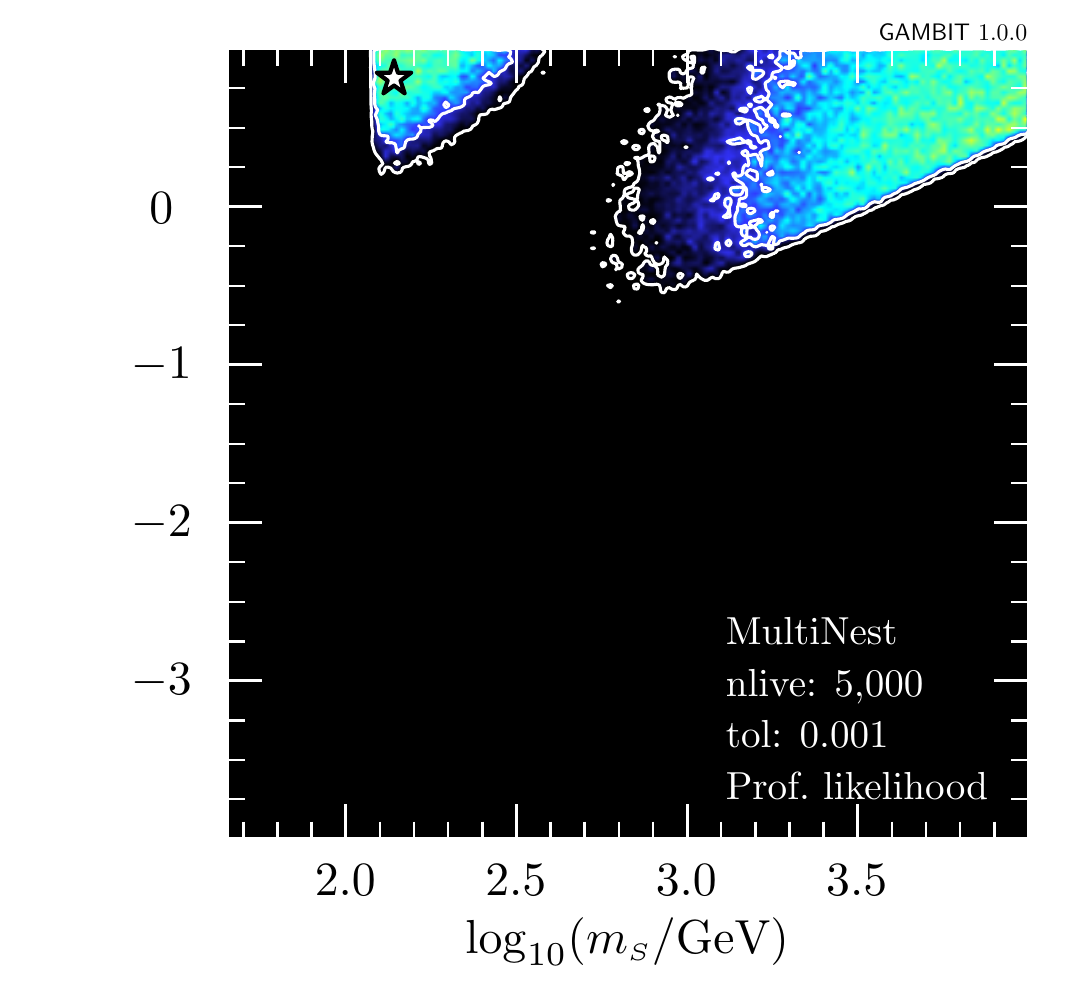}\hspace{-3mm}%
  \includegraphics[height=0.234\linewidth]{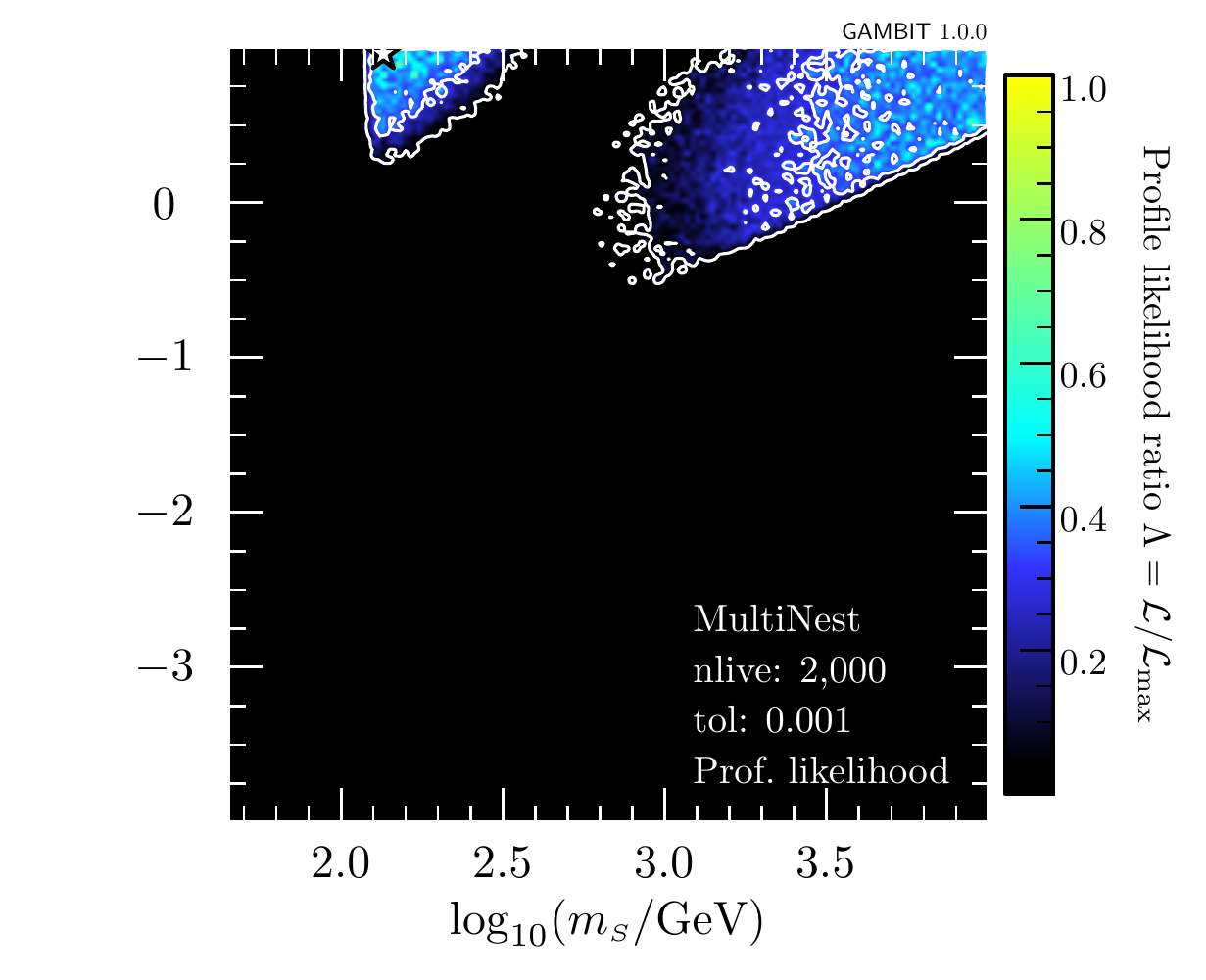}
  \caption{Profile likelihood ratio maps from a 15-dimensional scan of the scalar singlet parameter space, using the \multinest scanner with a selection of difference tolerances (\lstinline{tol}) and numbers of live points (\lstinline{nlive}).  The maximum likelihood point is shown by a white star.}
  \label{fig:multinest_plots}
\end{figure*}

\begin{figure*}[tp]
  \centering
  \includegraphics[height=0.234\linewidth]{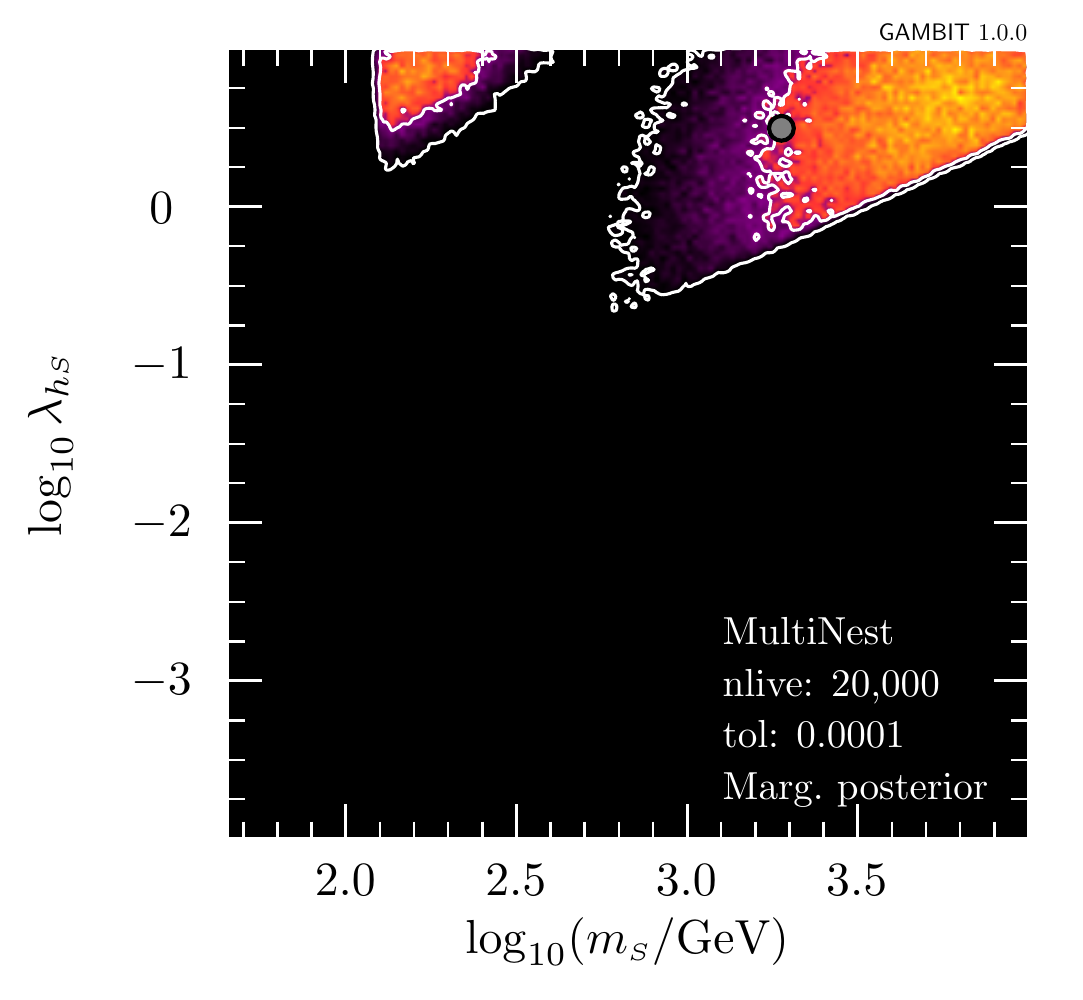}\hspace{-3mm}%
  \includegraphics[height=0.234\linewidth]{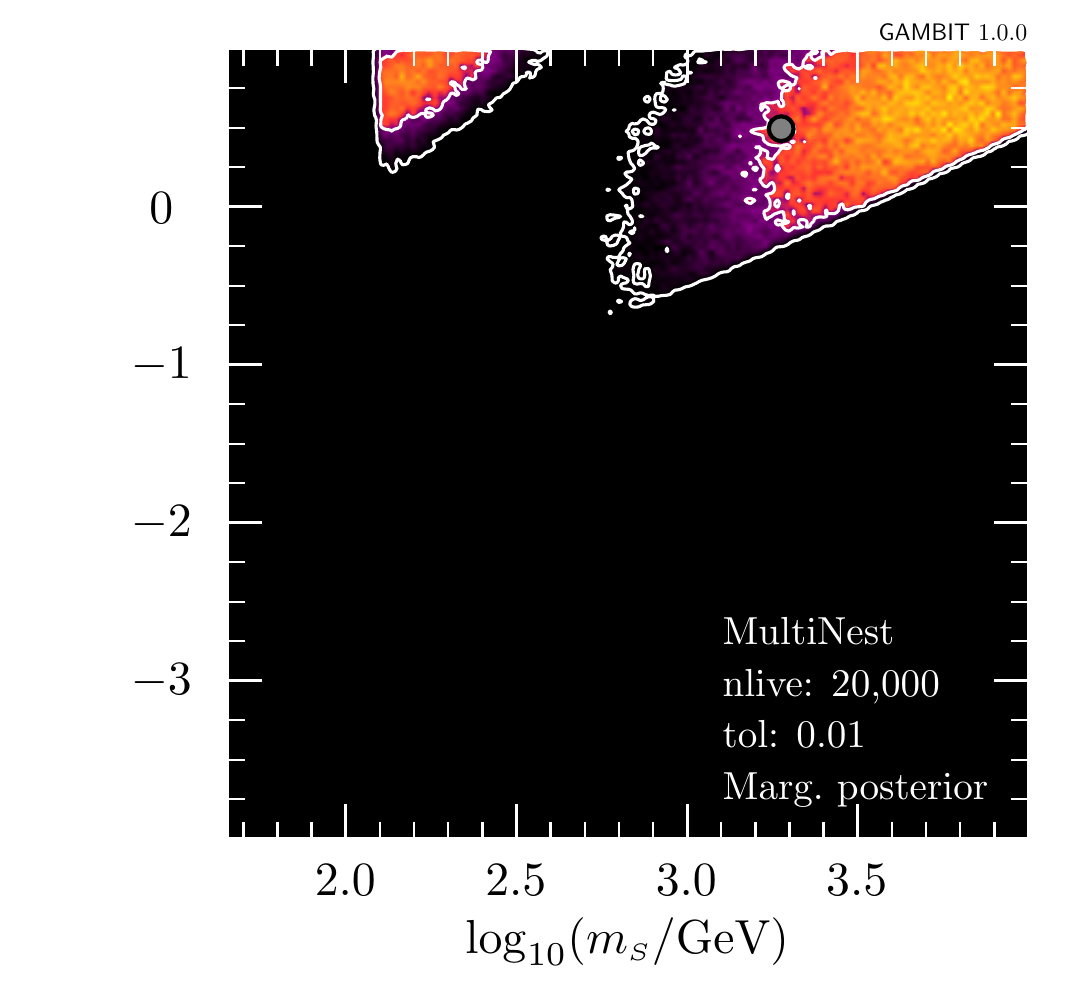}\hspace{-3mm}%
  \includegraphics[height=0.234\linewidth]{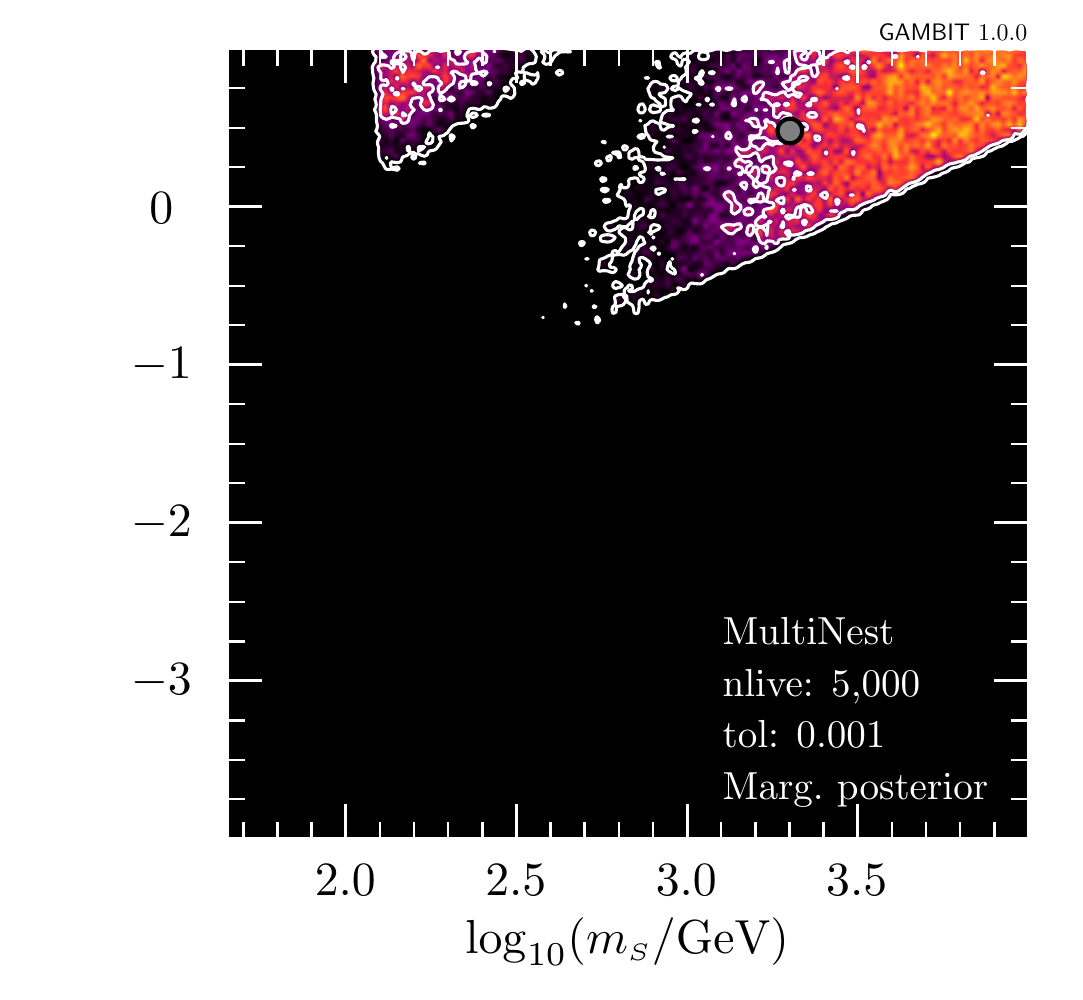}\hspace{-3mm}%
  \includegraphics[height=0.234\linewidth]{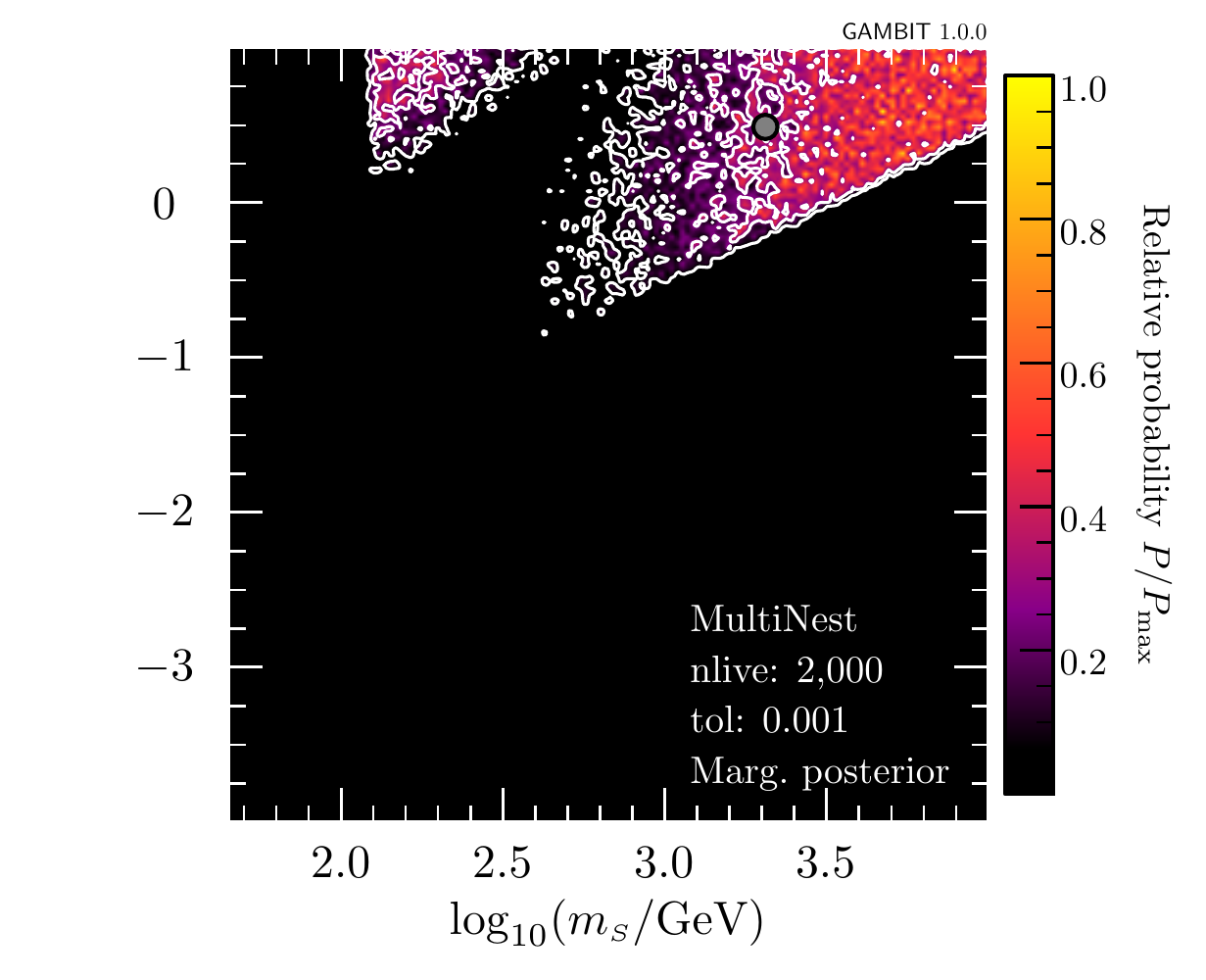}
  \caption{Marginalised posterior probability density maps from a 15-dimensional scan of the scalar singlet parameter space, using the \multinest scanner with a selection of difference tolerances (\lstinline{tol}) and numbers of live points (\lstinline{nlive}).  Note that the colourbar strictly only applies to the rightmost panel, and that colours map to the same enclosed posterior mass on each plot, rather than to the same iso-posterior density level (i.e.~the transition from red to purple is designed to occur at the edge of the $1\sigma$ credible region, and so on).  The posterior mean is shown with a grey bullet point.}
  \label{fig:multinest_plots_post}
\end{figure*}

\begin{figure*}[tp]
  \centering
  \includegraphics[height=0.234\linewidth]{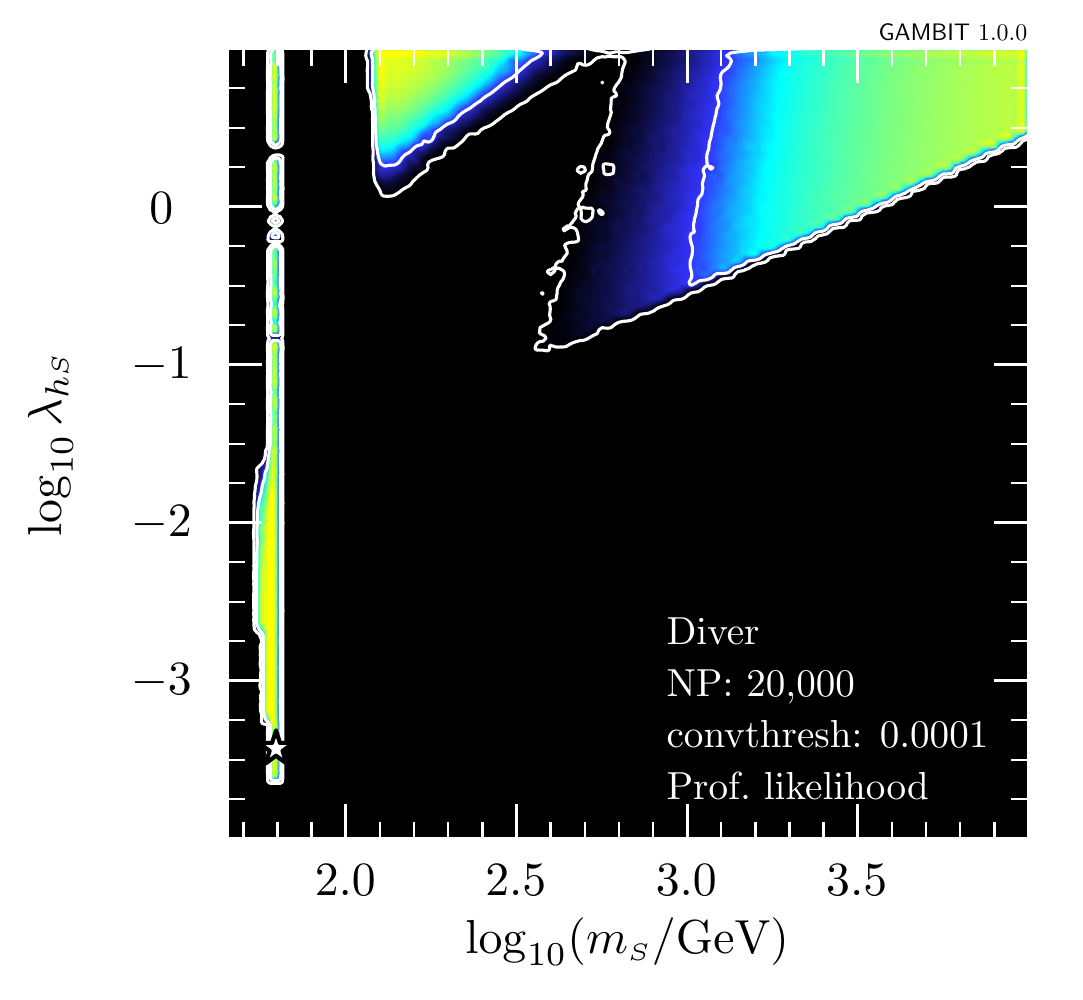}\hspace{-3mm}%
  \includegraphics[height=0.234\linewidth]{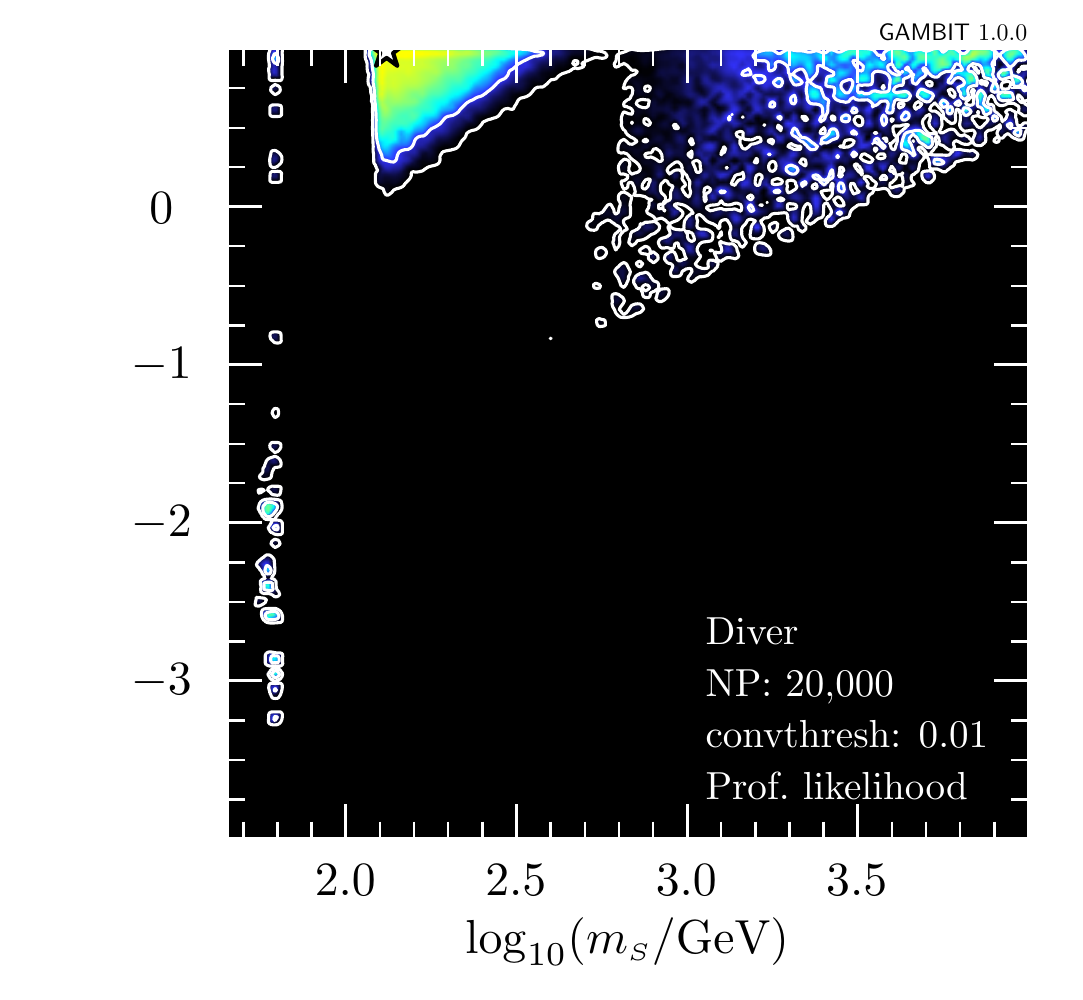}\hspace{-3mm}%
  \includegraphics[height=0.234\linewidth]{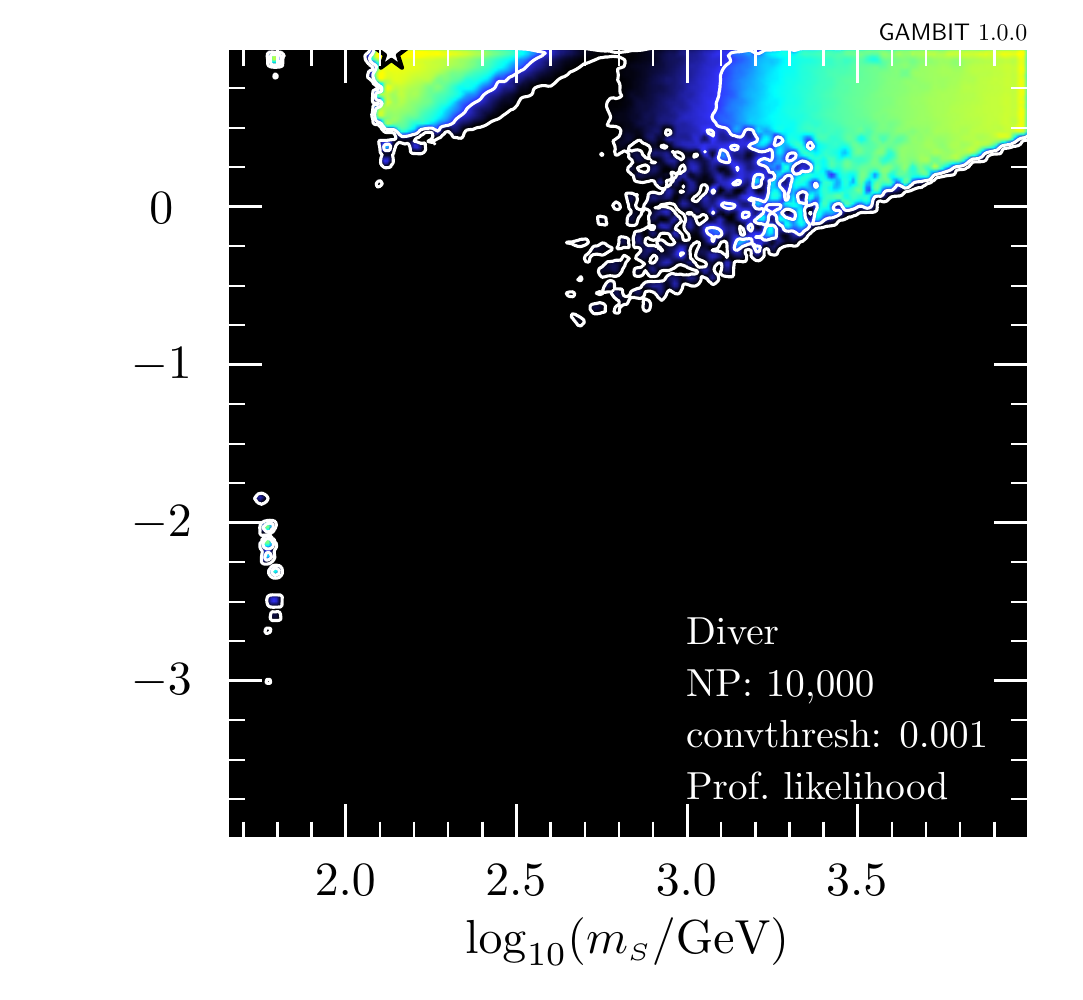}\hspace{-3mm}%
  \includegraphics[height=0.234\linewidth]{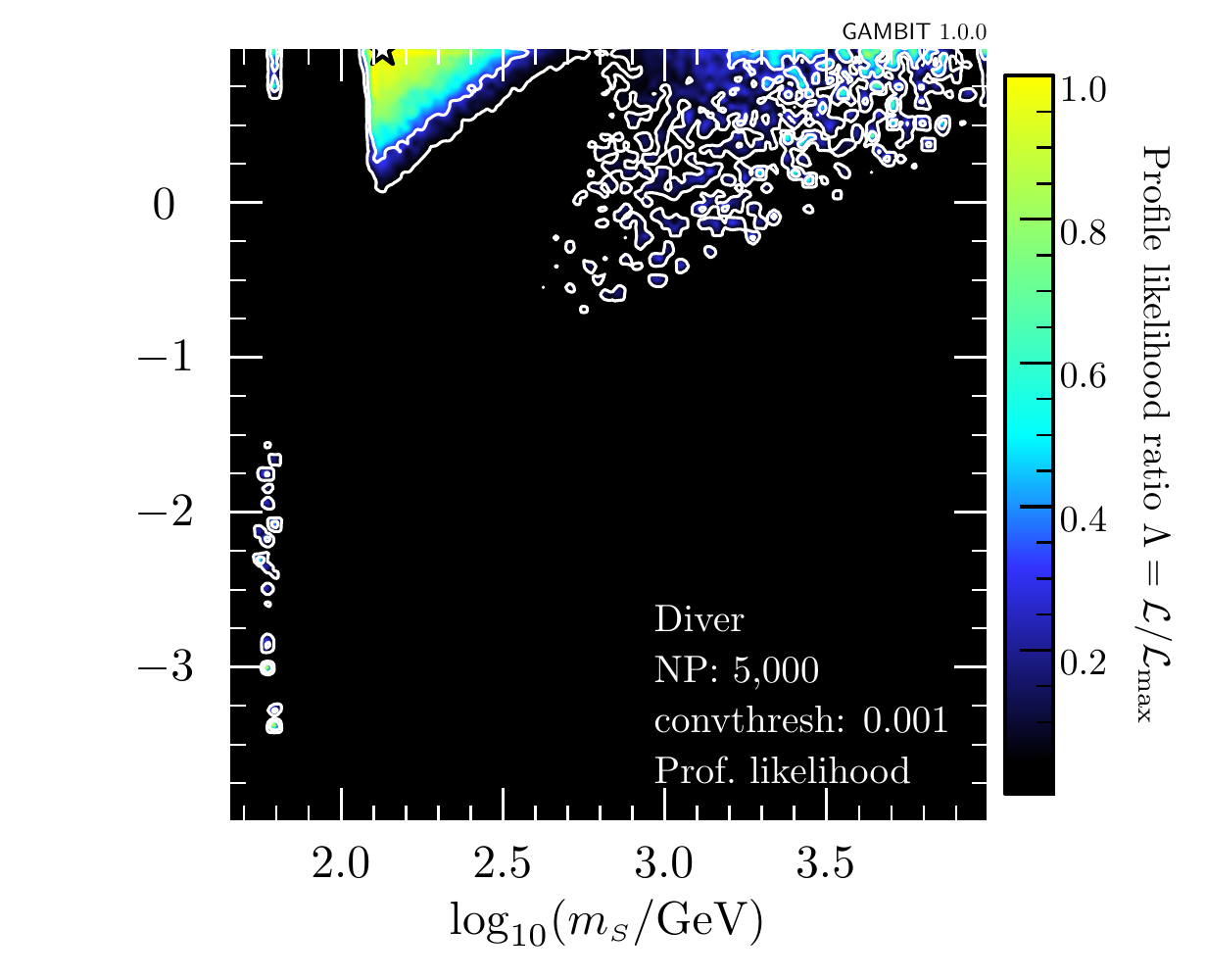}
  \caption{Profile likelihood ratio maps from a 15-dimensional scan of the scalar singlet parameter space, using the \diver scanner with a selection of difference convergence thresholds (\protect\fortran{convthresh}) and population sizes (\protect\fortran{NP}).  The maximum likelihood point is shown by a white star.}
  \label{fig:diver_plots}
\end{figure*}

In Secs. \ref{sec:multinest}--\ref{sec:great} we discuss the most appropriate choices of settings for \MultiNest, \diver, \twalk and \great, respectively.  In order to make comparisons, we require fair metrics with which to compare the outcomes of scans.  We first look at the best value of the log-likelihood found in each scan, which is crucial for the correct normalisation of the profile likelihood (Figs.\ \ref{fig:Diver_MultiNest_1}, \ref{fig:Diver_MultiNest_2}, \ref{fig:Twalk} and \ref{fig:GreAT}).  The results of this test favour algorithms primarily intended as optimisers, whilst disadvantaging those mainly designed to map the likelihood function or posterior.  We therefore also compare the visual quality of the profile likelihood maps (Figs.\ \ref{fig:multinest_plots}, \ref{fig:diver_plots}, \ref{fig:twalk_plots} and \ref{fig:great_plots}), and the corresponding posterior maps (Figs.\ \ref{fig:multinest_plots_post}, \ref{fig:twalk_plots_post} and \ref{fig:great_plots_post}).  This is a more qualitative approach, better suited for algorithms intended to explore the parameter space.

We also make some additional comparisons between the four sampling algorithms.  In the first two of these tests, we are interested in the relative performance as a function of parameter space dimensionality (Sec.\ \ref{sec:dimensional_performance}) and the total CPU time required to complete a scan (Sec.\ \ref{sec:time_performance}).  Here, we focus mostly on the value of the best-fit log-likelihood and the time taken to achieve it.  These sections are most relevant for evaluating profile likelihood performance; in Sec.\ \ref{sec:posterior}, we instead focus on the specific merits of different algorithms for mapping the Bayesian posterior.  We discuss the overall implications of these results in Sec.\ \ref{sec:compare}.

We performed all tests using a high-performance computing cluster, taking advantage of the ability to run \GB in parallel across multiple processors.  In the interests of making sensible use of computing resources and time, we ran the two-dimensional scans on a single 24-core compute node, using 24 \mpi processes.  For the seven- and fifteen-dimensional scans, we used 10 nodes, for a total of 240 \mpi processes.  For the scans where we compare performance with respect to dimensionality, a consistent computing environment is required; here we used 5 nodes for all scans, corresponding to 120 \mpi processes.\footnote{Although \GB is also able to use \omp threads for further  (likelihood-level) parallelisation within individual \mpi processes \cite{gambit}, here we limit ourselves to distributed-memory parallelisation with \mpi, seeing as this is the form of parallelisation employed by the scanning algorithms.}  The two-dimensional profile likelihood and marginalised posterior maps that we show in the following subsections were produced with \pippi \cite{pippi}, using 150 bins in each dimension.

\subsection{\MultiNest}\label{sec:multinest}

\multinest's ability to accurately evaluate the evidence and map the posterior is directly affected by the number of live points used in a scan, with more live points increasing the chance of finding all relevant modes of the posterior.  On the other hand, more live points means more likelihood evaluations, and requires greater computing resources.  The overall duration of the scan is also influenced by the stopping criterion, which is given by the tolerance on the final evidence (the estimate of the largest evidence contribution that can be made with the remaining portion of the posterior volume).  The sampling parameters that we vary are therefore the number of live points ($N_{live}$, \lstinline{nlive}) and the tolerance (\lstinline{tol}).

We perform runs with 2000, 5000, 10\,000 and 20\,000 live points, and tolerances of $10^{-4}$,  $10^{-3}$, $10^{-2}$ and $10^{-1}$.  The values of the best-fit log-likelihoods achieved for scans using these parameters are shown in Figs.\ \ref{fig:Diver_MultiNest_1} and \ref{fig:Diver_MultiNest_2}.  In Fig.\ \ref{fig:multinest_plots}, we present a selection of the profile likelihoods from \MultiNest scans in the full 15-dimensional parameter space; in Fig.\ \ref{fig:multinest_plots_post} we give corresponding marginalised posterior maps.

We see consistent best fits from all scans when \lstinline{tol} $\le$ $10^{-3}$.  A sufficiently small tolerance appears to provide a good best-fit value over a large range of \lstinline{nlive} values.  On the other hand, even with larger values of \lstinline{nlive}, setting \lstinline{tol} too large will still negatively impact the quality of the best-fit point; even with 20\,000 live points we still see a poor best-fit likelihood if the tolerance is greater than $10^{-3}$.  The number of live points has a more significant impact on the sampling of the parameter space, as can be seen in Figs.\ \ref{fig:multinest_plots} and \ref{fig:multinest_plots_post}.  In these plots, a significant difference in the quality of both profile likelihood and posterior sampling is evident even between runs done with 2000 and 5000 live points.

On the basis of these results, we recommend an upper bound on the tolerance of $10^{-3}$ if \multinest is to be relied upon for obtaining the appropriate normalisation for profile likelihoods.  The number of live points required will depend on the desired quality of the resultant profile likelihood or posterior contours, and the dimensionality of the parameter space.  In Fig.\ \ref{fig:multinest_plots}, it is clear that in fifteen dimensions a value of at least 20\,000 for \lstinline{nlive} is required to give fine-grained sampling of the profile likelihood.  Because in most cases one is interested in a global fit over many parameters, we recommend a value of 20\,000 live points as the lower limit. We note however that this may be reduced somewhat if dealing with a lower-dimensional parameter space, or if one is only interested in mapping the posterior at a lower resolution (less bins) than we have employed here.

\subsection{\diver}\label{sec:diver}

\diver is a differential evolution optimisation package that is also highly effective at sampling parameter spaces.  The size of the evolving population is determined by the \fortran{NP} parameter, and the threshold for convergence is controlled by the \fortran{convthresh} parameter.

We examine population sizes of \fortran{NP} = 2000, 5000, 10\,000 and 20\,000, and \fortran{convthresh} values of $10^{-4}$, $10^{-3}$, $10^{-2}$ and $10^{-1}$.  Although these parameters have different definitions to \lstinline{nlive} and \lstinline{tol} in \MultiNest, we take advantage of the similarity in the appropriate ranges for these and plot the scan results on the same axes in Figs.\ \ref{fig:Diver_MultiNest_1} and \ref{fig:Diver_MultiNest_2}.  We see that a \fortran{convthresh} value of less than $10^{-3}$ gives consistent results for the best-fit log-likelihood at all values of \fortran{NP}.

In two dimensions, both \MultiNest and \diver are able to find roughly the same or equivalently good best-fit points.  The differences in the algorithms become evident in seven and fifteen dimensions however, where \diver consistently outperform \MultiNest for equivalent parameter values.  This is somewhat expected, given that \diver is designed as an optimisation routine, whereas \multinest is intended to compute the Bayesian evidence and sample the posterior distribution.  In two dimensions, the sampling is dense enough that \MultiNest has been able to locate the best-fit point, but in higher dimensions the task is more suited to an optimisation-specific routine.  Because the maximum likelihood is located in the low-mass region in both two and fifteen dimensions, it is indeed a result of poor sampling that \multinest has not located the same best fit that \diver has achieved (see Appendix \ref{app:scanner_comparisons} for equivalent plots for two dimensional scans).  We return to this discussion in Sec.\ \ref{sec:compare}.

In Fig.\ \ref{fig:diver_plots}, we investigate the ability of \diver to accurately map the contours of the profile likelihood.  We see that both the \fortran{convthresh} and \fortran{NP} settings are relevant in reproducing the desired contours.  A \fortran{convthresh} of $10^{-3}$ appears appropriate in fifteen dimensions, along with an \fortran{NP} value of at least 20\,000.  However, these requirements become less stringent in a lower-dimensional parameter spaces (data not shown), where they can be reduced by at least an order of magnitude whilst still achieving a suitable mapping of the profile likelihood.

From these tests, we recommend similar settings as for \MultiNest for similar parameters: for a detailed picture of the profile likelihood a value of 20\,000 is recommended for \fortran{NP} (although this can be reduced for lower dimensional parameter spaces), and to consistently find the best-fit point an upper bound of $10^{-3}$ is recommended for the \fortran{convthresh} convergence tolerance.

\subsection{\twalk}\label{sec:twalk}

\twalk is an ensemble MCMC algorithm. The primary parameters of interest are the number of chains used during the scan and the stopping criterion. The latter is controlled by the parameter \yaml{sqrtR}, which is the square root of the Gelman-Rubin $R$ statistic, where 1 is perfect.  For comparison with other scanners, we define the equivalent tolerance of \twalk scans as $tol$ $\equiv$ \yaml{sqrtR} $-$ 1.  The \yaml{chain_number} is bounded below by 1 + \yaml{projection\_dimension} + the number of \mpi processes in use (see Sec.\ \ref{app:options:twalk}).  For two dimensions, we have a lower limit of 27 $(24 + 2 + 1)$, and therefore perform tests with 27, 54, 81 and 108 chains.  For higher-dimensional scans, the increase in the number of \mpi processes requires larger chain numbers, so we choose 256 and 512.  We consider $tol$ values of 0.3, 0.1, 0.03 and 0.01.

\begin{figure}[tp]
  \centering
  \includegraphics[width=1\linewidth]{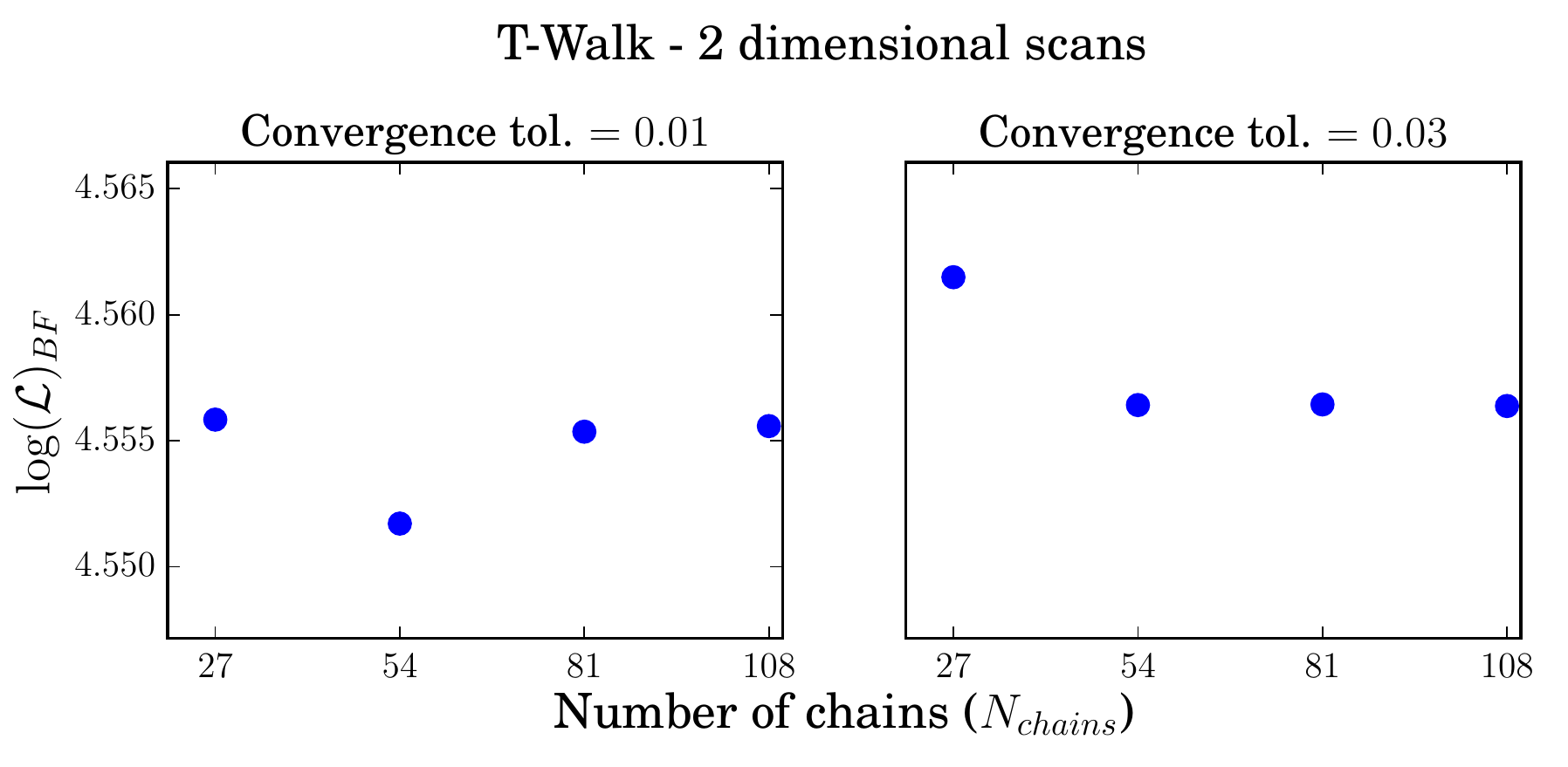}
  \includegraphics[width=1\linewidth]{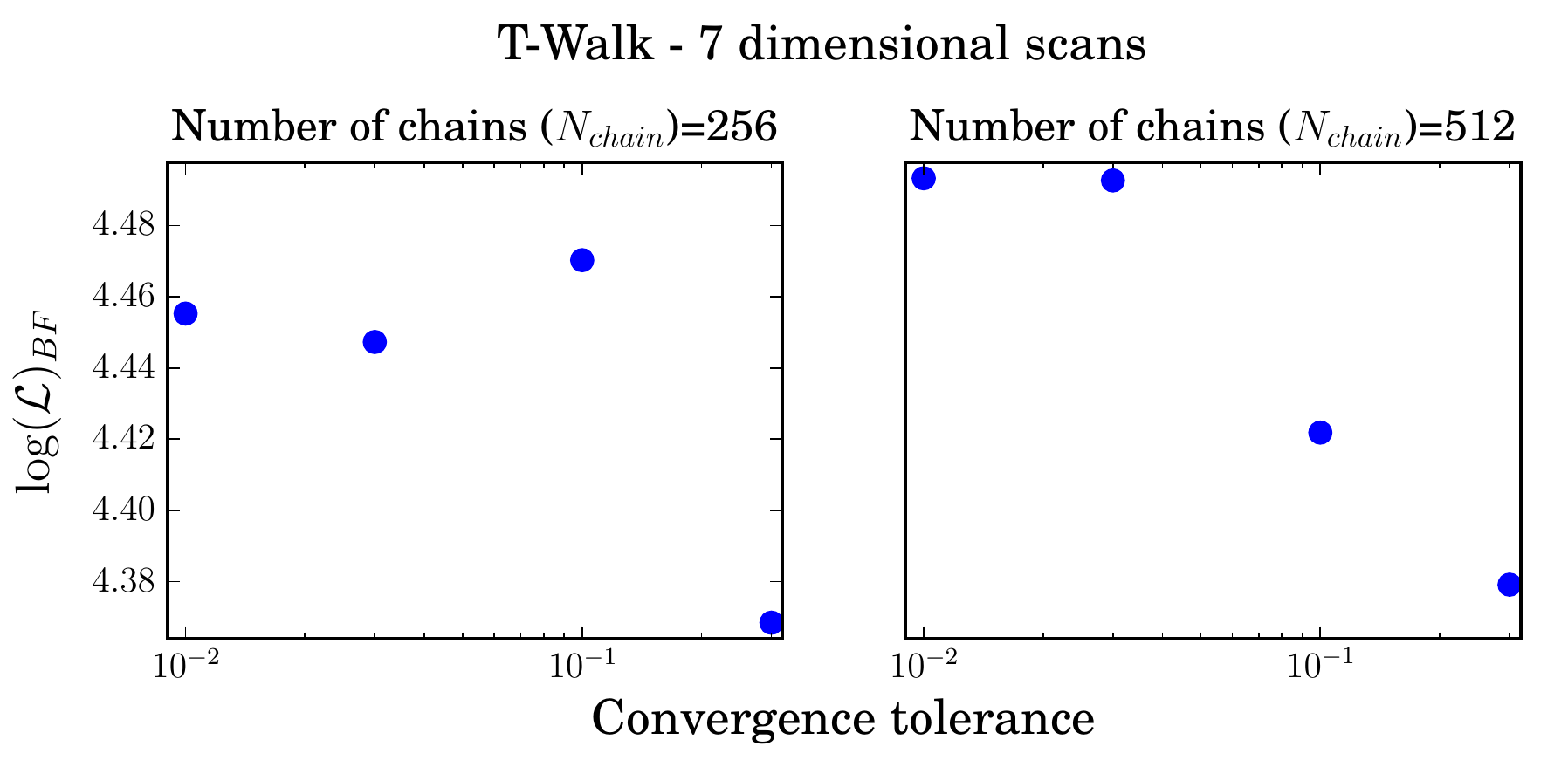}
  \includegraphics[width=1\linewidth]{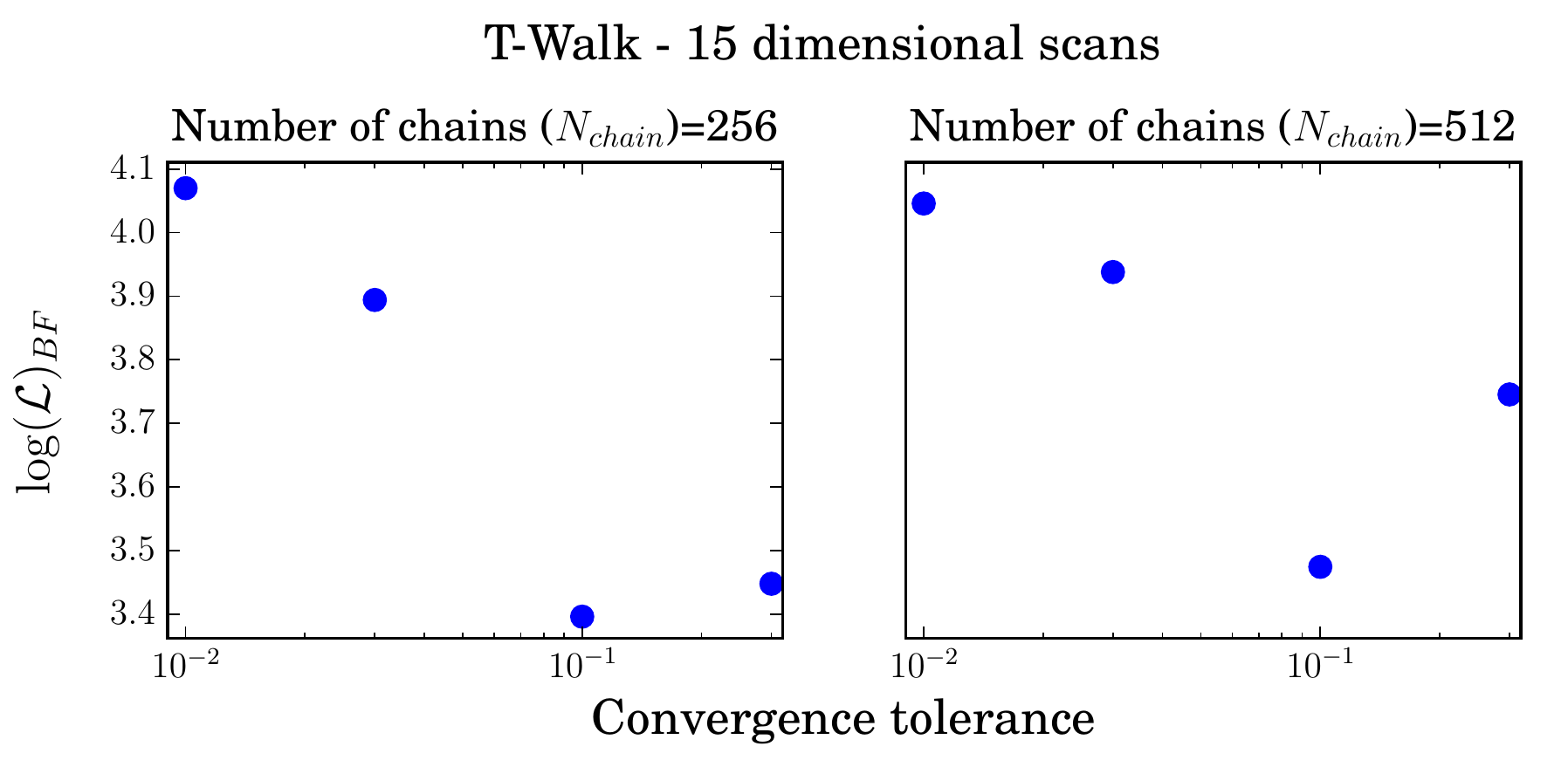}
  \caption{\textit{Top row:} Best-fit log-likelihoods for two-dimensional scans using the \twalk algorithm, as a function of the number of chains used, for two different convergence tolerances (\lstinline{tol}).  \textit{Middle and bottom panels:}  Best-fit log-likelihoods as a function of convergence tolerance (\lstinline{tol}), for \twalk scans in seven and fifteen dimensions with a fixed number of chains.  Note that the likelihood is dimensionful, leading to $\mathcal{L_{\rm BF}}>1$ \cite{gambit}.}
  \label{fig:Twalk}
\end{figure}

The best-fit log-likelihoods from scans using various \twalk settings are given in Fig.\ \ref{fig:Twalk}.  In two dimensions, we hold the tolerance fixed and investigate the effect of varying the chain number.  We see no notable trend with chain number, for either of the tolerance values.  For the seven and fifteen-dimensional scans, we therefore instead focus on varying the tolerance for a fixed number of chains.  This reveals the expected trend: smaller tolerances result in improvements to the best-fit log-likelihoods.  A significant improvement seems to occur when $tol\lesssim 0.1$.  We also notice no significant difference between the scans with 256 and 512 chains, consistent with what we saw in the two-dimensional scans.

\begin{figure*}[tp]
  \centering
  \includegraphics[height=0.234\linewidth]{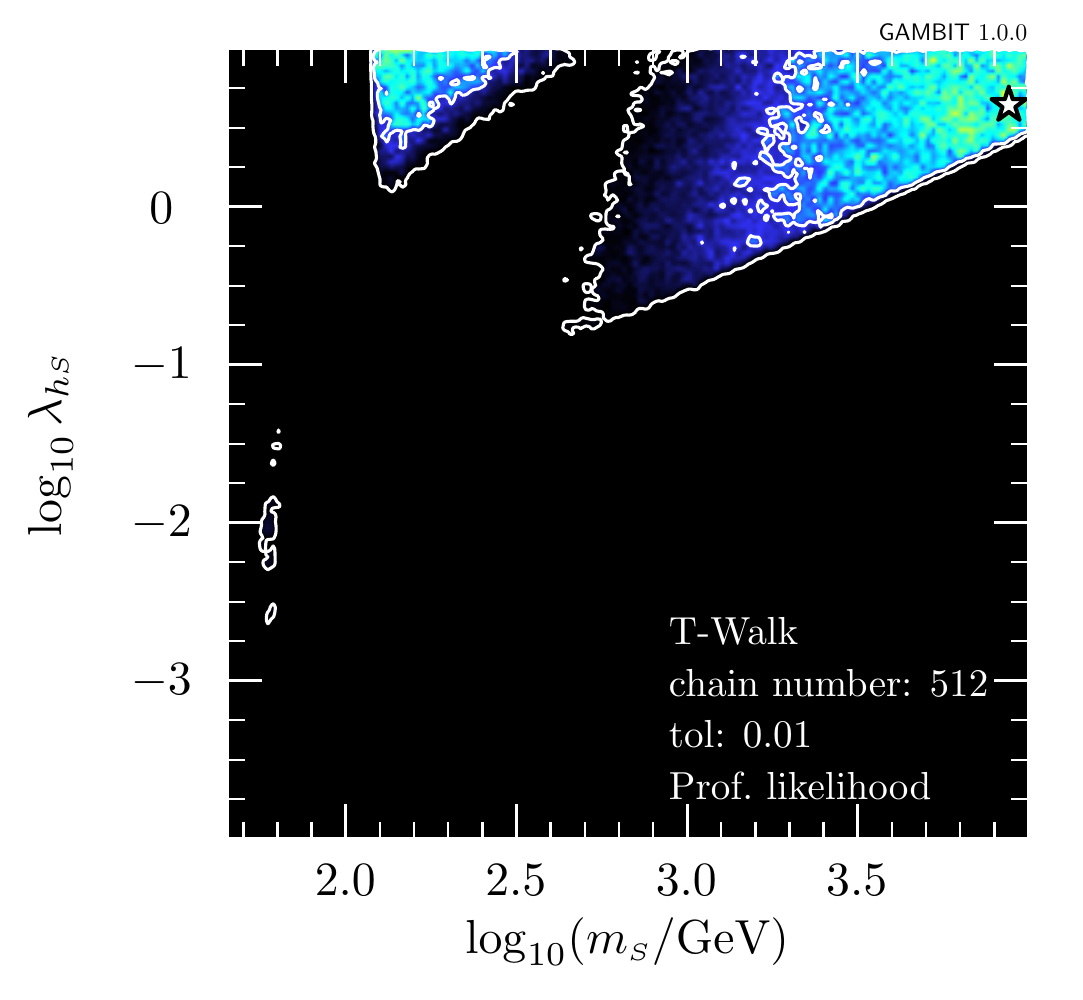}\hspace{-3mm}%
  \includegraphics[height=0.234\linewidth]{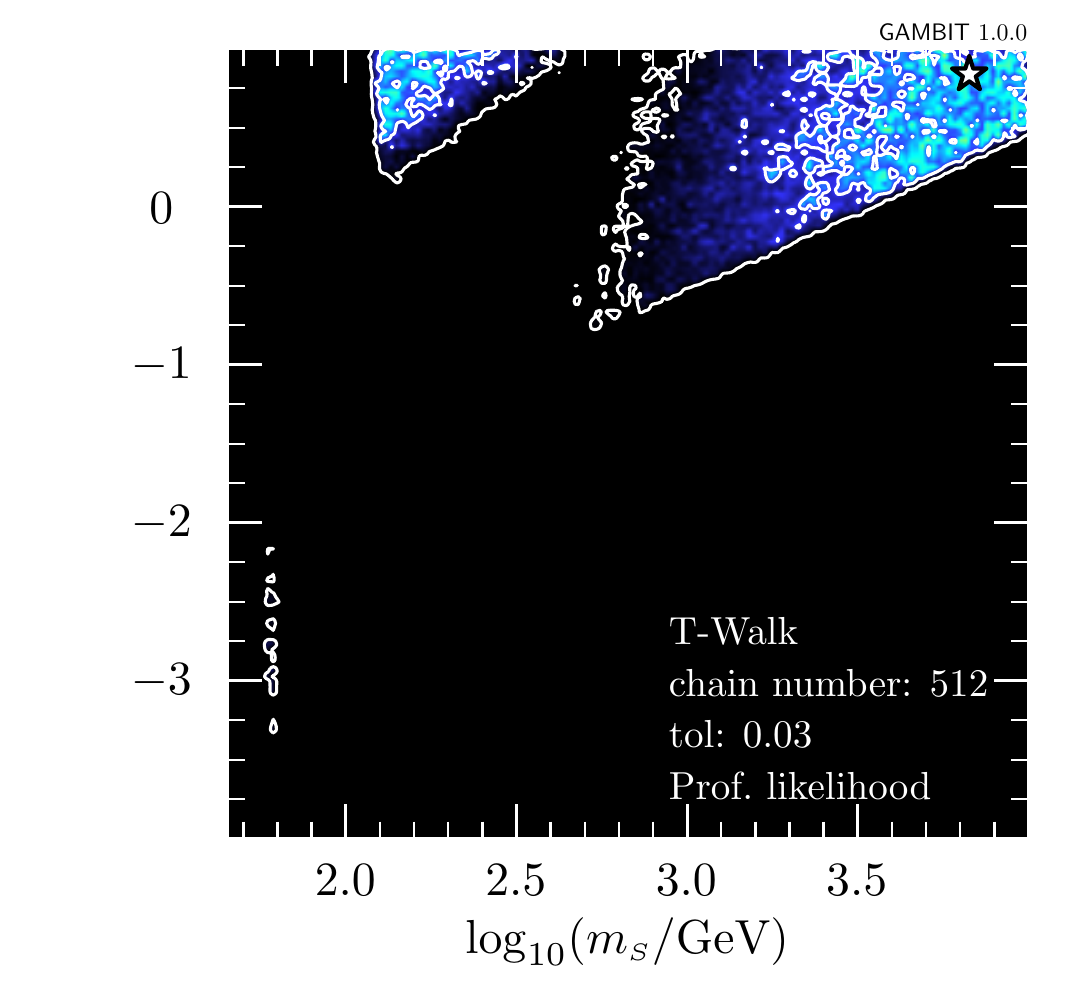}\hspace{-3mm}%
  \includegraphics[height=0.234\linewidth]{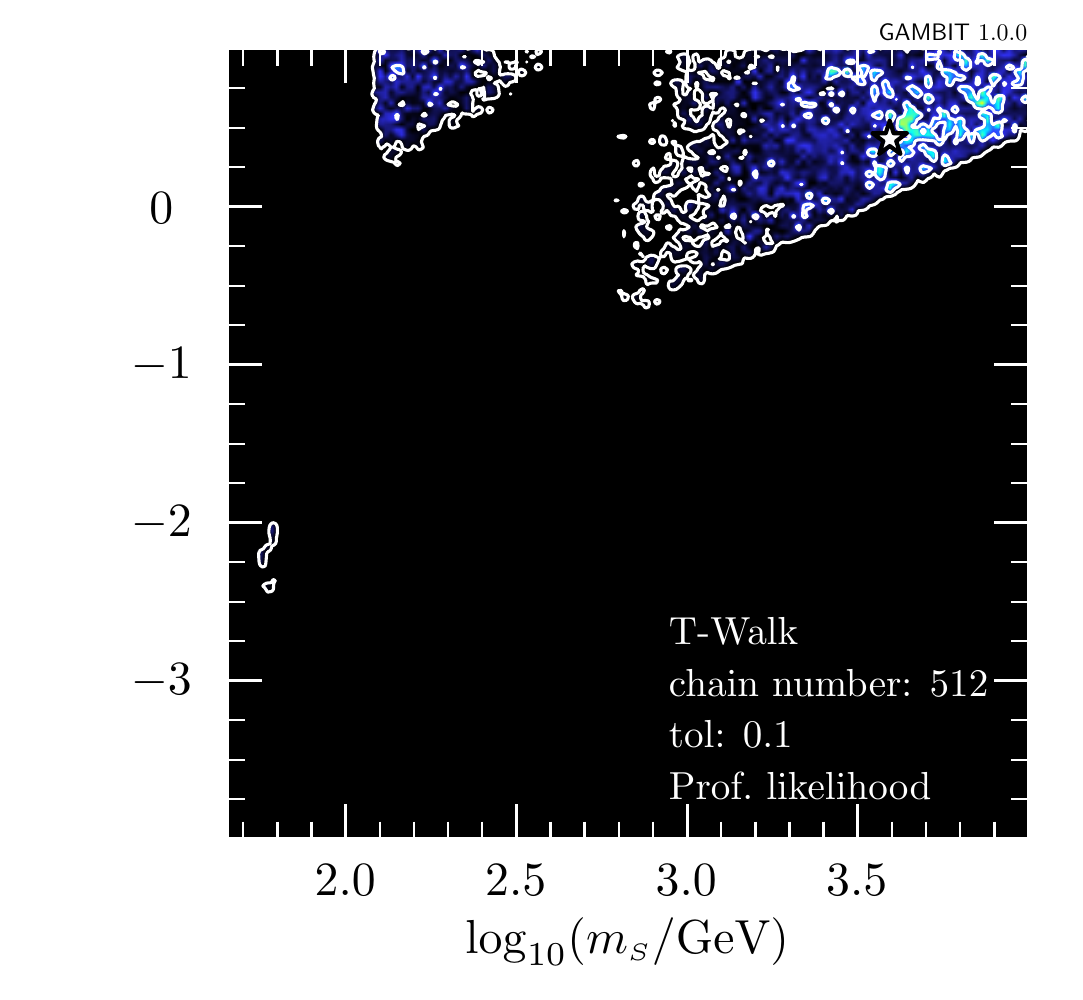}\hspace{-3mm}%
  \includegraphics[height=0.234\linewidth]{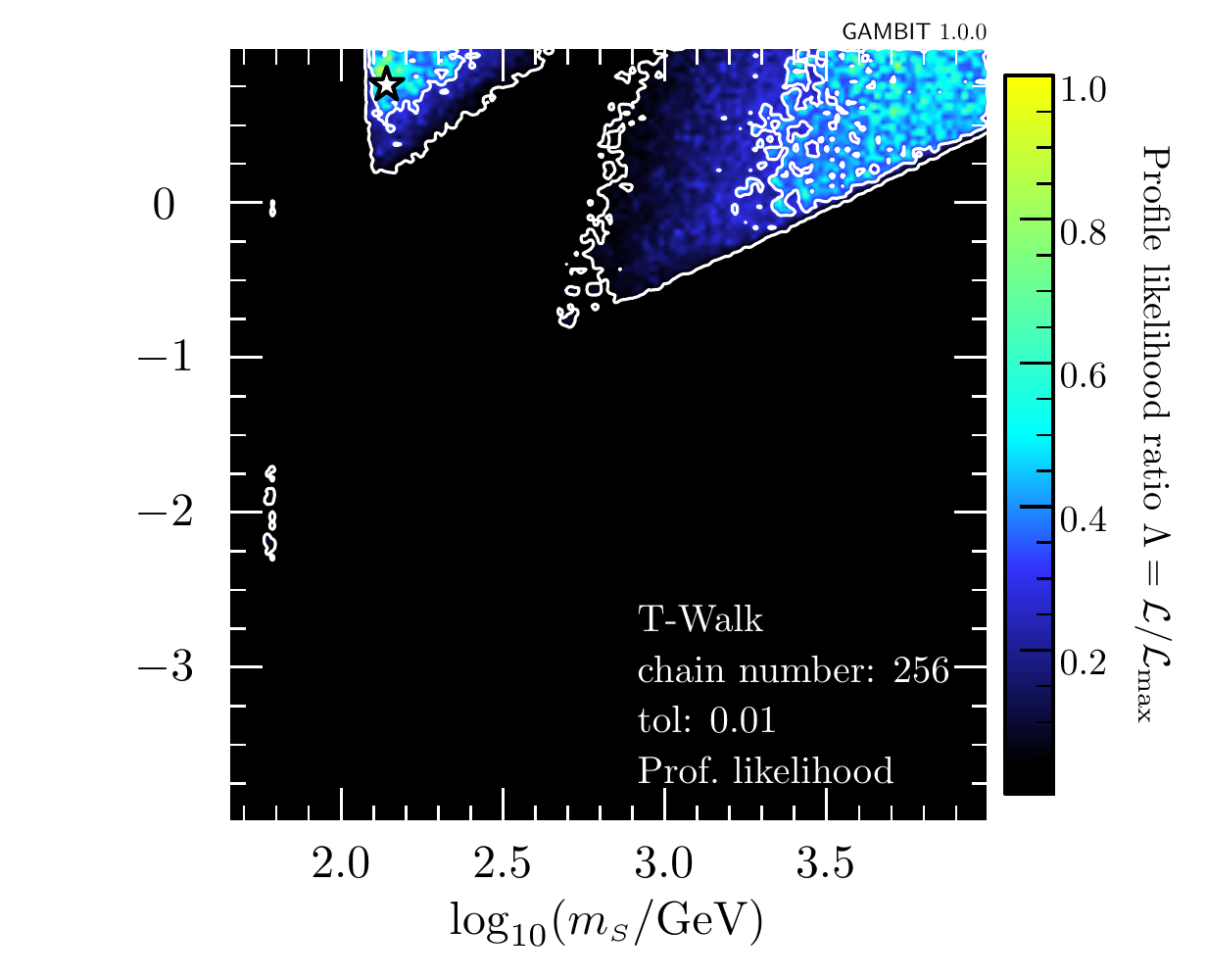}
  \caption{Profile likelihood ratio maps from a 15-dimensional scan of the scalar singlet parameter space, using the \twalk scanner with various numbers of chains and different tolerances.  The maximum likelihood point is shown by a white star.}  \label{fig:twalk_plots}
\end{figure*}

\begin{figure*}[tp]
  \centering
  \includegraphics[height=0.234\linewidth]{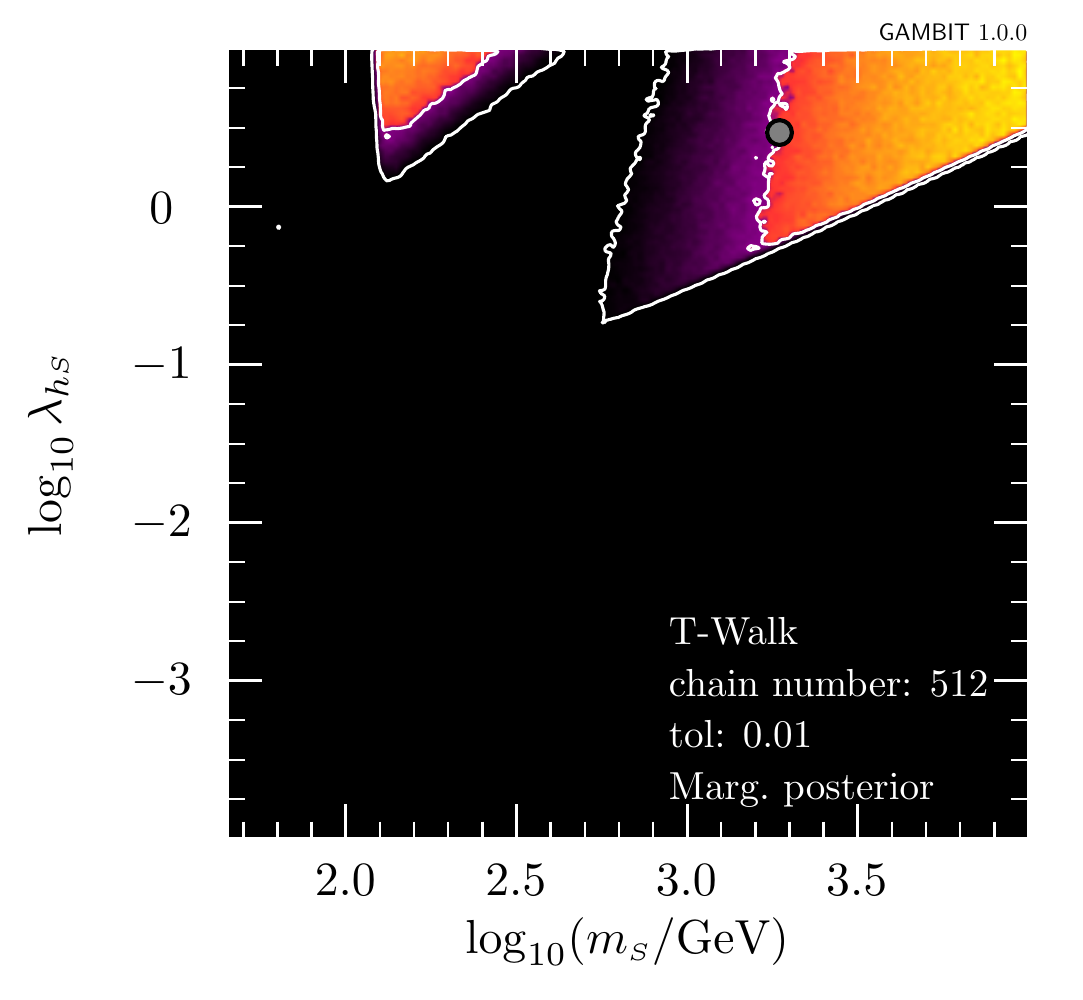}\hspace{-3mm}%
  \includegraphics[height=0.234\linewidth]{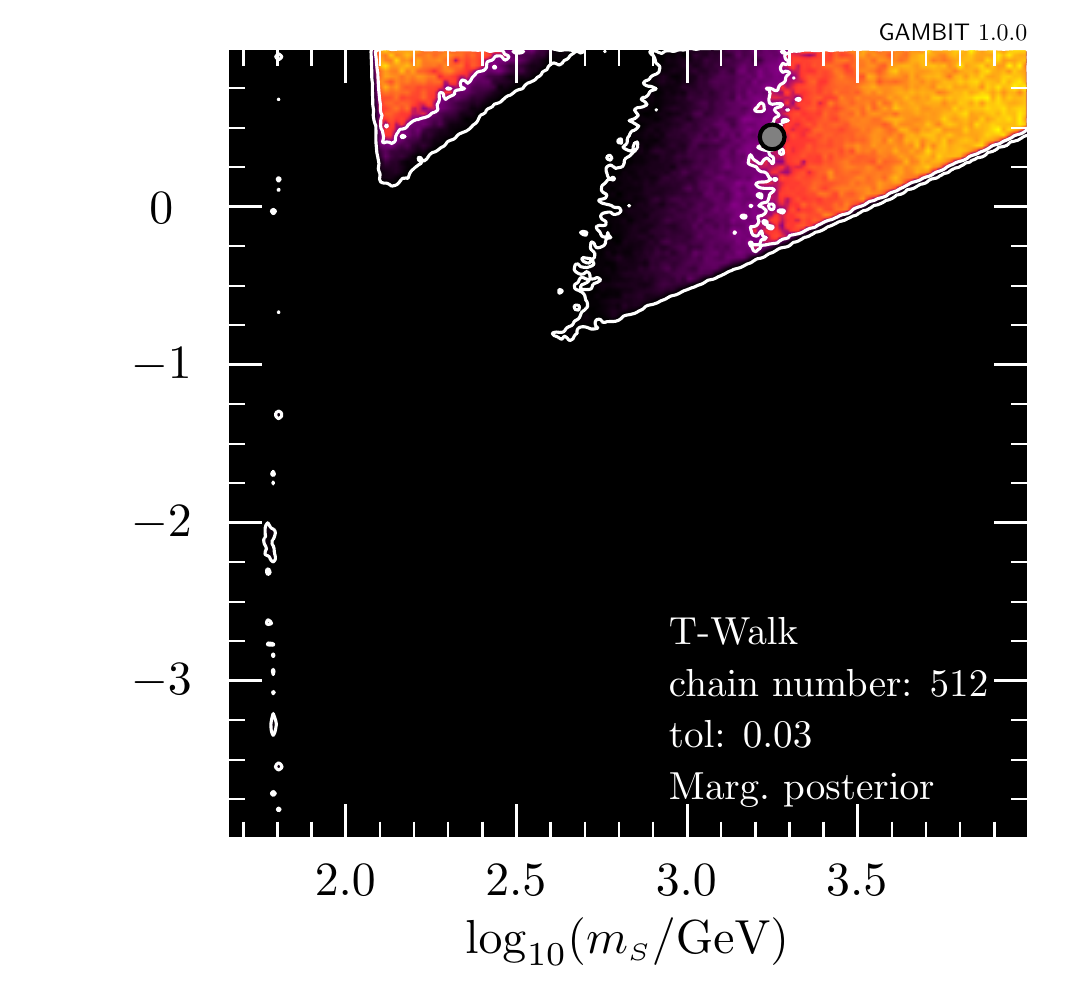}\hspace{-3mm}%
  \includegraphics[height=0.234\linewidth]{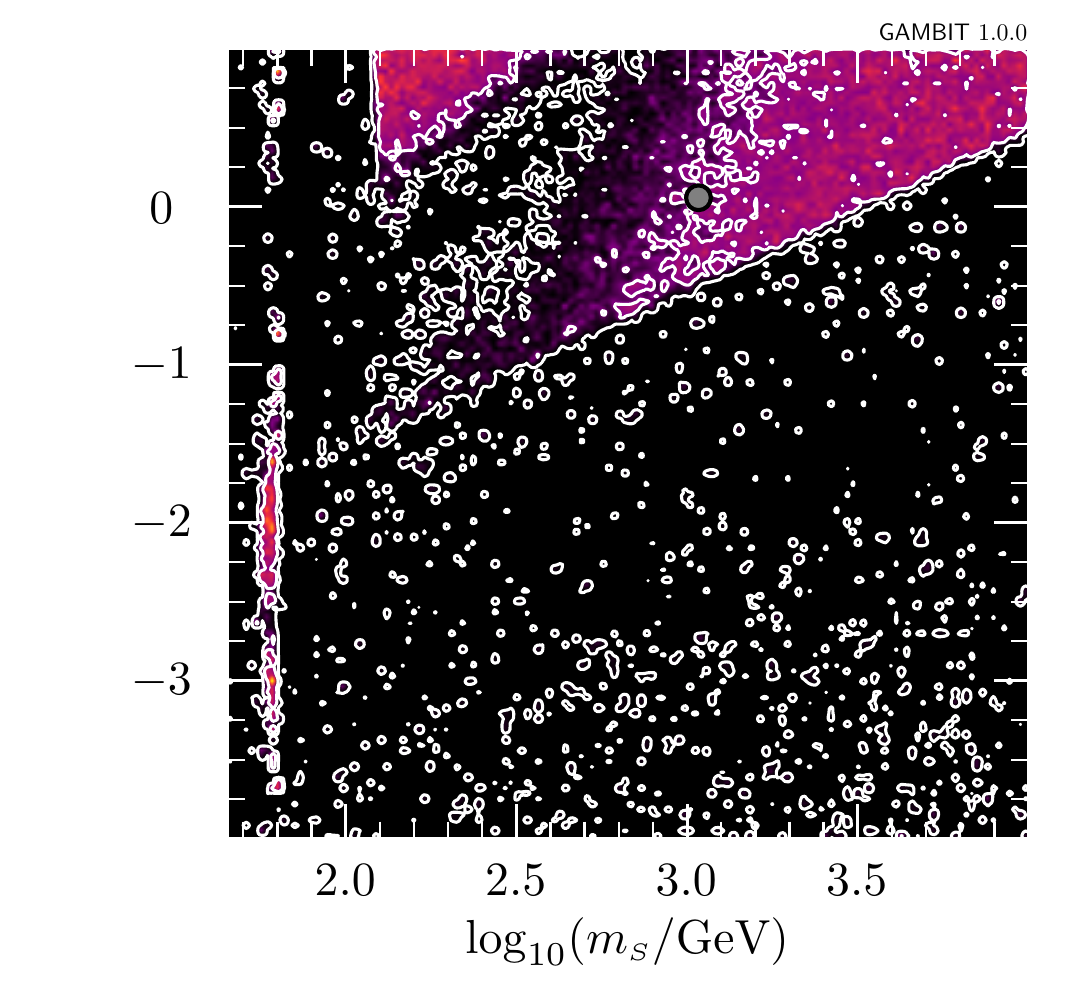}\hspace{-3mm}%
  \includegraphics[height=0.234\linewidth]{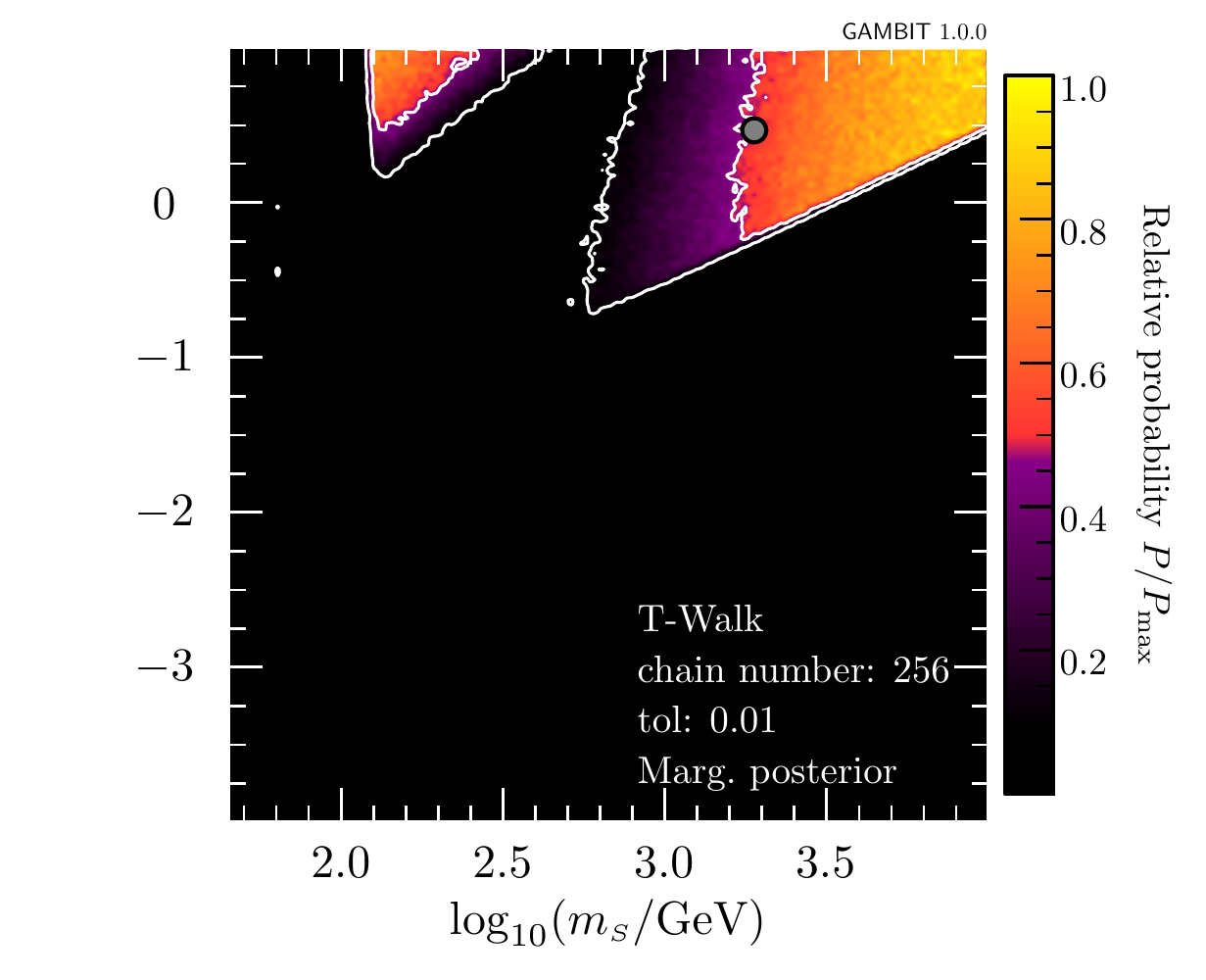}
  \caption{Marginalised posterior probability density maps from a 15-dimensional scan of the scalar singlet parameter space, using the \twalk scanner with various numbers of chains and different tolerances.  The second to rightmost panel is from a 512-chain scan with a tolerance of 0.1. Note that the colourbar strictly only applies to the rightmost panel, and that colours map to the same enclosed posterior mass on each plot, rather than to the same iso-posterior density level (i.e.~the transition from red to purple is designed to occur at the edge of the $1\sigma$ credible region, and so on).  The posterior mean is shown with a grey bullet point.}
  \label{fig:twalk_plots_post}
\end{figure*}

In Fig.\ \ref{fig:twalk_plots}, we show a selection of profile likelihood maps of the 15-dimensional scalar singlet parameter space.  We immediately see that smaller tolerances are preferable for a detailed sampling, and doubling the number of chains has no notable impact on the quality of the sampling.  In Fig.\ \ref{fig:twalk_plots_post}, we show a selection of the marginalised posterior maps of the 15-dimensional scalar singlet parameter space achieved by \twalk.  Here we see that whilst the main posterior modes appear to be better explored with smaller values of \yaml{tol}, leading to smoother, better-converged posterior contours, the presence of the minority mode at low mass would seem to be more evident in scans using a \textit{higher} tolerance.  This may appear counter-intuitive; why should poorer sampling apparently do better at uncovering small regions such as this?  In reality, this region has been sampled more carefully in the scans with lower \yaml{tol} values, despite appearing less prominently in the posterior maps.  That the sampling in these regions is better at lower tolerances can be seen from Fig.\ \ref{fig:twalk_plots}, where lower tolerances pick up better-fit points in this region.  Nevertheless, the additional samples retrieved in runs with lower tolerances provide a steadily more accurate indication of relative posterior weights of each of these modes, gradually leading to the low-mass solution to become reweighted and disfavoured in the better-sampled posterior maps of Fig.\ \ref{fig:twalk_plots_post}.

Recommending parameters for the \twalk algorithm is difficult, due to the sensitivity of the convergence to the $tol$ $=$ \fortran{sqrtR} $-$ 1 parameter.  However, values less than $\sim$0.1 appear to be safe for the scans we have conducted here.  Increasing the number of chains above the minimum value does not appear to result in any improvement in the quality of the best-fit, nor in the overall sampling.  As starting values for a study using the \twalk scanner, we therefore recommend setting $tol<0.1$ and leaving \yaml{chain_number} at the default (minimum) value.

\begin{figure*}[tp]
  \centering
  \includegraphics[height=0.3\linewidth]{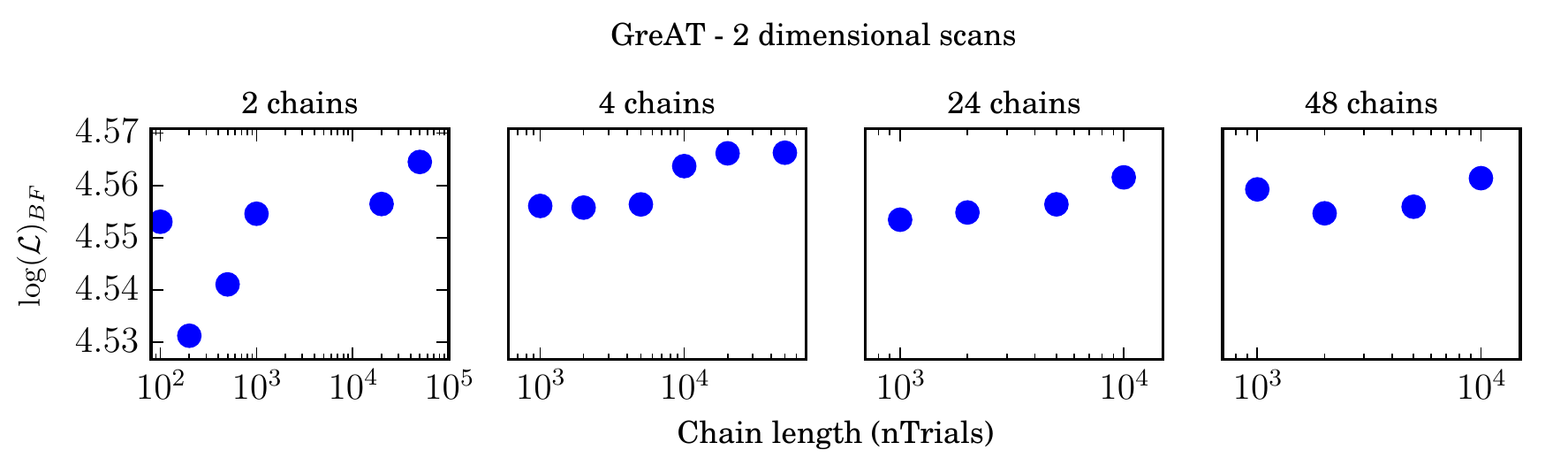}
  \includegraphics[height=0.345\linewidth]{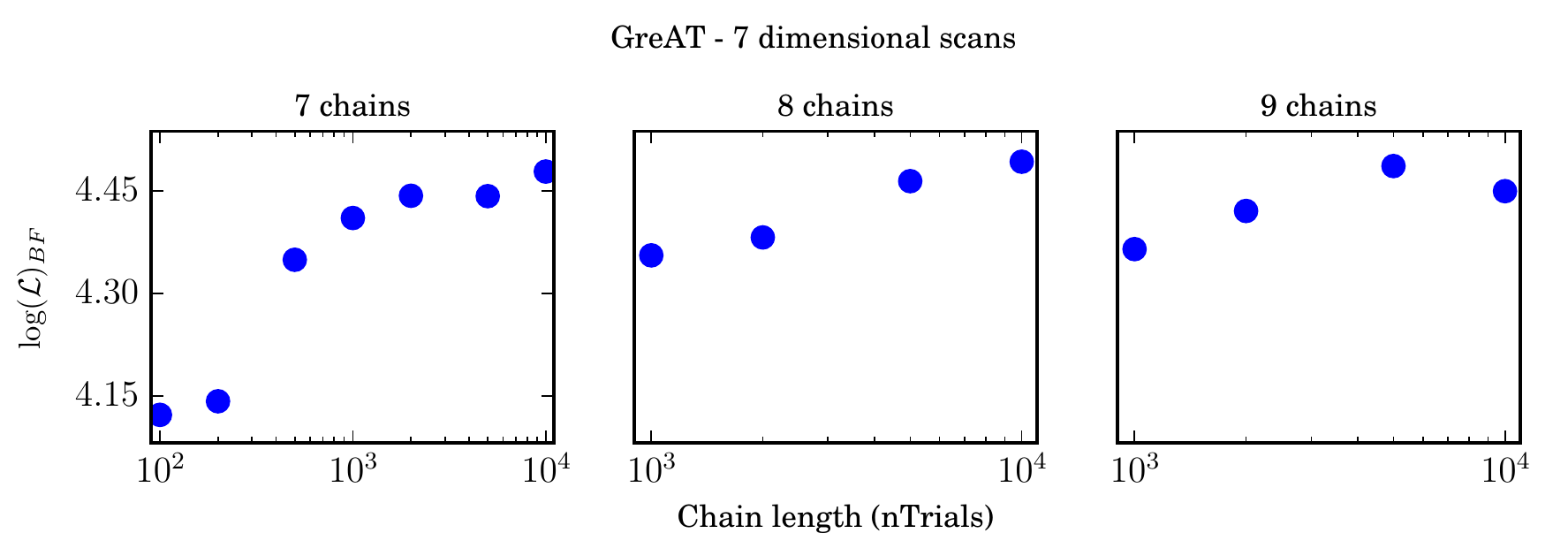}
  \includegraphics[height=0.345\linewidth]{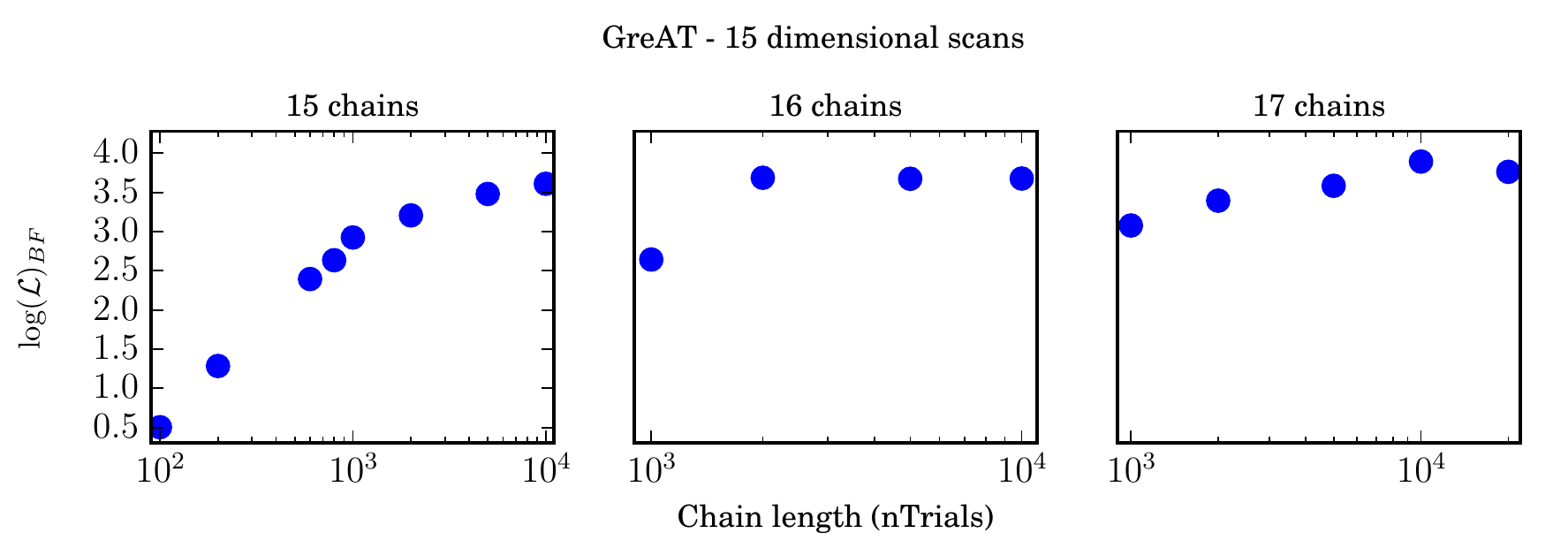}
  \caption{Best-fit log-likelihoods for scans using the \great sampler in two (\textit{top row}), seven (\textit{middle row}) and fifteen dimensions (\textit{bottom row}).  The number of chains is set by the \lstinline{nTrialLists} parameter.  Note that the likelihood is dimensionful, leading to $\mathcal{L_{\rm BF}}>1$ \cite{gambit}.}
  \label{fig:GreAT}
\end{figure*}

\begin{figure*}[tp]
  \centering
  \includegraphics[height=0.234\linewidth]{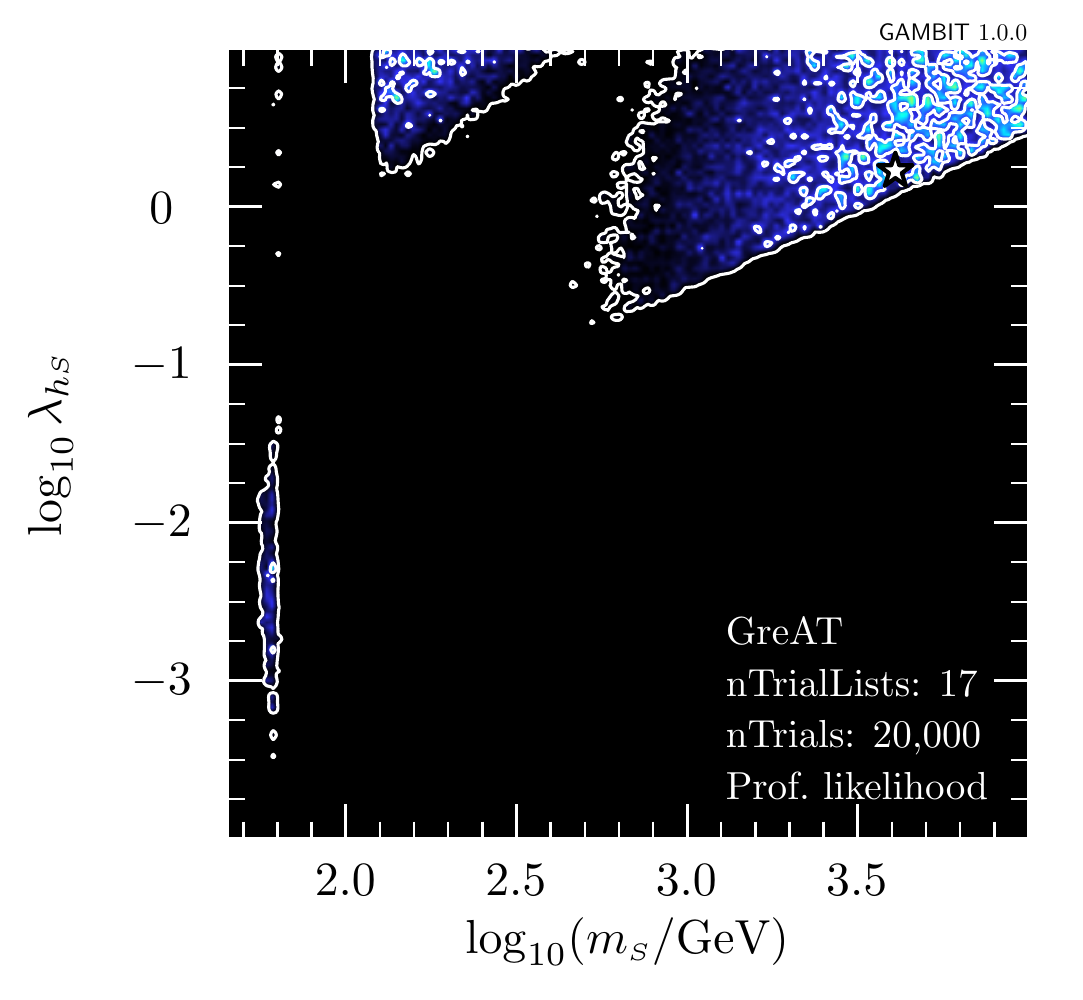}\hspace{-3mm}%
  \includegraphics[height=0.234\linewidth]{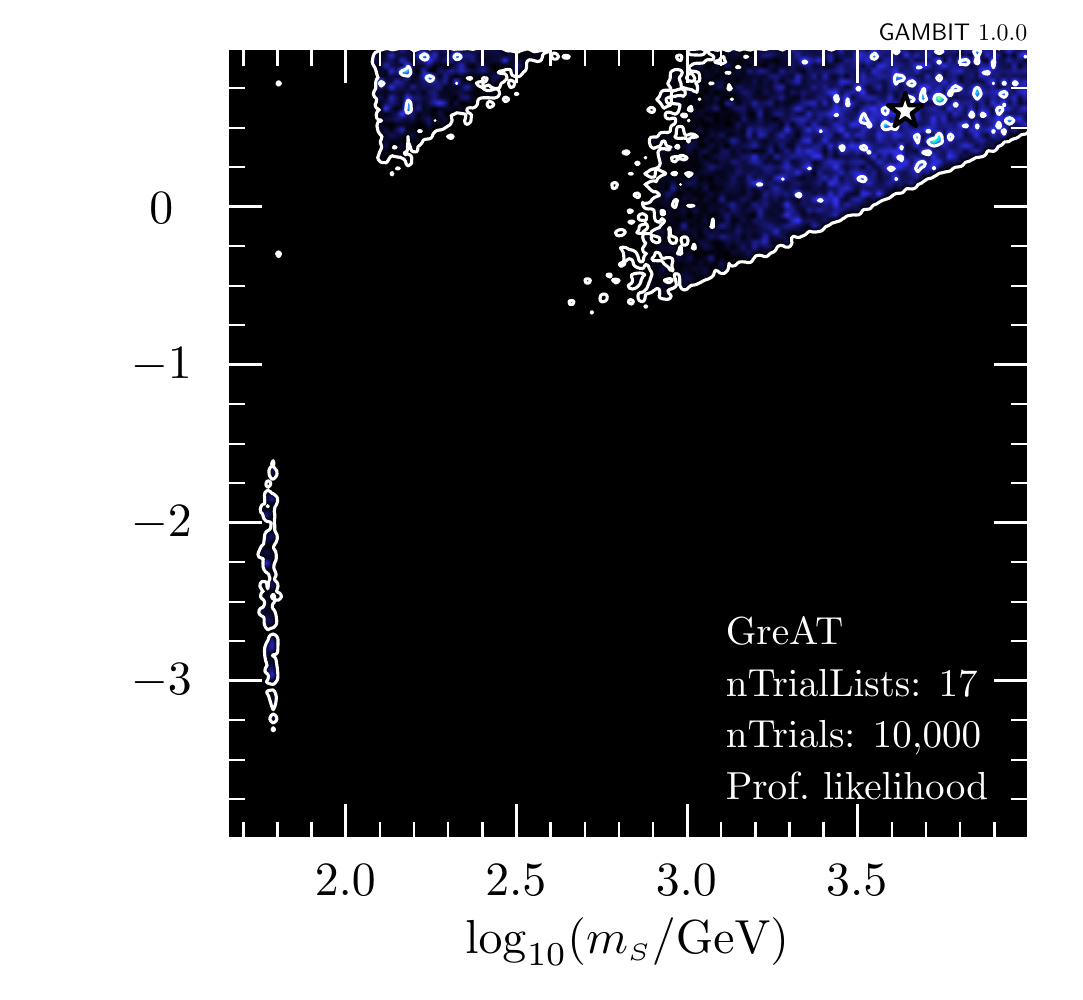}\hspace{-3mm}%
  \includegraphics[height=0.234\linewidth]{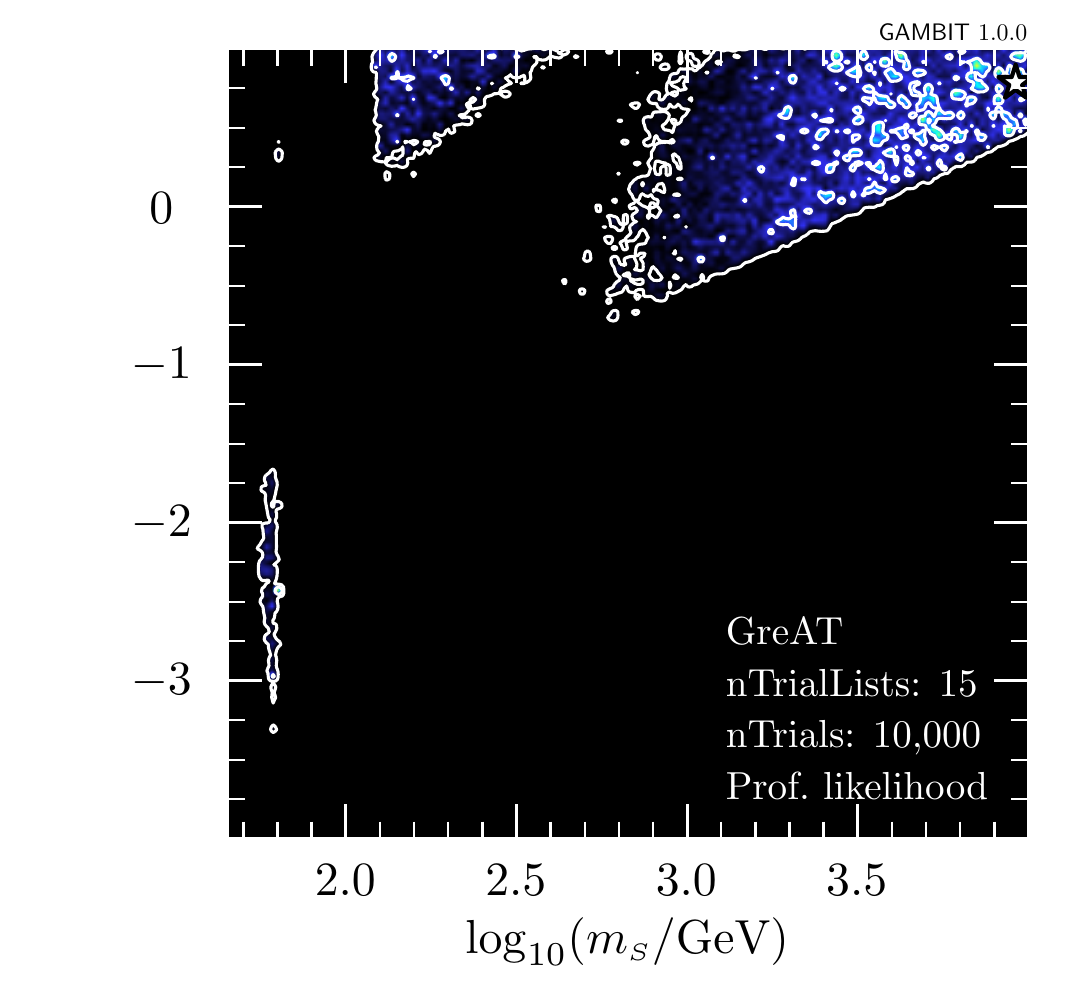}\hspace{-3mm}%
  \includegraphics[height=0.234\linewidth]{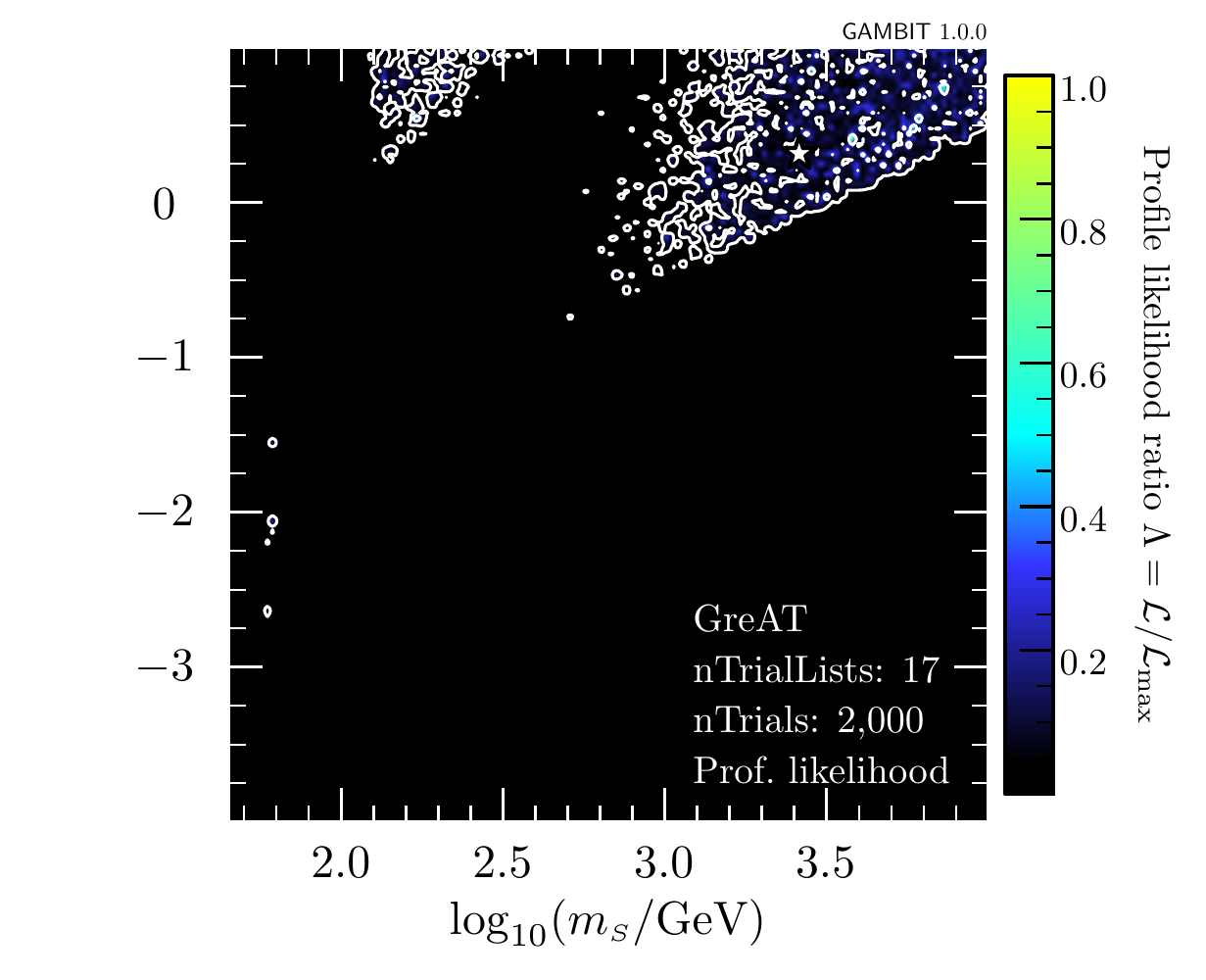}
  \caption{Profile likelihood ratio maps from a 15-dimensional scan of the scalar singlet parameter space, using the \great sampler with various numbers of chains (\cpp{nTrialLists}) and chain lengths (\cpp{nTrials}).  The maximum likelihood point is shown by a white star.}
  \label{fig:great_plots}
\end{figure*}

\begin{figure*}[tp]
  \centering
  \includegraphics[height=0.234\linewidth]{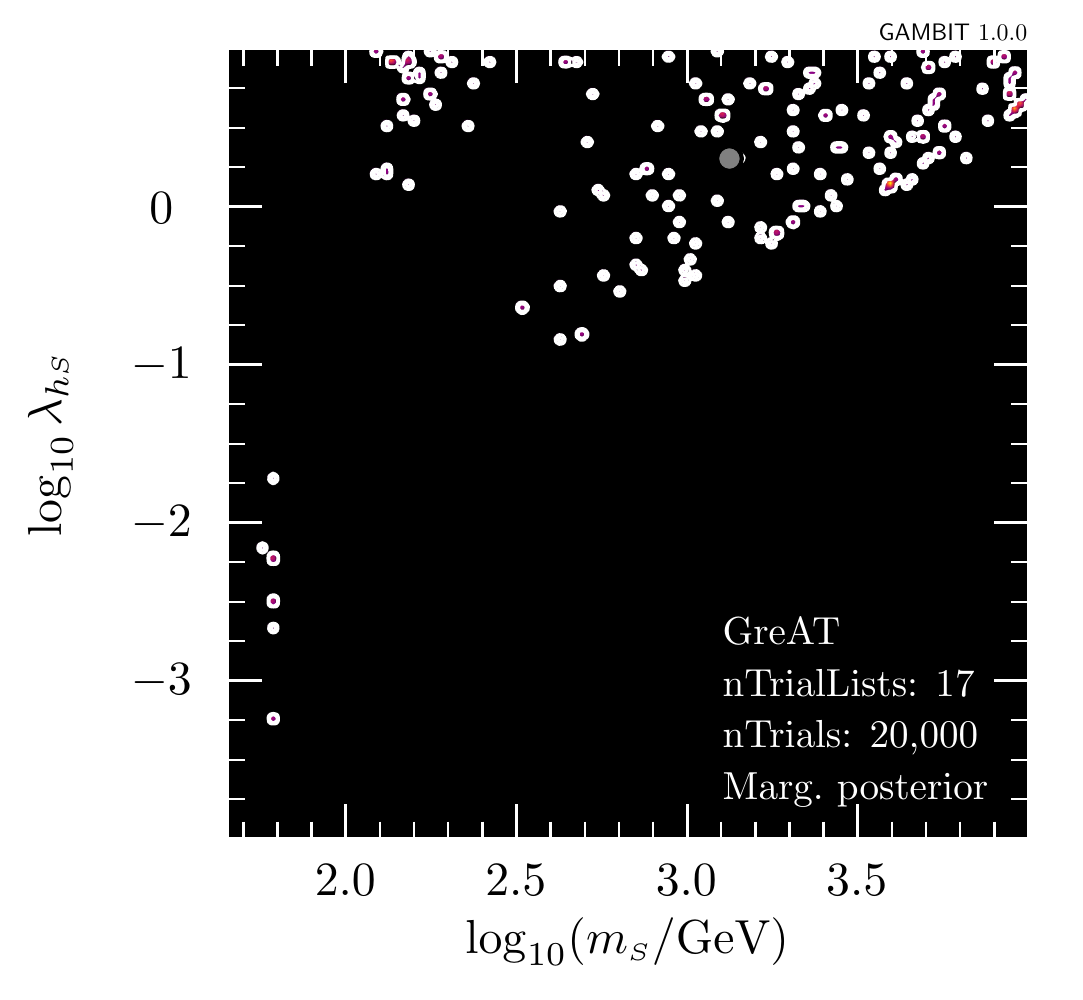}\hspace{-3mm}%
  \includegraphics[height=0.234\linewidth]{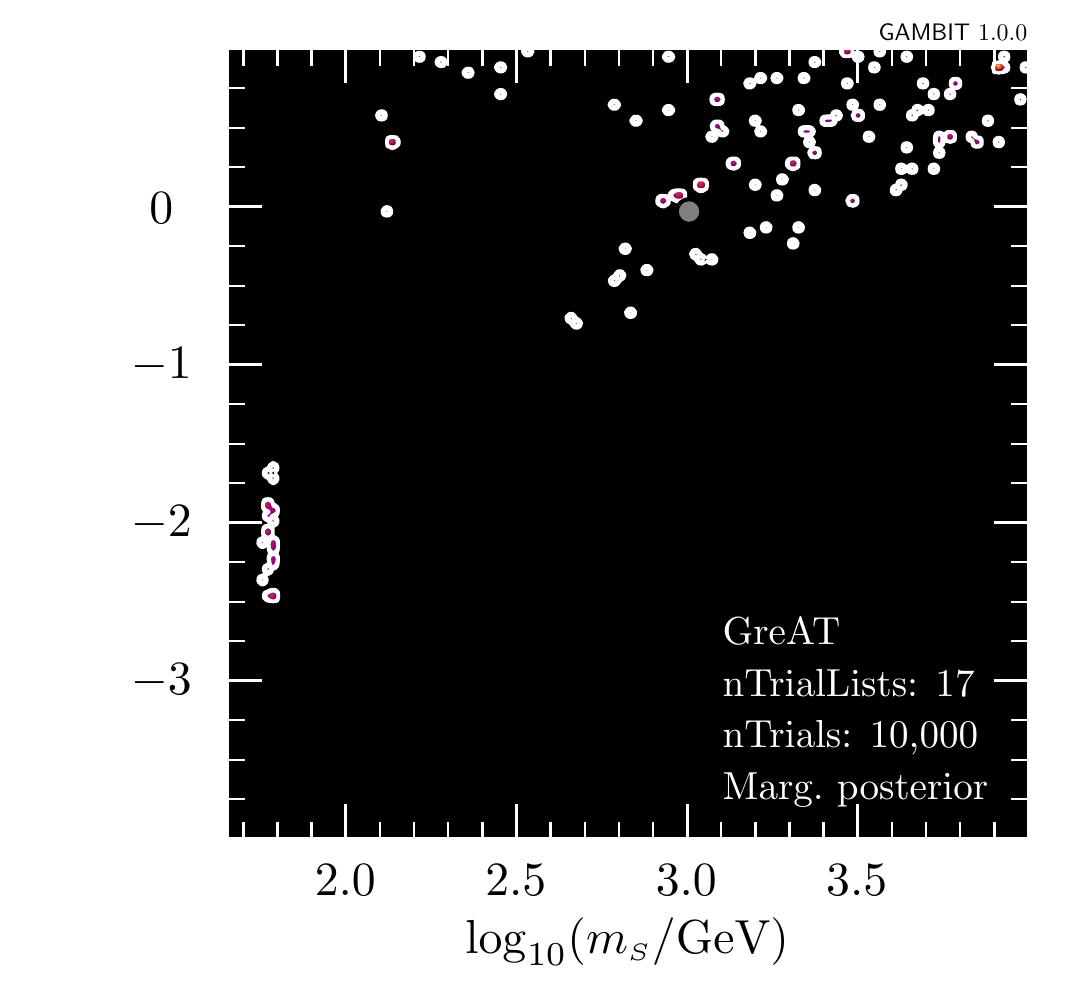}\hspace{-3mm}%
  \includegraphics[height=0.234\linewidth]{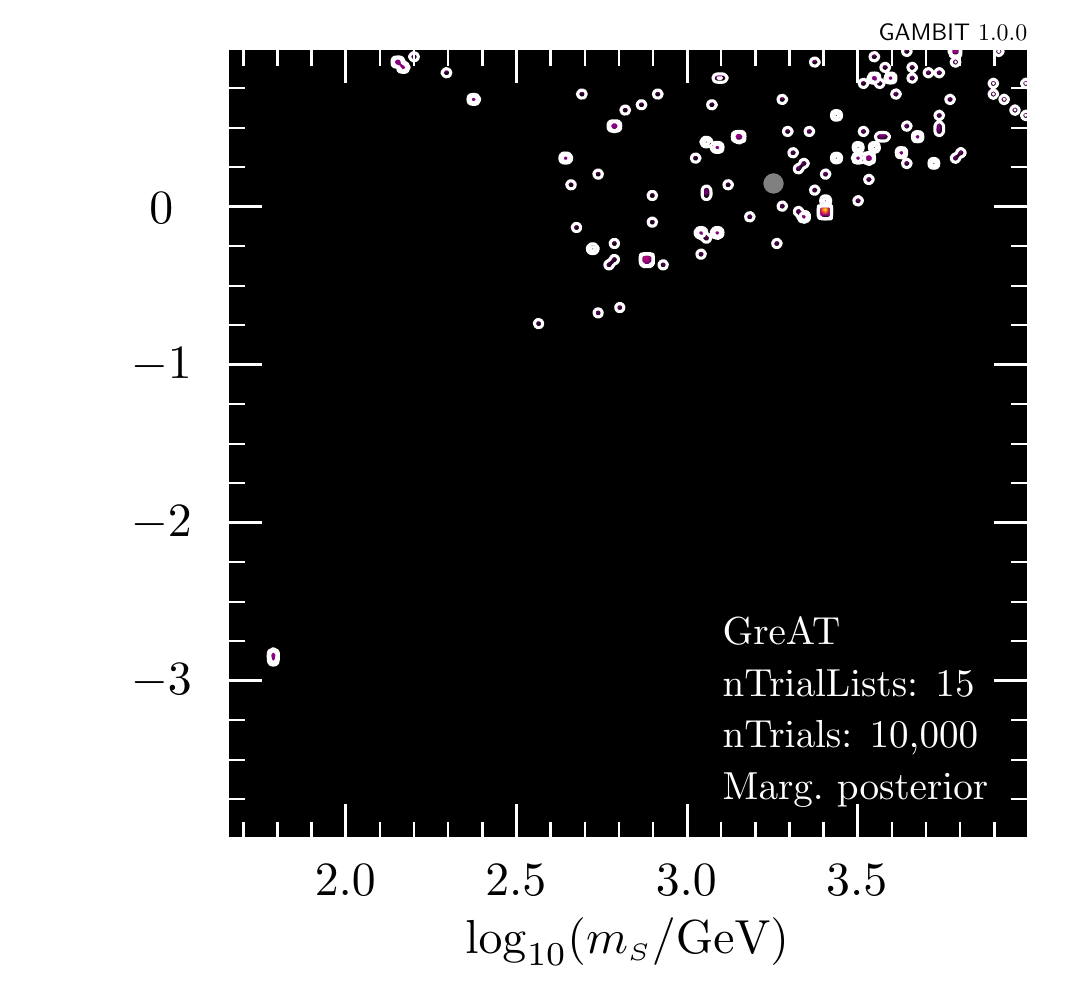}\hspace{-3mm}%
  \includegraphics[height=0.234\linewidth]{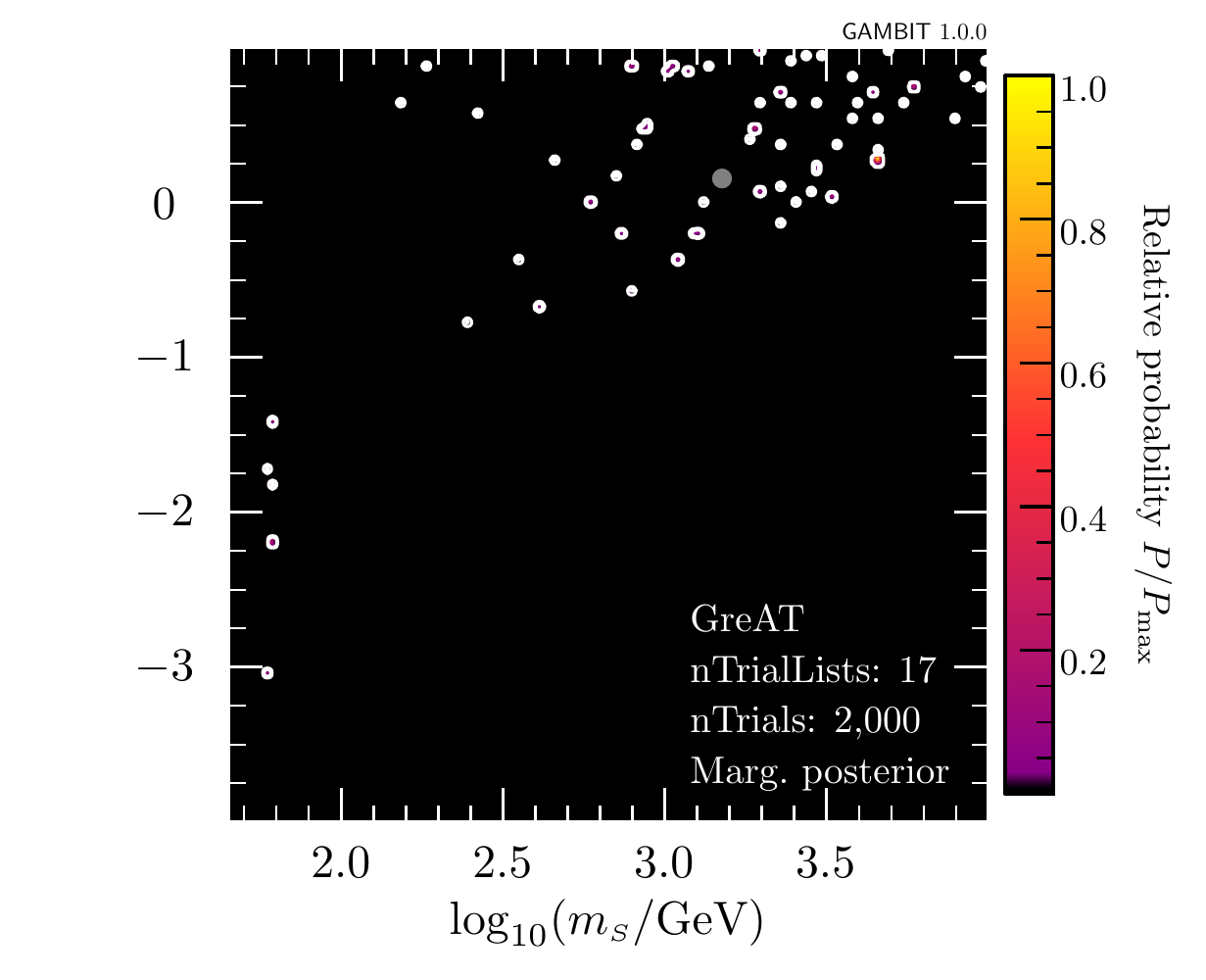}
  \caption{Marginalised posterior ratio maps from a 15-dimensional scan of the scalar singlet parameter space, using the \great sampler with various numbers of chains (\cpp{nTrialLists}) and chain lengths (\cpp{nTrials}).  Note that the colourbar strictly only applies to the rightmost panel, and that colours map to the same enclosed posterior mass on each plot, rather than to the same iso-posterior density level (i.e.~the transition from red to purple is designed to occur at the edge of the $1\sigma$ credible region, and so on).  The posterior mean is shown with a grey bullet point.}
  \label{fig:great_plots_post}
\end{figure*}

\subsection{\great}\label{sec:great}

The Grenoble Analysis Toolkit (\great \cite{great}) is a traditional Metropolis-Hastings MCMC able to sample parameters in parallel using multiple independent chains.  The number of chains is controlled by the \cpp{nTrialLists} parameter, and the number of points to run each chain for is controlled by \cpp{nTrials}.  No other convergence criteria are available.

For all dimensionalities, we consider \cpp{nTrials} values of 100, 200, 500, 1000, 2000, 5000 and 10\,000.  For scans in $N_{\text{dim}} = 7$ or 15 dimensions, we test \cpp{nTrialLists} values of $N_{\text{dim}}$, $N_{\text{dim}}+1$ and $N_{\text{dim}}+2$.  For the two-dimensional scans, we consider a larger range, setting \cpp{nTrialLists} to 2, 4, 24 and 48.  We plot a selection of these results in Fig.\ \ref{fig:GreAT}.

In two dimensions, we see that more chains result in some improvement in the reliability of the algorithm in uncovering competitive values of the best-fit likelihood. Unsurprisingly, Fig.\ \ref{fig:GreAT} also illustrates a tendency for longer chains to uncover slightly better fits.  These trends are both borne out substantially more strongly in seven and fifteen dimensions.  Visual inspection of the profile likelihood maps in Fig.\ \ref{fig:great_plots} indicates that beyond \cpp{nTrials} of about 1000, these improvements in best-fit likelihood with increasing numbers of chains do not come with any substantial impact on the overall quality of sampling across the rest of the parameter space.  We do notice a small runtime improvement, however. For example, two two-dimensional scans, each with 10\,000 samples per chain, took 119\,min to complete with \cpp{nTrialLists} = $48$, but 165\,min with \cpp{nTrialsLists} = 4.  The best-fit log-likelihoods returned by the two scans were equal to the third significant figure.  This timing difference reflects the improvement in acceptance that can be achieved when \great is able to draw on many different chains for constructing its correlation matrix.

In Fig.\ \ref{fig:great_plots_post}, we show the posterior maps resulting from the final set of independent samples returned by \great after its thinning process.  Clearly, none of the scans we have run produce enough independent samples for a convergent map of the posterior, at least at the relatively high bin resolution that we employ for these tests.

For all scans, we observe that a minimum value between 1000 and 10\,000 for \cpp{nTrials} is required in order to achieve a consistent value for the best-fit log-likelihood.  We also notice that very low values (below $\sim$1000) map the profile likelihood rather poorly.  The value of \cpp{nTrialLists} appears to be less crucial to the quality of the result; in general, values of $N_{\text{dim}}+1$ and above appear to give relatively stable results when coupled with \cpp{nTrials} $\gtrsim$ 10\,000.  Substantially longer chains (\cpp{nTrials} $\gg$ 10\,000) would probably be required to obtain high-resolution posterior maps.

\subsection{The effect of dimensionality on performance}\label{sec:dimensional_performance}

\begin{figure*}[tp]
  \centering
  \includegraphics[width=0.48\linewidth]{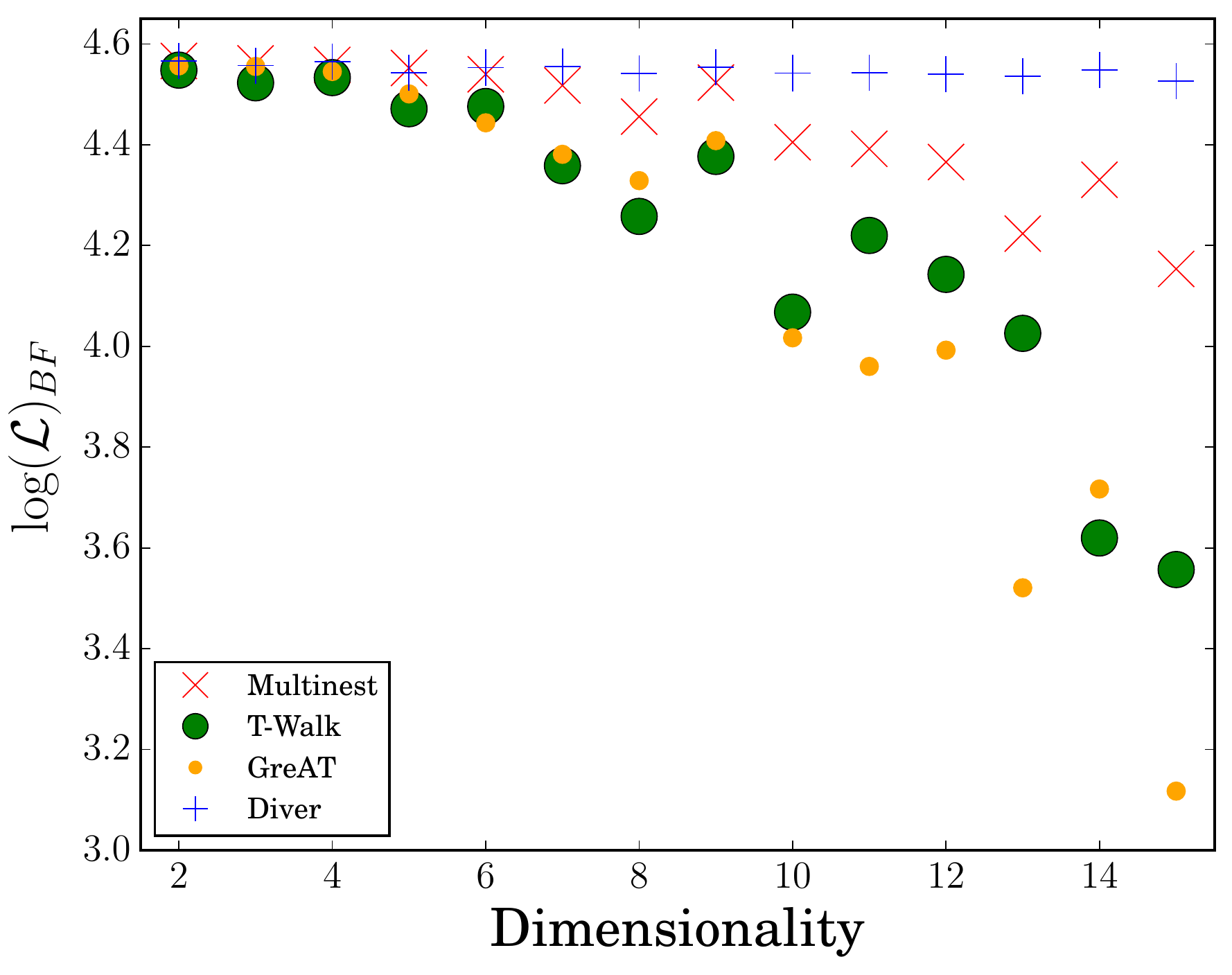}\hspace{0.04\linewidth}\includegraphics[width=0.48\linewidth]{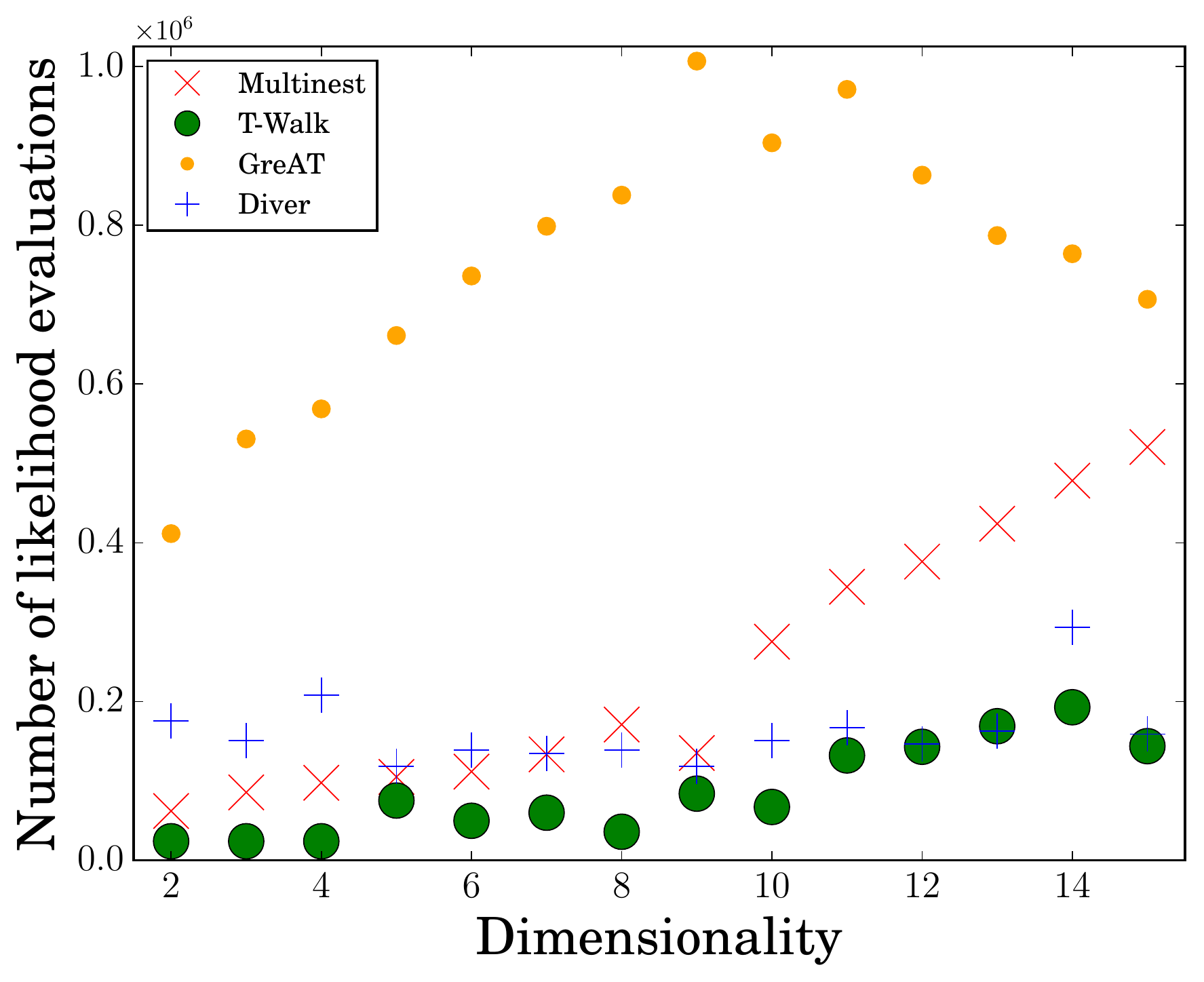}
  \caption{Best-fit log-likelihood (\textit{left}) and number of likelihood evaluations (\textit{right}) as a function of dimensionality, for all four scanning algorithms, using a fixed computing configuration and scanner settings.  Note that the likelihood is dimensionful, leading to $\mathcal{L_{\rm BF}}>1$ \cite{gambit}.}
  \label{fig:dimensionality}
\end{figure*}

We have studied scanner performance in detail for two, seven and fifteen-dimensional parameter spaces, by increasing the number of nuisance parameters; each additional parameter adds an additional Gaussian component to the likelihood, and modifies the existing components.  We now fix the computing configuration and scanner parameters (or apply a consistent scaling with dimensionality, where appropriate), and carry out scans for every possible dimensionality from two to fifteen.  The results of these tests are presented in Fig.\ \ref{fig:dimensionality}.  The scanner settings we use for these tests are:
\begin{description}
\item\diver: \fortran{NP} = 20\,000, \fortran{convthresh} = $10^{-3}$
\item\MultiNest: \lstinline{nlive} = 20\,000, \lstinline{tol} = $10^{-3}$
\item\twalk: \cpp{chain_number} = number of \mpi processes + $N_{\text{dim}}+1$,  $tol$ = \yaml{sqrtR} $-$ 1 = 0.05
\item\great: \cpp{nTrials} = 2000, \cpp{nTrialsList} = $N_{\text{dim}}+1$
\end{description}

To reach convergence, \great requires significantly more likelihood evaluations for a larger number of dimensions.  Although this is undoubtedly in part due to the increased number of chains used in higher dimensions, even with this increased number of evaluations, the best-fit log-likelihood is not competitive with that achieved by either \diver or \MultiNest.  If we demanded that all scanners must achieve the same quality of best fit, then it is clear that \great would require an even greater number of function evaluations to achieve this.  Judging from the quality of best fit, the decrease in the number of evaluations required for convergence by \great in higher dimensions is clearly the result of spurious early convergence, rather than any increase in performance.

\diver performs extremely well at all dimensionalities, out-performing the other three scanners in terms of quality of best fit at $N_\text{dim}\ge 10$.  It also achieves this using a consistent number of likelihood evaluations across the full dimensionality range.  \MultiNest is able to achieve a competitive best-fit log-likelihood up until $N_\text{dim}\sim10$, however this comes with a steady increase in the number of evaluations with respect to dimensionality.  \twalk runs for a consistent number of likelihood evaluations across all dimensions, despite the required increase in number of chains, yet the best-fit deteriorates significantly with respect to dimensionality, in much the same way as it does with \great.  The ensemble version of the MCMC algorithm implemented by \twalk essentially provides the same best-fit performance as the regular MCMC (\great), but with a significant improvement in efficiency with increasing dimension.  Overall, at least in this parameter space, \diver appears to be the scanner of choice for larger dimensions.

\begin{figure}[tp]
  \centering
  \includegraphics[width=1\linewidth]{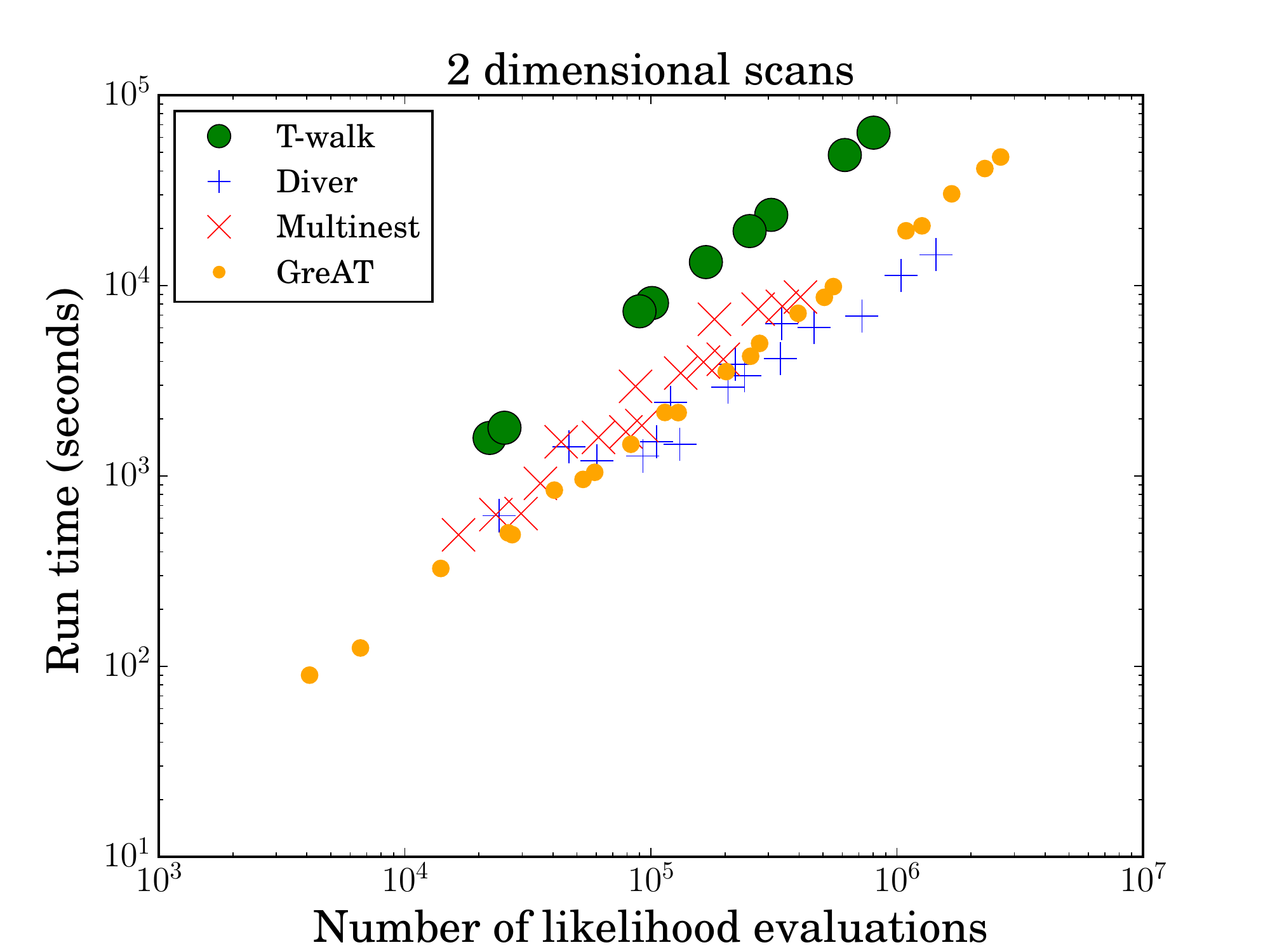}
  \includegraphics[width=1\linewidth]{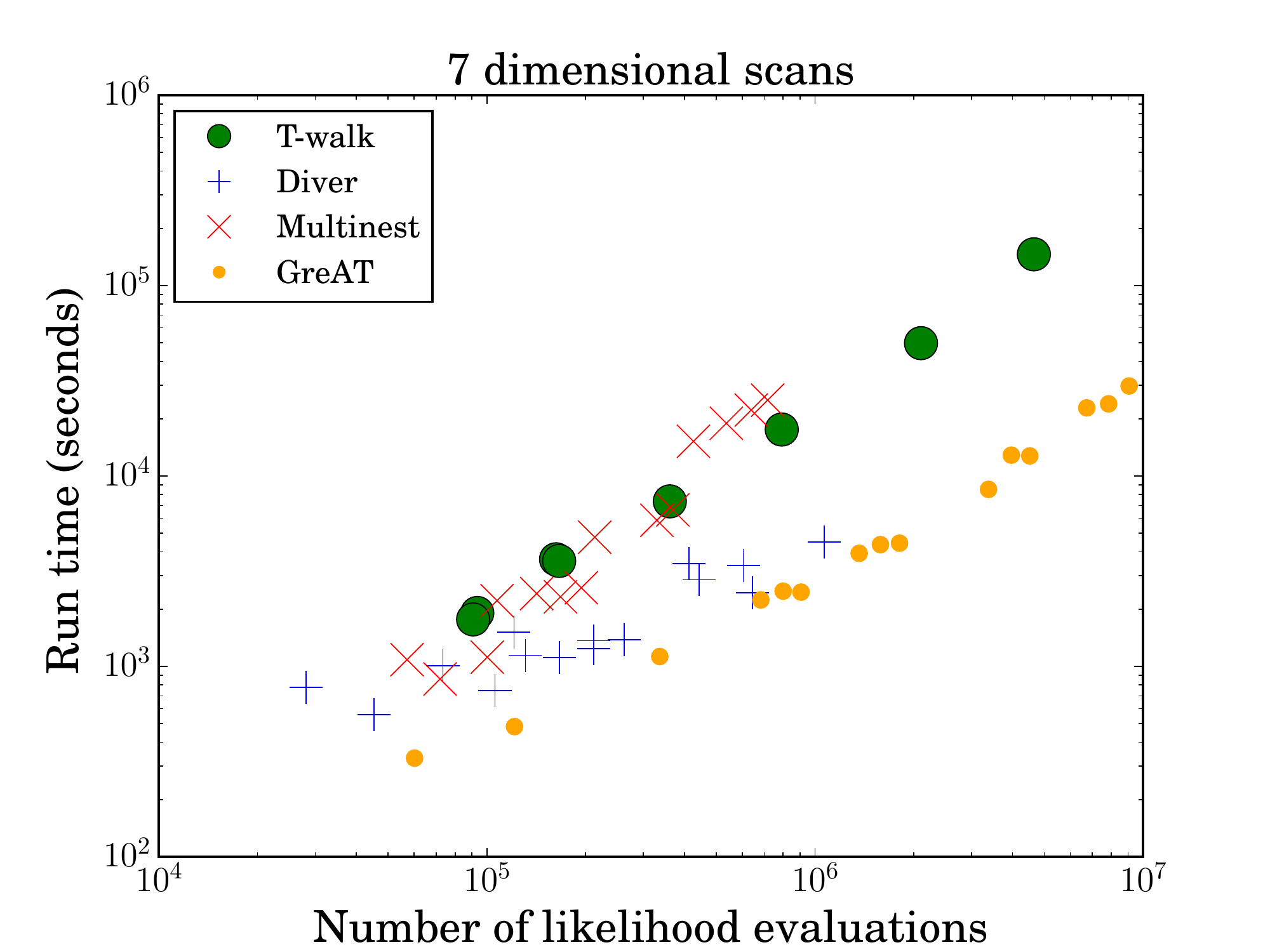}
  \includegraphics[width=1\linewidth]{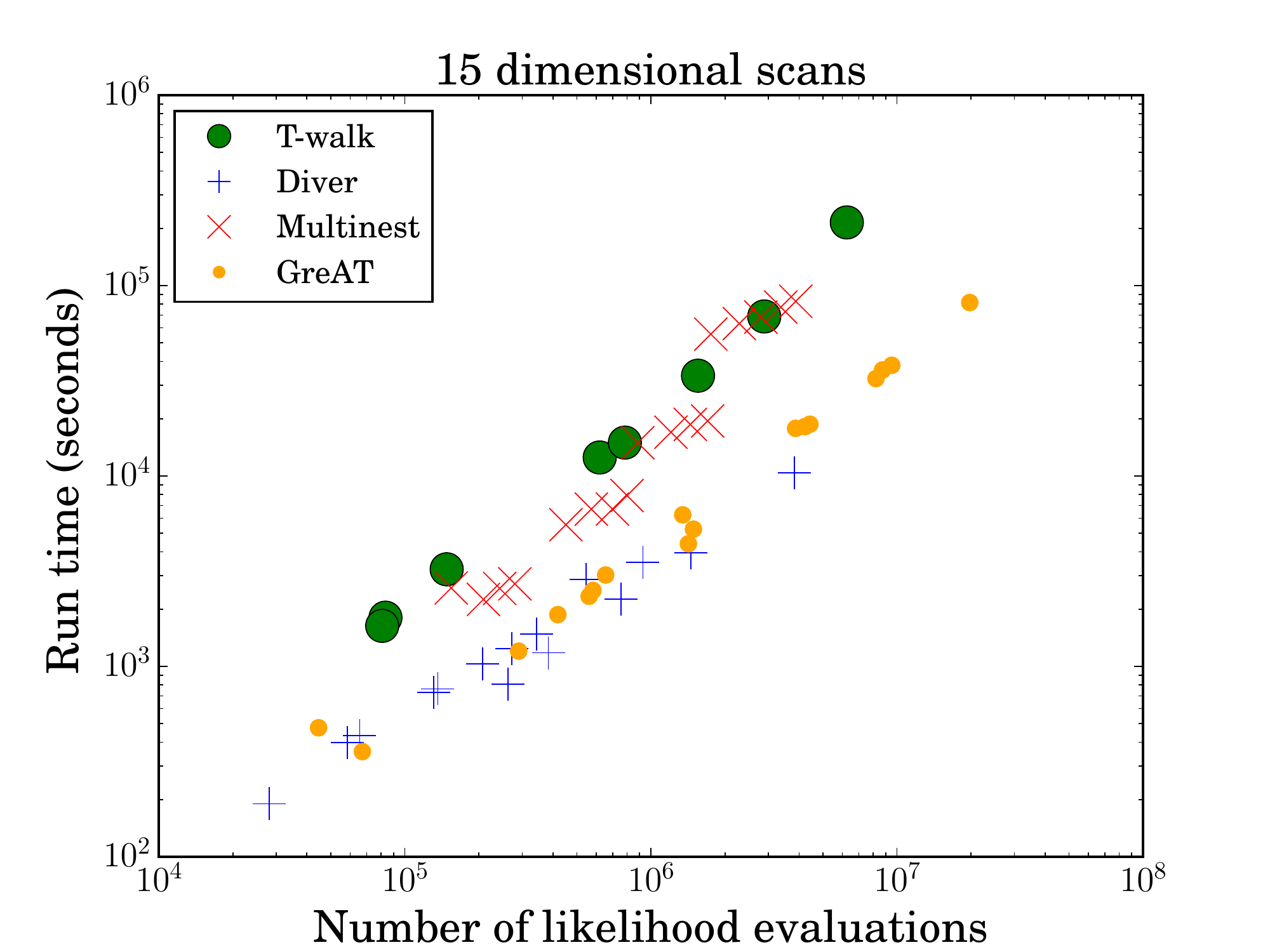}
  \caption{The real time required as a function of likelihood evaluations for two- (\textit{upper}), seven- (\textit{middle}) and fifteen-dimensional (\textit{lower}) scans.}
  \label{fig:timing}
\end{figure}

\begin{figure}[tp]
  \centering
  \includegraphics[width=1\linewidth]{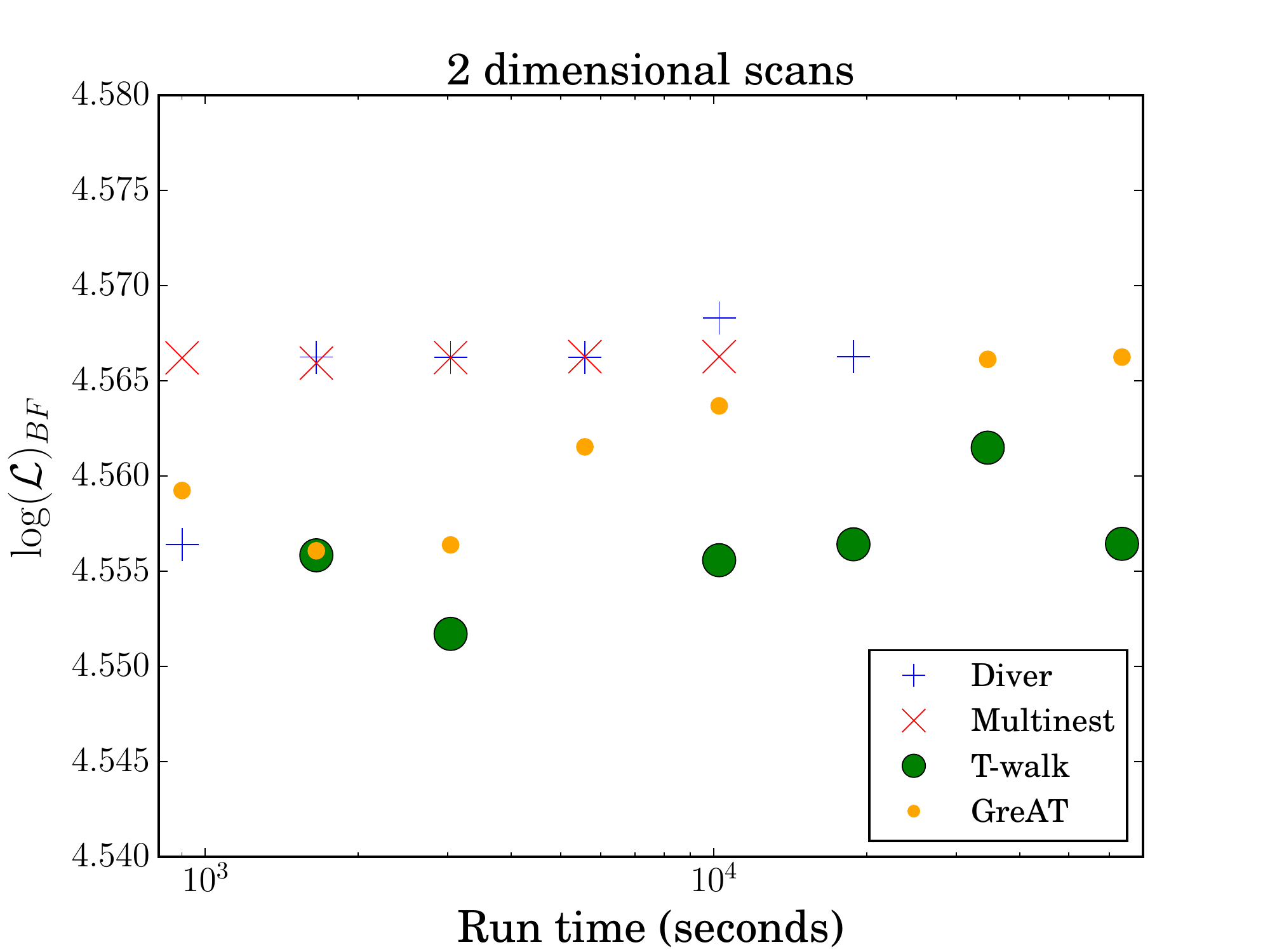}
  \includegraphics[width=1\linewidth]{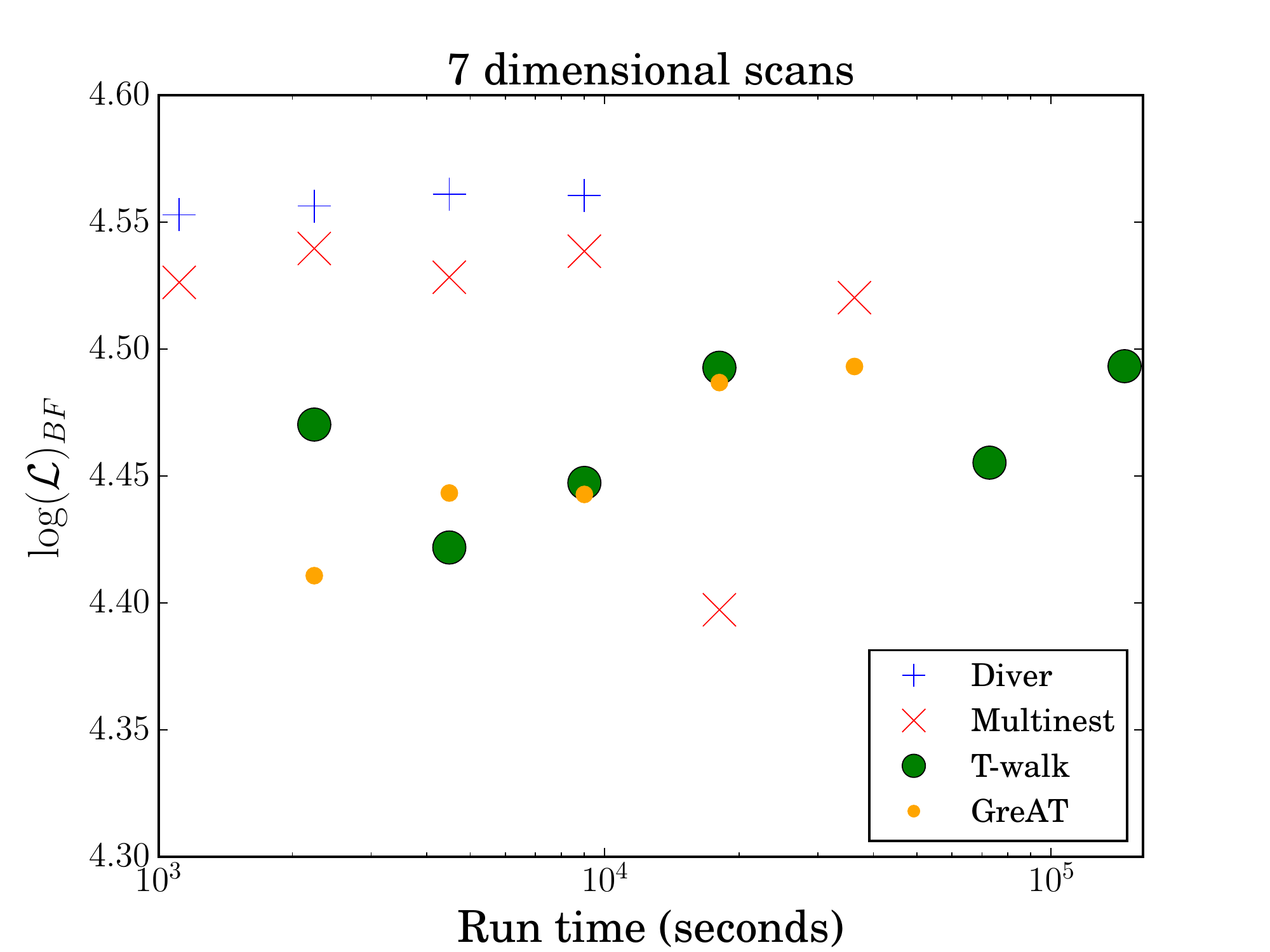}
  \includegraphics[width=1\linewidth]{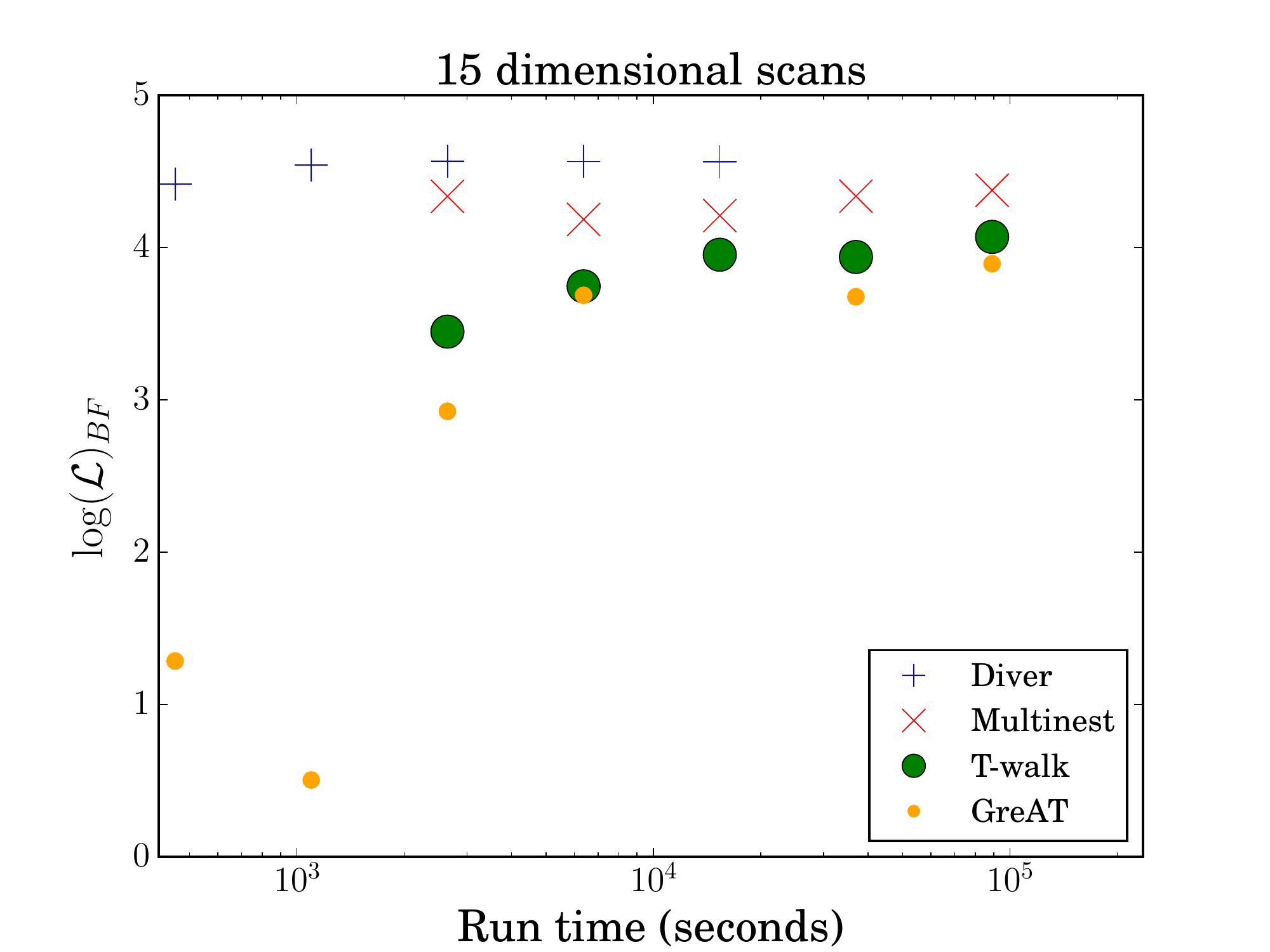}
  \caption{The best-fit likelihood achieved by each scanner within a given time limit, for two (\textit{upper}), seven (\textit{middle}) and fifteen-dimensional (\textit{lower}) scans.}
  \label{fig:timing_2}
\end{figure}

\subsection{Scanning efficiency}\label{sec:time_performance}

The number of likelihood evaluations required to reach convergence is not the only reasonable metric for scanner efficiency. In general the number of evaluations is used as a proxy for time, as the likelihood evaluations are generally expected to be the bottleneck in most scans -- but it is also illustrative to look directly at actual runtime.  The efficiency of a scanner can be degraded by poor use of parallel processing capabilities, or by complicated calculations performed between likelihood evaluations.  This can lead to a divergence between the apparent performance assessed purely by number of function evaluations, and the true walltime needed.  We therefore record the actual CPU time used for all scans, and compare with the total number of likelihood evaluations in Figure \ref{fig:timing}.\footnote{Here we use 24 processes for the two dimensional scans, and 240 processes for the seven and fifteen-dimensional scans, so time comparisons should not be drawn between the two-dimensional plots and the seven/fifteen-dimensional ones.}

Fig.\ \ref{fig:timing} shows that dimensionality has a significant impact on the relative efficiency per likelihood evaluation of each algorithm.  For two-dimensional scans, we see that \twalk performs the least efficiently, while the other algorithms are reasonably similar.  However, in the higher-dimensional parameter spaces, the efficiency of the nested sampling in \MultiNest becomes comparable to the MCMC in \twalk, whereas \great and \diver remain relatively efficient.  The reduction in performance by \multinest in higher dimensions is probably due to the complicated calculations required to perform its ellipsoidal sampling of multi-dimensional modes.  These calculations must be performed between each generation of live points.  Another potential cause of the performance reduction in \twalk and \MultiNest is the intrinsic level of parallelisability of their algorithms, relative to the other scanners. For problems with larger numbers of parameters, we observe that the most efficient sampling algorithms are \great and \diver, with both exhibiting the lowest average latency between likelihood evaluations.

In Fig.\ \ref{fig:timing_2}, we summarise the overall performance of the algorithms in terms of time and fit quality at each dimensionality.  We bin all completed test scans logarithmically in the total convergence time, and for each sampler, choose the scan in each bin with the best fit.  There are no \diver points in the longer bins, simply because the longest \diver scans took less time than the longest scans with other samplers.  \diver clearly outperforms the other algorithms in high dimensions by this metric as well, finding a better fit in a shorter runtime than the other three algorithms.  It is also important to note the vertical scales in Fig.\ \ref{fig:timing_2}, where the likelihood values span a much wider range in seven and fifteen dimensions than in two. On close inspection however, we can see even in two dimensions that \diver and \MultiNest obtain better fits in less time than either \twalk or \great.

We also notice that in higher dimensions, although \twalk takes less evaluations than \great, both take a similar amount of runtime to reach convergence, suggesting that \twalk's reduced sampling is offset by additional algorithmic complexity requiring more extended calculations \textit{between} samples.

\subsection{Posterior sampling}
\label{sec:posterior}

Figs.\ \ref{fig:multinest_plots_post}, \ref{fig:twalk_plots_post} and \ref{fig:great_plots_post} show the posterior sampling abilities of \twalk, \multinest and \great, respectively.  The best-quality posterior in \twalk took 9\,hr, while in \multinest the best posterior we show took over 21\,hr. The highest-quality \great posterior we show took even longer, and is clearly a poorer result than what was achieved by \twalk and \multinest.

Comparing the quality of the posterior maps achieved by \twalk and \multinest reveals some interesting trends.  Firstly, despite taking less than half the runtime, the best posterior map returned by \twalk appears to have given a better-converged map of the posterior than the best effort by \multinest.

We can also see a distinct tendency for the shapes of the contours returned by \multinest to erroneously `smooth away' sharper features in the posterior, which are mapped far more carefully and accurately by \twalk.  This is most likely due to the ellipsoidal sampling method intrinsic to \multinest, which biases the algorithm towards finding new live points within elliptically-shaped regions encompasing its current population of points.  This makes it rather easy for the algorithm to miss sharp features in the posterior, such as the low-coupling tip of the highest-mass mode in the scaler singlet parameter space, which would protrude beyond the approximate contour defined by the bounding ellipsoids in \multinest.

We also see that posterior maps become poorer for shorter scans with both \twalk and \multinest, but in quite distinct ways.  In \multinest, a scan performed with too few live points or too high a tolerance will give a poorly-sampled posterior with few favoured regions, essentially because the algorithm has only managed to locate the most dominant modes of the posterior at the outset.   In contrast, a poorly-converged \twalk scan, particularly one with a large \yaml{tol} value, will typically instead result in a map that includes all relevant modes across the parameter space, but with their relative contributions poorly determined, such that they appear alongside a number of other, spurious, favoured regions.  When inspecting a posterior map, particularly from brief scans, it is important to be aware of these differences between the algorithms.

\begin{table}[tp]
\caption{The recommended starting parameters for each scanner available in \GB \textsf{1.0.0}.  Here $N_\text{dim}$ is the dimensionality of the scan and $N_{\text{\mpi}}$ is the number of (distributed-memory) parallel processes available to \GB.}\label{table:ideal_params}
\centering
\begin{tabular}{l l l}
\hline
Scanner & Parameter & Recommendation\\
\hline
\MultiNest & \cpp{nlive} & $2\times 10^{4}$ \\
  & \cpp{tol} & $10^{-3}$ \\
\hline
\diver & \fortran{NP} & $2\times 10^{4}$ \\
  & \fortran{convthresh} & $10^{-3}$ \\
  \hline
\twalk & \cpp{chain_number} &$ N_\text{dim}+N_{\text{\mpi}}+1$ \\
  & \cpp{sqrtR} & $<1.01$ \\
    \hline
\great & \cpp{nTrialLists} &$ N_\text{dim}+1$\\
  & \cpp{nTrials} & $10^{4}$ \\
\hline\end{tabular}
\end{table}

\subsection{Discussion}\label{sec:compare}

We have investigated the performance of the four major samplers available in \scannerbit as part of \GB \textsf{1.0.0}, over a range of algorithmic settings and parameter space dimensionalities.  In Table \ref{table:ideal_params}, we summarise our recommended values for the two most important settings of each scanner.  These are intended as starting values that will give reasonably robust results.  However, every parameter space is different and a publication quality results may require significantly more stringent settings, in order for final results to be sufficiently robust.  See Secs.\ \ref{sec:diver}--\ref{sec:great} for more detailed recommendations.

We are also able to make detailed comparisons between the four scanning algorithms.  In Secs.\ \ref{sec:dimensional_performance} and \ref{sec:time_performance} it became evident that differential evolution, as implemented in \diver, consistently out-performs the other algorithms in the computation of profile likelihoods.  This becomes particularly clear in high dimensions, where \diver leads the other algorithms in likelihood mapping, the quality of the best fit found, and overall efficiency.

The true best-fit point for this likelihood is located in the low-mass region, regardless of the number of additional free parameters.  The scanners did not always locate this point, and in many cases located a best-fit in one of the high-mass modes.  Although locating this point in two dimensions is less challenging (see Appendix \ref{app:scanner_comparisons}), once the dimensionality is increased, only \diver (with most stringent convergence criteria) was able to successfully locate the best fit in the low-mass mode.  All other scans converged to a best fit in a completely different mode, demonstrating the value of using alternative algorithms to fully understand the parameter space.

For careful mapping of the posterior, we find that \twalk is the most effective algorithm, followed by \MultiNest and \great.  \twalk manages to sample the posterior distribution at higher resolution in less time than the other two scanners, and avoids the ellipsoidal biases that appear to afflict \multinest.  For computing low-resolution posteriors however, \multinest has the advantage that it requires less parameter tuning than \twalk, and can more quickly identify which are the most relevant posterior modes.

In many cases, having both Bayesian and frequentist interpretations of results is desirable.  This makes it necessary to use a sampler able to effectively sample the posterior, such as \MultiNest or \twalk.  However, our tests show that this is best performed \textit{after} the likelihood function has been carefully mapped with another sampler, in order to find all modes.  For example, in Figure \ref{fig:multinest_plots}, \multinest has completely missed the likelihood mode at low mass.  This mode was successfully found by all three of the other samplers.  If \MultiNest were to be used exclusively, then this region --- which contains best-fit points degenerate with those in the other modes --- would be completely unexplored.  However, with the knowledge gained from the other scanners, a localised study can be performed using \MultiNest around the low-mass region (a technique used in Refs.\ \cite{CMSSM,SSDM}), in order to correctly evaluate the full posterior.  In this way, the ability to use complementary scanners significantly improves the statistical robustness of results.

For lower-dimensional problems where both posterior distribution and profile likelihood are required, \MultiNest could potentially be used solo, to save repeating analyses with multiple scanners.  We find that it is able to locate all modes when scanning only the two-dimensional parameter space, and that it is reasonably efficient compared with the other algorithms.  In general though, relying on only a single sampling algorithm is risky.

The two MCMC-based scanners available in \GB \textsf{1.0.0}, \twalk and \great, provide the user with a somewhat more traditional class of sampling methods.  Although these algorithms are demonstrably less effective scanners in higher-dimensional profile likelihood problems, they may suit lower-dimensional studies better.

Notably, our tests here are based on only one physical problem; although this is intended as a realistic example, no single example could ever represent the full diversity of problems that might be encountered.  Other parameter spaces and likelihood functions may therefore reveal different trends to those we have observed with the scalar singlet model.

\section{Conclusions}
\label{sec:conclusions}

In this paper we have presented \scannerbit, the statistical and sampling module for the new global fitting package \gambit.  \scannerbit manages the overhead associated with choosing parameter combinations and applying prior transforms, and offers an extremely flexible framework into which any existing sampling code can be easily integrated.  It is able to perform sampling in standard random, grid and raster patterns, or employ more sophisticated statistical methods including nested sampling, differential evolution, Markov Chain Monte Carlo and ensemble Monte Carlo. It interfaces seamlessly with the \gambit printer system to allow statistical and physical outputs of parameter scans to be saved to a common format of choice, entirely independent of the model under investigation or the sampling algorithm in use.  It can also post-process existing sets of samples previously computed and saved with \gambit.  \scannerbit can be used from within \gambit, or as a standalone package independent of \gambit, allowing the user to connect to an arbitrary likelihood function and sample it using their desired algorithm.

In addition to \scannerbit itself, we have presented a new standalone sampling package based on differential evolution: \diver.  \diver features a full suite of differential evolution variants, from standard rand/1/bin to adaptive and discrete versions, and additional operation modes designed to provide approximate Bayesian results.  We have also presented a new implementation of the \twalk algorithm, implemented natively in \scannerbit.

We compared the performance of the four main sampling algorithms interfaced to \scannerbit in \gambit \textsf{1.0.0}: \diver, \multinest, \twalk and \great.  We found that for profile likelihood analyis at low dimensionality, \diver and \multinest outperform \twalk and \great, and provide roughly equivalent performance to each other.  At higher dimensions (10 and above), \diver substantially outperforms the other three algorithms on all metrics.  \twalk provides a more accurate, timely and complete mapping of the Bayesian posterior than \multinest, although \multinest identifies the primary posterior mode more quickly.

\scannerbit and \gambit can be obtained from \href{gambit.hepforge.org}{gambit.hepforge.org}, and are both released under the terms of the standard 3-clause BSD license.\footnote{\href{http://opensource.org/licenses/BSD-3-Clause}{http://opensource.org/licenses/BSD-3-Clause}.  Note that \textsf{fjcore} \cite{Cacciari:2011ma} and some outputs of \flexiblesusy \cite{Athron:2014yba} (incorporating routines from \SOFTSUSY \cite{Allanach:2001kg}) are also shipped with \GB \textsf{1.0}.  These code snippets are distributed under the GNU General Public License (GPL; \href{http://opensource.org/licenses/GPL-3.0}{http://opensource.org/licenses/GPL-3.0}), with the special exception, granted to \GB by the authors, that they do not require the rest of \GB to inherit the GPL.}  \diver can be downloaded from \href{diver.hepforge.org}{diver.hepforge.org}, or installed automatically from within \gambit by simply typing \term{make} \term{diver}; it is released under a license that makes it free to use and distribute for academic and non-profit purposes.

\section{Acknowledgements}
We thank the other members of the \gambit Collaboration for helpful discussions.  \gambitacknos

\appendix

\section{Sources, options and outputs of the \diver package}
\label{app:diver}

\subsection{Sources}
\label{sec:sources}

Each of the source files located in \term{diver/src/} contains a single eponymous \Fortran module:
\begin{description}
	\item\term{de.f90}: the main module of \diver, containing the function \protect\fortran{diver()}, by which the package is invoked.
	\item\term{init.f90}: contains routines to set all parameters for the run and to initialise the population every generation
	\item\term{mutation.f90}: contains routines to allow standard DE mutation following equation~\ref{eq:mutation} and self-adaptive
		mutation using jDE or $\lambda$jDE (see Sec.~\ref{subsec:jDE}).
	\item\term{crossover.f90}: contains routines to allow binomial or exponential crossover, or self-adaptive crossover using jDE or $\lambda$jDE.
	\item\term{selection.f90}: performs selection of the next generation of vectors, applies boundary conditions, and removes duplicate
		vectors to ensure population diversity (see Sec.~\ref{subsec:duplicates}).  If \mpi is used, this is where most \mpi routines are called.
	\item\term{converge.f90}: checks whether the population has converged sufficiently to end the current DE run.
	\item\term{io.f90}: saves the parameters of the run as well as the population at regular intervals.  Contains routines to
		continue a run that was stopped partway through.
	\item\term{evidence.f90}: contains routines used for calculating approximate Bayesian evidence values.
	\item\term{posterior.f90}: contains routines used for calculating approximate Bayesian posterior probability density functions.
	\item\term{detypes.f90}: contains interfaces to the likelihood function and prior, as well as the definitions of the internal data types used by \diver.
	\item\term{deutils.f90}: contains utility routines.
	\item\term{cwrapper.f90}: acts as an interface between \plainC/\Cpp drivers and \term{de.f90}.
\end{description}

\subsection{Run options}
\label{sec:options}

Options for a \diver run are passed directly as arguments to \fortran{diver()} or \cpp{cdiver()}.  The required arguments are:
\begin{description}
	\item\fortran{double precision func()}: The function to optimise, assumed to be positive definite; should generally correspond to the negative log-likelihood for statistical scans.  See example driving programs for suggested use.  Must take the following arguments:
		\begin{description}
			\item\fortran{double precision params}: An array of size equal to the sum of $D$ (the dimensionality of the parameter space) and \fortran{nDerived}, the number of derived quantities to be output in the run.
			\item\fortran{integer fcall}: The total number of calls to \fortran{func}; should be incremented appropriately by the objective function.
			\item\fortran{logical quit}: A flag set by the objective function.  If this is ever set to \fortran{true}, \diver will save and quit at the end of the current generation.
			\item\fortran{logical validvector}: A flag set by \diver.  If this is \fortran{false}, the point in parameter space represented by \fortran{params} is outside the specified parameter boundaries, and should not be evaluated.
      \item\fortran{c_ptr context}: A context pointer, allowing the driving program to pass arbitrary information to \fortran{func}.  Can be modified in a call to \fortran{func}, and will retain its value the next time the function is called.
		\end{description}
	\item\fortran{double precision lowerbounds}:  An array of size $D$, giving the desired lower bounds of the parameter space.
	\item\fortran{double precision upperbounds}: An array of size $D$, giving the desired upper bounds of the  parameter space.
	\item\fortran{character path}: The path to which output files should be saved.
\end{description}

Other arguments are optional and default to sensible values if left unspecified.  Here we list these in the format \metavar{option}\fortran{[}\metavar{default}\fortran{]}:
\begin{description}
	\item\fortran{integer nDerived[0]}: The number of derived quantities to be calculated by the likelihood/objective function.  If \fortran{outputSamples} is \fortran{true}, these are saved in human-readable format along with the original parameters in a \term{.sam} file.
  \item\fortran{integer discrete[empty]}: A vector listing all dimensions of the parameter space that should be treated as discrete parameters.  See Sec.\ \ref{subsec:discrete} for details.
  \item\fortran{logical partitionDiscrete[false]}: Evolve discrete parameters as separate populations. See Sec.\ \ref{subsec:discrete} for details.
  \item\fortran{integer maxciv[2000 if doBayesian else 1]}: The maximum number of `civilisations' to run.  A civilisation is a full DE run with multiple generations, which terminates either because it has converged or reached generation number \fortran{maxgen}.  If \fortran{doBayesian} is \fortran{true}, \diver will run additional civilisations up to \fortran{maxciv} until the approximate Bayesian evidence has converged; if \fortran{doBayesian} is \fortran{false}, \diver will simply repeat DE optimisation \fortran{maxciv} times, and save the results as a single set of samples.
	\item\fortran{integer maxgen[300]}: The maximum number of generations for the DE run.  Usually the default convergence criterion will
		cause \diver to end the DE run before this number has been reached.
	\item\fortran{integer NP[10*size(lowerbounds)]}: The population size.  Larger populations take longer to run but are less likely to become trapped in local
		minima.  Small populations run more quickly because they require fewer likelihood/objective evaluations per generation, but they
		lack diversity and may converge prematurely.  The default is set to $10D$; we recommend that \fortran{NP} never be set to less than $D$. If \diver is invoked using \mpi, the actual population size will be increased from the requested size until it is a multiple of the number of \mpi processes to be used.
	\item\fortran{double precision F[0.7]}: The mutation scale factor(s); see Secs.~\ref{subsec:mutation} and \ref{subsec:mutation_strategies}.  This should
		be supplied as an array.  The scale factor, and the degree to which the population is spread out, together determine the
		radius around the population in which new points can be proposed.  For this reason, \fortran{F} should be smaller than 1,
		to help convergence, but not too small, to prevent premature convergence.
		This option is ignored when \fortran{jDE} or \fortran{lambdajDE} is \fortran{true}.
	\item\fortran{double precision Cr[0.9]}: The crossover rate; see Secs.~\ref{subsec:crossover} and \ref{subsec:mutation_strategies}.  This option encourages mixing between the trial and target
		vectors, and can encourage search along individual dimensions.  This parameter should be set between 0 and 1, inclusive.
		If it is set to 0, trial vectors will differ from the target vector along only one dimension.  If it is set to 1, trial vectors will be entirely
		unrelated to their target vectors.  This option is ignored when \fortran{jDE} or \fortran{lambdajDE} is \fortran{true}.
	\item\fortran{double precision lambda[0]}: A scale factor linking the best target vector in the population to the initial vector chosen for mutation; see
		Sec.~\ref{subsec:mutation_strategies}.  This may take any value between 0 and 1, inclusive.  If \mbox{\fortran{lambda = 0}}, the best
		vector is not used for mutation.  If \mbox{\fortran{lambda = 1}}, mutation will use the best vector as the starting point for all new vectors.
		As a result, setting \fortran{lambda} $>$ 0 will cause DE to optimise more aggressively.  This option is ignored when \fortran{jDE} or \fortran{lambdajDE} is \fortran{true}.
	\item\fortran{logical current[false]}: Use the current target vector as a base for mutation; see Sec.~\ref{subsec:mutation_strategies}.
    This option is ignored when \fortran{jDE} or \fortran{lambdajDE} is \fortran{true}.
	\item\fortran{logical expon[false]}: Use exponential crossover instead of binomial;
		see Sec.~\ref{subsec:mutation_strategies}.  This option is ignored when \fortran{jDE} or \fortran{lambdajDE} is \fortran{true}.
	\item\fortran{integer bndry[1]}: Controls the behaviour when trial vectors are outside the allowed boundaries of the parameter space.
		Should be set to an integer between 1 and 4:
		\begin{description}
			\item\fortran{1} (`brick wall'): points outside the boundaries are rejected during the selection phase.
			\item\fortran{2} (`random reinitialisation'): For each point outside the bounds, a random new valid point is chosen.
			\item\fortran{3} (`reflection'): points outside the boundaries are reflected across the limits so that they land inside.  This option is
				recommended if full exploration of the edges of parameter space is desired.
			\item$\ge$\fortran{4} (none): boundary conditions are not enforced.  This may lead to the population drifting away from the initially
				specified region of parameter space, and should be used with caution.
		\end{description}
	\item\fortran{logical jDE[true]}: Use self-adaptive rand/1/bin DE, as described in section \ref{subsec:jDE}.
		If this option is \fortran{true}, the values set for \fortran{F}, \fortran{Cr}, \fortran{lambda}, \fortran{current}, and \fortran{expon} are ignored.  This option is ignored when \fortran{lambdajDE} is  \fortran{true}.
	\item\fortran{logical lambdajDE[true]}: Use self-adaptive rand-to-best/1/bin DE, as described in section
		\ref{subsec:jDE}.  If this option is \fortran{true}, the values set for \fortran{F}, \fortran{Cr}, \fortran{lambda}, \fortran{current},  \fortran{expon},
		and \fortran{jDE} are ignored.  If less aggressive optimisation is required, we recommend that this be turned off, and jDE used instead.
	\item\fortran{double precision convthresh[0.001]}: The threshold for convergence of one DE population (a `civilisation').  The smoothed fractional improvement in the population over successive generations must drop below this value for a population to achieve convergence. Assuming that the likelihood/objective function \fortran(func()) has been chosen to return  $\ln\mathcal{L}$, the smoothed fractional improvement in the mean is defined as
  \begin{equation}
    \delta_\mathrm{smooth} = \frac{1}{n}\sum_{i=j}^{j-n+1} \left[ 1 - \frac{\sum_\mathrm{population} \ln\mathcal{L}_{i-1}}{\sum_\mathrm{population} \ln\mathcal{L}_i}\right],
  \end{equation}
  where $i$ is the generation index, $j$ is the current generation number, and $n$ is the population smoothing length, given by \fortran{convsteps}.
	\item\fortran{integer convsteps[10]}: The number of generations over which to smooth the fractional improvement of the mean population value of the likelihood/objective function when testing for convergence.
	\item\fortran{logical removeDuplicates[}{\footnotesize see Sec.\ \ref{subsec:duplicates}}\fortran{]}: Remove duplicate vectors within a single generation.  Turning this on is generally good for population diversity.  Duplicates are however exceedingly rare when either \fortran{jDE} or \fortran{current} is \fortran{true}, so keeping \fortran{removeDuplicates = true} in these cases is not necessary, but can be a useful debug check against \mpi problems.
	\item\fortran{logical doBayesian[false]}: Estimate posterior weights of population members, and the natural log of the Bayesian evidence $\ln Z$; see Sec.~\ref{subsec:posterior}.
	\item\fortran{double precision prior()}: The prior function to be accounted for in approximate Bayesian computations; see Sec.~\ref{subsec:posterior}.  Required if \fortran{doBayesian} is \fortran{true}, ignored otherwise.
	\item\fortran{integer maxNodePop[1.9]}: The population above which to perform node division in the binary spanning tree used to estimate posterior weights; see Sec.~\ref{subsec:posterior}. Ignored unless \fortran{doBayesian} is \fortran{true}.
  \item\fortran{double precision Ztolerance[0.01]}: The fractional uncertainty in $\ln Z$ taken to indicate convergence of the evidence; Sec.~\ref{subsec:posterior}.  Ignored unless \fortran{doBayesian} is \fortran{true}.
	\item\fortran{integer savecount[1]}: The number of generations that should pass between periodic saves of the population.
	\item\fortran{logical resume[false]}: Resume from a previous run.
	\item\fortran{logical outputSamples[true]}: Write samples and derived quantities in an output \term{.sam} file.  Even if this is \fortran{false}, the \term{.sam} file will still be written if \fortran{discrete} is non-empty.
  \item\fortran{integer init_population_strategy[0]}: Strategy to employ when initialising the first generation.  Should be set to an integer between 0 and 2:
    \begin{description}
      \item\fortran{0} (`one-shot'): initialise each member of the first generation to a different random point drawn from between the stated \fortran{lowerbounds} and \fortran{upperbounds}, without regard to its fitness.
      \item\fortran{1} (`$n$-shot'): draw candidate initial population members randomly from between \fortran{lowerbounds} and \fortran{upperbounds}. Accept a candidate if its function value is below \fortran{max_acceptable_value}, otherwise attempt to draw an alternative candidate.  Continue until \fortran{max_initialisation_attempts} is reached, then if a good candidate has still not been found, accept the next candidate without regard to its fitness.
      \item\fortran{2} (`fatal $n$-shot'): as per \fortran{1}, but throw a hard error if \fortran{max_initialisation_attempts} is reached when initialising any member of the first generation.
    \end{description}
  \item\fortran{integer max_initialisation_attempts[10000]}:  Maximum number of times to try to find a valid vector when initialising each member of the initial population if \fortran{init_population_strategy > 0}; ignored otherwise.
  \item\fortran{double precision max_acceptable_value[10}{\footnotesize\ttfamily$^\mathtt{6}$}\fortran{]}:  The cutoff value of the objective function below which to consider a candidate initial population member `acceptable' if \fortran{init_population_strategy > 0}; ignored otherwise.
	\item{\footnotesize\ttfamily\color{blue} c\_ptr} \fortran{context[C_NULL_PTR]}: A raw \cpp{void} callback pointer, used to pass information from the driver program to the objective function.  This is typically used to pass an external function address, which the objective function then uses to help with its evaluation.
	\item\fortran{integer verbose[1]}: The amount of information to print to screen.  Recognised values are:
		\begin{description}
			\item\fortran{0}(`Quiet'):  Only error messages will be printed.
			\item\fortran{1}(`Laconic'): Prints warning messages and a summary at the beginning and end.
			\item\fortran{2}(`Chatty'): Prints civilisation-level and basic gener-ation-level information.
			\item\fortran{3+}(`Verbose'): Prints detailed information for each generation.
		\end{description}
\end{description}

\subsection{Output formats}
\label{sec:outputs}

\diver produces up to four different output files, in plain ASCII format.  The first three of these are always generated, and are needed for resuming a run.

\begin{description}

\item\metavar{path}\term{.rparam}: the complete range of \diver settings in use in the current run, including optional parameters.  The meaning of each entry in this file can be read off the comments provided in the routine \fortran{save_run_params} in \term{io.f90}.  This file is created during the first save operation, which takes place after \fortran{savecount} generations have been completed (see Sec.\ \ref{sec:options}).

\item\metavar{path}\term{.devo}: convergence and other dynamic runtime information.  This is the file to check for evaluating the progress of a given run.  Its contents are as follows:
\begin{lstterm}
civilisation number, generation number
@$\mathcal{Z}$@, @$\langle P^2 \rangle$@, @$\Delta\mathcal{Z}$@, unpolished @$\mathcal{Z}$@
@$N_\mathrm{s}$@, individuals saved, number of calls to @\cpppragma{func}@

fitness at best fit @$\theta_\mathrm{best}$@
raw (non-discretised) parameter values at @$\theta_\mathrm{best}$@
parameter values and derived quantities at @$\theta_\mathrm{best}$@

fitnesses of current population
raw parameters of current population
parameters & derived quantities of current pop.

if @\cpppragma{jDE}@ or @\cpppragma{lambdajDE}@:
  @$F$@ values of current population
  @$Cr$@ values of current population
  @$\lambda$@ values of current population

@$\delta_\mathrm{smooth}$@
individual contributions to @$\delta_\mathrm{smooth}$@ from each of
 the last @\cpppragma{convsteps}@ generations
\end{lstterm}

Further information can be found in the routine \fortran{save_state} of \term{io.f90}.  Like the \term{.rparam} file, this file is created during the first save operation.

\item\metavar{path}\term{.raw}: the posterior weight, fitness, civilisation number, generation number and raw parameter values (in this order), for every individual so far generated in a scan.  The data for each individual occupies a single line in the file.  In order to allow proper resumption of the run, the sampled values of any discrete parameters appear as they are used internally for mutation, i.e.~as values of a continuous parameter.  This file is created before the initial population is generated.

\item\metavar{path}\term{.sam}: all parameter samples, in a similar format to the \term{.raw} file, but with additional columns for each derived quantity calculated in a scan.  The sampled values of any discrete parameters are also given rounded to their true discrete values in this file, unlike in the \term{.raw} file. This file is only generated if \fortran{outputSamples =} \fortran{true} and either \fortran{discrete} is non-empty or \fortran{nDerived} $\ne$ \fortran{0}.  This file is created immediately after the \term{.raw} file.

\end{description}

\section{Scanner options and outputs}
\label{app:options}

For quick reference, here we provide the \scannerbit\ \YAML file options and output formats for all five of the major scanners mentioned in this paper: the postprocessor (Sec.\ \ref{sec:postprocessor}), \great (Sec.\ \ref{sec:mcmc}), \twalk (Sec.\ \ref{sec:ensemble}) MultiNest (Sec.\ \ref{sec:nested sampling}) and Diver (Sec.\ \ref{sec:de}).

\subsection{Postprocessor}
\label{app:options:postprocessor}

The \YAML setup required to run the \textsf{postprocessor} spans two sections of the master \YAML file: the usual \yaml{Scanner} section, plus also the \yaml{Parameters} section.\footnote{Some of the requirements of the \yaml{Parameters} section can be optionally implemented in the \yaml{Priors} section instead.} In the \yaml{Scanner} section, the options [and defaults] are as follows:
\begin{description}
 \item\yaml{like}: The purpose to use as the objective; should generally match the purpose set for likelihood components (e.g. in the \yaml{ObsLike} section of a \GB \YAML file).
 \item\yaml{reweighted_like}: The output label used for the final result of \yaml{add_to_like} and \yaml{subtract_from_like} operations.
 \item\yaml{add_to_like[}empty\yaml{]}: A vector of names of datasets present in the input samples, presumably log-likelihood values, to be added to the newly computed \yaml{like} and output as \yaml{reweighted_like}.  (Note that the `newly-computed' \yaml{like} may be zero if no entries in the \GB\ \yaml{ObsLike} section have been assigned a \yaml{purpose} that matches \yaml{like}). For example, if the combined likelihood of a previous scan were labelled \yaml{"LogLike"}, and one were to choose \yaml{like: New_LogLike} as the new composite (log-)likelihood for a new `scan', then the way to ensure that the old and the new composite log-likelihoods were automatically summed for every model point would be to set \yaml{add_to_like: [LogLike]}. The results of this summation would appear in the new output with the label by \yaml{reweighted_like}.
 \item\yaml{subtract_from_like[}empty\yaml{]}: As per \yaml{add_to_like}, except the old output is {\em subtracted} from \yaml{like}.
 \item\yaml{permit_discard_old_likes[false]}: When set to \yaml{false}, this option forbids the purpose chosen for \yaml{like} from clashing with any data label in the input samples. For example: if the original purpose was \yamlvalue{LogLike}, a different purpose must be chosen for \yaml{like}, or an error will be thrown. If this option is set \yaml{true}, then clashes are permitted, and will be resolved in the new output by replacing the old data with the newly-computed data (as occurs automatically for all other clashes between old and new dataset names). This option also applies to likelihood components listed in \yaml{add_to_like}, \yaml{subtract_from_like}, and \yaml{reweighted_like}. If set to \yaml{false} then these names may not be recomputed during postprocessing.
 \item\yaml{update_interval[1000]}: Defines the number of iterations between messages reporting on the progress of the postprocessing.
 \item\yaml{reader}: Options under this item specify the format of the old output file to be read, along with e.g.\ the path at which the file is located. The required options differ depending on which \GB \cross{printer} was used to save the results of the previous scan.
\end{description}

The final option, \yaml{reader}, is used to inform the \textsf{postprocessor} of the format and location of the old data that needs to be reprocessed. In this first release of \GB there are only two possible \cross{printer} formats, \yaml{ascii} and \yaml{hdf5}, as described in \cite{gambit}. There are therefore at present only two sets of options that may be specified for the \yaml{reader}. For files created with the \yaml{hdf5} printer:
\begin{description}
  \item\yaml{type}: \yaml{hdf5}
  \item\yaml{file}: Path to the \textsf{HDF5} file containing the data to be parsed
  \item\yaml{group}: Group within the \textsf{HDF5} file containing datasets to be parsed.
\end{description}
For \yaml{ascii} output:
\begin{description}
  \item\yaml{type}: \yaml{ascii}
  \item\yaml{data_filename}: Path to the ASCII file containing the data to be parsed
  \item\yaml{info_filename}: Path to the ASCII `header' file that contains the labelling information for the columns of \yaml{data_filename}.
\end{description}
Note that the \yaml{reader} need not match the chosen \cross{printer} in a postprocessing run; reading samples in \yaml{ascii} and outputting updated samples in \yaml{hdf5}, or vice versa, is permitted.  This allows \GB samples produced in one format to be easily converted into any other format.

Note also that where new print overloads have been defined for one or more printers, as described in Sec.\ 9.3 of Ref.\ \cite{gambit}, users wishing to postprocess the resulting data must \textit{also} overload the equivalent \cpp{_retrieve} method of the reader in use, so that it can successfully read the new type in from the existing scan output.  To do this, one needs to follow the instructions for adding a new print overload in Sec.\ 9.3 of Ref.\ \cite{gambit}, and then also add the body of the new \cpp{_retrieve} function to the file \term{Printers/src/printers/}\metavar{printer\_name} \cpp{/retrieve_overloads.cpp}.

Using the \textsf{postprocessor} scanner also places some special requirements on the \yaml{Parameters} and/or \yaml{Priors} sections of the \YAML file. First, the \doublecross{models}{model} chosen in the \yaml{Parameters} section must be a subset of the models that were used for the original scan. Secondly, the \yaml{prior_type} for all the parameters in those models must be set to \textsf{none}. This disables the standard \GB prior system and allows the postprocessor to manually set parameter values (see Sec.\ \ref{sec:other_priors} for details).

\subsection{\great}
\label{app:options:great}

The following options (with defaults in brackets) set the chain length and number of steps taken used by the \great sampler:
\begin{description}
    \item\yaml{nTrialLists[10]}: Number of Markov chains to be run.
    \item\yaml{nTrials[20000]}: Number of steps in each Markov chain.
\end{description}

At the end of the run, the complete statistics for all chains run (burn-in length, correlation length, number of independent samples) are printed out in \great's native format. The independent samples and their multiplicities are stored and outputed to the \gambit printer system.

\subsection{\twalk}
\label{app:options:twalk}

The options available for \twalk in \scannerbit (with defaults in square brackets) are:
\begin{description}
    \item\yaml{ kwalk\_ratio[0.9836]}:  ratio of \textsf{walk} and \textsf{traverse} to \textsf{hop} and \textsf{blow} moves.  The default is to strongly prefer \textsf{walk} and \textsf{traverse} moves.
    \item\yaml{ projection\_dimension[4]}: dimension of the projection subspace in which \textsf{walk} and \textsf{traverse} moves are performed.
    \item\yaml{ walk\_distance[2.5]}: width of the distribution function for the distance of the walk move ($a_w$; see Eq.\ \ref{eq:walk}).
    \item\yaml{ traverse\_distance[6]}: width of the distribution function for the distance of the traverse move ($a_t$, see Eq.\ \ref{eq:traverse}).
    \item\yaml{ gaussian\_distance[2.4]}: Gaussian jump parameter $d$ for the \textsf{hop} and \textsf{blow} moves.  See Eq.\ \ref{eq:hopblow}.
    \item\yaml{ chain\_number[1+projection\_dimension+}number of \mpi processes\yaml{]}: total number of MCMC chains. \twalk will be highly inefficient if this parameter is set to anything less than the default.
    \item\yaml{ hyper\_grid[true]}: confines the search to the hypercube defined by the priors.
    \item\yaml{ sqrtR[1.001]}: the version of \twalk in \scannerbit uses the Gelman-Rubin convergence diagnostic $\sqrt{R}$ \cite{GelmanRubin} to determine when a scan has converged. This compares the inter-chain dispersion to the total dispersion of each parameter.  Values closer to 1 are better converged; when $\sqrt{R}$ drops below the value given for \yaml{ sqrtR}, the scan terminates.
\end{description}

The \twalk scanner also outputs various variables associated with the scan to the \GB printer system:
\begin{description}
    \item\term{mult}:     Multiplicity (posterior weight) of each sample.
    \item\term{chains}:   Chain  number for each sample.  Rejected proposal points are assigned the number $-1$.
\end{description}

\subsection{MultiNest}
\label{app:options:multinest}
The \scannerbit plugin that runs the \MultiNest sampler takes the following \YAML options, which it passes directly through to the external \MultiNest library (defaults are given in square brackets):
\begin{description}
 \item\yaml{IS[true]}: do nested importance sampling?
 \item\yaml{mmodal[true]}: do mode separation?
 \item\yaml{ceff[false]}: run in constant efficiency mode? Setting this \yaml{true} can result in poor evidence estimates.
 \item\yaml{nlive[1000]}: number of live points.
 \item\yaml{efr[0.8]}: required efficiency (only relevant if \yaml{ceff =} \yaml{true}).
 \item\yaml{tol[0.5]}: stopping tolerance; the scan halts when the ratio of the estimated remaining unsampled evidence to the current estimate of the evidence drops below this value.
 \item\yaml{nClsPar[}{\CPPidentifierstyle ndims}\yaml{]}: number of parameters to do mode separation on. The default is to do separation on all parameters being scanned.
 \item\yaml{updInt[1000]}: update interval; this sets the number of iterations between output file updates and any feedback passed to standard output.  The \multinest\ \cpp{dumper} function, which handles the calls to the \GB printer, runs every \cpp{10*updInt} iterations.
 \item\yaml{Ztol[-10}{\footnotesize\ttfamily$^{\mathtt{90}}$}\yaml{]}: the threshold in the logarithm of the evidence below which to ignore modes of the posterior.
 \item\yaml{maxModes[100]}: expected maximum number of modes (used only for memory allocation).
 \item\yaml{seed[-1]}: seed to use for the internal \multinest random number generator. If this is negative, the seed is taken from the system clock.
 \item\yaml{fb[true]}: provide feedback on run progress to standard output?
 \item\yaml{outfile[true]}: write native \MultiNest output files? \scannerbit does \textit{not} add prior-transformed parameter values nor auxilliary observable values to the native \MultiNest output, so this output is not very useful for analysis purposes. However, the native outputs are required for \multinest to be able to resume scans that were previously interrupted. We recommend leaving this option set unless running scans that will definitely not need to be resumed.
 \item\yaml{maxiter[0]}: maximum iterations permitted; a non-positive value is interpreted to mean infinity.
\end{description}

There are several other options that \MultiNest ordinarily requires when run outside of \scannerbit, but for which \scannerbit can infer appropriate values and set automatically. These \textit{cannot} be set in the \yaml{Scanner} section of the \YAML file (although some can be changed indirectly by modifying the scan setup elsewhere):

\begin{description}
 \item[Number of parameters ({\CPPidentifierstyle ndims}):] \scannerbit sets this option according to the number of varying parameters that exist in the model being scanned.
 \item[Size of `cube' array ({\CPPidentifierstyle nPar}):] This is set to \cpp{ndims+2}. The first \cpp{ndims} slots contain the hypercube parameters, and in the extra two slots \scannerbit stores an ID number for each point, plus the \mpi rank of the process that produced it. Together these two numbers uniquely identify every point sampled in a scan. These numbers are also stored in the \GB printer system output, so they can be used to correlate the native \MultiNest output with the \GB printer output.
 \item[Resume mode ({\CPPidentifierstyle resume}):] \scannerbit activates resume mode by default unless the \term{-r} switch (for `restart scan') is given at the command line.
 \item[Minimum loglike ({\CPPidentifierstyle logZero}):] points with $\ln\mathcal{L} < $ \cpp{logZero} will be ignored by \MultiNest. This is set to 0.9999 times the value of \yaml{model_invalid_for_lnlike_below} in the \yaml{likelihood} node of the \yaml{KeyValues} section of the main \YAML file.
 \item[Initialise \mpi ({\CPPidentifierstyle initMPI}):] This is set to \cpp{false} because \scannerbit handles the initialisation of \mpi.
\end{description}

Note that \GB sets \cpp{logZero} to slightly more than \yaml{model_invalid_for_lnlike_below}.  This is so that invalid points, assigned $\ln\mathcal{L}$ = \yaml{model_invalid_for_lnlike_below} by the likelihood container \cite{gambit}, are treated as having zero likelihood by \multinest.  This is the desired behaviour during live point generation, as it prevents any of the initial live points being invalid.

During live point replacement however, this can prevent efficient parallelisation, as \multinest requires all \mpi nodes to continue testing proposed points until they each find one with $\ln\mathcal{L} >$ \cpp{logZero}.  In complicated parameter spaces, where the ellipsoids encompass large regions of invalid parameter space, this can lead to many nodes idling whilst they wait for a small number of nodes to find their valid points, even if one of the points already found has a high enough likelihood to use for live point replacement.  To circumvent this, following live point generation, when the  \multinest\ \cpp{dumper} function first runs, the \multinest plugin communicates to \scannerbit and \GB that likelihoods for invalid points should no longer be set to \yaml{model_invalid_for_lnlike_below}, but instead to the value of the alternative option \yaml{model_invalid_for_lnlike_below_alt}.  This key can also found in the \yaml{likelihood} node of the \yaml{KeyValues} section of the main \YAML file.  The value of \yaml{model_invalid_for_lnlike_below_alt} defaults to half \yaml{model_invalid_for_lnlike_below}.  Whenever it is set to more than \cpp{logZero} (i.e.~0.9999 times \yaml{model_invalid_for_lnlike_below}), \multinest considers all samples found to be valid, and does not demand additional samples before evaluating whether those found are appropriate for live point replacement.  We find that this often results in more than an order of magnitude improvement in performance when running \multinest with $\mathcal{O}(100)$ or more \mpi processes.

\subsection{Diver}
\label{app:options:diver}

The \YAML entry \yaml{KeyValues::}\yaml{likelihood::}\yaml{lnlike_offset} can be used to set the offset to be applied to the log-likelihood function passed to \diver, in order to maintain positive definiteness of the fitness function; this defaults to $10^{-4}$ times \yaml{KeyValues::}\yaml{likelihood::}\yaml{model_invalid_for_lnlike_below}.

The \diver interface in \scannerbit provides almost all of the run options mentioned in Sec.\ \ref{sec:options}, configurable directly from the \diver entry in the main \YAML file. With a few exceptions, these options have the same names and default values as in \diver itself.  The exceptions are:\begin{itemize}
\item \fortran{NP} has no default, and must be specified in the \YAML file
\item \fortran{maxgen} defaults to \fortran{5000}, not \fortran{300}
\item \fortran{bndry} defaults to \fortran{3}, not \fortran{1}
\item \fortran{removeDuplicates} defaults to \fortran{true}, regardless of other options
\item \fortran{outputSamples} is instead referred to by the \YAML option \fortran{full_native_output}
\item \fortran{init_population_strategy} defaults to \fortran{2}, not \fortran{0}
\item \fortran{max_acceptable_value} defaults to 0.9999 times the value of \yaml{model_invalid_for_lnlike_below} in the \yaml{likelihood} node of the \yaml{KeyValues} section of the main \YAML file
\item \fortran{verbose} is instead referred to by the \YAML option \fortran{verbosity}, and defaults to \yaml{0} instead of \fortran{1}.
\end{itemize}

Note that \fortran{doBayesian} is not available as a \YAML option, and is hard-coded to \fortran{false}; there are multiple other scanners available in \scannerbit more efficient and accurate at scanning the Bayesian posterior than \diver.  Correspondingly, \fortran{maxNodePop} and \fortran{Ztolerance} are not offered as \YAML parameters either.  Any user especially interested in obtaining posteriors from \diver running within \scannerbit should find this relatively easy to recode by comparison with e.g. the \multinest or \great interface.

\section{Custom priors}
\label{app:priors}

\scannerbit allows for users to add their own priors.  These should be declared inside the \cpp{priors} namespace, in new headers placed in \term{ScannerBit/include/gambit/} \term{ScannerBit/priors}, and new source files placed in \term{ScannerBit/src/priors}.

Declaration of a new prior \metavar{prior\_name}, arising from a new class \metavar{prior\_class}, takes the form:
\begin{lstcpp}
class @\metavar{prior\_class}@ : public BasePrior
{
  public:

    @\metavar{prior\_class}@(const std::vector<std::string>&
     params, const Options& options)
    : BasePrior(params, @\metavar{cube\_size}@)
    {//insert optional initialisation code}

    void transform(const std::vector<double>&
     unitpars, std::unordered_map<std::string,
     double>& outputMap) const
    {//insert non-optional transformation code}
};

LOAD_PRIOR(@\metavar{prior\_name}@, @\metavar{prior\_class}@)
\end{lstcpp}
Given this recipe, the only real input required of a user when implementing a new prior is to decide on its dimensionality (\metavar{cube\_size}), and to write the body of its transformation function (\metavar{prior\_class}\cpp{::transform}).

The class defining the user-specified prior inherits from the abstract base class \cpp{BasePrior}.  This class has the following members:
\begin{description}
\item \cpp{BasePrior(std::vector<std::string>, int)}:
  Base class constructor.  Takes in a vector of strings that defines the parameter names, and an integer that specifies the dimension of the unit hypercube to be operated on by this prior.  Typically this will be \cpp{1}, or the entire parameter space, available by simply calling the \cpp{size()} method on the vector passed as the first argument.

\item \cpp{void transform(std::vector<double>,} \\
\cpp{std::unordered_map<std::string, double>)}:
  A pure virtual function that defines the prior transformation.  Takes as input a vector of doubles with the input unit hypercube values, converts them to the actual model parameters, and stores them in the unordered map passed (by reference) as the second argument.

\item \cpp{unsigned int size(), unsigned int\& sizeRef()}:
  Returns the dimension of the input unit hypercube.

\item \cpp{void setSize(int)}:
  Set the unit hypercube dimension.

\item \cpp{std::vector<std::string> param\_names}:
  A protected member variable (i.e.\ accessible from derived classes only), which contains the names of the parameters as passed to the constructor.

\end{description}
A user-defined prior is registered in the \scannerbit prior database by invoking the following macro after the class declaration:
\begin{description}
 \item[{\CPPidentifierstyle LOAD\_PRIOR(\metavar{prior\_name}, \metavar{prior\_class})}]
   Macro that loads the prior defined in class \metavar{prior\_class}, and assigns it the internal name \metavar{prior\_name}.
\end{description}

Here we give a worked example of the declaration of a custom prior.  This prior is contained in the \scannerbit source file \term{ScannerBit/include/gambit/} \term{ScannerBit/priors/dummy.hpp}.  This prior simply transforms the unit hypercube to the same unit hypercube.

\begin{lstcpp}
namespace Gambit
{

  namespace Priors
  {

    class Dummy : public BasePrior
    {

      public:

        Dummy(const std::vector<std::string>&
         param, const Options&)
        : BasePrior(param, param.size())
        {}

        void transform(const std::vector<double>&
         unitpars, std::unordered_map<std::string,
         double>& outputMap) const
        {
          auto it_vec = unitpars.begin();
          for (auto it = param_names.begin(),
           end = param_names.end(); it != end;
           it++)
          {
            outputMap[*it] = *(it_vec++);
          }
        }

    };

    LOAD_PRIOR(dummy, Dummy)
  }

}
\end{lstcpp}

Here, the \lstinline|Dummy| class inherits from the \lstinline|BasePrior| class.  The constructor passes the entered parameter names to the \cpp{BasePrior} constructor, as well as the hypercube size.  The \lstinline|transform| function transforms a \lstinline|vector<double>| representing the unit hypercube into actual parameter values, which are stored in the output map.  In this case, the hypercube values are directly stored in the output map.  Lastly, the \cpp{Dummy} prior is loaded into the prior system and given the name \cpp{dummy}, by calling the macro \lstinline|LOAD_PRIOR(dummy, Dummy)|.

\section{Plugin Declaration and Interface}
\label{app:plugin_decl}

In the following subsections, we go through the definition, design, and operation of plugins in detail, starting with their declaration in Sec.\ \ref{plugin_setup}.  \scannerbit provides a broad suite of utility functions that can be called from plugins.  We first deal with the functions available to all plugins, for accessing information in the initialisation file of a scan (Appendix\ \ref{sec:plugin_to_ini}), the chosen prior transformation (Appendix\ \ref{sec:plugin_to_prior}), and the \GB printers (Appendix\ \ref{sec:printers}).  We then list utility functions available only to scanner (Appendix\ \ref{sec:scanner_plugins}) or objective (Appendix \ref{sec:objective_plugins}) plugins.

\subsection{Plugin declaration}
\label{plugin_setup}

Source code for a plugin \metavar{plugin\_name} is located within a directory \term{ScannerBit/src/}\metavar{plugins\_kind/}\metavar{plugin\_name}. Headers are found in \term{ScannerBit/include/gambit/} \term{ScannerBit/}\metavar{plugins\_kind}\term{/}\metavar{plugin\_name}.  Here \metavar{plugins\_kind} is either \term{scanners} or \term{objectives}.

Code for all plugins follows the same basic layout (with \metavar{plugin\_kind} either \term{scanner} or \term{objective}):
\begin{lstcpp}
#include "@\metavar{plugin\_kind}@_plugin.hpp"

@\metavar{plugin\_kind}@_plugin(@\metavar{plugin\_name}@, version(...))
{
  @\metavar{environmental\_macros}@

  plugin_constructor {...}

  @\metavar{return\_type}@ plugin_main(@\metavar{args}@) {...}

  plugin_deconstructor {...}
}
\end{lstcpp}
The plugin body can contain three blocks of code: a \cpp{plugin_constructor}, a \cpp{plugin_main}, and a \cpp{plugin_destructor}.  The utility functions detailed in the following subsections can be accessed from within any of these three blocks. The \cpp{plugin_constructor} and \cpp{plugin_deconstructor} blocks will run when the plugin is loaded and unloaded, respectively. The code here can used to initialise, allocate, or deallocate variables needed by the plugin. The \cpp{plugin_main} block defines the function that will be run by the plugin. The form of the arguments for \cpp{plugin_main} required by \scannerbit depends on whether the plugin is a \cross{scanner plugin} or an objective \doublecross{(test function) plugin}{test function plugin}.

For scanner plugins, \cpp{plugin_main} must take the form
\begin{lstcpp}
void plugin_main() { @\metavar{code}@ }
\end{lstcpp}
where \metavar{code} is the code that actually drives a statistical sampling algorithm.  We give a full example of a minimal scanner plugin in Appendix\ \ref{app:scanner}.

Objective plugins can be further categorised into `likelihood' plugins, which compute likelihoods, and `prior' plugins, which provide the transformation function needed to implement a \scannerbit prior (see Sec.\ \ref{sec:stats}).  For likelihood plugins, \cpp{plugin_main} must be of the form
\begin{lstcpp}
double plugin_main(const std::vector<double>&)
\end{lstcpp}
whereas for prior plugins, the required form is
\begin{lstcpp}
void plugin_main(const std::vector<double>&,
 std::unordered_map<std::string, double>&)
\end{lstcpp}
We give a worked example of a minimal likelihood-oriented objective plugin in Appendix\ \ref{app:objective}.

Each plugin is built in a separate programming environment, with its own user-specified library dependencies and compile-time options.  A set of \metavar{environmental\_macros} that define the compilation environment can be declared at the beginning of a plugin.  These macros can be used to define additional compilation flags, required libraries, required headers, or required entries in the input \YAML file of a scan.  The following macros are available:
\begin{description}
  \item \cpp{reqd\_inifile\_entries("}\metavar{X}\cpp{"}\cpp{,}\cpp{"}\metavar{Y}\cpp{"}\cpp{,...)}: Indicates that the plugin will not be permitted to load unless the \YAML node corresponding to the plugin in question, in the \YAML input file of the scan, contains the options \metavar{X} and \metavar{Y}.  Any number of required entries can be given as a comma-separated list.
  \item \cpp{cxx\_flags(}\metavar{flag\_string}\cpp{)}: Additional flags to append to the compilation commands for this plugin.
  \item \cpp{reqd\_libraries("}\metavar{A}\cpp{"}\cpp{,}\cpp{"}\metavar{B}\cpp{"}\cpp{,...)}: Tells \scannerbit to search for and link the libraries \metavar{A} and \metavar{B} if using this plugin. Any number of libraries can be given as a comma-separated list.
  \item \cpp{reqd\_headers("}\metavar{C}\cpp{"}\cpp{,}\cpp{"}\metavar{D}\cpp{"}\cpp{,...)}: Specifies that the headers \metavar{C} and \metavar{D} must exist for the plugin to compile; any number of headers can be given in a comma-separated list. Like libraries, \scannerbit will automatically search for the specified headers.
\end{description}

If a library or a header listed in \cpp{reqd\_libraries} or \cpp{reqd\_headers} is in a non-standard location, or if \scannerbit is unable to locate it, the location can be specified in the \term{config/scanner_locations.yaml} or \term{config/objective_locations.yaml} configuration files.\footnote{Note that the current version of \scannerbit locates both libraries and headers at \textsf{cmake} time, not at runtime.  This means that \textsf{cmake} must be run (or re-run) and \scannerbit rebuilt after scanners are built or moved. This is in contrast to the \GB backend system, which locates and loads backend libraries entirely at runtime.  It is expected that future versions of \scannerbit will dynamically load the shared plugin libraries, in line with \GB backend practice.}  Entries in the configuration files follow the format
\begin{lstyaml}
@\metavar{plugin\_name}@:
  @\metavar{plugin\_version}@:
    - inc: @\metavar{include\_dir}@
    - lib: @\metavar{library\_path}@
\end{lstyaml}
This entry gives the locations of the libraries and headers needed for version \metavar{plugin\_version} of the plugin \metavar{plugin\_name}.  Note that libraries require full paths, whereas headers require only an include directory.  The plugin version can be given as ``\term{any\_version}'', in which case the indicated library and/or header locations will be applied to every version of the plugin.  If the \term{config/scanner\_locations.yaml} or \term{config/objective\_locations.yaml} configuration files do not exist, or a relevant entry is missing from them for a given plugin, then \scannerbit will use any relevant entry it can find in the files \term{config/scanner\_locations.yaml.default} and \term{config/objective\_locations.yaml.default}. These \term{.default} files ship with \scannerbit and should not be modified; it is up to the user to create \term{config/scanner\_locations.yaml} and/or \term{config/objective\_locations.yaml} if they wish to override or add to any of the defaults.

\subsection{Interface to input file}\label{sec:plugin_to_ini}

Detailed instructions on how to construct and format a \YAML input file for a scan are given in Sec.\ \ref{sec:interface}.  To extract entries from this file, the following functions are provided to both scanner and objective plugins:
\begin{description}
\item \metavar{ret\_type} \cpp{ get\_inifile\_value<}\metavar{ret\_type}\cpp{>(std::string key,} \metavar{ret\_type} \cpp{default_value)}:
   Retrieves the value assigned to the \YAML key \cpp{key}.
   If \cpp{key} is not present in the relevant part of the \YAML file, an optional \cpp{default_value} to be returned can be specified.  If
   no default is given, and the \cpp{key} is absent from the \YAML file, \scannerbit will throw an error.  The return value obtained will
   be interpreted as a quantity of type \metavar{ret\_type}.  Note that the key \cpp{default\_output\_path}
   will always return a value; if this key is not set in the \YAML file, the output defaults to
   \term{scanner_plugins/}\metavar{plugin\_name} (where \metavar{plugin\_name} is
   the name of the plugin calling \cpp{get\_inifile\_value}).  This is true for both scanner and objective plugins, although only scanner plugins are typically expected to generate output files.
\item \cpp{YAML::Node get\_inifile\_node(std::string key)}:
   Retrieves an entire \YAML node with a given \metavar{key} from the input \YAML file.
\end{description}

\subsection{Interface to prior object}\label{sec:plugin_to_prior}

Both scanner and objective plugins can directly access the prior transformation object used in any given scan, via the function \cpp{get\_prior()}.  See Appendix\ \ref{app:priors} for details of how to use this object.

\subsection{Interface to \GB printer system}\label{sec:printers}

Within the body of a \scannerbit plugin, the \cpp{get_printer()} function can be called to obtain an object that acts as an interface to the \GB the \cross{printer} system.  \GB's printer system removes the need for scanners or their plugins to directly output sampled parameter values, as this responsibility is taken on by \scannerbit itself.  The printer system also removes any need for scanners to output total likelihoods, individual likelihood components or observables; these are to be printed by objective plugins themselves, or in the case of \GB, by the \cross{likelihood container} (which is in effect just a very sophisticated likelihood plugin).  This arrangement is designed to increase modularity, by allowing individual likelihoods to print their own --- potentially highly model-specific --- results, without the need to modify any scanner or scanner plugin code.  Printing of scanner-specific quantities (such as posterior weights or chain multiplicities) must be handled by the scanner plugins themselves, and these quantities must be uniquely associated with specific parameter combinations. This is accomplished by assigning each parameter combination a unique point ID number via which the printer can associate any future outputs with a specific parameter combination.

The basic interface is contained within the \lstinline|printer_interface| object returned by \lstinline|get_printer()|.  This object offers the following useful member functions:
\begin{description}
\item \cpp{printer* get\_stream(std::string name)}:
   Gets a pointer to the printer stream \cpp{name}. If no name is specified, the main printer is returned.\footnote{Note that \cpp{printer} is just a local \scannerbit\ \cpp{typedef} of the \GB printer base class.}
\item \cpp{void new\_stream(std::string name, YAML::Node option)}:
   Create a new printer stream named \cpp{name}, using the options contained in a \YAML node \cpp{option} (which is itself optional).  This typically only needs to be done on the \mpi master process (See Ref.\ \cite{gambit}).  To then ensure that all \mpi processes are aware of the new streams, the helper function \cpp{void assign_aux_numbers} \cpp{(std::string name1, std::string name2, ...)} should be called by all \mpi processes.
\item \cpp{bool resume\_mode()}:
   Returns true if the printers have resumed writing to the outputs of a previous scan.  Generally, scanner plugins should take their cue on whether or not to resume a previous run from the printers.
\end{description}
At the heart of the printer system are the \cpp{printer} stream objects.  These objects provide the necessary methods for printing values and associating them with a given point ID.  The printer stream is manipulated using the following member functions:
\begin{description}
\item \cpp{void reset(bool force)}: Deletes output that was already in the stream.  By default, the main printer cannot be reset; to override this behaviour, set \cpp{force} to \cpp{true}.
\item \cpp{void print(}\metavar{value\_type} \cpp{value, std::string name,}\\
   \cpp{int rank, unsigned long long int id)}:
   This function prints the actual output, sending a single datum of the given \cpp{value} and \metavar{value\_type} to the printer.  The output is identified as being the quantity \cpp{name}, and corresponding to the parameter combination uniquely identified by the point \cpp{id} and \mpi\ \cpp{rank}.
\end{description}
Scanner-specific output files not associated with the \GB printer system should typically be saved in the default scanner output path, which is accessed with \cpp{get_inifile_value<std::string>("default_output_path")}, and set to \term{scanner_plugins/}\metavar{plugin\_name} by default.

\subsection{Scanner plugins}
\label{sec:scanner_plugins}

Scanner plugins receive access to an additional pair of utility functions and a class, for obtaining likelihood functors and scanner information:

\begin{description}
\item \cpp{unsigned int get\_dimension()}:
   Gets the dimension of the unit hypercube being explored.

\item \cpp{void* get\_purpose(std::string purpose)}:
   Gets a pointer to a functor that is able to compute the quantity corresponding to \cpp{purpose}.  In \GB scans, purpose is conventionally \cpp{"LogLike"}, and the functor returned will be a direct conduit to the \cross{likelihood container}.

\item \cpp{like\_ptr}:
   A functor class used to contain the output of \cpp{get_purpose}, primarily designed to act as the local representation of the likelihood function within a plugin.  A \cpp{like_ptr} can be called as if it were a function with signature \lstinline|double (const| \cpp{std::vector<double>&)}.  Typically, within a  scanner plugin, the scanner passes a vector of unit hypercube parameter values to the \cpp{like_ptr}.  This functor automatically performs any required prior transformation, computes the quantities corresponding to its \cpp{purpose}, and sends the corresponding quantities and hypercube parameters to the printer.  The \cpp{like_ptr} member function \cpp{disable_external_shutdown()} can also be used from the plugin constructor to tell the objective function not to carry out its own shutdown procedure, but to simply set an internal \cpp{quit} flag (referred to in Ref.~\cite{gambit}) and rely on the scanner to terminate the scan itself.
\end{description}

\subsubsection{Scanner plugin example}
\label{app:scanner}

Here we give a simple example of a scanner plugin declaration, which closely follows one contained in the \scannerbit source code (\term{ScannerBit/src/scanners/random.cpp}).
The example declares a scanner plugin named \textsf{random}, version \textsf{1.0.0-example}.
This scanner enters \cpp{number} random points in the functor corresponding to the purpose specified by the \cpp{like} \YAML file option.

\begin{lstcpp}
#include "scanner_plugin.hpp"

scanner_plugin(random, version(1, 0, 0, example))
{
  reqd_inifile_entries("number");

  like_ptr loglike;
  int num, dim;

  plugin_constructor
  {
    std::string purpose =
     get_inifile_value<std::string>("like")
    loglike = get_purpose(purpose);
    num = get_inifile_value<int>("number");
    dim = get_dimension();
  }

  int plugin_main(void)
  {
    std::vector<double> a(dim);
    for (int j = 0; j < num; j++)
    {
      for (int i = 0; i < dim; i++)
      {
        a[i] = Gambit::Random::draw();
      }
      loglike(a);
    }
    return 0;
  }

  plugin_deconstructor
  {
    std::cout << "no more plugin" << std::endl;
  }
}
\end{lstcpp}

The actual scanner code is declared within the \lstinline|plugin_main| function, and randomly draws a parameter point from the hypergrid via the line
\begin{lstcpp}
a[i] = Gambit::Random::draw();
\end{lstcpp}
When the plugin is loaded, the \lstinline|plugin_constructor| function is run, initialising the variables \lstinline|loglike|, \lstinline|num|, and \lstinline|dim|.  The likelihood calculations and printing are done by the line \lstinline|loglike(a)|.  When the plugin is unloaded, the \lstinline|plugin_deconstructor| function runs, and indicates to \term{stdout} that the plugin has been unloaded.  At the top of the plugin declaration, the \lstinline|reqd_inifile_entries("number")| macro indicates that that the inifile entry \yaml{number} is required in order to use this scanner (see Sec.~\ref{sec:plugins}).

\subsection{Objective plugins}
\label{sec:objective_plugins}

In addition to the general plugin functions described in Secs.\ \ref{sec:plugin_to_ini}--\ref{sec:printers}, objective functions are provided with utility functions that can be used to probe the parameters being scanned, set the hypercube dimension and print parameters:
\begin{description}
\item \cpp{std::vector<std::string>& get\_keys()}:
   Retrieve the names of all the parameters being scanned over.
\item \cpp{void set\_dimension(unsigned int dim)}:
   For plugins that will be used as priors. Sets the hypercube dimension that will be operated on by the prior to \cpp{dim}.
\item \cpp{void print\_parameters(std::unordered_map<std::string,}
   \cpp{double> map)}:
   Prints the contents of a map from strings to double-precision floating-point variables. Typically used to print a set of parameters, where the map associates parameter names with their values.
\end{description}

\subsubsection{Objective plugin example}
\label{app:objective}

Here we give a simple example of an objective plugin declaration contained in the \scannerbit source code (\term{ScannerBit/src/objectives/examples.cpp}).
This example declares a scanner plugin \textsf{EggBox}, version \textsf{1.0.0}.  It returns a likelihood of the form:
\begin{equation}
 \mathcal{P}(x, y) = \left[2 + \cos\left(\tfrac{\pi}{2}x\right)\cos\left(\tfrac{\pi}{2}y\right)\right]^5.
\end{equation}

\begin{lstcpp}
#include "objective_plugin.hpp"

objective_plugin(EggBox, version(1, 0, 0))
{
  std::pair <double, double> length;
  unsigned int dim;

  plugin_constructor
  {
    dim = get_keys().size();

    if (dim != 2)
    {
      scan_err << "EggBox: Need two parameters."
               << scan_end;
    }
    length = get_inifile_value<std::pair<double,
     double> > ("length", std::pair<double,
     double>(10.0, 10.0));
  }

  double plugin_main(std::unordered_map
   <std::string,double> &map)
  {
    print_parameters(map);

    double params[2];
    params[0] = map[get_keys()[0]]*length.first;
    params[1] = map[get_keys()[1]]*length.second;

    return 5.0*std::log(2.0 +
     std::cos(params[0]*M_PI_2) *
     std::cos(params[1]*M_PI_2));
  }
}
\end{lstcpp}

In the \lstinline|plugin_constructor|, the hypercube dimension is obtained by testing how many parameters are returned from the \lstinline|get_keys()| function.  If the hypercube dimension does not match expectations, a runtime error is thrown with the \lstinline|scan_err| and \lstinline|scan_end| macros.  The constructor initialises the scale length for each of the hypercube dimensions with the values assigned to the \yaml{length} key in the input \YAML file.  If no values are specified, both lengths default to 10.  The \lstinline|plugin_main| function does the actual likelihood calculation, as it is the function run by the scanner for every parameter combination.  For each likelihood evaluation, the \cpp{plugin_main} receives an \lstinline|unordered_map| with the parameter names and values, which it uses to compute the value of the likelihood.  The contents of the map are printed with the command \lstinline|print_parameters(map)|.

\section{Scanner comparisons in a two-dimensional parameter space}
\label{app:scanner_comparisons}

The scanner comparisons presented in Sec.\ \ref{sec:comparison} are based on about 16 separate scans for each scanner in two, seven and fifteen dimensions.  We also included results from 52 more scans to cover each dimensionality between two and fifteen. However, for clarity we only displayed two-dimensional profile likelihoods for the 15-dimensional scans (Figs.\ \ref{fig:multinest_plots}--\ref{fig:diver_plots}, \ref{fig:twalk_plots}, \ref{fig:twalk_plots_post}, \ref{fig:great_plots} and \ref{fig:great_plots_post}).  In this Section we present the equivalent plots to these for the two-dimensional scans.  In some cases, where the optimal settings depends strongly on dimensionality, we have chosen different sampler settings in two than in fifteen dimensions, so as to allow a meaningful comparison.

 \begin{figure*}[tp]
  \centering
  \includegraphics[height=0.234\linewidth]{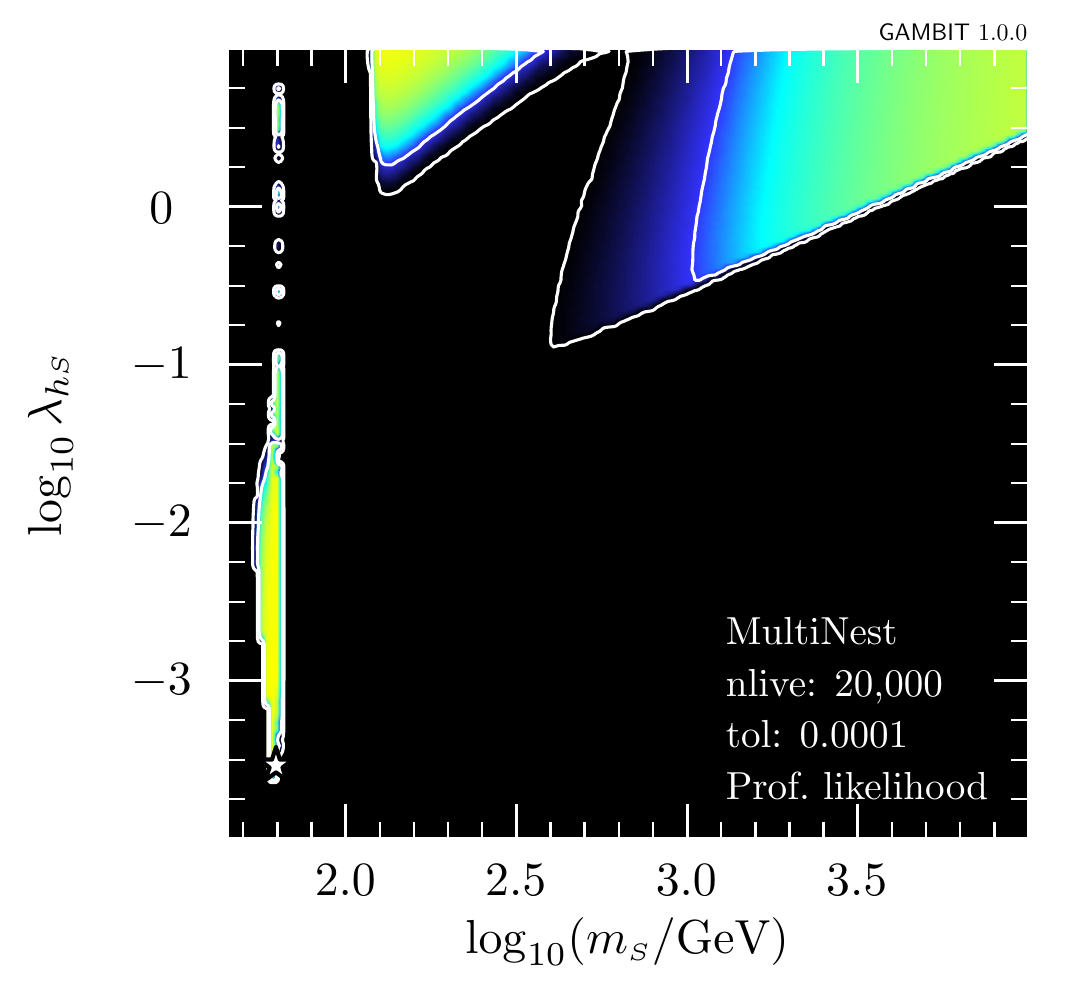}\hspace{-3mm}%
  \includegraphics[height=0.234\linewidth]{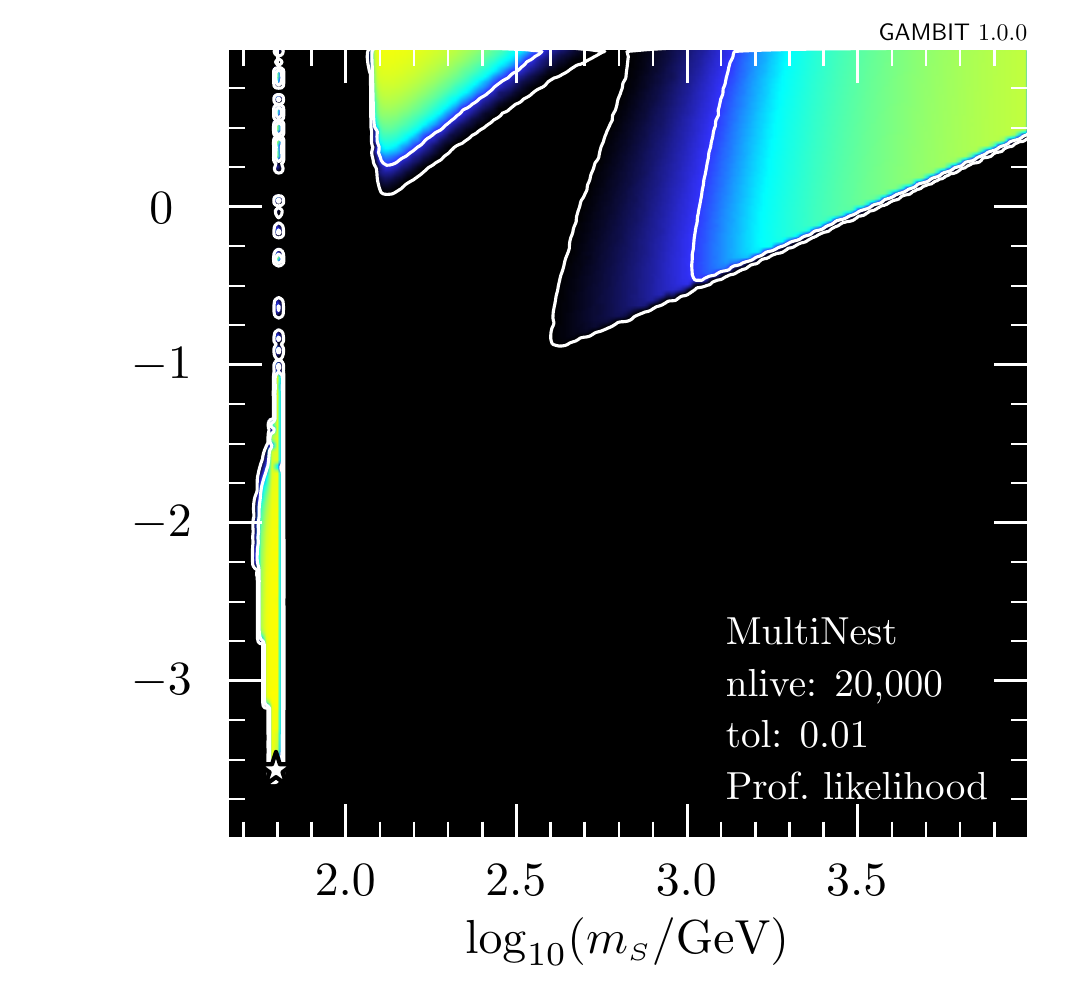}\hspace{-3mm}%
  \includegraphics[height=0.234\linewidth]{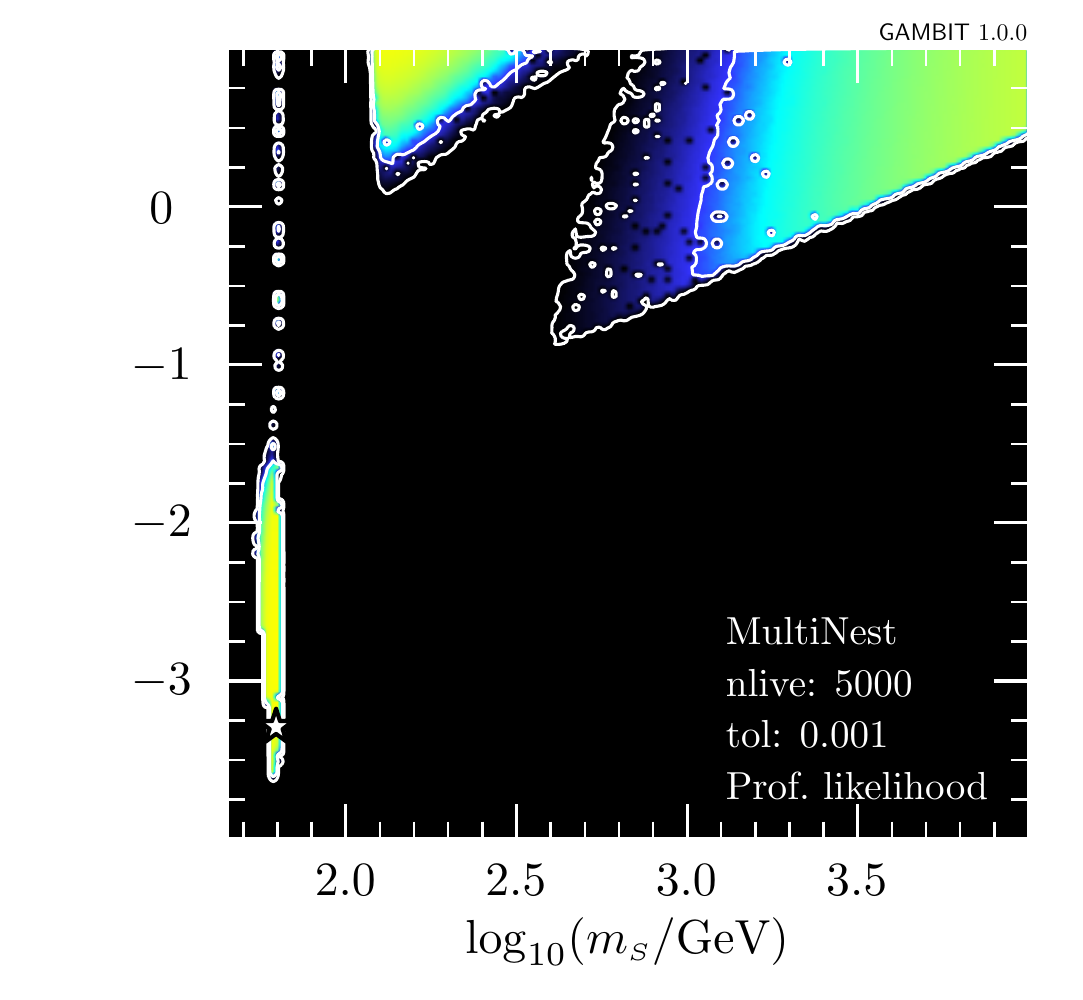}\hspace{-3mm}%
  \includegraphics[height=0.234\linewidth]{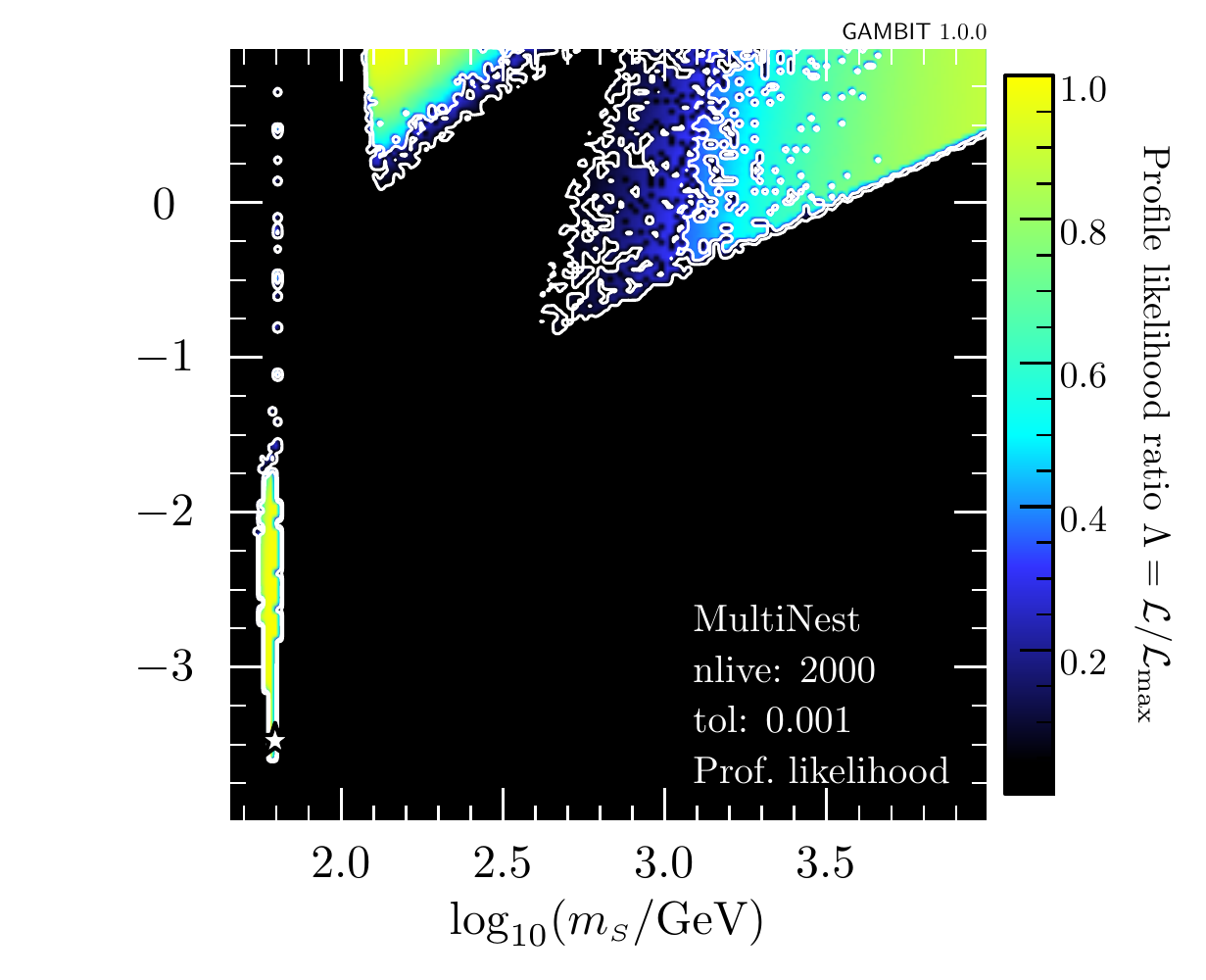}
  \caption{Profile likelihood ratio maps from a 2-dimensional scan of the scalar singlet parameter space, using the \multinest scanner with a selection of difference tolerances (\lstinline{tol}) and numbers of live points (\lstinline{nlive}).  The maximum likelihood point is shown by a white star.}
  \label{fig:multinest_plots2}
\end{figure*}

\begin{figure*}[tp]
  \centering
  \includegraphics[height=0.234\linewidth]{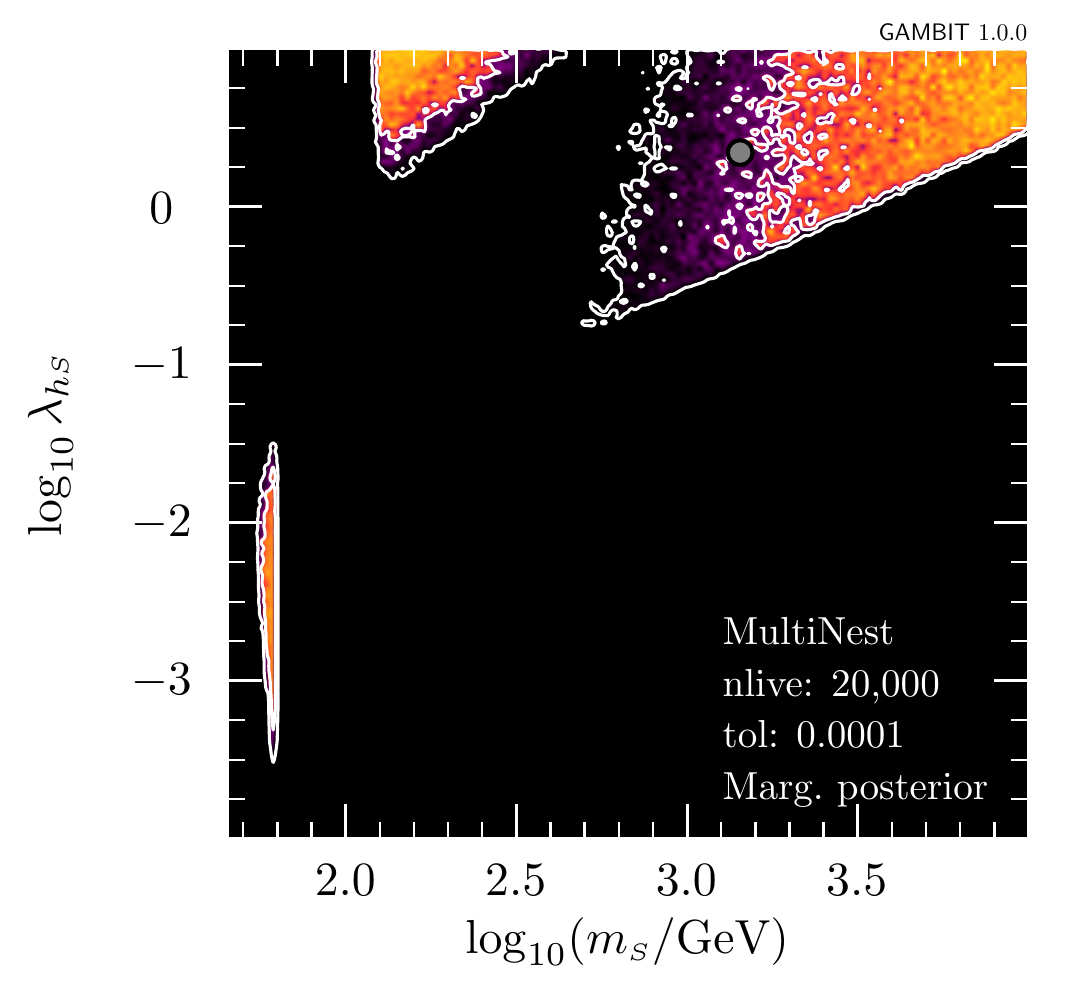}\hspace{-3mm}%
  \includegraphics[height=0.234\linewidth]{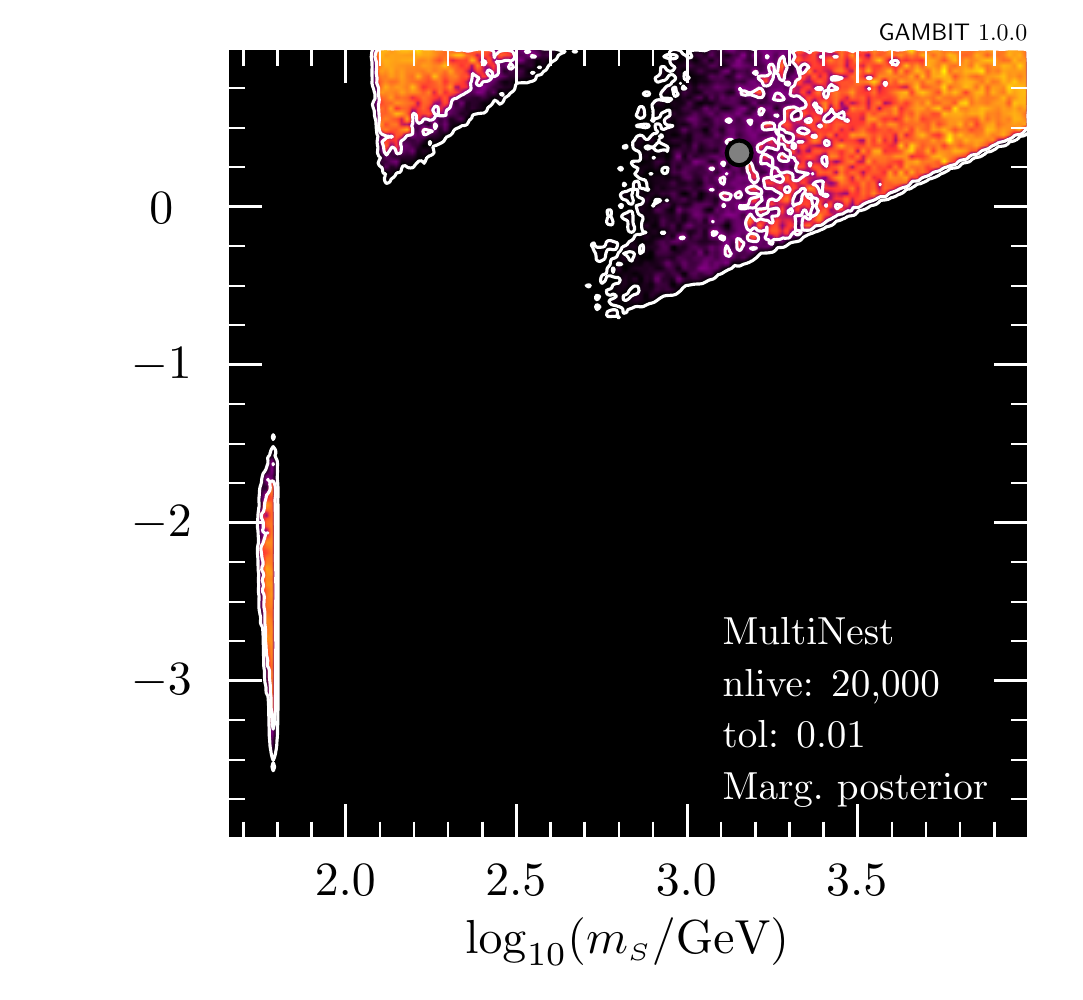}\hspace{-3mm}%
  \includegraphics[height=0.234\linewidth]{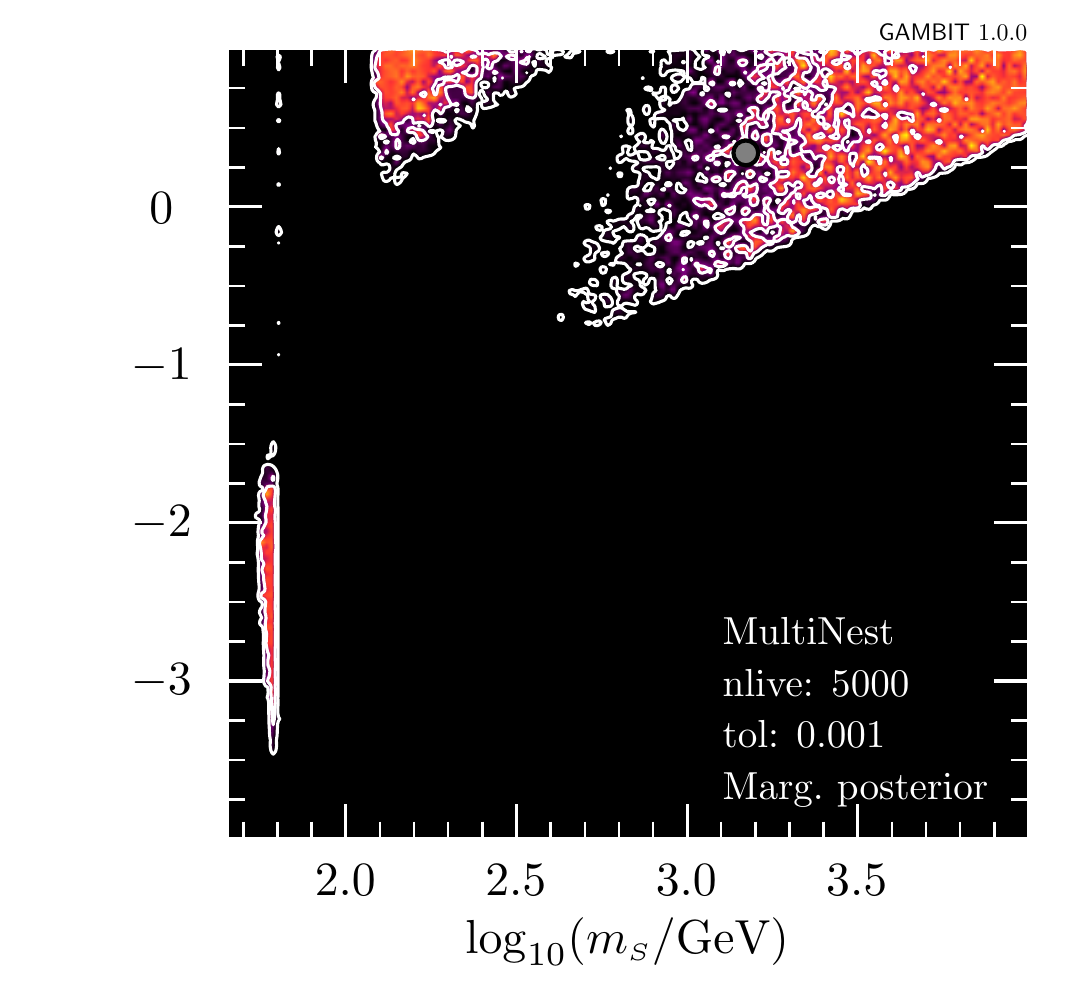}\hspace{-3mm}%
  \includegraphics[height=0.234\linewidth]{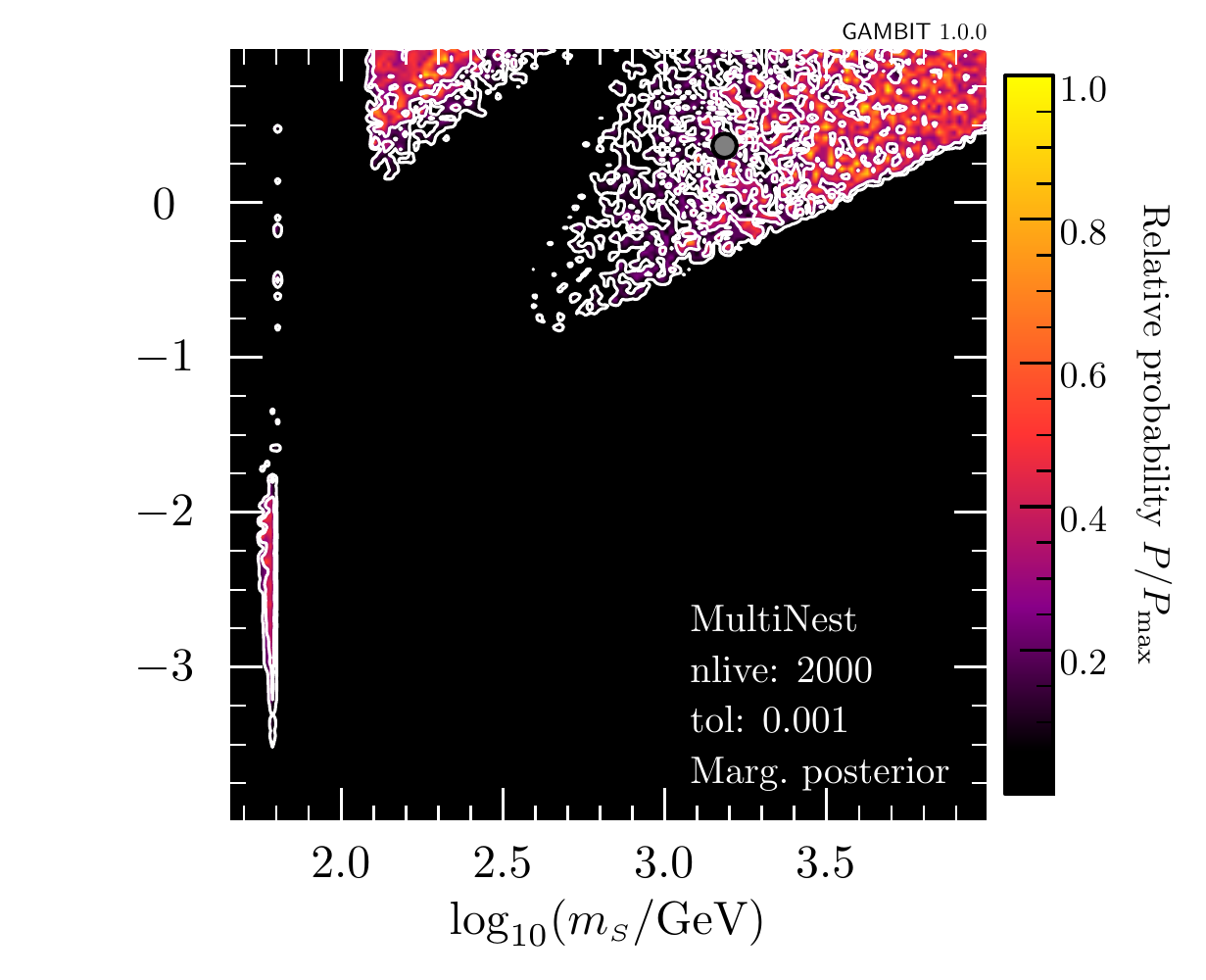}
  \caption{Marginalised posterior probability density maps from a 2-dimensional scan of the scalar singlet parameter space, using the \multinest scanner with a selection of difference tolerances (\lstinline{tol}) and numbers of live points (\lstinline{nlive}).  Note that the colourbar strictly only applies to the rightmost panel, and that colours map to the same enclosed posterior mass on each plot, rather than to the same iso-posterior density level (i.e.~the transition from red to purple is designed to occur at the edge of the $1\sigma$ credible region, and so on).  The posterior mean is shown with a grey bullet point.}
  \label{fig:multinest_plots_post2}
\end{figure*}

\begin{figure*}[tp]
  \centering
  \includegraphics[height=0.234\linewidth]{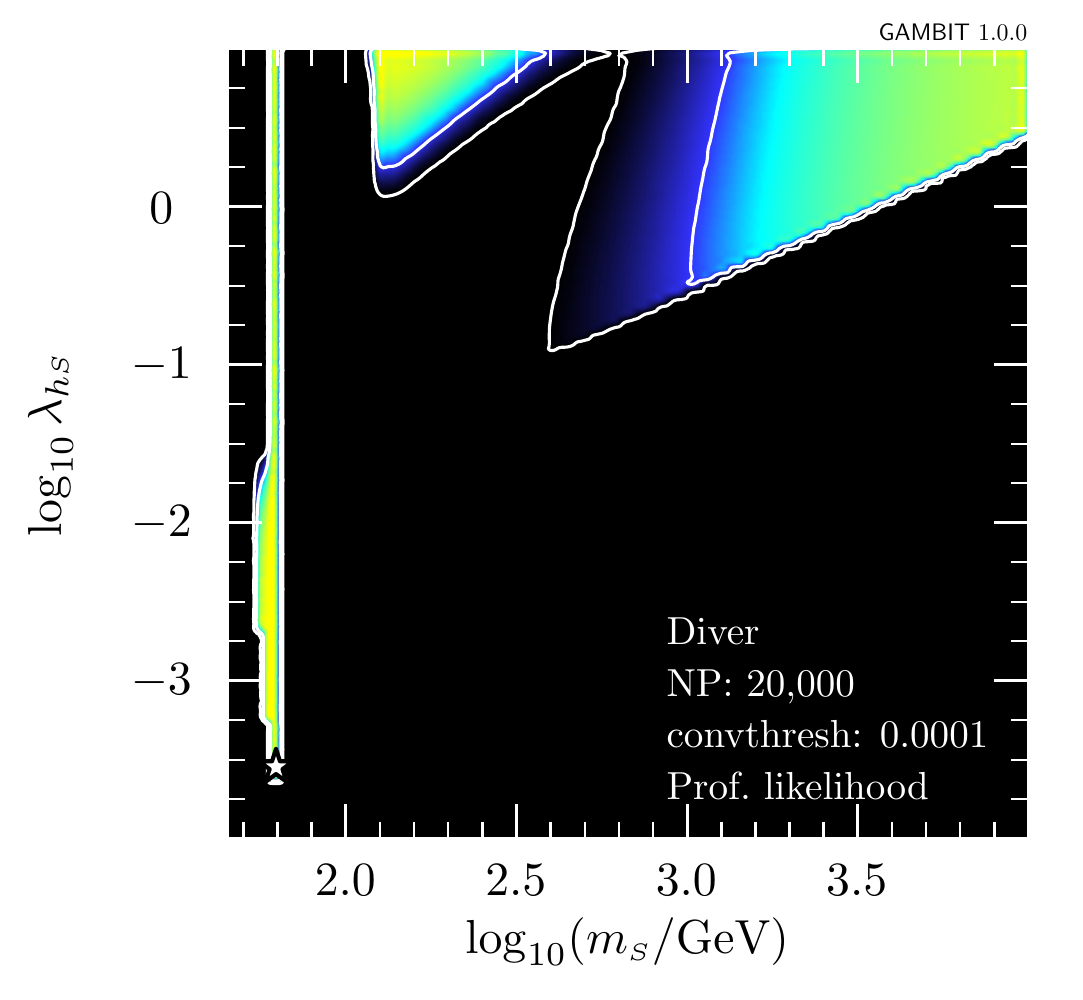}\hspace{-3mm}%
  \includegraphics[height=0.234\linewidth]{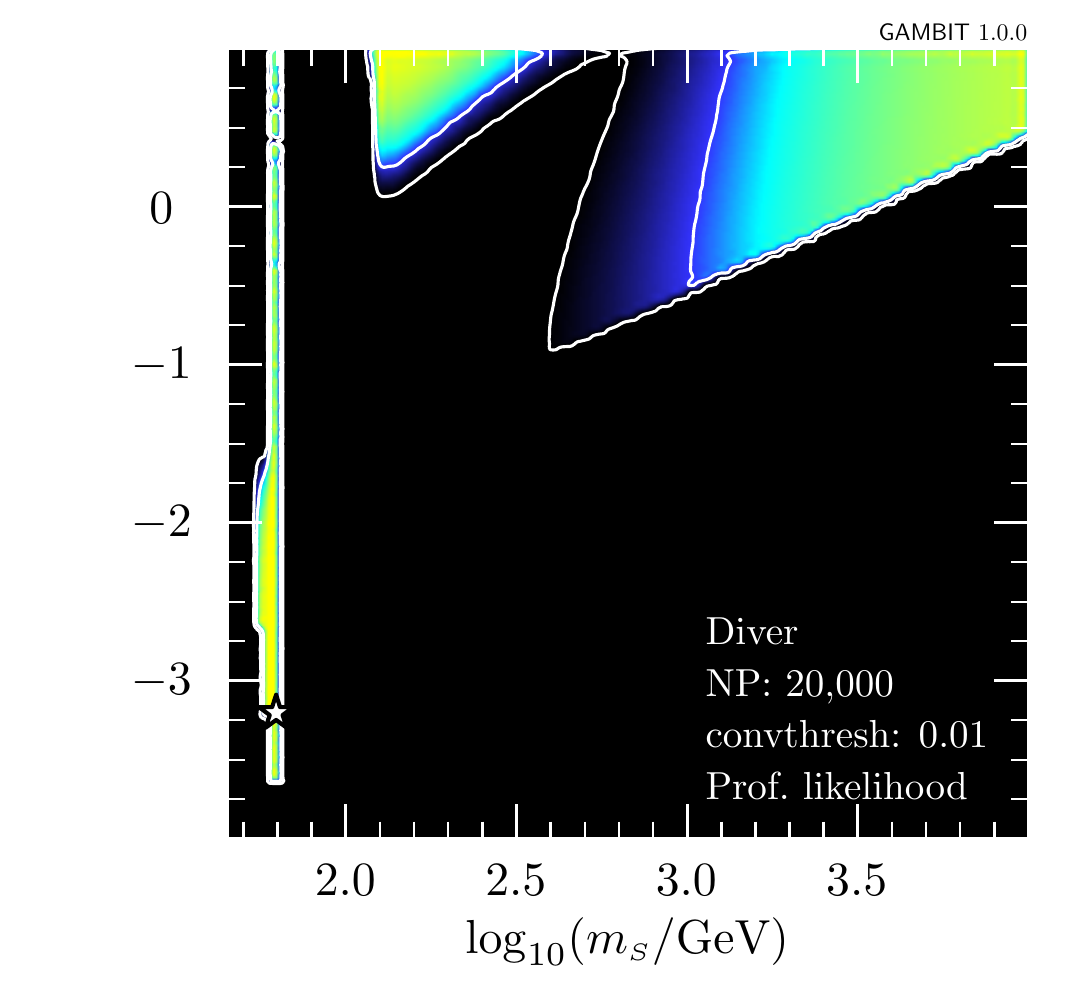}\hspace{-3mm}%
  \includegraphics[height=0.234\linewidth]{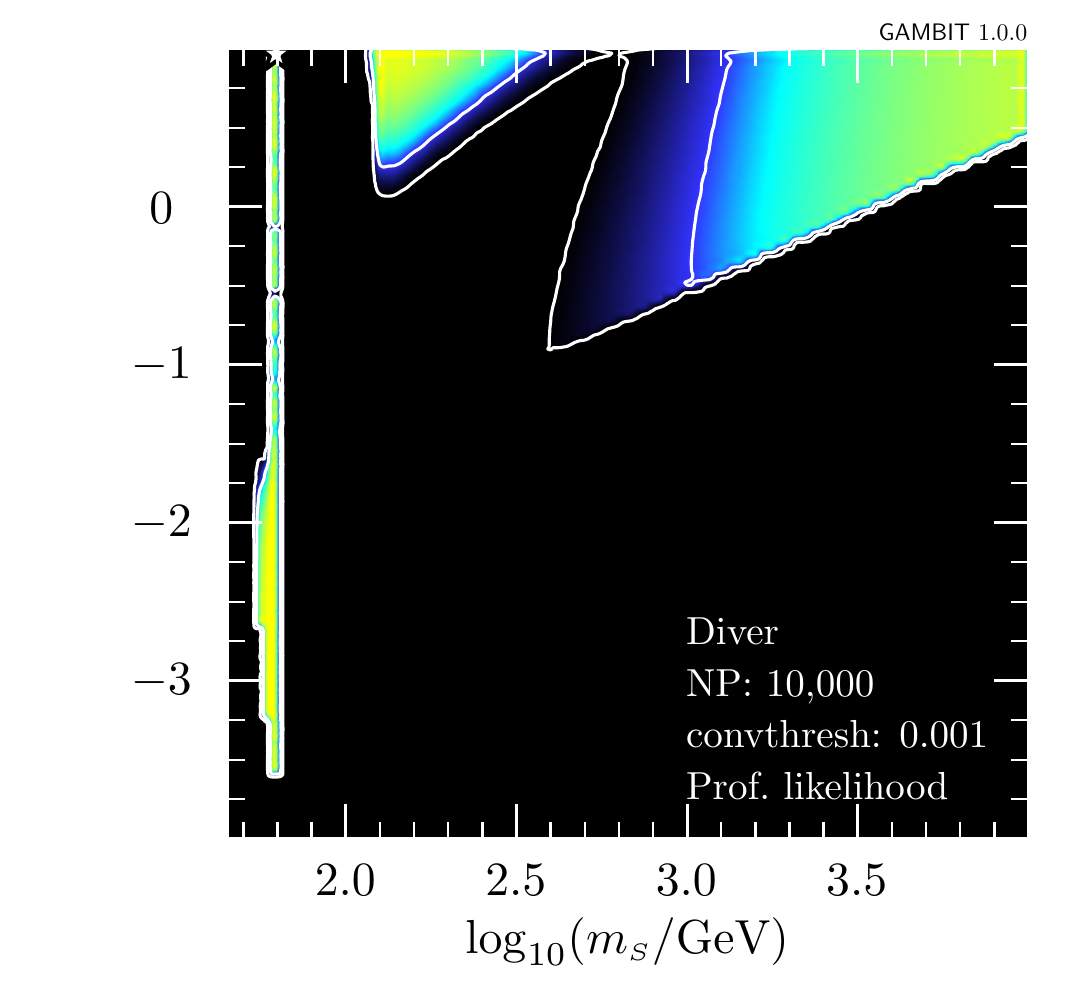}\hspace{-3mm}%
  \includegraphics[height=0.234\linewidth]{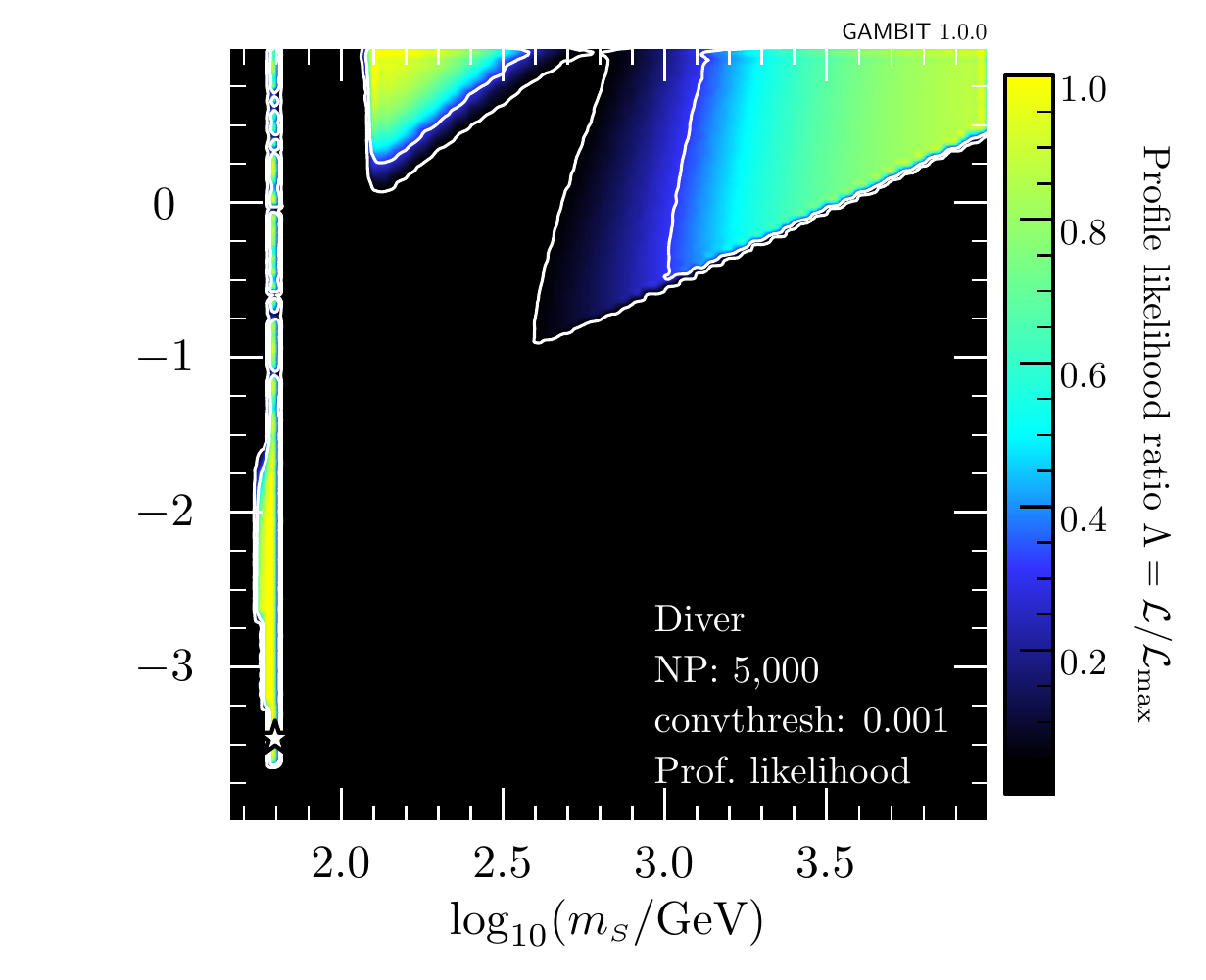}
  \caption{Profile likelihood ratio maps from a 2-dimensional scan of the scalar singlet parameter space, using the \diver scanner with a selection of different convergence thresholds (\protect\fortran{convthresh}) and population sizes (\protect\fortran{NP}).  The maximum likelihood point is shown by a white star.}
  \label{fig:diver_plots2}
\end{figure*}

\subsection{\multinest \& \diver}\label{app:multinest}

The profile likelihoods for \multinest and \diver are presented in Figs.\ \ref{fig:multinest_plots2} and \ref{fig:diver_plots2} respectively.  The marginalised posterior for \multinest is given in Fig. \ref{fig:multinest_plots_post2}.  For both \multinest and \diver, we present scans with the same settings as used for the 15-dimensional equivalent (Figs.\ \ref{fig:multinest_plots}, \ref{fig:multinest_plots_post} and \ref{fig:diver_plots}).

The quality of the profile likelihood is dramatically better in the two-dimensional scans than in the fifteen-dimensional equivalents.  Although \multinest did not sample the low-mass region at all in fifteen dimensions, it has been well sampled in two.  The maximum likelihood point is located in the low-mass mode in all scans presented in Figs. \ref{fig:multinest_plots2} and \ref{fig:diver_plots2}.  This is in good agreement with the analysis in Figs. \ref{fig:Diver_MultiNest_1} and \ref{fig:Diver_MultiNest_2}, in which the maximum likelihood was easily achieved in two dimensions with less stringent scanner settings.

The marginalised posteriors in Fig. \ref{fig:multinest_plots_post2} show some qualitative differences to their 15-dimensional counterparts in Fig.\ \ref{fig:multinest_plots_post}.  The primary difference is that the low-mass region shows in two dimensions, but not in fifteen.  This is because in two dimensions, the low-mass region does not suffer from the same fine-tuning penalty (imposed by the integration over the nuisance parameters) as in fifteen dimensions.  This penalty is due to the dependence of the exact location and shape of the low-mass region on the values of the 13 nuisance parameters included in the 15-dimensional scan. This reduces the ratio of the posterior mass of the low-mass mode to the posterior mass of the high-mass mode in the fifteen-dimensional scan.

\begin{figure*}[tp]
  \centering
  \includegraphics[height=0.234\linewidth]{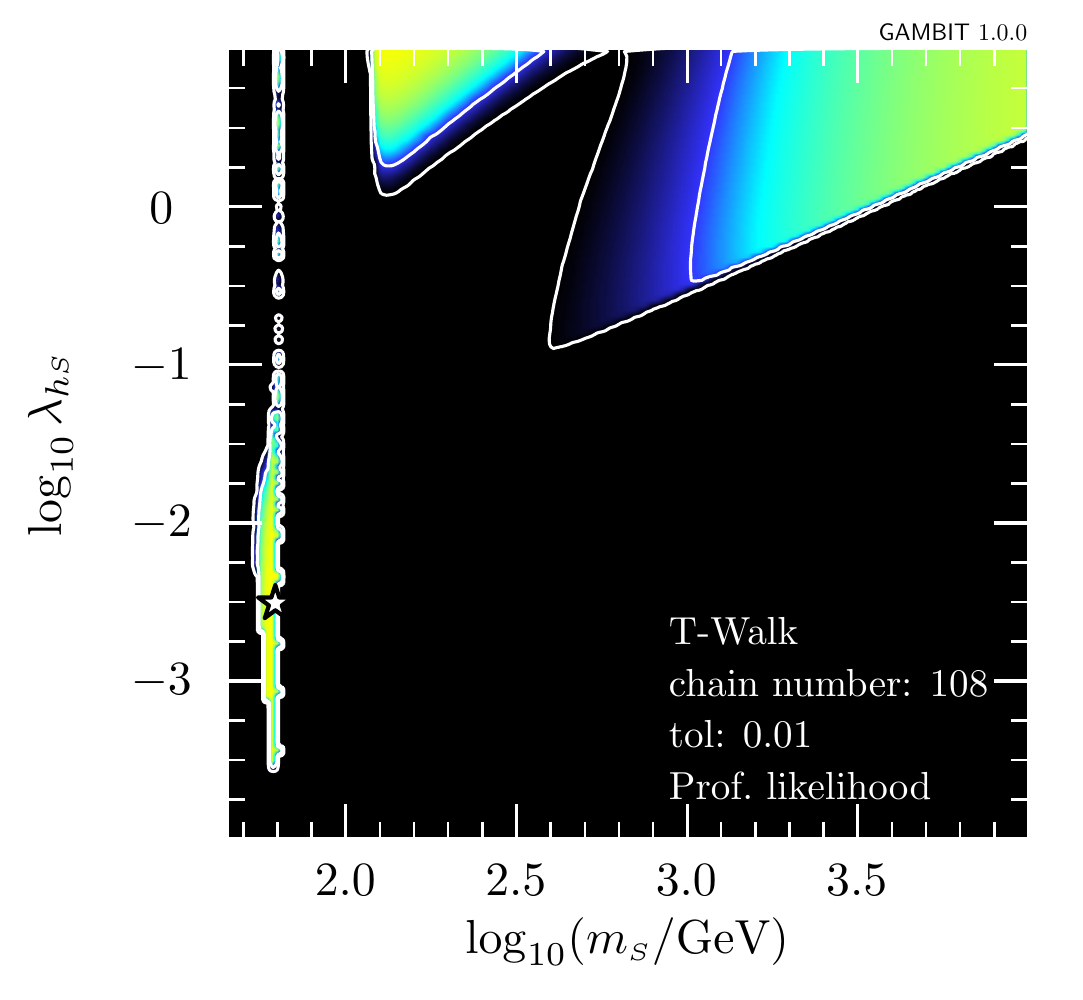}\hspace{-3mm}%
  \includegraphics[height=0.234\linewidth]{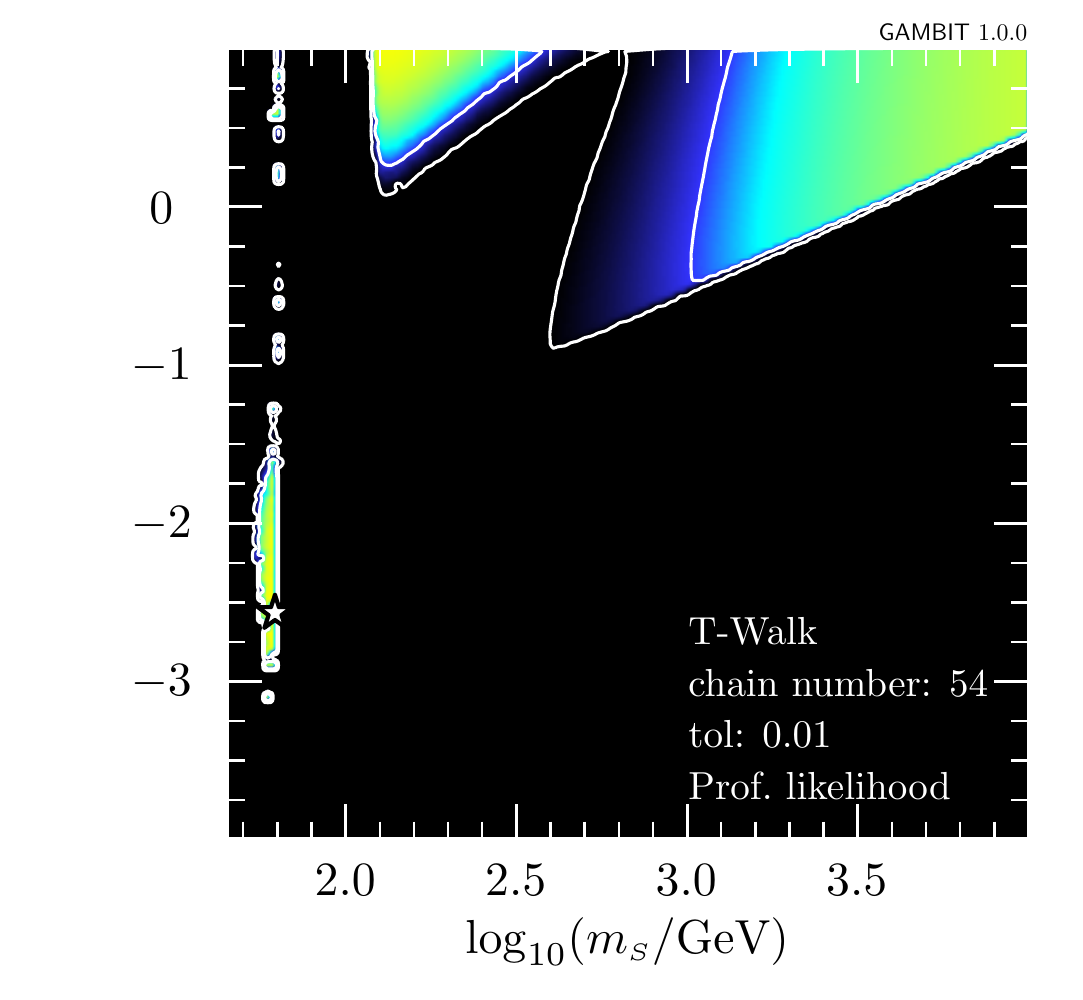}\hspace{-3mm}%
  \includegraphics[height=0.234\linewidth]{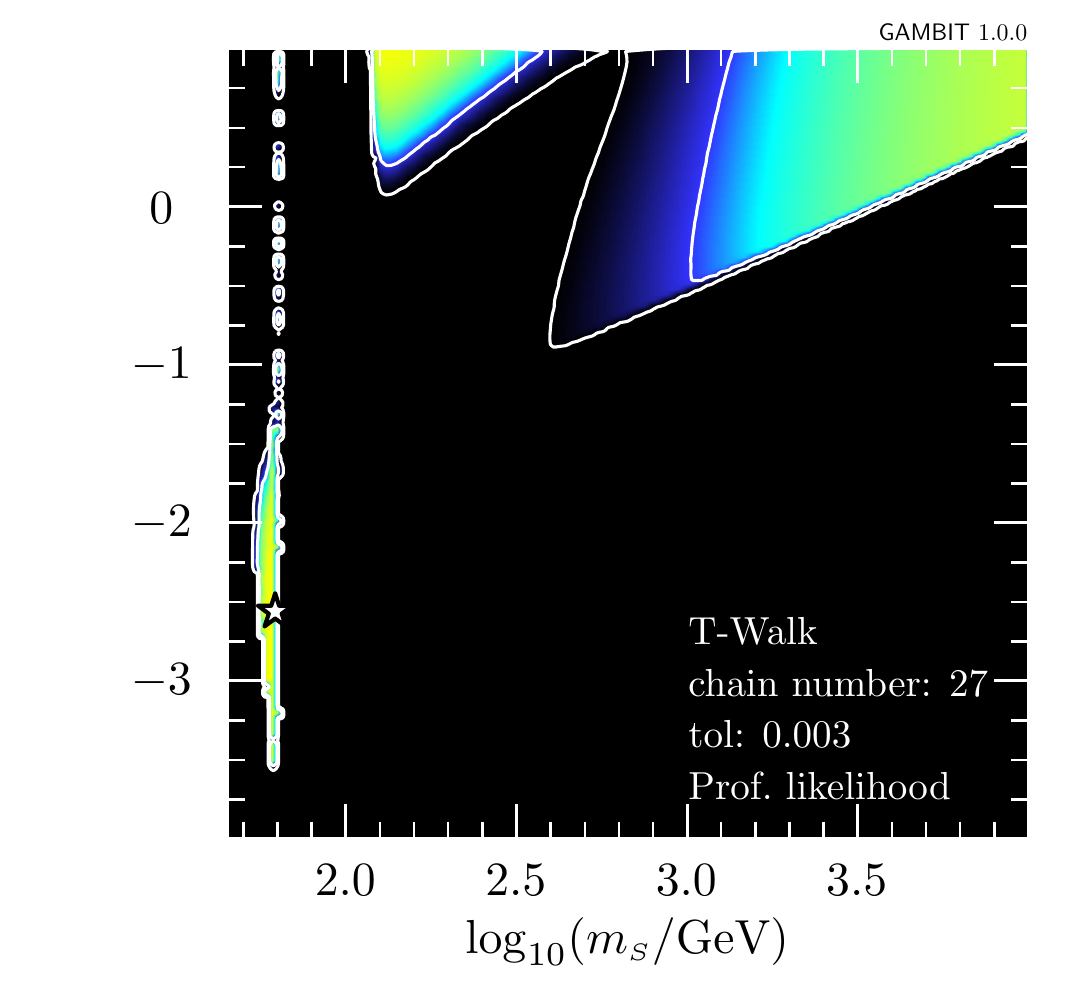}\hspace{-3mm}%
  \includegraphics[height=0.234\linewidth]{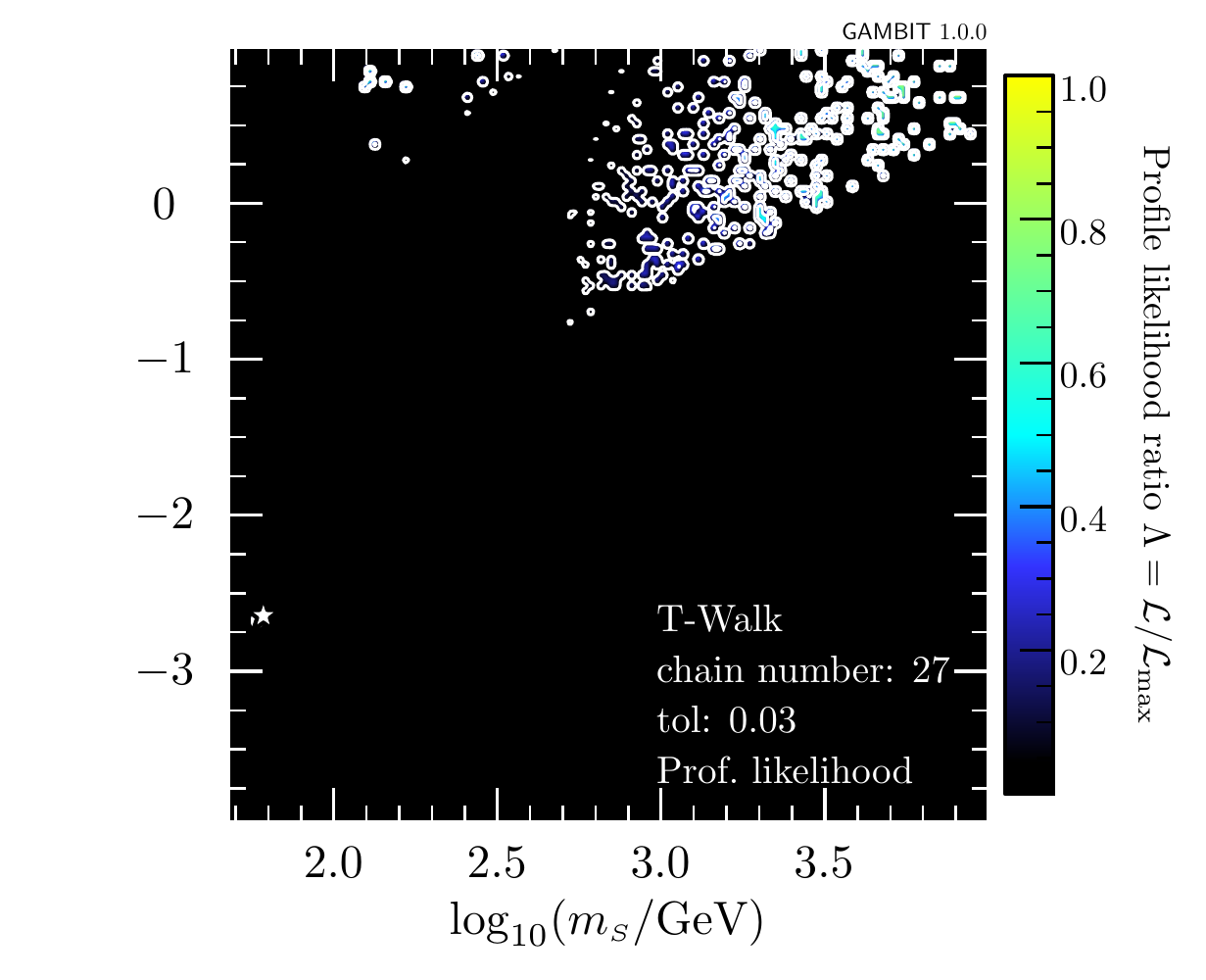}
  \caption{Profile likelihood ratio maps from a 2-dimensional scan of the scalar singlet parameter space, using the \twalk scanner with various numbers of chains and different tolerances.  The maximum likelihood point is shown by a white star.}  \label{fig:twalk_plots2}
\end{figure*}

\begin{figure*}[tp]
  \centering
  \includegraphics[height=0.234\linewidth]{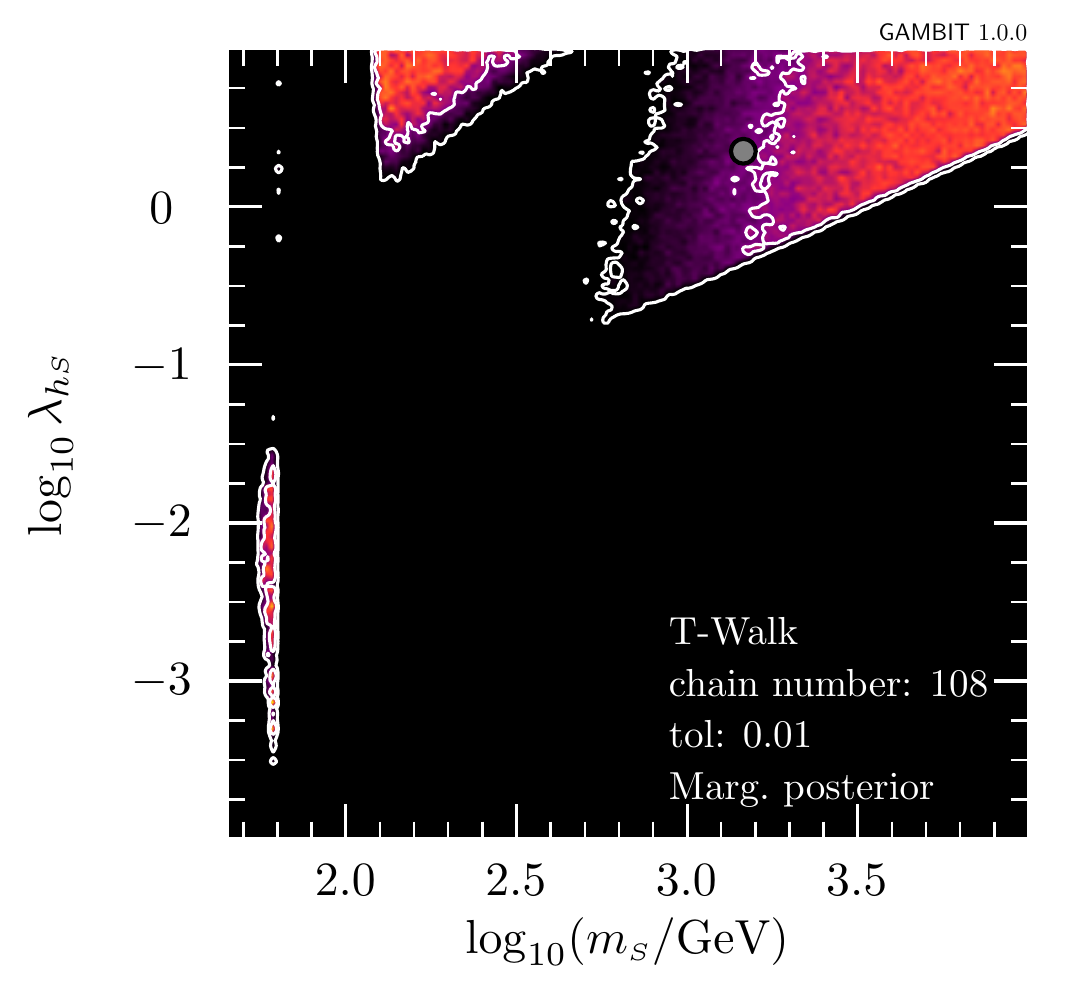}\hspace{-3mm}%
  \includegraphics[height=0.234\linewidth]{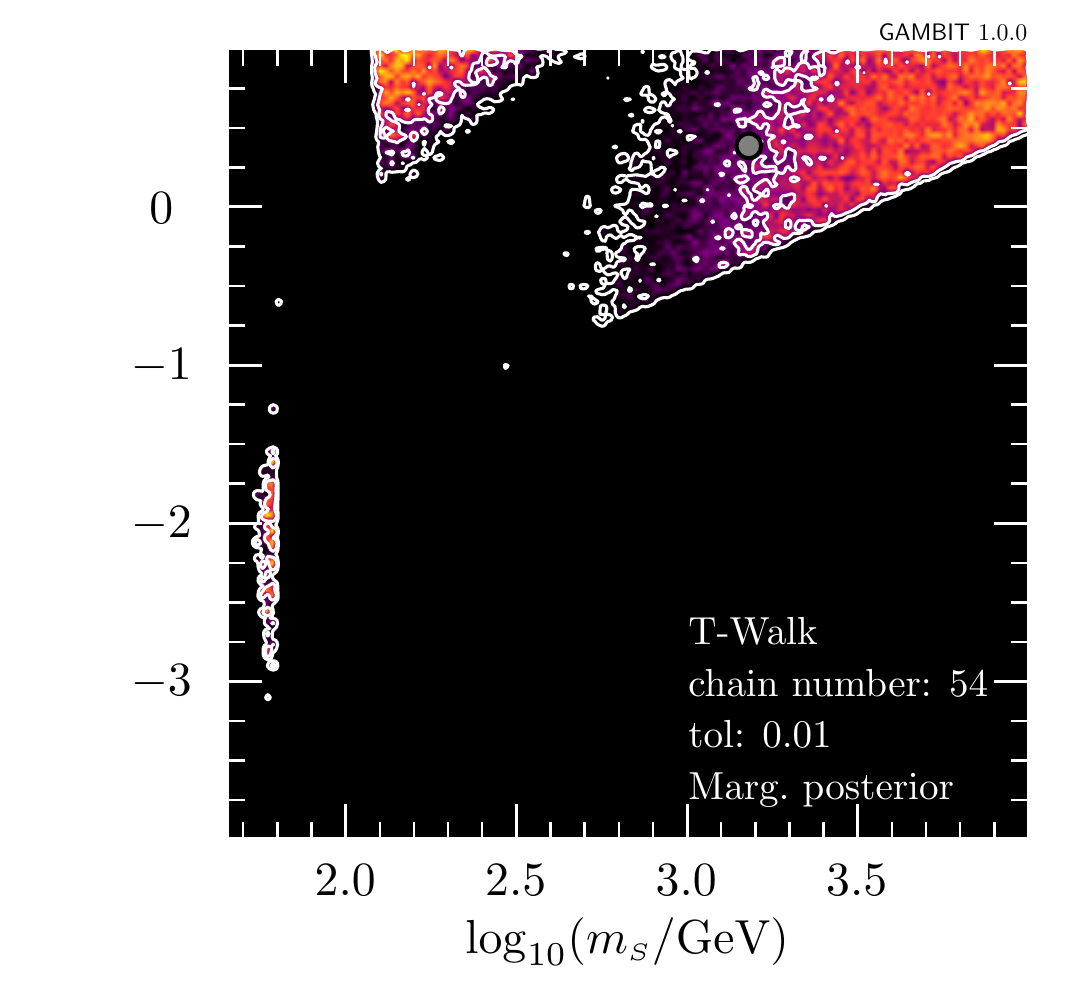}\hspace{-3mm}%
  \includegraphics[height=0.234\linewidth]{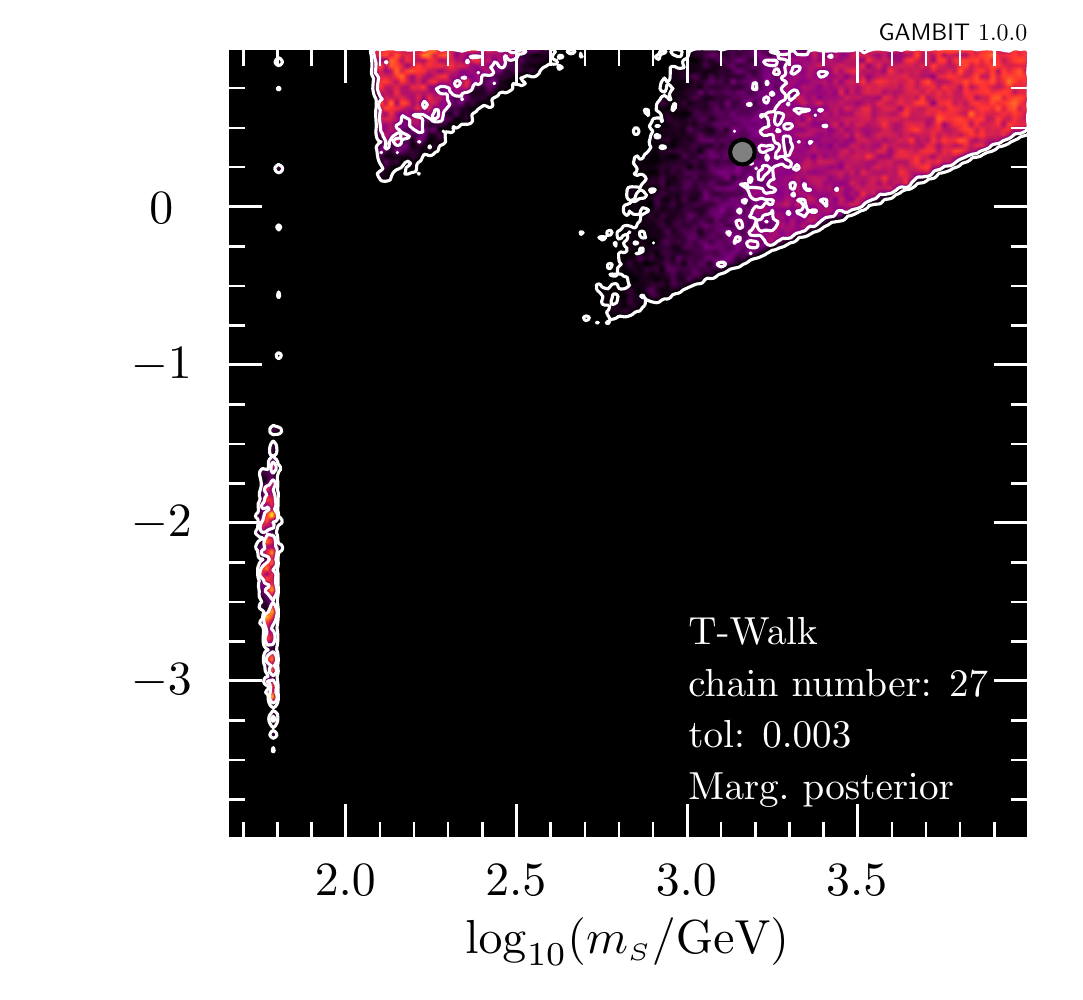}\hspace{-3mm}%
  \includegraphics[height=0.234\linewidth]{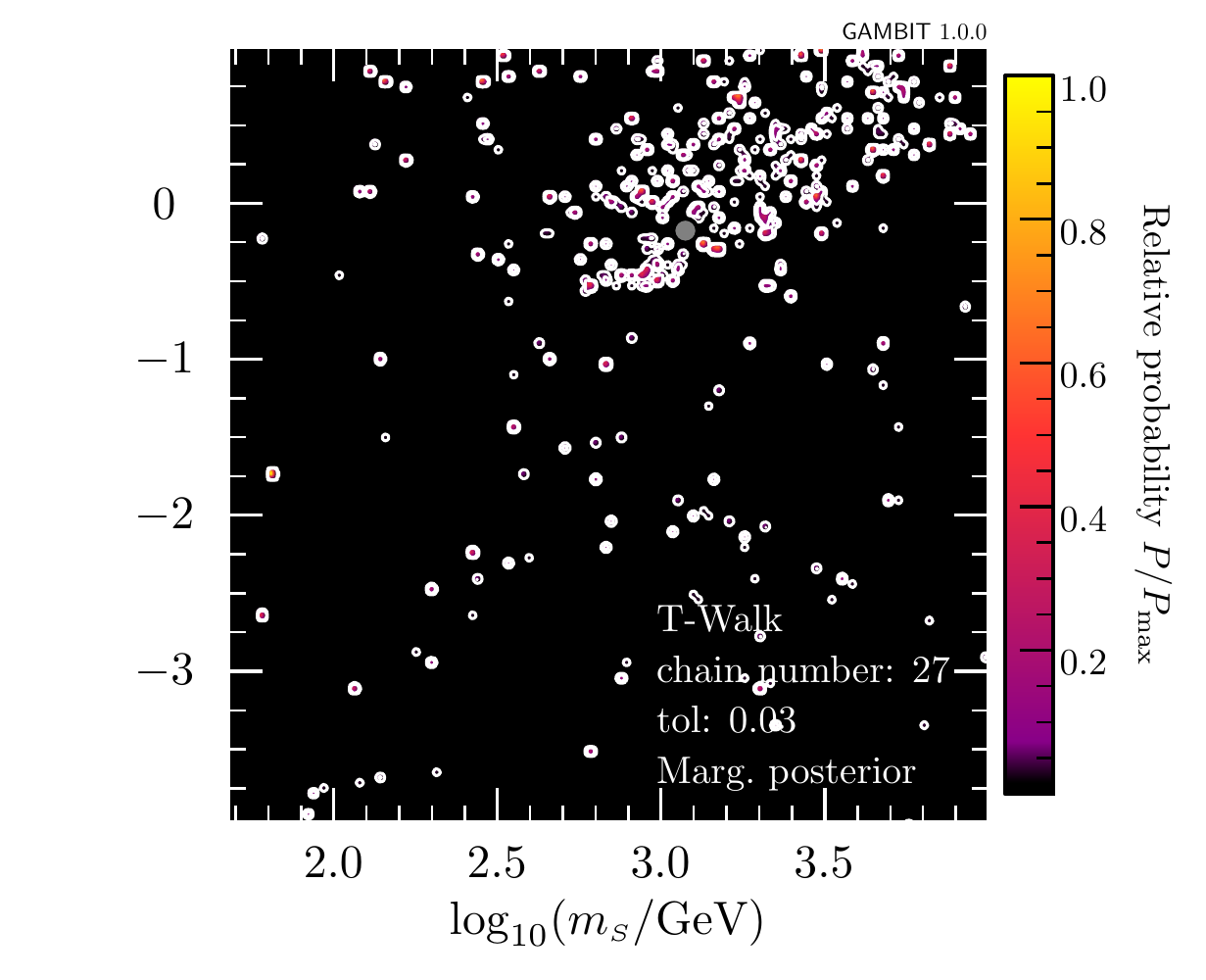}
  \caption{Marginalised posterior probability density maps from a 2-dimensional scan of the scalar singlet parameter space, using the \twalk scanner with various numbers of chains and different tolerances.  Note that the colourbar strictly only applies to the rightmost panel, and that colours map to the same enclosed posterior mass on each plot, rather than to the same iso-posterior density level (i.e.~the transition from red to purple is designed to occur at the edge of the $1\sigma$ credible region, and so on).  The posterior mean is shown with a grey bullet point.}
  \label{fig:twalk_plots_post2}
\end{figure*}

\subsection{\twalk}

The profile likelihoods and marginalised posteriors for two-dimensional \twalk scans are presented in Figs. \ref{fig:twalk_plots2} and \ref{fig:twalk_plots_post2}, respectively.  We use different \twalk settings compared to Figs. \ref{fig:twalk_plots} and \ref{fig:twalk_plots_post}.  This is primarily dictated by the dimensional dependence of the optimal number of chains, \yaml{chain_number}, as discussed in Secs.\ \ref{sec:twalk} and \ref{app:options:twalk}.  We find that values of \yaml{tol} $\sim$ 0.1 causes very rapid convergence in two dimensions, even before any meaningful sampling can occur. This behaviour can be seen in the right-most plot of Fig.\ \ref{fig:twalk_plots2}, where \yaml{tol} = 0.03.  We therefore use different settings, more appropriate for the two-dimensional parameter space.

We find that \twalk samples the profile likelihood very well in two dimensions when \yaml{tol} $\lesssim$ 0.01.  The number of chains appears to have less influence on the quality of the sampling, but dramatically increases the runtime.  The scans of the two left-most plots in Fig.\ \ref{fig:twalk_plots2} took $\sim$4 hours (\yaml{chain_number} = 54) and $\sim$18 hours (\yaml{chain_number} = 108).

Although the sampling of the profile likelihood is much more complete in these two-dimensional scans than in the 15 dimensional case, there is no significant improvement in the marginalised posteriors (Fig. \ref{fig:twalk_plots_post2}).  However, we do see that the low-mass region appears within the two-sigma contours (as discussed in Sec.\ \ref{app:multinest}).

\begin{figure*}[tp]
  \centering
  \includegraphics[height=0.234\linewidth]{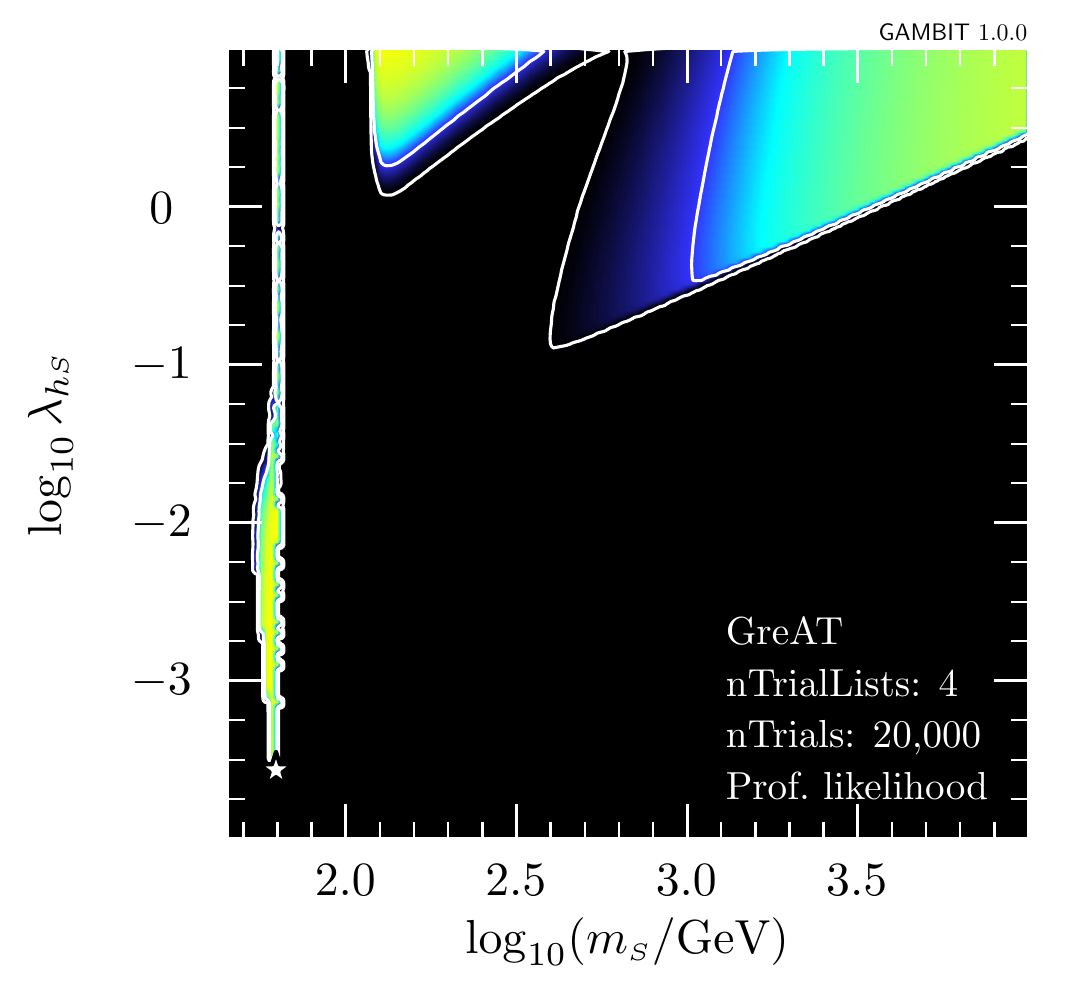}\hspace{-3mm}%
  \includegraphics[height=0.234\linewidth]{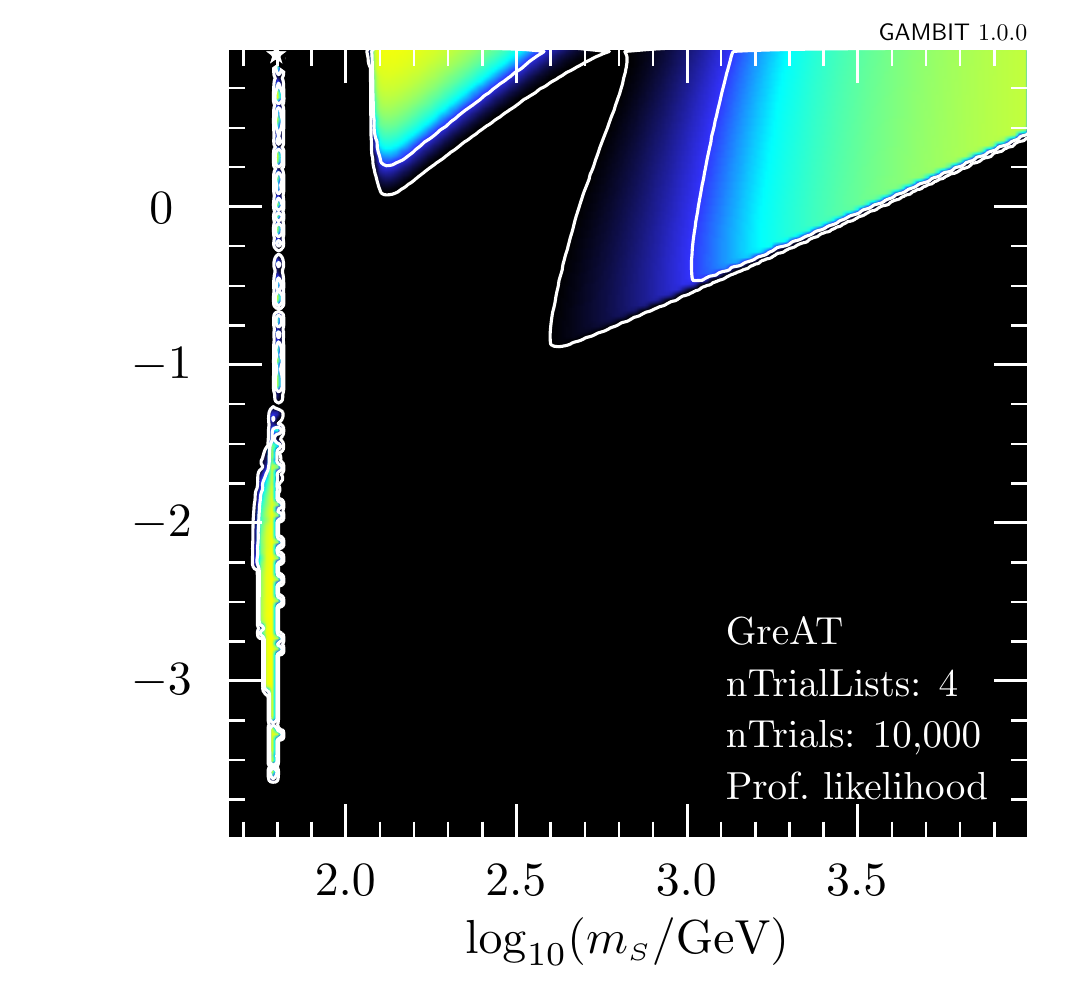}\hspace{-3mm}%
  \includegraphics[height=0.234\linewidth]{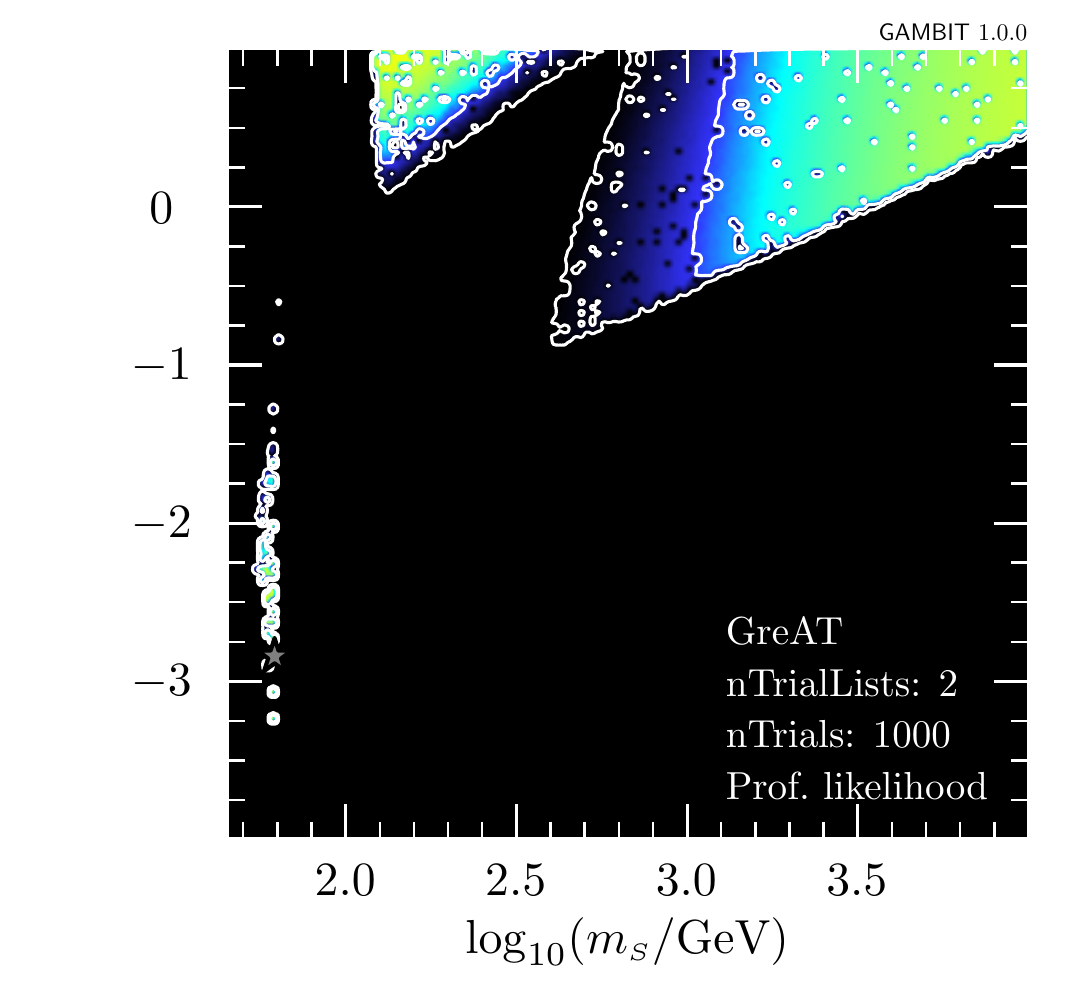}\hspace{-3mm}%
  \includegraphics[height=0.234\linewidth]{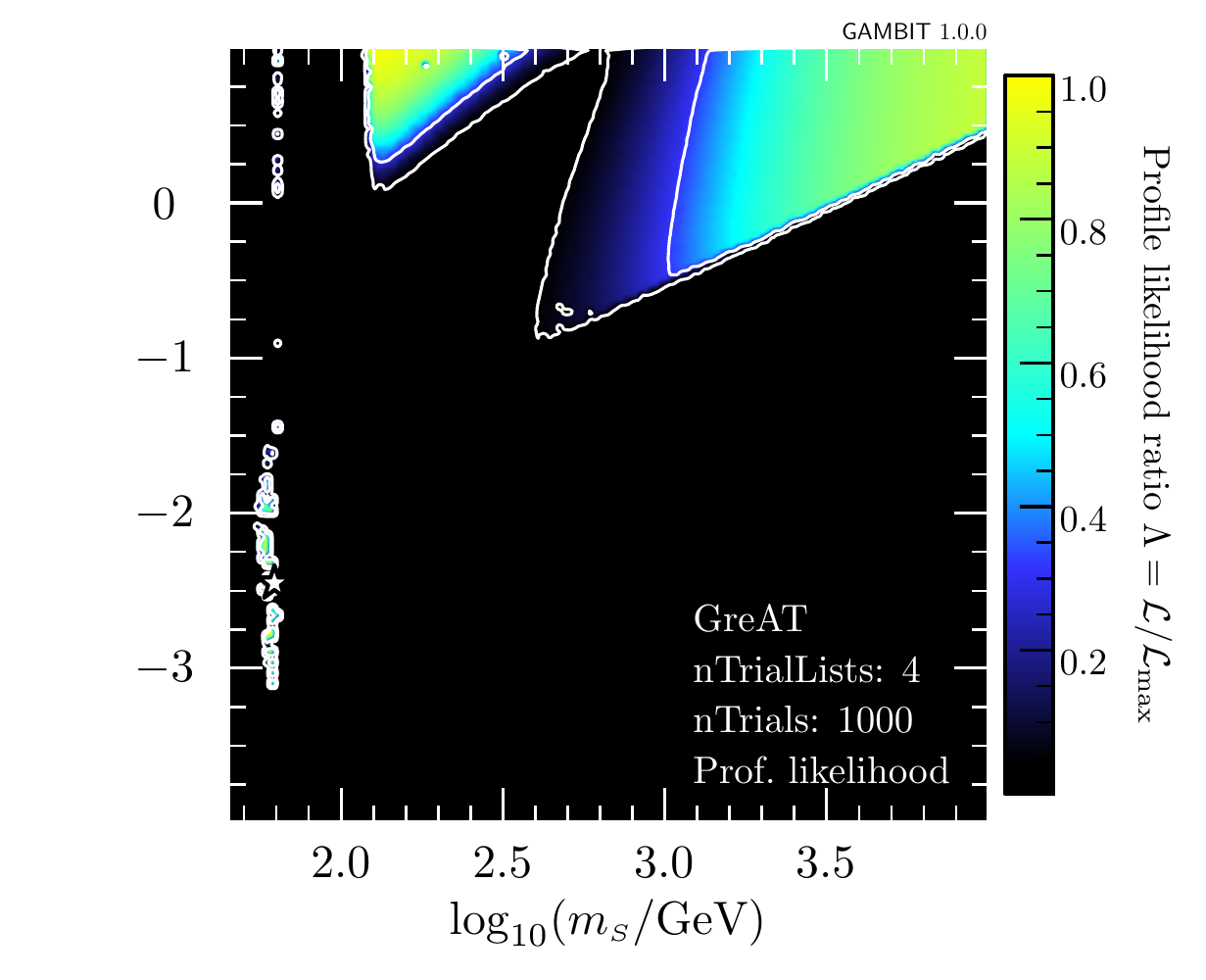}
  \caption{Profile likelihood ratio maps from a 2-dimensional scan of the scalar singlet parameter space, using the \great sampler with various numbers of chains (\cpp{nTrialLists}) and chain lengths (\cpp{nTrials}).  The maximum likelihood point is shown by a white star.}
  \label{fig:great_plots2}
\end{figure*}

\begin{figure*}[tp]
  \centering
  \includegraphics[height=0.234\linewidth]{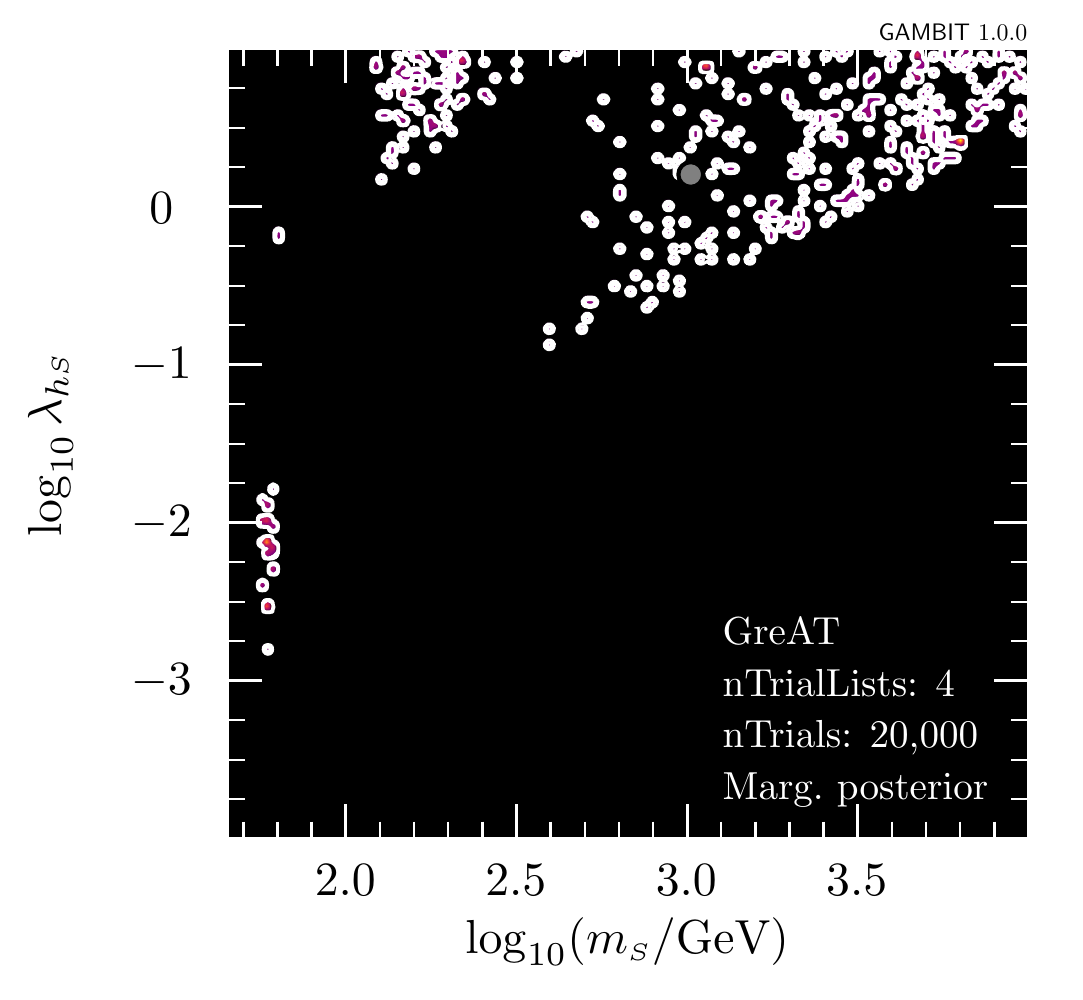}\hspace{-3mm}%
  \includegraphics[height=0.234\linewidth]{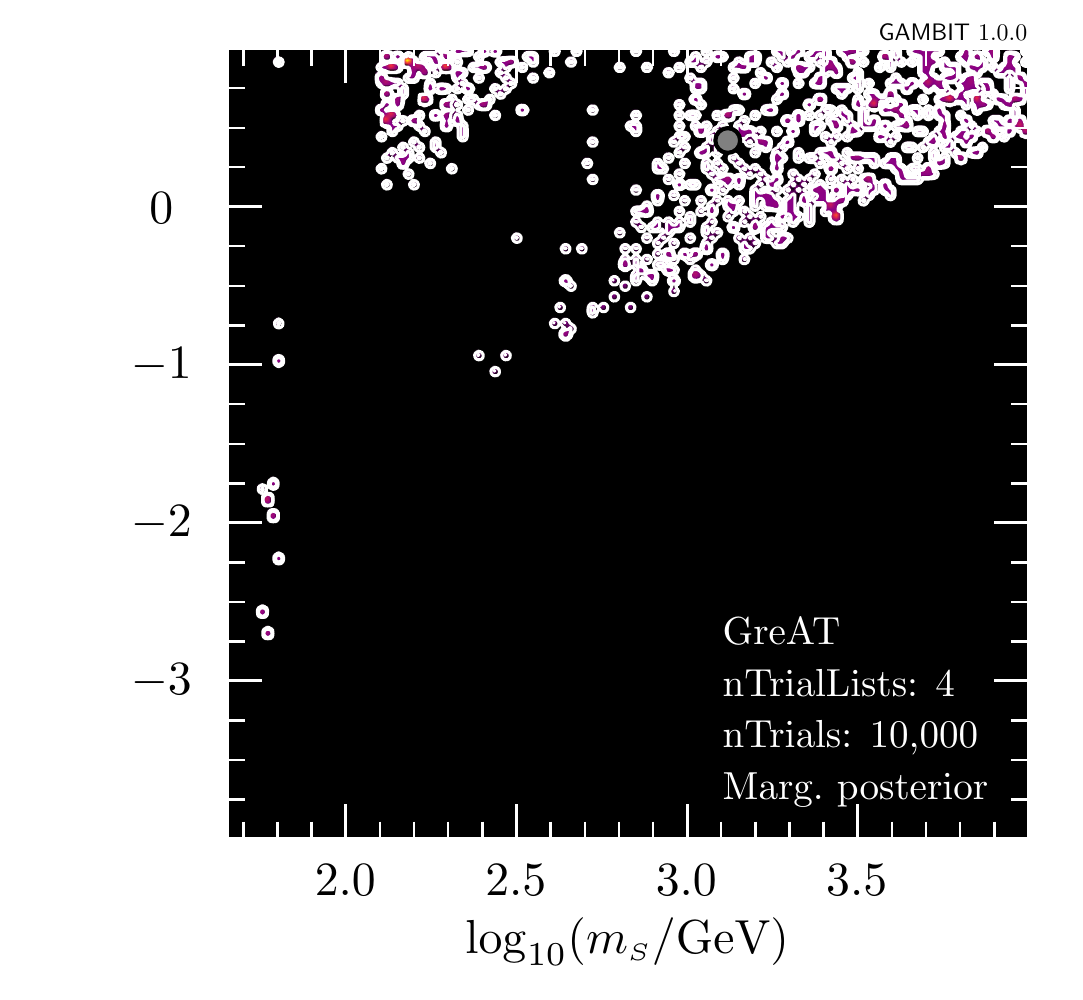}\hspace{-3mm}%
  \includegraphics[height=0.234\linewidth]{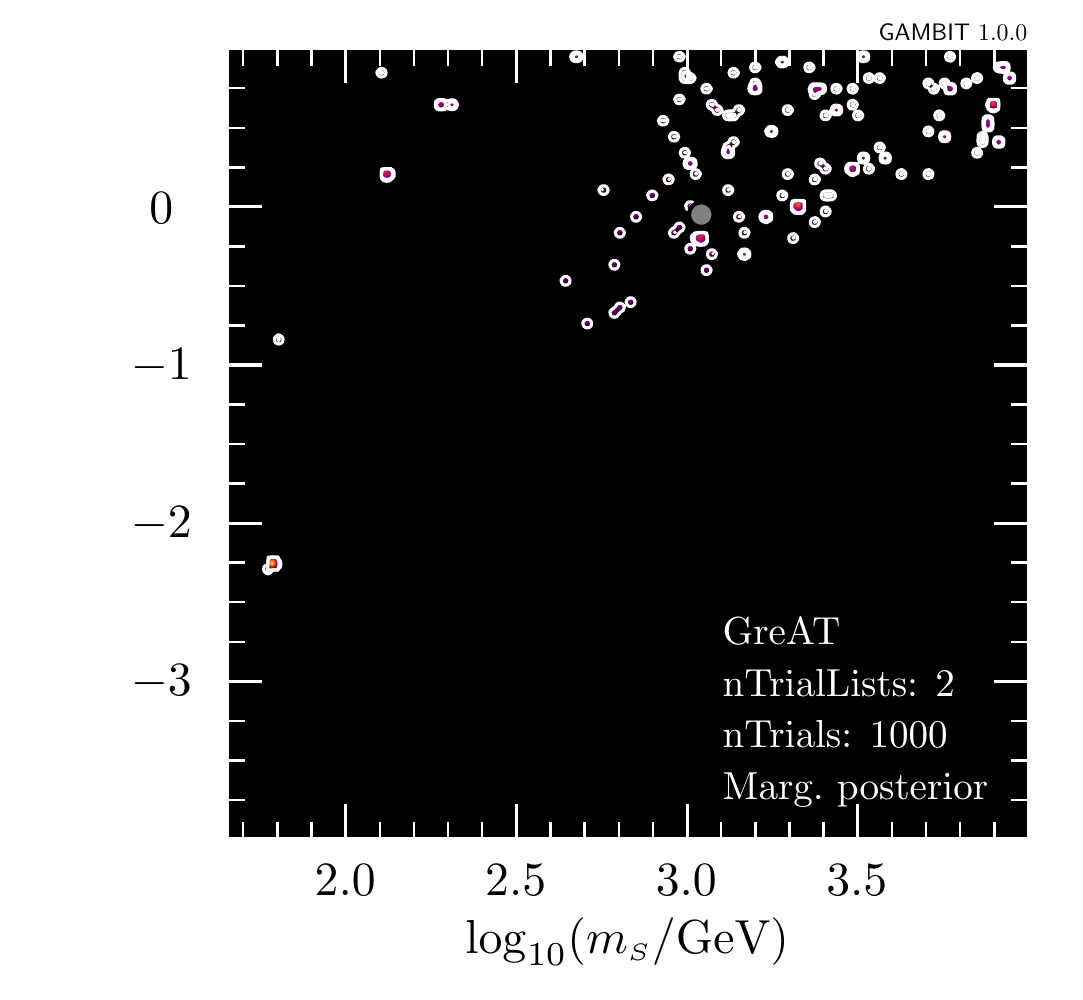}\hspace{-3mm}%
  \includegraphics[height=0.234\linewidth]{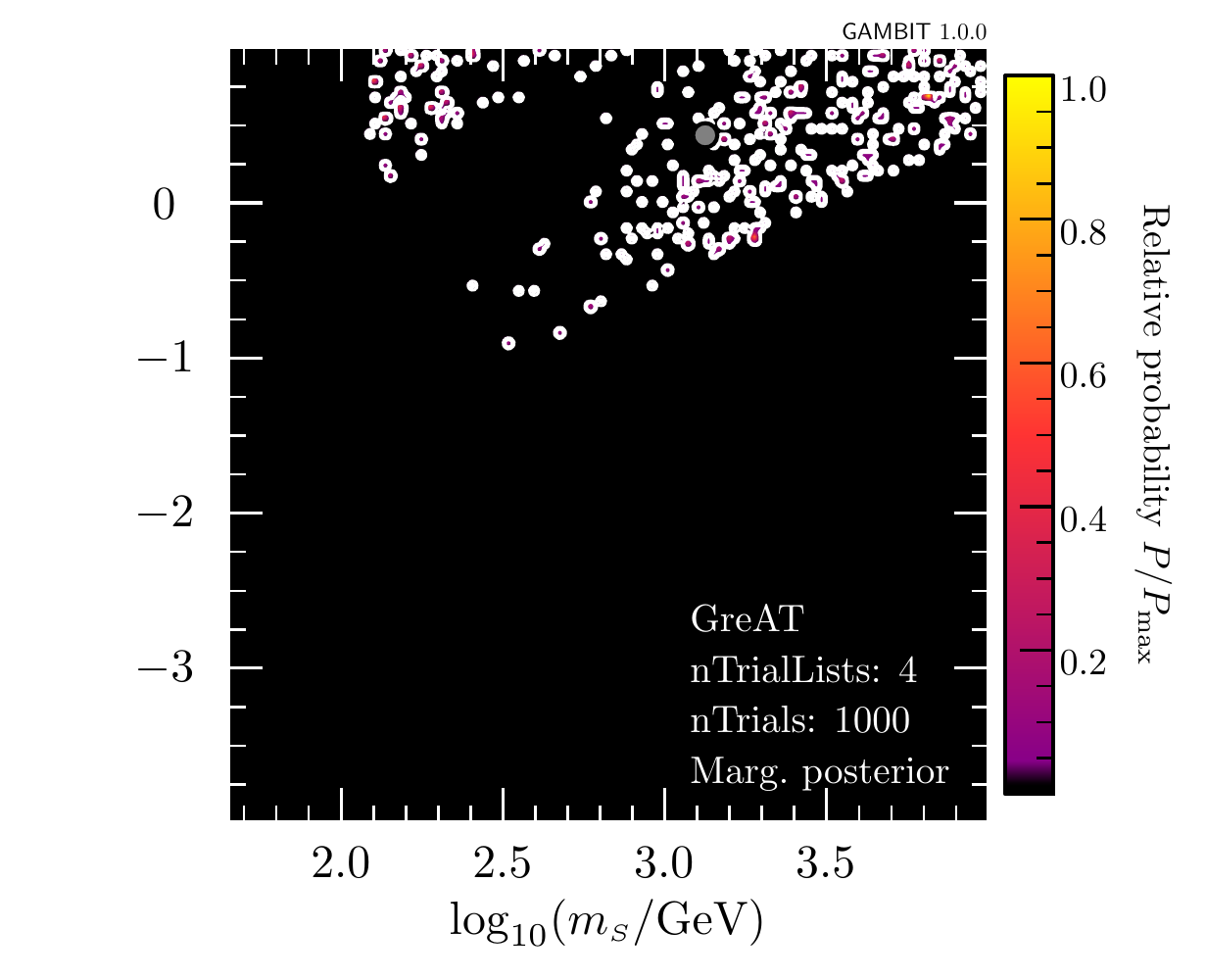}
  \caption{Marginalised posterior ratio maps from a 2-dimensional scan of the scalar singlet parameter space, using the \great sampler with various numbers of chains (\cpp{nTrialLists}) and chain lengths (\cpp{nTrials}).  Note that the colourbar strictly only applies to the rightmost panel, and that colours map to the same enclosed posterior mass on each plot, rather than to the same iso-posterior density level (i.e.~the transition from red to purple is designed to occur at the edge of the $1\sigma$ credible region, and so on).  The posterior mean is shown with a grey bullet point.}
  \label{fig:great_plots_post2}
\end{figure*}

\subsection{\great}

The profile likelihoods and marginalised posteriors for \great scans in a two-dimensional parameter space are presented in Figs. \ref{fig:great_plots2} and \ref{fig:great_plots_post2}, respectively.  The scanner settings in these plots are equivalent to those in Figs. \ref{fig:great_plots} and \ref{fig:great_plots_post}, except for \yaml{nTrialLists}, which is set to $N_{\text{dim}} = 2$ or $N_{\text{dim}}+2 = 4$.

The two left-most plots of Fig. \ref{fig:great_plots2} clearly show that these settings are excessive for sampling the profile likelihood in two dimensions.  Even though all panels in Fig. \ref{fig:great_plots2} exibit well-sampled profile likelihoods, one can make an optimal choice when considering the computing time taken.  From left to right, the scans took $\sim$5\,hr, 3\,hr, 8\,min and 17\,min.  Only in the quickest two scans does some degradation of the contours and sampling begin to appear.  In contrast to the quality of the profile likelihood, we see in Fig. \ref{fig:great_plots_post2} that even with a long scan, in two dimensions the marginalised posterior is not well sampled by \great.

\subsection{Summary}

We have presented profile likelihoods and marginalised posteriors for scans of a two-dimensional parameter space, directly comparable to the 15-dimensional case presented in Sec.\ \ref{sec:comparison}.  These plots show that the inclusion of the additional 13 nuisance parameters does not significantly alter the joint profile likelihood of $\lhs$ and $\ms$.  We find that sampling performance is significantly improved, demonstrating that although the additional 13 parameters are well constrained by unimodal likelihoods, their inclusion creates a significant challenge for the sampling algorithms.

\section{\YAML input file example}
\label{app:inifile}

Below is an example \YAML input file for \textsf{ScannerBit\_standalone} that uses the custom prior defined in Appendix \ref{app:priors}, and the scanner and objective plugins declared in Appendices \ref{app:scanner} and \ref{app:objective}.

\begin{lstyaml}
Parameters:
  EggBox:
    param_0:
      range: [@\yamlvalue{0, 1}@]
    param_1:
      prior_type: dummy

Scanner:
  use_objectives: eggbox_like
  use_scanner: random_scanner

  objectives:
    eggbox_like:
      plugin: EggBox
      purpose: loglike
      length: [@\yamlvalue{12, 12}@]

  scanners:
    random_scanner:
      plugin: random
      point_number: 2000
      like: loglike

Printer:
  printer: ascii
  options:
    output_file: "results.txt"

KeyValues:
  likelihood:
    model_invalid_for_lnlike_below: -1e5
\end{lstyaml}

Here, the model chosen for scanning is actually given as \yamlvalue{EggBox}, which is the name of an objective plugin.  Although we have not discussed such usage earlier in this paper, an objective plugin can in fact even be listed as a model when it provides the likelihood that is to be scanned, as is done here.  This can be useful for avoiding any need to explicitly define a new model in \GB format when all one is interested in is computing some external function provided by an objective plugin. In this case, the names given to parameters in the \YAML file are entirely arbitrary.  Here, the parameter \yaml{param\_0} is defined to have a flat prior in the range \yaml{[}\yamlvalue{0,1}\yaml{]}, and parameter \yaml{param\_1} is defined to use the custom prior \yaml{dummy}.  Next, the objective and scanner plugins are defined in the \yaml{Scanner} section.  The \yaml{eggbox_like} objective is selected with the \yaml{use_objectives} directive, and the \yaml{random\_scanner} scanner is selected as the chosen scanner via the \yaml{use_scanner} directive.  The \yaml{eggbox\_like} objective is defined to use the \yaml{EggBox} plugin, with purpose set to \yaml{loglike}, and the option \yaml{length} set to \yaml{[}\yamlvalue{0, 1}\yaml{]}.  The \yaml{random_scanner} scanner is set to use the \yaml{random} plugin, with functions assigned the purpose \yaml{loglike} used to make up the likelihood function that it will call.  The \yaml{point\_number} option is set to ensure that \yaml{2000} samples are taken.

\startglossary

\gitem{backend}\input{"glossary/backend.glossentry"}
\gitem{backend function}\input{"glossary/backend_function.glossentry"}
\gitem{backend requirement}\input{"glossary/backend_requirement.glossentry"}
\gitem{backend variable}\input{"glossary/backend_variable.glossentry"}
\newcommand{\seecompdatabase}{see Sec.\ 10.7 of Ref.\ \cite{gambit}}
\gitem{capability}\input{"glossary/capability.glossentry"}
\gitem{dependency}\input{"glossary/dependency.glossentry"}
\gitem{dependency resolver}\input{"glossary/dependency_resolver.glossentry"}
\gitem{dependency resolution}\input{"glossary/dependency_resolution.glossentry"}
\gitem{frontend}\input{"glossary/frontend.glossentry"}
\gitem{frontend header}\input{"glossary/frontend_header.glossentry"}
\gitem{likelihood container}\input{"glossary/likelihood_container.glossentry"}
\gitem{model}\input{"glossary/model.glossentry"}
\gitem{module}\input{"glossary/module.glossentry"}
\gitem{module function}\input{"glossary/module_function.glossentry"}
\gitem{physics module}\input{"glossary/physics_module.glossentry"}
\newcommand{\inifilesec}{Sec. 6.6 of Ref.\ \cite{gambit}}
\gitem{printer}\input{"glossary/printer.glossentry"}
\gitem{purpose}\input{"glossary/purpose.glossentry"}
\gitem{scanner plugin}\input{"glossary/scanner_plugin.glossentry"}
\gitem{test function plugin}\input{"glossary/test_function_plugin.glossentry"}
\gitem{rollcall header}\input{"glossary/rollcall_header.glossentry"}
\gitem{type}\input{"glossary/type.glossentry"}

\finishglossary

\bibliography{R1}

\providecommand{\href}[2]{#2}\begingroup\raggedright\begin{thebibliography}{10}

\bibitem{Berger09}
C.~F. {Berger}, J.~S. {Gainer}, J.~A.~L. {Hewett}, and T.~G. {Rizzo}, {\it
  {Supersymmetry without prejudice}},  {\em \jhep} {\bf 2} (2009) 23,
  [\href{http://arxiv.org/abs/0812.0980}{{\tt arXiv:0812.0980}}].

\bibitem{ATLAS15}
ATLAS Collaboration: {ATLAS Collaboration}, {\it {Summary of the ATLAS
  experiment's sensitivity to supersymmetry after LHC Run 1 -- interpreted in
  the phenomenological MSSM}},  {\em \jhep} {\bf 10} (2015) 134,
  [\href{http://arxiv.org/abs/1508.06608}{{\tt arXiv:1508.06608}}].

\bibitem{Christensen01}
N.~{Christensen}, R.~{Meyer}, L.~{Knox}, and B.~{Luey}, {\it {Bayesian methods
  for cosmological parameter estimation from cosmic microwave background
  measurements}},  {\em Classical and Quantum Gravity} {\bf 18} (2001)
  2677--2688, [\href{http://arxiv.org/abs/astro-ph/0103134}{{\tt
  astro-ph/0103134}}].

\bibitem{Dunkley05}
J.~{Dunkley}, M.~{Bucher}, P.~G. {Ferreira}, K.~{Moodley}, and C.~{Skordis},
  {\it {Fast and reliable Markov chain Monte Carlo technique for cosmological
  parameter estimation}},  {\em \mnras} {\bf 356} (2005) 925--936,
  [\href{http://arxiv.org/abs/astro-ph/0405462}{{\tt astro-ph/0405462}}].

\bibitem{CosmoMC}
A.~{Lewis} and S.~{Bridle}, {\it {Cosmological parameters from CMB and other
  data: A Monte Carlo approach}},  {\em \prd} {\bf 66} (2002) 103511,
  [\href{http://arxiv.org/abs/astro-ph/0205436}{{\tt astro-ph/0205436}}].

\bibitem{CosmoMC++}
A.~{Lewis} and S.~{Bridle}, {\it {CosmoMC++}},  {\em unpublished note} (2006).
  \href{http://cosmologist.info/notes/CosmoMC.pdf}{http://cosmologist.info/notes/CosmoMC.pdf}.

\bibitem{Baltz04}
E.~A. {Baltz} and P.~{Gondolo}, {\it {Markov Chain Monte Carlo Exploration of
  Minimal Supergravity with Implications for Dark Matter}},  {\em \jhep} {\bf
  10} (2004) 52, [\href{http://arxiv.org/abs/hep-ph/0407039}{{\tt
  hep-ph/0407039}}].

\bibitem{Allanach06}
B.~C. {Allanach} and C.~G. {Lester}, {\it {Multidimensional mSUGRA likelihood
  maps}},  {\em \prd} {\bf 73} (2006) 015013,
  [\href{http://arxiv.org/abs/hep-ph/0507283}{{\tt hep-ph/0507283}}].

\bibitem{Fittino06}
P.~{Bechtle}, K.~{Desch}, and P.~{Wienemann}, {\it {Fittino, a program for
  determining MSSM parameters from collider observables using an iterative
  method}},  {\em \cpc} {\bf 174} (2006) 47--70,
  [\href{http://arxiv.org/abs/hep-ph/0412012}{{\tt hep-ph/0412012}}].

\bibitem{Ruiz06}
R.~{Ruiz de Austri}, R.~{Trotta}, and L.~{Roszkowski}, {\it {A Markov chain
  Monte Carlo analysis of CMSSM}},  {\em \jhep} {\bf 5} (2006) 2,
  [\href{http://arxiv.org/abs/hep-ph/0602028}{{\tt hep-ph/0602028}}].

\bibitem{Buchmueller08}
O.~{Buchmueller}, R.~{Cavanaugh}, {\em et.~al.}, {\it {Predictions for
  supersymmetric particle masses using indirect experimental and cosmological
  constraints}},  {\em \jhep} {\bf 9} (2008) 117,
  [\href{http://arxiv.org/abs/0808.4128}{{\tt arXiv:0808.4128}}].

\bibitem{Skilling04}
J.~{Skilling}, {\it {Nested Sampling}},  in {\em American Institute of Physics
  Conference Series} (R.~{Fischer}, R.~{Preuss}, and U.~V. {Toussaint}, eds.)
  {\bf 735} (2004) 395--405.

\bibitem{Trotta08}
R.~{Trotta}, F.~{Feroz}, M.~{Hobson}, L.~{Roszkowski}, and R.~{Ruiz de Austri},
  {\it {The impact of priors and observables on parameter inferences in the
  constrained MSSM}},  {\em \jhep} {\bf 12} (2008) 24,
  [\href{http://arxiv.org/abs/0809.3792}{{\tt arXiv:0809.3792}}].

\bibitem{Scott09c}
P.~{Scott}, J.~{Conrad}, {\em et.~al.}, {\it {Direct constraints on minimal
  supersymmetry from Fermi-LAT observations of the dwarf galaxy Segue 1}},
  {\em \jcap} {\bf 1} (2010) 31, [\href{http://arxiv.org/abs/0909.3300}{{\tt
  arXiv:0909.3300}}].

\bibitem{Planck15cosmo}
{Planck Collaboration}, P.~A.~R. {Ade}, {\em et.~al.}, {\it {Planck 2015
  results. XIII. Cosmological parameters}},  {\em \aap} {\bf 594} (2016) A13,
  [\href{http://arxiv.org/abs/1502.01589}{{\tt arXiv:1502.01589}}].

\bibitem{MasterCodeMSSM10}
K.~J. {de Vries}, E.~A. {Bagnaschi}, {\em et.~al.}, {\it {The pMSSM10 after LHC
  run 1}},  {\em \epjc} {\bf 75} (2015) 422,
  [\href{http://arxiv.org/abs/1504.03260}{{\tt arXiv:1504.03260}}].

\bibitem{MultiNest}
F.~{Feroz}, M.~P. {Hobson}, and M.~{Bridges}, {\it {MULTINEST: an efficient and
  robust Bayesian inference tool for cosmology and particle physics}},  {\em
  \mnras} {\bf 398} (2009) 1601--1614,
  [\href{http://arxiv.org/abs/0809.3437}{{\tt arXiv:0809.3437}}].

\bibitem{IC79_SUSY}
IceCube Collaboration: M.~G. {Aartsen} {\em et.~al.}, {\it {Improved limits on
  dark matter annihilation in the Sun with the 79-string IceCube detector and
  implications for supersymmetry}},  {\em \jcap} {\bf 04} (2016) 022,
  [\href{http://arxiv.org/abs/1601.00653}{{\tt arXiv:1601.00653}}].

\bibitem{ColliderBit}
\GB Collider Workgroup: C.~{Bal{\'a}zs}, A.~{Buckley}, {\em et.~al.}, {\it
  {ColliderBit: a GAMBIT module for the calculation of high-energy collider
  observables and likelihoods}},  {\em \epjc\ in press} (2017)
  [\href{http://arxiv.org/abs/1705.07919}{{\tt arXiv:1705.07919}}].

\bibitem{Akrami09}
Y.~{Akrami}, P.~{Scott}, J.~Edsj{\"o}, J.~{Conrad}, and L.~{Bergstr{\"o}m},
  {\it {A profile likelihood analysis of the Constrained MSSM with genetic
  algorithms}},  {\em \jhep} {\bf 4} (2010) 57,
  [\href{http://arxiv.org/abs/0910.3950}{{\tt arXiv:0910.3950}}].

\bibitem{DEnu}
M.~{Ghulam}, A.~{Faisal}, and M.~{Bilal}, {\it {Optimization of Neutrino
  Oscillation Parameters Using Differential Evolution}},  {\em Communications
  in Theoretical Physics} {\bf 59} (2013) 324--330,
  [\href{http://arxiv.org/abs/1109.2431}{{\tt arXiv:1109.2431}}].

\bibitem{SBSpike}
F.~{Feroz}, K.~{Cranmer}, M.~{Hobson}, R.~{Ruiz de Austri}, and R.~{Trotta},
  {\it {Challenges of profile likelihood evaluation in multi-dimensional SUSY
  scans}},  {\em \jhep} {\bf 6} (2011) 42,
  [\href{http://arxiv.org/abs/1101.3296}{{\tt arXiv:1101.3296}}].

\bibitem{Akrami11coverage}
Y.~{Akrami}, C.~{Savage}, P.~{Scott}, J.~{Conrad}, and J.~{Edsj{\"o}}, {\it
  {Statistical coverage for supersymmetric parameter estimation: a case study
  with direct detection of dark matter}},  {\em \jcap} {\bf 7} (2011) 2,
  [\href{http://arxiv.org/abs/1011.4297}{{\tt arXiv:1011.4297}}].

\bibitem{SBcoverage}
M.~{Bridges}, K.~{Cranmer}, {\em et.~al.}, {\it {A coverage study of CMSSM
  based on ATLAS sensitivity using fast neural networks techniques}},  {\em
  \jhep} {\bf 3} (2011) 12, [\href{http://arxiv.org/abs/1011.4306}{{\tt
  arXiv:1011.4306}}].

\bibitem{Strege12}
C.~{Strege}, R.~{Trotta}, G.~{Bertone}, A.~H.~G. {Peter}, and P.~{Scott}, {\it
  {Fundamental statistical limitations of future dark matter direct detection
  experiments}},  {\em \prd} {\bf 86} (2012) 023507,
  [\href{http://arxiv.org/abs/1201.3631}{{\tt arXiv:1201.3631}}].

\bibitem{Fittinocoverage}
P.~{Bechtle}, J.~E. {Camargo-Molina}, {\em et.~al.}, {\it {Killing the cMSSM
  softly}},  {\em \epjc} {\bf 76} (2016) 96,
  [\href{http://arxiv.org/abs/1508.05951}{{\tt arXiv:1508.05951}}].

\bibitem{great}
A.~{Putze} and L.~{Derome}, {\it {The Grenoble Analysis Toolkit (GreAT)-A
  statistical analysis framework}},  {\em Physics of the Dark Universe} {\bf 5}
  (2014) 29--34.

\bibitem{cosmoabc}
E.~E.~O. {Ishida}, S.~D.~P. {Vitenti}, {\em et.~al.}, {\it {COSMOABC:
  Likelihood-free inference via Population Monte Carlo Approximate Bayesian
  Computation}},  {\em Astronomy and Computing} {\bf 13} (2015) 1--11,
  [\href{http://arxiv.org/abs/1504.06129}{{\tt arXiv:1504.06129}}].

\bibitem{gambit}
\GB Collaboration: P.~{Athron}, C.~{Balazs}, {\em et.~al.}, {\it {GAMBIT: The
  Global and Modular Beyond-the-Standard-Model Inference Tool}},  {\em \epjc\
  in press} (2017) [\href{http://arxiv.org/abs/1705.07908}{{\tt
  arXiv:1705.07908}}].

\bibitem{DarkBit}
\GB Dark Matter Workgroup: T.~{Bringmann}, J.~{Conrad}, {\em et.~al.}, {\it
  {DarkBit: A GAMBIT module for computing dark matter observables and
  likelihoods}},  {\em \epjc\ in press} (2017)
  [\href{http://arxiv.org/abs/1705.07920}{{\tt arXiv:1705.07920}}].

\bibitem{SDPBit}
\GB Models Workgroup: P.~{Athron}, C.~{Bal{\'a}zs}, {\em et.~al.}, {\it
  {SpecBit, DecayBit and PrecisionBit: GAMBIT modules for computing mass
  spectra, particle decay rates and precision observables}},  {\em submitted to
  \epjc} (2017) [\href{http://arxiv.org/abs/1705.07936}{{\tt
  arXiv:1705.07936}}].

\bibitem{FlavBit}
\GB Flavour Workgroup: F.~U. {Bernlochner}, M.~{Chrz\k{a}szcz}, {\em et.~al.},
  {\it {FlavBit: A GAMBIT module for computing flavour observables and
  likelihoods}},  {\em \epjc\ in press} (2017)
  [\href{http://arxiv.org/abs/1705.07933}{{\tt arXiv:1705.07933}}].

\bibitem{CMSSM}
\GB Collaboration: P.~{Athron}, C.~{Bal{\'a}zs}, {\em et.~al.}, {\it {Global
  fits of GUT-scale SUSY models with GAMBIT}},  {\em \epjc\ in press} (2017)
  [\href{http://arxiv.org/abs/1705.07935}{{\tt arXiv:1705.07935}}].

\bibitem{MSSM}
\GB Collaboration: P.~{Athron}, C.~{Bal{\'a}zs}, {\em et.~al.}, {\it {A global
  fit of the MSSM with GAMBIT}},  {\em \epjc\ in press} (2017)
  [\href{http://arxiv.org/abs/1705.07917}{{\tt arXiv:1705.07917}}].

\bibitem{SSDM}
\GB Collaboration: P.~{Athron}, C.~{Bal{\'a}zs}, {\em et.~al.}, {\it {Status of
  the scalar singlet dark matter model}},  {\em \epjc} {\bf 77} (2017) 568,
  [\href{http://arxiv.org/abs/1705.07931}{{\tt arXiv:1705.07931}}].

\bibitem{pippi}
P.~{Scott}, {\it {Pippi -- painless parsing, post-processing and plotting of
  posterior and likelihood samples}},  {\em \epjp} {\bf 127} (2012) 138,
  [\href{http://arxiv.org/abs/1206.2245}{{\tt arXiv:1206.2245}}].

\bibitem{Metropolis}
N.~Metropolis, A.~W. Rosenbluth, M.~N. Rosenbluth, A.~H. Teller, and E.~Teller,
  {\it Equation of state calculations by fast computing machines},  {\em \jcp}
  {\bf 21} (1953) 1087--1092.

\bibitem{2003it...book....M}
D.~{MacKay}, {\em {Information Theory, Inference, and Learning Algorithms}}.
\newblock Publisher: Cambridge University Press.~ ISBN: 0521642981, 2003.

\bibitem{1993Neal}
R.~M. {Neal}, {\it {Probabilistic Inference Using Markov Chain Monte Carlo
  Methods}},  {Technical Report CRG-TR-93-1}, 1993.

\bibitem{Hastings}
W.~K. Hastings, {\it Monte carlo sampling methods using markov chains and their
  applications},  {\em Biometrika} {\bf 57} (1970) 97--109.

\bibitem{ChristenFox10}
J.~A. {Christen} and J.~{Weare}, {\it {A general purpose sampling algorithm for
  continuous distributions (the t-walk)}},  {\em Bayesian\ Anal.} {\bf 5}
  (2010) 263.

\bibitem{GoodmanWeare10}
J.~{Goodman} and J.~{Weare}, {\it {Ensemble samplers with affine invariance}},
  {\em Comm.\ App.\ Math.\ Comp.\ Sci.} {\bf 5} (2010) 65.

\bibitem{GelmanRubin}
A.~{Gelman} and D.~B. {Rubin}, {\it Inference from iterative simulation using
  multiple sequences},  {\em Statistical Science} {\bf 7} (1992) 457--472.

\bibitem{StornPrice95}
R.~Storn and K.~Price, {\it Differential evolution: A simple and efficient
  heuristic for global optimization over continuous spaces},  {\em Journal of
  Global Optimization} {\bf 11} (1997) 341--359.

\bibitem{Price05wholebook}
K.~Price, R.~M. Storn, and J.~A. Lampinen, {\em Differential evolution: a
  practical approach to global optimization}.
\newblock Springer, 2005.

\bibitem{DasSuganthan11}
S.~Das and P.~Suganthan, {\it Differential evolution: A survey of the
  state-of-the-art},  {\em Evolutionary Computation, IEEE Transactions on} {\bf
  15} (2011) 4--31.

\bibitem{Price13}
K.~Price, {\it Differential evolution},  in {\em Handbook of Optimization}
  (I.~Zelinka, V.~Sn\'{a}\v{s}el, and A.~Abraham, eds.), vol.~38 of {\em
  Intelligent Systems Reference Library}, pp.~187--214.
\newblock Springer Berlin Heidelberg, 2013.

\bibitem{Price05chp2}
K.~Price, R.~M. Storn, and J.~A. Lampinen, {\it The differential evolution
  algorithm},  in {\em Differential Evolution: A Practical Approach to Global
  Optimization}, Natural Computing Series, pp.~37--134.
\newblock Springer Berlin Heidelberg, 2005.

\bibitem{Zaharie07}
D.~Zaharie, {\it A comparative analysis of crossover variants in differential
  evolution},  {\em Proceedings of IMCSIT 2007} (2007) 171--181.

\bibitem{Zaharie08}
D.~Zaharie, {\it Statistical properties of differential evolution and related
  random search algorithms},  in {\em COMPSTAT 2008} (P.~Brito, ed.),
  pp.~473--485.
\newblock Physica-Verlag HD, 2008.

\bibitem{Mezura-Montes06}
E.~Mezura-Montes, J.~Vel\'{a}zquez-Reyes, and C.~A. Coello~Coello, {\it A
  comparative study of differential evolution variants for global
  optimization},  in {\em Proceedings of the 8th annual conference on Genetic
  and evolutionary computation}, GECCO '06, (New York, NY, USA), ACM (2006)
  485--492.

\bibitem{Zaharie09}
D.~Zaharie, {\it Influence of crossover on the behavior of differential
  evolution algorithms},  {\em Applied Soft Computing} {\bf 9} (2009) 1126 --
  1138.

\bibitem{Brest06}
J.~Brest, S.~Greiner, B.~Boskovic, M.~Mernik, and V.~Zumer, {\it Self-adapting
  control parameters in differential evolution: A comparative study on
  numerical benchmark problems},  {\em Evolutionary Computation, IEEE
  Transactions on} {\bf 10} (2006) 646--657.

\bibitem{NeriTirronen10}
F.~Neri and V.~Tirronen, {\it Recent advances in differential evolution: a
  survey and experimental analysis},  {\em Artificial Intelligence Review} {\bf
  33} (2010) 61--106.

\bibitem{Cuoco:2016jqt}
A.~{Cuoco}, B.~{Eiteneuer}, J.~{Heisig}, and M.~{Kr{\"a}mer}, {\it {A global
  fit of the {$\gamma$}-ray galactic center excess within the scalar singlet
  Higgs portal model}},  {\em \jcap} {\bf 6} (2016) 050,
  [\href{http://arxiv.org/abs/1603.08228}{{\tt arXiv:1603.08228}}].

\bibitem{Beniwal}
A.~{Beniwal}, F.~{Rajec}, {\em et.~al.}, {\it {Combined analysis of effective
  Higgs portal dark matter models}},  {\em \prd} {\bf 93} (2016) 115016,
  [\href{http://arxiv.org/abs/1512.06458}{{\tt arXiv:1512.06458}}].

\bibitem{Cline13b}
J.~M. {Cline}, K.~{Kainulainen}, P.~{Scott}, and C.~{Weniger}, {\it {Update on
  scalar singlet dark matter}},  {\em \prd} {\bf 88} (2013) 055025,
  [\href{http://arxiv.org/abs/1306.4710}{{\tt arXiv:1306.4710}}].

\bibitem{Cheung:2012xb}
K.~Cheung, Y.-L.~S. Tsai, P.-Y. Tseng, T.-C. Yuan, and A.~Zee, {\it {Global
  Study of the Simplest Scalar Phantom Dark Matter Model}},  {\em \jcap} {\bf
  1210} (2012) 042, [\href{http://arxiv.org/abs/1207.4930}{{\tt
  arXiv:1207.4930}}].

\bibitem{Mambrini11}
Y.~{Mambrini}, {\it {Higgs searches and singlet scalar dark matter: Combined
  constraints from XENON 100 and the LHC}},  {\em \prd} {\bf 84} (2011) 115017,
  [\href{http://arxiv.org/abs/1108.0671}{{\tt arXiv:1108.0671}}].

\bibitem{Burgess01}
C.~P. {Burgess}, M.~{Pospelov}, and T.~{ter Veldhuis}, {\it {The Minimal Model
  of nonbaryonic dark matter: a singlet scalar}},  {\em \nphysb} {\bf 619}
  (2001) 709--728, [\href{http://arxiv.org/abs/hep-ph/0011335}{{\tt
  hep-ph/0011335}}].

\bibitem{McDonald94}
J.~{McDonald}, {\it {Gauge singlet scalars as cold dark matter}},  {\em \prd}
  {\bf 50} (1994) 3637--3649, [\href{http://arxiv.org/abs/hep-ph/0702143}{{\tt
  hep-ph/0702143}}].

\bibitem{SilveiraZee}
V.~{Silveira} and A.~{Zee}, {\it {Scalar Phantoms}},  {\em \plb} {\bf 161}
  (1985) 136--140.

\bibitem{PDB}
Particle Data Group: K.~A. Olive {\em et.~al.}, {\it {Review of Particle
  Physics}},  {\em Chin.\ Phys.\ C} {\bf 38} (2014) 090001.

\bibitem{PDG15}
Particle Data Group: K.~A. Olive {\em et.~al.}, {\it {Review of Particle
  Physics}},  {\em \normalfont{update to Ref.~\cite{PDB}}} (2015).
  \href{http://pdg.lbl.gov/2015/tables/rpp2015-sum-gauge-higgs-bosons.pdf}{http://pdg.lbl.gov/2015/tables/rpp2015-sum-gauge-higgs-bosons.pdf}.

\bibitem{Cacciari:2011ma}
M.~Cacciari, G.~P. Salam, and G.~Soyez, {\it {FastJet User Manual}},  {\em
  \epjc} {\bf 72} (2012) 1896, [\href{http://arxiv.org/abs/1111.6097}{{\tt
  arXiv:1111.6097}}].

\bibitem{Athron:2014yba}
P.~Athron, J.-h. Park, D.~St{\"o}ckinger, and A.~Voigt, {\it {FlexibleSUSY - A
  spectrum generator generator for supersymmetric models}},  {\em \cpc} {\bf
  190} (2015) 139--172, [\href{http://arxiv.org/abs/1406.2319}{{\tt
  arXiv:1406.2319}}].

\bibitem{Allanach:2001kg}
B.~C. Allanach, {\it {SOFTSUSY: a program for calculating supersymmetric
  spectra}},  {\em \cpc} {\bf 143} (2002) 305--331,
  [\href{http://arxiv.org/abs/hep-ph/0104145}{{\tt hep-ph/0104145}}].

\end{thebibliography}\endgroup

\end{document}